RESEARCH ARTICLE

# Mitigating Molecular Aggregation in Drug Discovery with Predictive Insights from Explainable AI

Hunter Sturm,[a],† Jonas Teufel,[b],† Kaitlin A. Isfeld,[a],† Pascal Friederich[b],[c],*, Rebecca L. Davis[a],*

[a] H. Sturm, K. A. Isfeld, Dr. R. L. Davis
Department of Chemistry, University of Manitoba
Winnipeg, Canada
E-mail: rebecca.davis@umanitoba.ca

[b] J. Teufel, Dr. P. Friederich
Institute of Theoretical Informatics, Karlsruhe Institute of Technology
Karlsruhe, Germany
Email: pascal.friederich@kit.edu

[c] J. Teufel, Dr. P. Friederich
Institute of Nanotechnology, Karlsruhe Institute of Technology
Karlsruhe, Germany

† These authors contributed equally

Supporting information for this article is provided at the end of the document.

**Abstract:** Herein, we present the application of MEGAN, our explainable AI (xAI) model, for the identification of small colloidally aggregating molecules (SCAMs). This work offers solutions to the long-standing problem of false positives caused by SCAMs in high throughput screening for drug discovery and demonstrates the power of xAI in the classification of molecular properties that are not chemically intuitive based on our current understanding. We leverage xAI insights and molecular counterfactuals to design alternatives to problematic compounds in drug screening libraries. Additionally, we experimentally validate the MEGAN prediction classification for one of the counterfactuals and demonstrate the utility of counterfactuals for altering the aggregation properties of a compound through minor structural modifications. The integration of this method in high-throughput screening approaches will help combat and circumvent false positives, providing better lead molecules more rapidly and thus accelerating drug discovery cycles.

## Introduction

Interest in the application of machine learning (ML) in lead discovery has grown substantially in recent years, driven by academic and industrial initiatives to apply ML methods during early-stage drug discovery.[1,2] This trend is largely attributed to the availability of extensive datasets containing activity data generated through high-throughput screening (HTS) campaigns. The activity data produced by HTS has long been essential for hit identification in early-stage drug discovery and is becoming even more critical with the growing interest in ML approaches for predicting lead compounds.

A persistent challenge in HTS-based hit identification is the prevalence of false hits. While large-scale HTS campaigns typically generate numerous initial hits, only a small proportion represents the desired interactions between compounds and their target biomolecules. Many screening libraries contain a significant number of false positive and negative data points, with up to 80–95% of the hits from initial screening representing artifacts.[3,4] This long-standing challenge in medicinal chemistry now extends to ML-based approaches for drug discovery. Models trained on datasets containing large numbers of false hits are prone to predicting compounds that are not viable as leads. Addressing these issues in data quality is essential to enhance the hit discovery efforts of medicinal chemists as well as the predictive prowess of ML models.

Colloidal aggregation represents a significant source of false positives in HTS.[5] Aggregation occurs when molecules form supramolecular complexes, or colloids, at or above a critical aggregation concentration.[6] Small colloidally aggregating molecules (SCAMs) can interact nonspecifically with proteins, leading to local unfolding and functional disruption, or they can interfere through mechanisms such as aggregation-induced emission, where self-assembled molecules fluoresce upon reaching their critical aggregation concentration.[6,7] Estimates suggest that 15–20% of small molecules in public chemogenomic databases aggregate under standard screening conditions, underscoring the need for accurate prediction of aggregation to mitigate its impact on drug discovery.[8]

The experimental detection of SCAMs is both expensive and time-consuming which has led to numerous *in silico* methods being developed to screen aggregating compounds from HTS datasets (Figure 1).[8-15] One of the earliest tools, Aggregator Advisor, evaluates molecules represented as SMILES strings and determines their similarity to known aggregators based on LogP and Tanimoto similarity (Figure 1b).[9] In addition to providing a rule-based *in silico* method for aggregation screening, Aggregator Advisor has also provided the field with a valuable database of **a** Experimental Aggregation Detection Methods

- Dynamic light scattering (DLS)
- Detergent-based biological assays
- UV-Vis spectroscopy

***Pros***
✓ *Accurate detection*
✓ *Conditions variable*

***Cons***
✗ *Slow*
✗ *Expensive*
✗ *Non-predictive*
✗ *Individual molecule testing*

**b** *In silico* Aggregation Prediction

**Early Computational Approach**

**Aggregator Advisor**
*Similarity/Rule Based*
*Aggregator Advisor*

- Based on logP and similarity to experimentally known aggregators
- Can't extrapolate chemical patterns
- Source of data used to train ML models

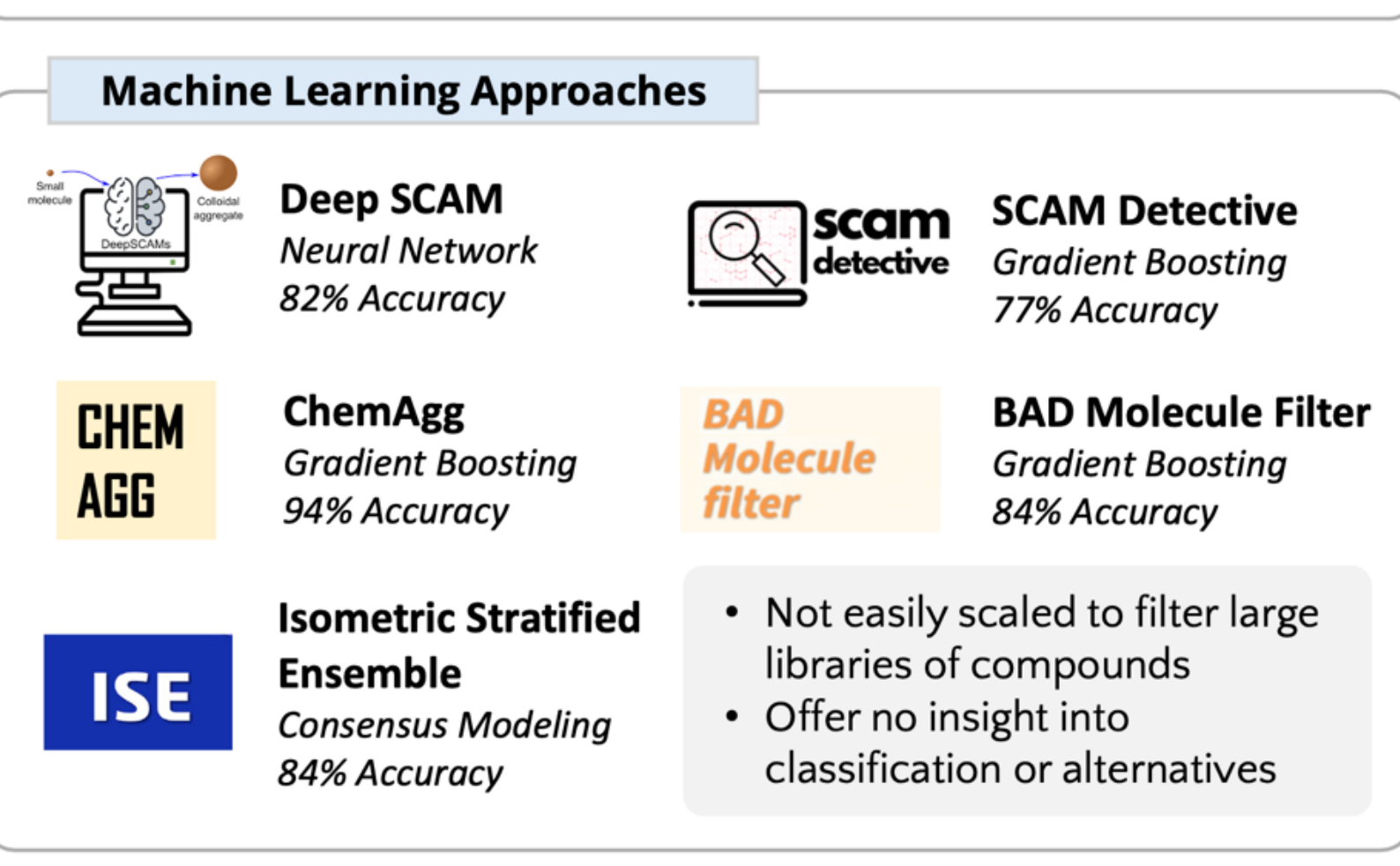

**c** This Work

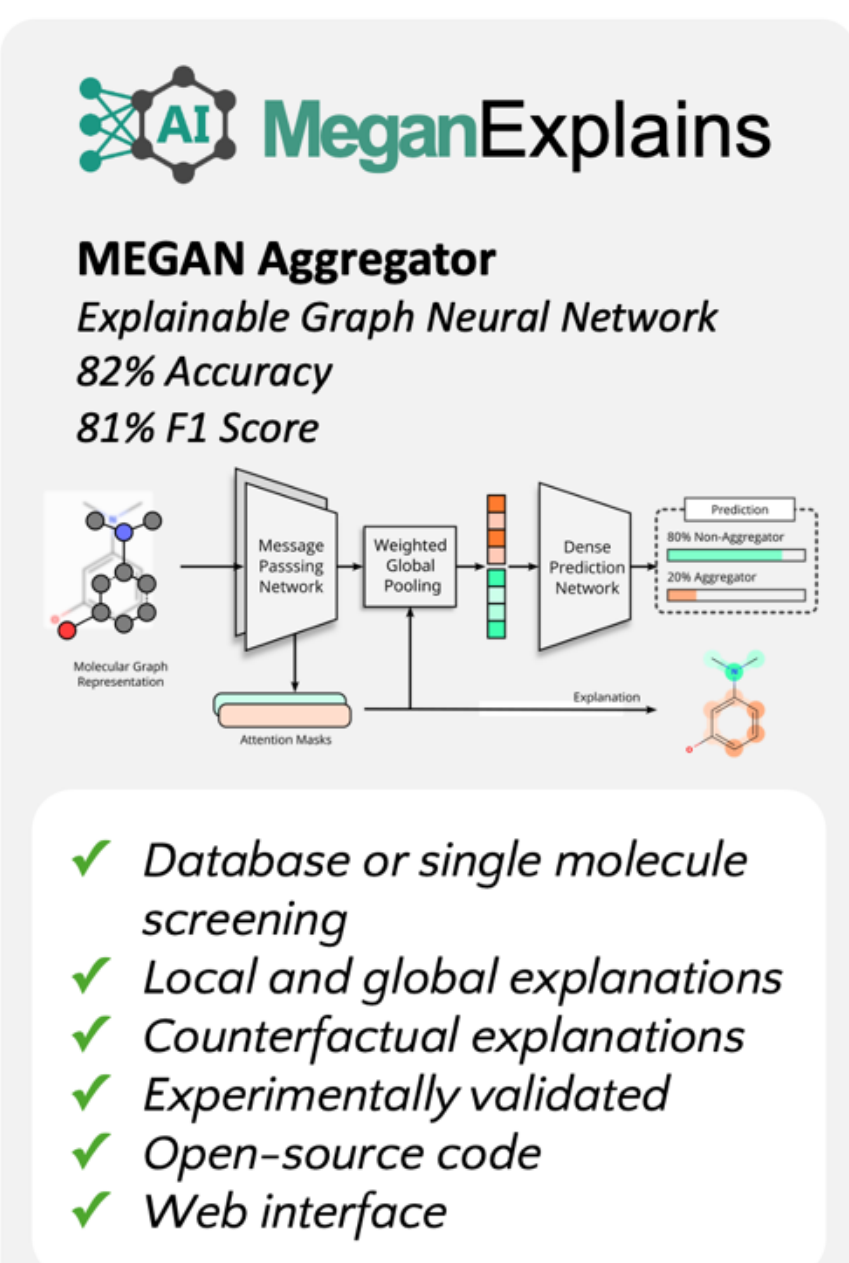

**Figure 1.** Summary of (a) experimental and (b) *in silico* approaches to aggregation determination (c) compared with our work. Accuracies are those reported in the respective publications (ref. 8-13) and are thus not directly comparable due to differences in training and testing datasets.

experimentally validated aggregators.[9] This dataset has been pivotal in the development and training of most of the ML models for aggregation prediction, enabling these models to classify small molecules as aggregators with greater accuracy and reliability (Figure 1b).[8, 10-13]

ML-based methods for the classification of aggregation have demonstrated accuracies exceeding 80% in predicting aggregators. However, these methods often do not provide scalable implementations or open-source code that allows for the filtering of large molecular libraries.[11] Scalable approaches, such as SCAM Detective and DeepSCAM, typically achieve accuracies in the range of 70-80%.[8, 10] Consequently, there remains a critical need for an accurate, scalable, and interpretable method capable of efficiently screening large libraries of compounds.

Aggregation is a complex phenomenon that is influenced by many variables, including concentration, pH, temperature, buffer, and solvent.[16,17] While numerous studies have explored the molecular features that contribute to small molecule aggregation, no clear consensus has been reached on the key factors driving this phenomenon.[18] Features such as logP, the number of hydroxyl groups, the number of sulfur atoms, and the number of aromatic rings are often proposed to contribute to aggregate formation; however, the complexity of molecular aggregation has led to difficulty in identifying reliable, universal trends distinguishing aggregating and non-aggregating compounds.[8, 9, 11] Given the absence of generalizable trends for aggregation prediction, there is a growing need for predictive models that can learn complex, nonlinear relationships. In this context, machine learning offers a powerful framework for the prediction of molecular aggregation and revealing insights into the structural patterns found and used by the machine learning models.

This study addresses the need for an accurate and scalable model capable of detecting SCAMs while providing interpretable explanations and the ability to create non-aggregating counterfactuals (Figure 1c). The explainable AI (xAI) model employed in this study, a multi-channel graph attention network (MEGAN), achieves an accuracy of 82% in predicting SCAMs and is suitable for screening both large molecular libraries and individual compounds.[19] Furthermore, the model generates explanations for its classifications, offering insights into why a compound is predicted as a SCAM or non-SCAM. The accompanying web server allows users to screen individual molecules and provides a user-specified number of counterfactual explanations. These counterfactuals are structurally similar to the query molecule but possess flipped classification labels (e.g., counterfactuals for a molecule predicted to aggregate are structurally similar molecules predicted to be non-aggregating, and vice versa). To validate our model and its application of counterfactuals, we synthesized and experimentally tested a non-aggregating derivative of clioquinol—an established aggregator—proposed by the model. The experimental results confirm the model's prediction and demonstrate the experimental relevance of our model in the informed design of molecules with tailored aggregation properties.

## Results and Discussion

### The MEGAN Model

To develop an accurate and scalable model for detecting SCAMs and elucidating the structure-property relationships underlying molecular aggregation, we leverage xAI techniques applied to predictions generated by graph neural network models. To apply a graph neural network to the task of chemical property prediction, each molecule is first converted into a molecular graph where atoms are represented as nodes and bonds are represented as edges. Based on this graph-structured input information, the graph neural network is trained on the dataset to predict the binary classification label of a given molecule as either an aggregator or non-aggregator (Figure 2).

In this work, we apply the multi-explanation graph attention network (MEGAN) to the task of aggregation prediction.[19] MEGAN is a self-explaining graph neural network model architecture for which node and edge attributional explanations are directly derived from the model's internal attention and masking mechanism. These attributional explanations assign an importance value between 0 and 1 to each node and edge of a given graph to indicate which substructures of a given graph are especially influential for the predicted outcome. The MEGAN model specifically generates one such attributional explanation for each of the possible classification outcomes—one explanation highlighting the structural evidence in favor of an aggregator classification (orange) and the other explanation highlighting substructures associated with the non-aggregator class (green) (Figure 2b).

### Prediction Accuracy and Benchmarking

We trained a MEGAN model on a dataset comprised of 12,338 aggregating and 177,048 non-aggregating molecules (see Methods in Supporting Information). For a quantitative evaluation of our trained model, we used a separate test set of 1500 aggregators and 1500 non-aggregators. The dataset was largely derived from a single experimental screen conducted under consistent conditions (e.g., phosphate buffer, pH = 7). To align molecular representations with the experimental conditions, protonation states were assigned based on physiological pH. Similar to most previously reported ML models, we achieved an accuracy of 82%. However, as we are using a training set with a class imbalance, accuracy can be misleading as it may achieve high accuracy simply by predicting the majority class. To further evaluate the MEGAN model's performance, we employed the F1 score, which, even on balanced test sets, offers a perspective on the balance between precision and recall beyond the accuracy metric. The MEGAN model achieved an F1 score of 81%, indicating that the model has a good balance between precision and recall. The similarity in the accuracy (82%) and F1 score (81%) also suggests that the model's performance is well-balanced across both classes (aggregators and non-aggregators). The performance of the MEGAN model was compared against the XGboost model of Yang et al. (ChemAgg), using our balanced test set. Attempts to compare to many of the other ML models listed in Figure 1 were unsuccessful due to the inaccessibility of the code used for these models. It was found that the MEGAN model provided a higher accuracy and F1 score than the ChemAgg model (acc = 73%, F1 = 74%), demonstrating the superior performance of graph neural networks in capturing molecular structures (see Supporting Information for details).

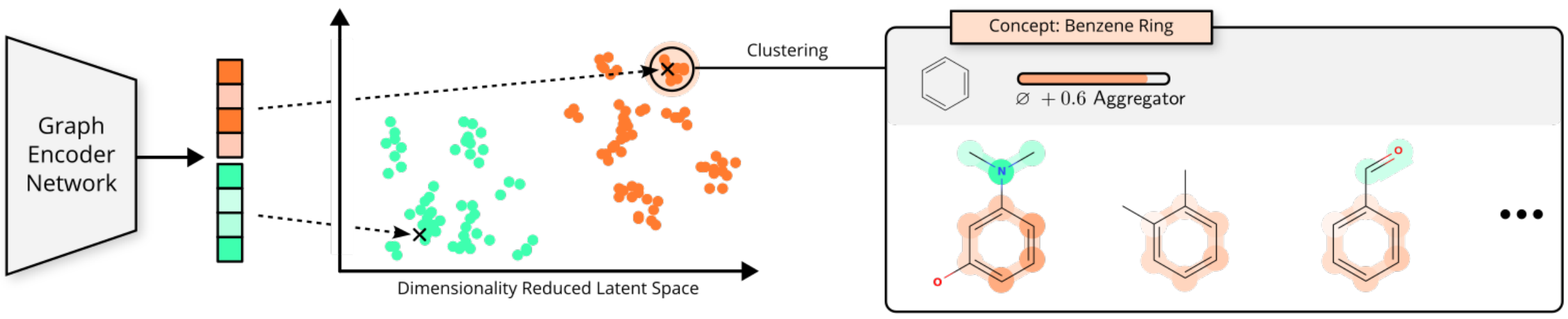

**Figure 2.** MEGAN model overview. (a) Molecular graph structures are used as input to multiple attention-based message-passing layers. Node representations are aggregated and passed to a fully connected network to output the predicted class. Explanation masks are derived from the internal attention values. (b) Attention-based explanations are explicitly split into separate channels for each possible output class. (c) Overarching structural explanations can be found by identifying clusters in the latent space of subgraph embeddings. Analyzing all members of a concept cluster yields general trends associated with certain structural motifs. Explicit explanation masks and values are constructed for illustration purposes.

### Sensitivity of the Model to Small Structural Modifications

When evaluating the performance of the MEGAN model in predicting molecular aggregation, its predictions were observed to exhibit a high degree of sensitivity to subtle modifications in molecular structure. To illustrate this sensitivity, two groups of compounds with experimentally validated aggregation behavior are presented: one comprising molecules from the training dataset (**A**, **B**) and the other consisting of molecules external to the training dataset (**C**, **D**).

Experimental studies reported in the literature have previously identified azacarbazole **A** as an aggregator and azacarbazole **B** as a non-aggregator, despite the two compounds differing by only a methyl group. The MEGAN model correctly predicted **A** as an aggregating and **B** as a non-aggregating compound with high confidence, likely due to their inclusion in the training dataset. However, more notable are the distinct explanation masks generated by the model for each compound. Examination of the explanation mask for **A** indicates that the methyl group does not significantly influence the model's prediction. Instead, the presence of the methyl group appears to cause the model to discriminate between different parts of the shared heterocyclic core of **A** and **B** in its classifications of each compound. The fact that the addition of a methyl group to the azacarbazole results in aggregation is not chemically intuitive, suggesting that we cannot directly derive trends in aggregation from the structures. This example shows the need for models like MEGAN to find edge cases like this and provide insight that can be further explored where traditional chemical intuition fails.

In comparing efonidipine (**C**) and manidipine (**D**), we again observe the MEGAN model's ability to correctly classify compounds based on changes in substructure. In both **C** and **D**, the model highlights the aromatic ring of the nitrobenzyl group as evidence for aggregation. In **C**, the bulky phosphonate group is identified as providing evidence against aggregation. From a chemical perspective, the explanation masks for **C** seem reasonable, as the phosphonate group likely reduces the molecule's capacity to form closely packed arrangements or stable intermolecular interactions that promote aggregate formation. In contrast to **C**, the methyl ester group of **D** decreases the attributional explanations for nonaggregation resulting from the dihydropyridine. This would suggest that the methyl ester is less disruptive of favourable intermolecular interactions that induce aggregation in **D**. These findings suggest that the model effectively identifies key structural features that either promote or inhibit aggregation, highlighting the potential for targeted modifications to reduce aggregation behavior.

### DFT Assessment of Physical Relevance for MEGAN Model Structural Sensitivity

To examine the physical relevance of the MEGAN model's sensitivity to small changes in molecular structure, quantum chemical modeling of a group of structurally similar compounds having different prediction labels was performed. As a significant portion of the molecules in the training dataset contain aromatic heterocycles, pyridine derivatives were selected as the primary focus of this study. This scaffold serves both as a simple example of an aromatic heterocycle and is easily modified to examine the effects of different intermolecular interactions. Although the 12 selected pyridine derivatives have not been experimentally evaluated for aggregation, their analysis provides insights into the relationship between aggregation prediction confidence and interaction energies. Based on the MEGAN model's prediction accuracy of 82%, approximately 2 of the 12 pyridine derivatives may be expected to be mislabeled; however, the functional groups that appear in the explanations are potentially still meaningful and likely trend with the actual aggregation-enhancing or aggregation-inhibiting behavior of these groups. The reason for that is that it is easier for the model to identify which groups trend with aggregation than to quantify the subtle relative influences of these groups on the final aggregation prediction. Therefore, even when the MEGAN model fails to predict the correct final label, it is very likely that the explanation masks are still correctly identifying and labeling relevant groups, but the model misjudges their exact relative influence on the overall aggregation likelihood. Interaction energies for the pyridine derivatives were calculated using density functional theory (DFT) and compared with the MEGAN predictions and explanation masks (Figure 4).[20-22]

The intermolecular interactions of pyridine derivatives predicted with high confidence as aggregators (**E**–**H**) were calculated to be at least 1.5 kcal/mol more energetically favorable than those of compounds predicted with high confidence as non-aggregators (**I**–**P**). All 2-hydroxypyridine derivatives were classified as aggregators with high confidence. Conformational searches of the dimers revealed that the strongest interactions between these molecules involved the formation of two intermolecular hydrogen bonds. The only other compound to form two intermolecular hydrogen bonds was 2-aminopyridine (**K**), which was classified as a non-aggregator. However, its interaction energy was only -6 kcal/mol, likely due to the relatively weaker nature of N-H-N hydrogen bonds compared to O-H-N hydrogen bonds. Hydroxypyridines **M**, **N**, **O**, and **P** were classified as non-aggregators and were capable of forming only one hydrogen bond, leading to weak interaction energies of greater than -9 kcal/mol. The remaining compounds, **I**, **J**, and **L**, classified as non-aggregators by the MEGAN model, all preferred conformations that promoted π-π interactions and had interaction energies of greater than -8 kcal/mol. To further investigate the electronic nature of the interactions studied, energy decomposition analysis was completed using Psi4's implementation of symmetry-adapted perturbation theory (SAPT) on the dimers of **E-P**.[23, 24] The SAPT results demonstrate trends that are consistent with the DFT interaction energies and show a clear distinction between the strengths of the interactions in the predicted aggregating and non-aggregating molecules (see Supporting Information for details). Overall, this demonstrates the ability of the model to discriminate based on intermolecular interaction strength.

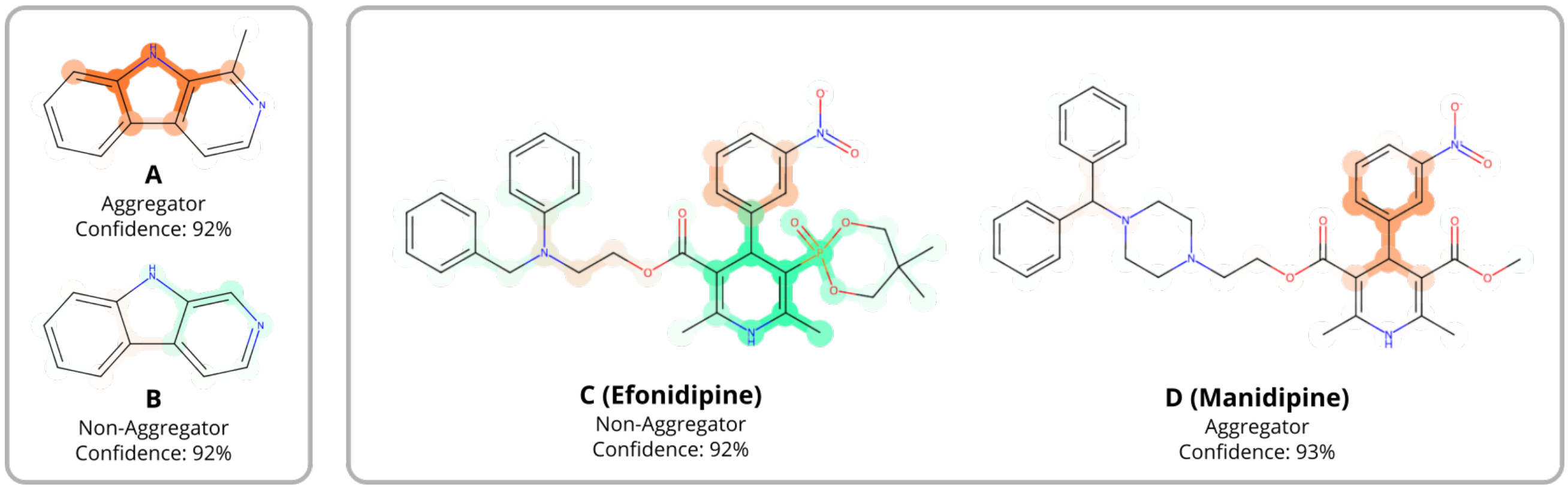

**Figure 3.** MEGAN predictions and explanation masks for two pairs of structurally similar molecules that exhibit experimentally contrasting aggregation behavior.

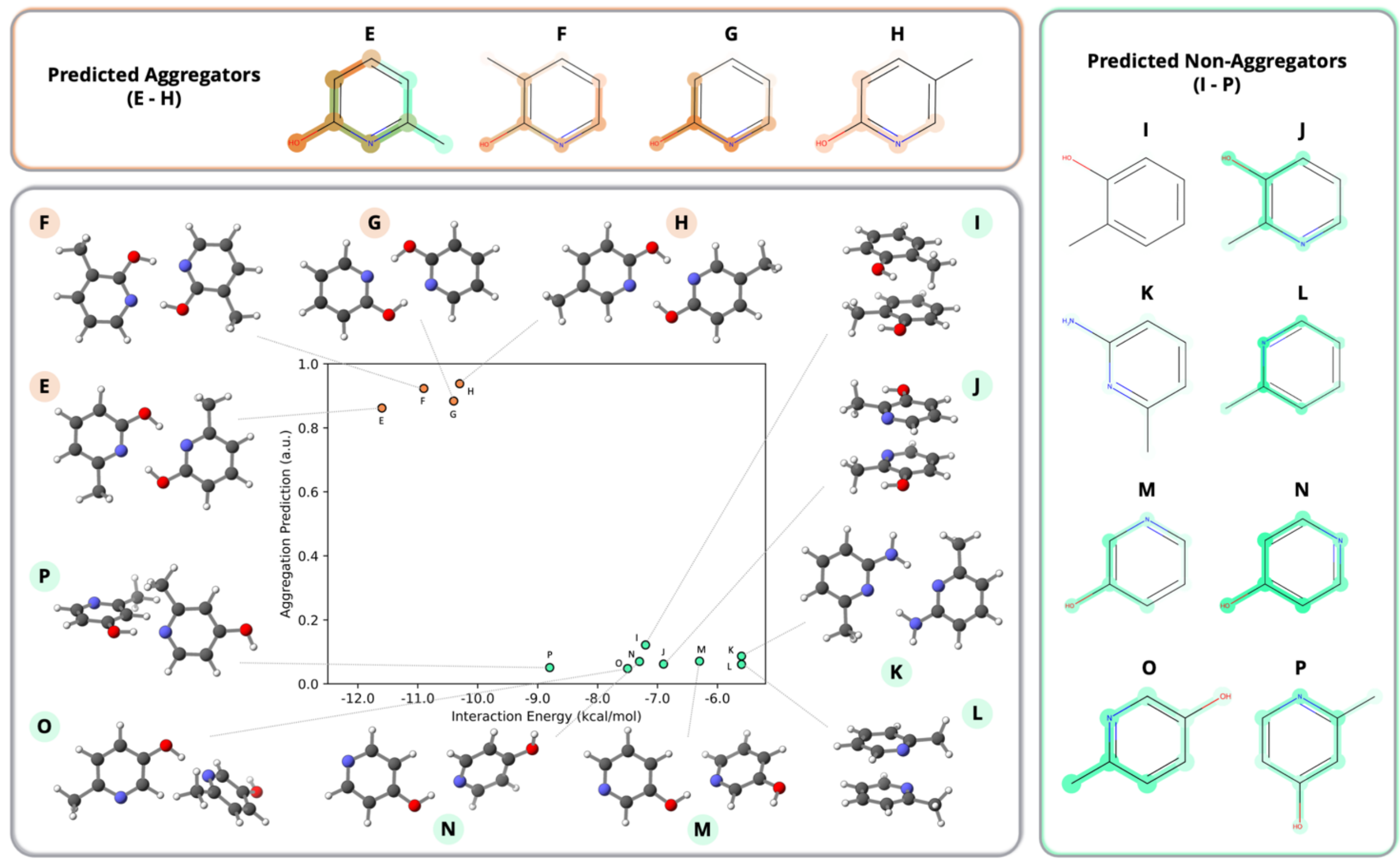

**Figure 4.** MEGAN predictions compared with DFT interaction energies of dimers for compounds **E** - **P**. Predictions are reported on a scale from 0 to 1, with 0 indicating a prediction of non-aggregator and 1 indicating a prediction of aggregator. Also shown are the 3D structures of the optimized dimer geometries and MEGAN explanation masks for each molecule. The light-colored explanation masks for **I** indicate that the MEGAN model detects no pattern systematically associated with either class.

In examining the MEGAN model's explanation masks for the hydroxypyridines, the model is found to distinguish between the various OH substitution patterns. For the predicted aggregators (**O**-**R**), the MEGAN model finds the HO-C-N substructure, which is responsible for the hydrogen bonding interactions in the dimers, as the key contributor to the aggregator classification (Figure 4). In contrast, for the isomers with weaker interaction energies (**J**, **M**, **N**, **O**, **P**) the model identifies the hydroxyl group as contributing to the classification of these compounds as non-aggregators, demonstrating that the model's prediction is influenced by the substitution pattern. The MEGAN model is also observed to be sensitive to the strength of the hydrogen bond donating abilities of the substituents. While the strongly hydrogen-bonding hydroxyl group of 2-hydroxypyridine (**G**) is highlighted as contributing to the aggregation prediction, the slightly weaker hydrogen-bonding amino group of 2-aminopyridine (**K**) is highlighted as contributing to a non-aggregator prediction. Therefore, not only do the interaction energies correlate with the MEGAN predictions, but the MEGAN model is able to identify the structural features of these molecules, which may be contributing to intermolecular interactions and aggregation.

### Global Explanations

To investigate the molecular substructures commonly contributing to the MEGAN model's prediction of molecules as aggregators or non-aggregators, a global concept extraction was performed.[25] In this process, the pooled graph explanations for each channel undergo dimensionality reduction and clustering to provide clusters of molecules that share structural explanations (Figure 2c). Each cluster represents one specific molecular substructure, which occurs as an important explanation in many individual samples in the training data. All clusters from the concept extraction are presented in the Supporting Information. The concept extraction produced 159 clusters, with 25 associated with the non-aggregator channel and the remaining 134 clusters associated with the aggregator channel.

Visual analysis of the clustered concepts suggests that flexible molecules and molecules containing groups that have the potential to disrupt π-stacking through sterics often contribute to the non-aggregation prediction. Alternatively, flat, and rigid molecules, as well as molecules with functional groups that can act as both hydrogen bond donors and acceptors, contribute to the prediction of a molecule as an aggregator. Many of the trends that were identified in our clustering analysis are consistent with those that have been identified previously in literature for classifying molecular aggregation potential in small molecules.[8-13]

One notable trend identified in our cluster analysis, not previously reported in the literature, is the influence of thioureas and ureas on the classification of molecular aggregation. The model frequently associates ureas with aggregation, whereas the thiourea moiety is strongly linked to non-aggregator labels. Interestingly, while the model has identified the urea and thiourea substructures as important to the classification of molecules as aggregating or non-aggregating, there is no notable difference in the relative frequency of the urea and thiourea substructures between the aggregating and non-aggregating datasets used for model training (relative frequencies provided in the Supporting Information (SVII and SXVI)).

It was further identified that when thioureas are adjacent to an electron-withdrawing group, such as a carbonyl, they instead contribute to the aggregator label. This indicates that the local electronic environment of the urea and thiourea substructures plays a decisive role in the aggregation tendencies of molecules containing these functional groups. Traditional fingerprint-based feature attribution methods would miss these subtleties, as they focus on substructure presence or absence. In contrast, MEGAN explanations reveal how specific atom environments and functional group contexts affect predictions. The insight provided by analysis of the global explanations has practical implications as it identifies new functional groups that correlate with aggregation but also highlights the critical importance of the local electronic and steric environment in controlling aggregation tendencies. This emphasizes the need for the MEGAN model and its explanations to capture the complexity of molecular aggregation.

To analyze whether the structural explanations generated based on the MEGAN model can be analysed and interpreted in terms of physicochemical concepts in an automated way, we used an approach based on the prior chemical knowledge and the pattern-recognition abilities of large language models, specifically GPT-4o (see Supporting Information (SIX) for prompts and results). The objective was to connect structural graph explanations to broader chemical concepts and human-understandable chemical trends in order to stimulate further ideas for more detailed analysis by experts and to potentially derive design rules.

When prompted to explain why molecular motifs derived from the global explanation analysis trend with aggregation, the LLM outputs referred to relevant concepts such as potential interactions with water molecules and dimer interactions. Specifically, the output of the GPT-4o model referred to hydrogen bonding effects, planar structures that promote π-stacking, as well as steric hindrance of π-stacking - effects that are commonly associated with aggregation and thus true but not novel.

However, in a subsequent blind test, we prompted the GPT-4o model with all structural motifs identified by the MEGAN model through the global explanation analysis, but did not reveal the context of aggregation or the specific role of the motifs in enhancing or reducing aggregation, in order to reduce the bias of the model to just repeat already known prior knowledge of aggregation. The results agreed surprisingly well with the results of the previous text, with hydrogen bonding, π-stacking, and sterics being important characteristics that separate the two groups of motifs. Some physicochemical characteristics and thus possible explanations were additionally mentioned that we did not consider before as relevant descriptors, e.g., electron-donating effects, which lead to increased electron density in the π-system of an aryl ring and facilitate solvation over intermolecular π-π stacking interactions. This provides insight that can be validated and quantified in further experiments. Overall, this demonstrates how specialized models, such as MEGAN, can potentially be interfaced with general-purpose models such as LLMs to provide insight for the analysis of complex and not-well-understood datasets.[26]

### Counterfactuals

In addition to MEGAN's attributional explanations, we also employ counterfactual explanations to gain further insights into the model's behavior and decision-making process. For a given original input molecule, we define a counterfactual as a molecule with a minimal structural change from the original molecule which causes the greatest deviation in the model's prediction. Counterfactuals explain which kinds of local perturbations to chemical structure the model, and by extension the underlying aggregation property, is most sensitive to. To probe some of the structural features and modifications influencing the model's classification predictions, counterfactuals were generated for azacarbazoles **A** and **B** (Figure 5). Recalling from the previous discussion, the model correctly classifies **A** as aggregating and **B** as non-aggregating despite their subtle structural differences (Figure 3).

In general, the counterfactuals generated for aggregator **A** include molecules with electron-withdrawing groups added to the pyridine ring of the azacarbazole as well as molecules with additional steric bulk added to the carbon of the methyl group at the C1 position of the azacarbazole. The addition of an electron-withdrawing group, such as an aldehyde (**A-C1**), to the pyridine ring was found to switch the classification label from aggregator to non-aggregator. Based on this observation, we suggest that electron-withdrawing groups could be capable of disrupting or altering the π-stacking arrangements that may produce aggregation in **A** by modifying the electronics of the azacarbazole ring system. Additionally, steric bulk added to the methyl group of the azacarbazole (**A-C1**, **A-C2**, **A-C3**) results in larger, more flexible substituents, which sit out of the plane of the ring system, potentially disrupting favourable π-stacking arrangements and preventing aggregation. The non-aggregator channel's explanation masks for the counterfactuals of **A** are nearly identical to those of the non-aggregator **B**, with the only substantial difference being that the additional substituent is also highlighted in the counterfactuals.

The counterfactuals generated for non-aggregator **B** included molecules containing either a hydroxyl (**B-C1**, **B-C2**) or thiol group (**B-C3**) adjacent to the nitrogen atom of the pyridine ring. These analogous substructures, which are ideally suited for forming hydrogen bonding interactions that may contribute to aggregation, are frequently observed in molecules classified by the model as aggregators. These findings are consistent with trends observed in the study of pyridine derivatives (**E**-**H**) where the HO-C-N substructure was found to be critical for aggregation.

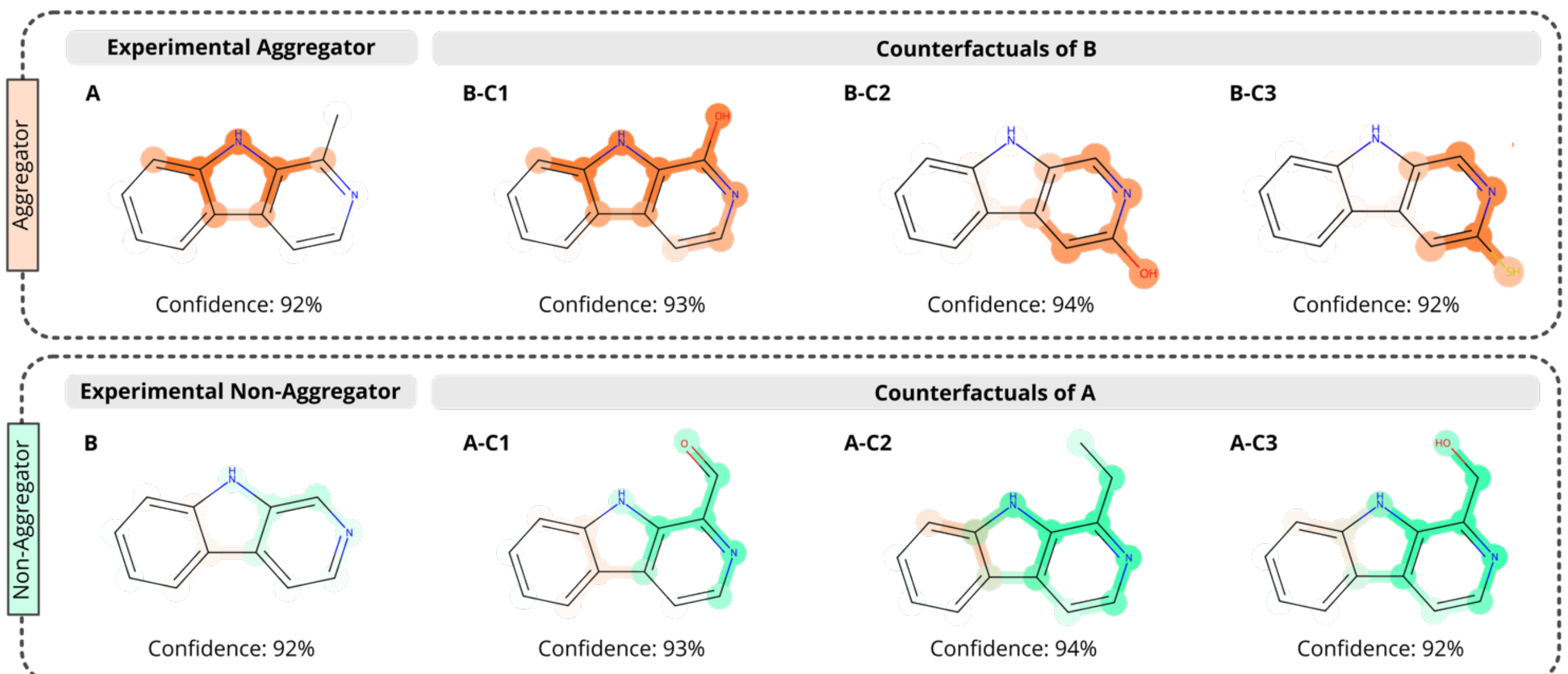

**Figure 5.** Counterfactuals of experimentally validated aggregator **A** and non-aggregator **B** with MEGAN model's prediction confidence and explanation masks.

The explanation masks for each of the counterfactuals of **B** (**B-C1**, **B-C2**, **B-C3**) highlight the hydroxy/thiol pyridine substructures in red, indicating that these substructures provide the model with evidence for the aggregation class. The identification of these functional groups is also consistent with the results of other studies, in which thiols and hydroxyl groups have been identified as functional groups commonly inducing aggregation.[8-13] Interestingly, compound **A-C3**, which contains a hydroxyl group not directly bound to the pyridine, is predicted as non-aggregating, suggesting that the hydroxyl group alone is not a predictor of aggregation for these compounds and further emphasizing the importance of the hydroxy/thio pyridine substructure.

### Experimental Validation of Predictions and Counterfactuals

In addition to serving as a means for identifying the structural features of molecules contributing to their classification as aggregators or non-aggregators by the MEGAN model, counterfactuals also allow for the prediction of new molecules highly similar to an input molecule with alternate aggregation properties. Applying the model in such a manner provides a method for the design of new compounds that maintain the desired structural aspects of a molecule, which may be crucial to forming necessary interactions within a biological target while modifying its aggregation tendencies. The following example provides experimental validation of the MEGAN prediction classification for one of the counterfactuals and demonstrates the utility of counterfactuals for altering the aggregation properties of a compound through minor structural modifications.

We investigated clioquinol, which has been previously reported to aggregate and is correctly predicted by the MEGAN model, and its flipped label counterfactual, methylclioquinol, which has never previously been experimentally examined for its aggregation properties (Figure 6). To compare the aggregation behaviour of these two molecules, dynamic light scattering (DLS) experiments were performed for each compound. For the DLS experiments, solutions of methylclioquonol and clioquinol at concentrations of 100, 75, 50, 25, 10, 8, 5, 3, 1 and 0.5 µM from were prepared by dilution of a 1 mM stock solution of the respective compound in DMSO with a 40 mM sodium phosphate buffer at pH 7.4.

Consistent with previous reports, clioquinol was found to aggregate, as demonstrated by a significant increase in its hydrodynamic radius at higher concentrations. In contrast, methylclioquinol exhibited minimal fluctuation in its hydrodynamic radius, indicating that aggregation does not occur in the concentration range studied. These findings highlight the potential power of the MEGAN model's ability to aid in the rational design of molecules with tailored aggregation properties, offering a versatile approach for optimizing compound behaviour in biological systems while preserving critical structural features.

To understand the model's rationale for suggesting methylclioquinol as a counterfactual, the explanation masks for both molecules were examined. The explanation masks for clioquinol highlight the majority of the quinoline scaffold, with a focus on the area containing the nitrogen and alcohol, as providing evidence for the aggregator class. In contrast, for methylclioquinol the methoxy group as well as the quinolin scaffold are both highlighted, albeit to a lesser extent, as providing evidence for the non-aggregator class. The methylation of clioquinol at the oxygen atom eliminates its ability to engage in hydrogen bonding interactions. These findings suggest that the model recognizes hydrogen bonding motifs and the steric bulk introduced by substituents as key factors influencing aggregation.

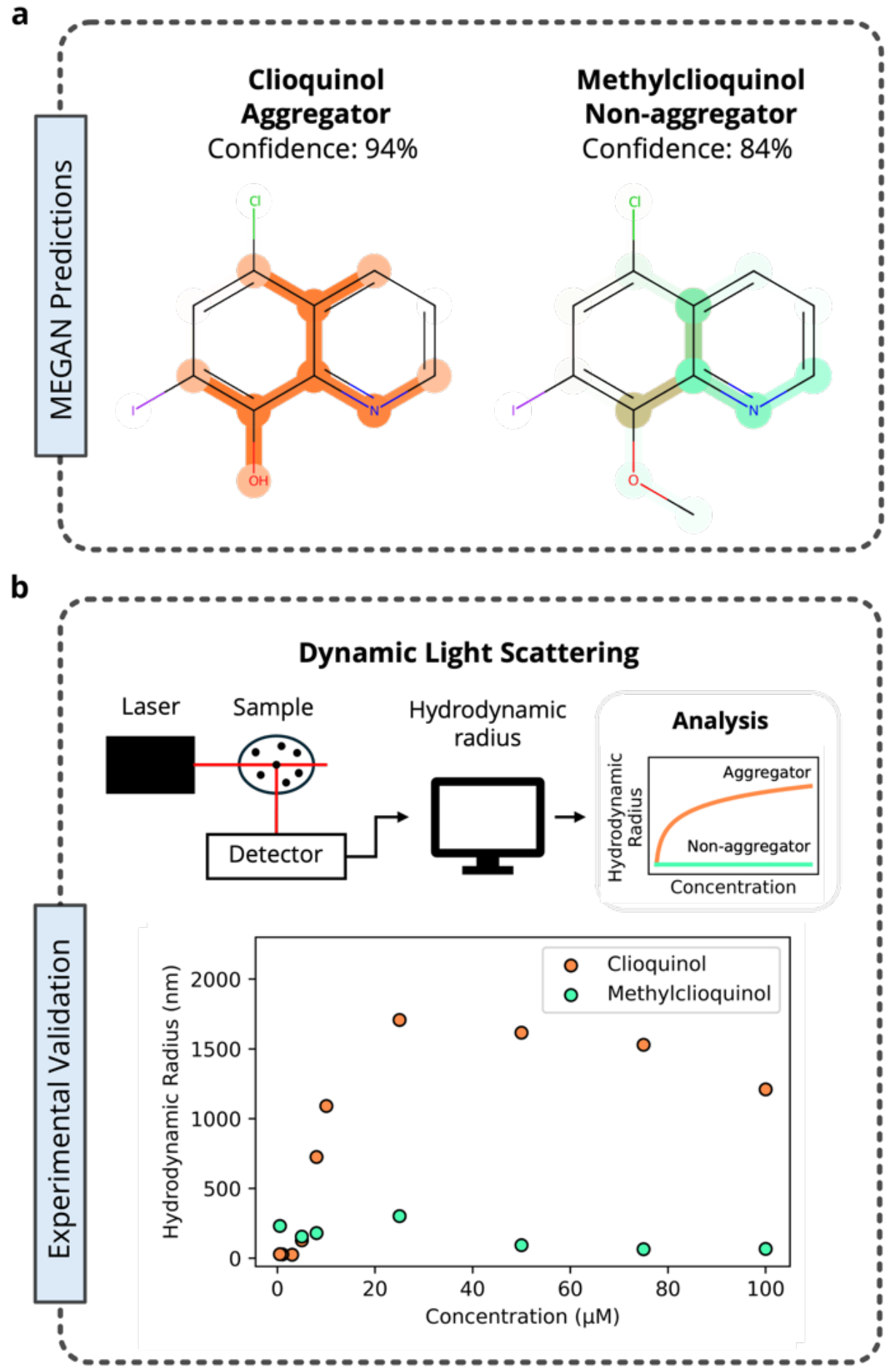

**Figure 6.** (a) MEGAN model's prediction confidence and explanation masks for clioquinol and methylclioquinol. (b) Overview of DLS method and experimental DLS results for clioquinol (orange) and methylclioquinol (green).

### External Validation

To further validate the predictive accuracy of the MEGAN model, a dataset of 58 structurally diverse drugs that have been experimentally characterized as aggregators or non-aggregators was compiled and assessed using the model. This external validation dataset was sourced from two literature references, where the compounds had been experimentally screened to evaluate the impact of colloidal aggregation on SARS-CoV-2 drug repurposing efforts.[27, 28] The dataset includes 30 aggregating and 28 non-aggregating compounds, none of which were present in the training or test sets for the MEGAN model, ensuring that they were previously unseen by the model.

The MEGAN model was capable of correctly classifying 15 of the 30 aggregating compounds and 24 of the 28 non-aggregating compounds, resulting in an accuracy of 68%. For comparison, we evaluated the performance of CHEMAgg, a widely recognized model for aggregation prediction, using the same validation dataset. CHEMAgg achieved an accuracy of only 50%, significantly lower than the MEGAN model. While CHEMAgg accurately identified 27 of the 28 non-aggregating compounds, it only correctly classified 2 of the 30 aggregating compounds. CHEMAgg's classification of 55 of the 58 molecules as non-aggregators suggests a strong bias in this model towards non-aggregator classification. In contrast, the MEGAN model provides a more balanced classification and demonstrates superior performance in accurately predicting molecular aggregation. This result highlights the ability of our model to outperform existing state-of-the-art models, particularly in identifying aggregating compounds, which is a critical capability for accurately screening datasets for promiscuous molecules.

## Conclusion

The MEGAN model presented in this study demonstrates a high level of accuracy for predicting SCAMs, achieving 82% on a balanced in-distribution test dataset and similar performance on a smaller external dataset of drug molecules. While commonly accepted features such as logP values and the number of aromatic rings are relevant for detecting SCAMs, the model's high accuracy and explanations indicate that it learns more complex relationships between molecular structure and aggregating behavior.

Through local and global model explanations and counterfactual analysis, we identified and systematically examined specific molecular motifs associated with aggregation. By combining expert knowledge, quantum mechanical modeling, and automated interpretations using large language models, we provide deeper insights into molecular aggregation. The MEGAN model not only highlights molecular motifs strongly linked to aggregation but also identifies small structural modifications that can significantly alter aggregating behavior, providing valuable tools for molecular design.

The MEGAN model's accuracy, combined with the accessibility and extended functionality provided through a publicly available web interface, enables improved detection of SCAMs. Furthermore, the use of molecular counterfactuals offers a practical approach for designing alternatives to problematic compounds. This capability facilitates the filtering of HTS libraries, helping to reduce false positives in drug discovery databases and improving overall screening reliability.

## Supporting Information

The authors have cited additional references within the Supporting Information.[29-47]

## Acknowledgements

P.F. acknowledges support by the Federal Ministry of Education and Research (BMBF) under Grant No. 01DM21001B (German-Canadian Materials Acceleration Center), as well as support by the Helmholtz Association within the Helmholtz Core Informatics Initiative. R.D. acknowledges support from the National Science and Engineering Council Discovery Grant Program and the Canadian Institutes of Health Research.

## Entry for the Table of Contents

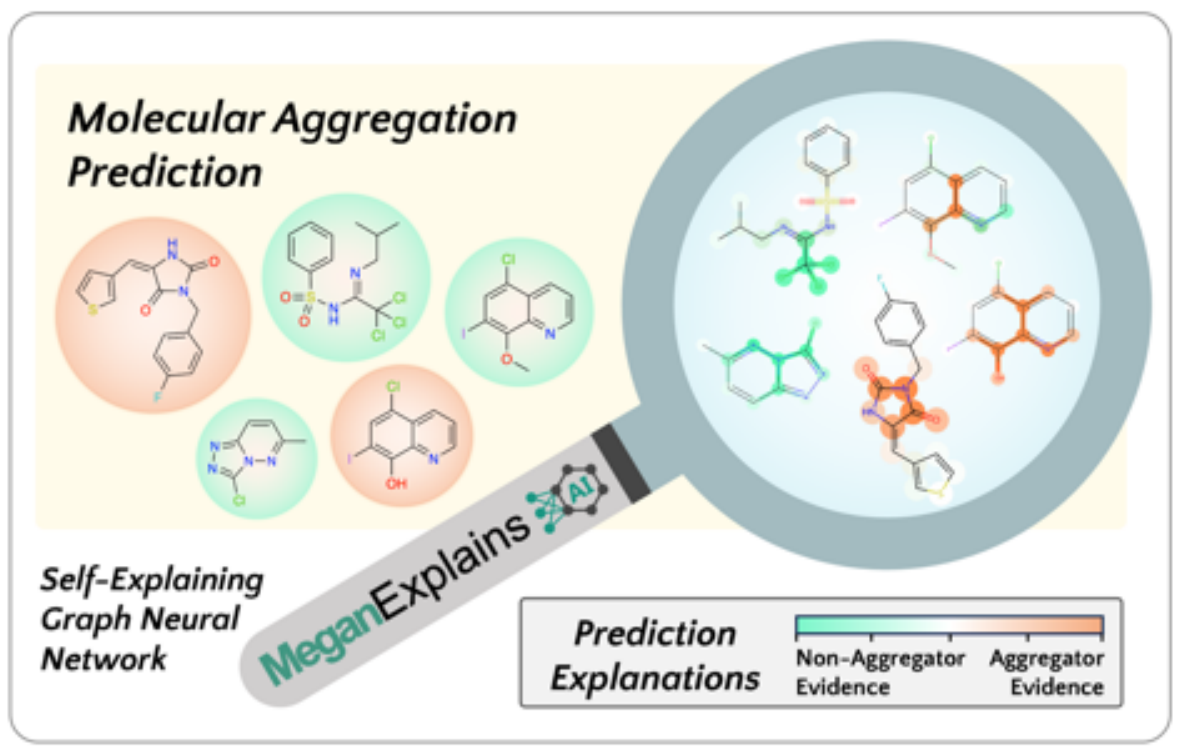

In this work, MEGAN, our explainable AI model, is applied to predict small colloidally aggregating molecules (SCAMs), a significant source of false positives in drug discovery. MEGAN provides interpretable predictions and proposes structurally similar, non-aggregating alternative molecules. Experimental validation of one such alternative demonstrates MEGAN's potential to guide molecular design and improve high-throughput screening reliability.

## Supporting Information

### Mitigating Molecular Aggregation in Drug Discovery with Predictive Insights from Explainable AI

Hunter Sturm, Jonas Teufel, Kaitlin A. Isfeld, Pascal Friederich, Rebecca L. Davis

## SI. Methods

### Dataset curation and preparation

For model training, a dataset of molecules that have been experimentally validated as either aggregating or non-aggregating was curated. In order to do so, data from the Shoichet laboratory utilized for the development of their Aggregator Advisor tool was used.[9] For the development of Aggregator Advisor, Stoichet et al. identified over 12,600 likely aggregators from their previously published experimental screens for colloidal aggregation. These compounds were used as our dataset of experimentally validated aggregators. To curate our dataset of non-aggregating molecules, one of the libraries screened for aggregation by Stoichet consisting of approximately 198,000 compounds was filtered to exclude any molecules in our dataset of aggregating molecules.[29] As the 198,000 compound dataset was utilized to populate the 12,600 aggregators in Aggregator Advisor, any compound in the dataset that was not identified by Stoichet as an aggregator has been experimentally validated as a non-aggregator.

Upon data curation, the resulting datasets consisted of 12,607 aggregating and 197,846 non-aggregating molecules represented as SMILES strings. In order to clean the data for use in training our machine learning model, the SMILES strings were canonicalized using RdKit.[30] If multiple fragments were present in the SMILES strings, the longest one was kept and the others (generally counterions or solvents) were removed. If there was more than one long (>12 characters) fragment, the data point was discarded. All duplicate molecules and molecules producing RdKit errors were also discarded from the datasets. Additionally, for the non-aggregating dataset, molecules containing metal atoms and inorganic molecules (i.e. containing zero carbon atoms were discarded. After filtering the molecules, the remaining datasets consisted of 12,338 aggregating and 177,048 non-aggregating molecules. Molecular descriptors were analyzed to examine the differences in structural features and properties of the compounds in the aggregating and non-aggregating datasets (see section SX of Supporting Information).

For the training of the graph neural network, the dataset was first pre-processed using the publicly available protonation software library DimorphiteDL to extract all possible protonation states at physiological pH for each molecule.[31] A value of 7.4 was used for both the min_ph and max_ph parameters of the DimorphiteDL software. The protonated molecule variants were then converted into graph representations using RDKit to extract numeric node and edge attribute vectors.[30] Node features include a one-hot representation of the atom type, as well as additional information such as the number of connected hydrogen and local charge distribution. Similarly, the edge features contain a one-hot encoding of the bond type and additional information such as whether the bond is part of a ring structure.

In case users are interested in different conditions than the ones listed above, fine-tuning of the trained model depending on conditions can be achieved in multiple ways: If condition information is available for all data points in a new dataset, then the information can be used as global information in the GNN model or as a condition on the prediction. In that way, during test time, condition-dependent prediction can be generated by the model. Alternatively, if conditions are only known for a subset of the datapoints, a general base-model can be trained with all datapoints and then fine-tuned for the subset of datapoints with desired target conditions.

## Explainable graph neural networks, MEGAN

The Multi-Explanation Graph Attention Network (MEGAN) is a graph neural network designed to make predictions while also explaining how it arrived at those predictions. Unlike many other models, MEGAN can provide multiple explanations for each prediction, such as highlighting features that contribute positively or negatively. It uses attention mechanisms to focus on important parts of the graph and a self-consistent, masking-based training process to ensure that its explanations align with the expected meanings. This makes MEGAN especially useful for tasks like predicting molecular properties, where understanding the role of specific substructures is critical.

Soley relying on attention masks to generate explanations has been demonstrated to lead to misleading results. For that reason, the MEGAN model[19] additionally aggregates attention scores across all layers to perform an attention-weighted graph pooling. This means that the global pooling step before the final regression MLP is weighted using the aggregated attention scores, which in this way receive additional training signal from the graph labels. We do this to ensure the faithfulness of the generated explanations: Only information from nodes with high attention values is used for the final prediction outcome. More information can be found in Teufel et al.[19] To ensure an interpretable alignment of the generated attention masks with human expectations, the training process also explicitly implements a semi-supervised explanation loss. This training loss promotes the model to approximately solve the primary classification task using only the explanation masks themselves. In this case, the "Aggregator" explanation channel, for example, is trained to highlight explanations only for true "Aggregator" elements. This effectively forces the model to recognize and highlight those substructures which occur more often in elements belonging to one class versus the other.

As demonstrated in Teufel et al.[19], the MEGAN model beats common graph explanation methods such as GNNExplainer and GNES in explanation accuracy, sparsity, and fidelity when trained on dedicated benchmark datasets with ground truth explanations. At the same time, the model matches the performance of other graph neural networks such as GAT, GIN, Schnet, and PAiNN in molecular regression and classification tasks. Thus, the MEGAN model is currently on the Pareto front of accuracy and explainability.

More details on the concrete implementation and benchmarks can be found in Teufel et al.[19]. An interactive interface to query the MEGAN model for aggregation prediction is available at https://megan.aimat.science/predict/megan_aggregator. Code and data are available on GitHub (See Data and Code Availability).

## Graph Representation

To obtain a prediction for a given molecule using the aforementioned MEGAN graph neural network, the molecule first has to be converted into a graph structure. In this graph structure, each node (atom) is represented by a numeric node feature vector and each edge is represented by a numeric edge feature vector. We construct the node feature vector by concatenating the following information:

- A one-hot representation of 15 common atom types, including carbon, nitrogen, oxygen, sulfur, chlorine, fluorine as well as a special encoding for "other" atom types taht are not explicitly included.
- Separate one-hot encoding of atom's hybridization state, its total degree and the total number of attached hydrogens.
- A single binary flag to indicate whether or not the atom is part of an aromatic ring.
- Continuous values for the atom's mass and it's formal charge.
- Individual atom contributions of the molecular descriptors LogP, TPSA and LabuteASA.

Likewise, we construct the edge feature vector by concatenating the following information:

- A one-hot encoding of the bond type and the stereo state of the bond.
- A single binary flag to indicate whether or not the bond is part of aromatic ring.

The conversion as well as the computation of the node and edge features is implemented using the RDKit software package which creates the molecular graph from its corresponding SMILES representation. The pre-processing and feature computation takes approximately 0.78±0.16 milliseconds (averaged over 1000 elements) and can therefore easily be considered real-time.

**Counterfactuals**

Counterfactuals are generated by exploring the immediate graph neighborhood of each original molecule and then selecting those that result in the highest prediction difference relative to the original prediction. We generate the local neighborhood of the molecular graphs with a procedure inspired by Riley et al. in which all chemically feasible atom and bond insertions and deletions are applied to the molecule recursively.[32] Depending on the recursion depth, this procedure typically produces 100-1000 perturbed graphs out of which we present the 10 graphs with the highest prediction difference as counterfactual explanations.

**Fingerprint and descriptor-based ensemble methods**

To compare the accuracy of our model to the aggregation prediction model reported by Yang et al., we re-implemented their computational approach and trained it on the dataset presented in this work. We used three types of descriptors, namely fingerprint features (circular fingerprints as well as ECFP4/Morgan fingerprints as implemented in RdKit), MACCS features as implemented in RdKit, and CATS features as implemented in https://github.com/alexarnimueller/cats-descriptor.[33-35] Within each set of descriptors, we used a feature importance analysis based on a Random Forest classifier to determine how many features are required to achieve the highest accuracy on the validation set. This resulted in a total selection of 33 features. After concatenation and hyperparameter optimization, we found the optimal hyperparameters of the Random Forest classifier to be an ensemble size of 500, with no limit on depth, and entropy as an impurity measure; of the Gradient Boosting classifier to be an ensemble size of 500, a maximum depth of 9, and a learning rate of 0.5; and of the xGBoost model to be an ensemble size of 500, a maximum depth of 8, and a learning rate of 0.1.

## DFT modeling of interaction energies

DFT modeling of pyridine derivatives **E - P** was performed to assess the energetic favorability of the intermolecular interactions formed between compounds as well as the favored geometries of the dimers. To locate the lowest energy geometries of the monomers and dimers, initial geometries were constructed, and conformer ensembles were generated using CREST based on GFN2-xTB calculations.[36,37] The lowest energy monomer and dimer geometries were then optimized using wB97xD/def2-TZVP with the SMD solvent model ($H_2O$) in Gaussian 16.[38-41] Vibrational analysis performed at 273.15 K confirmed that all optimized geometries corresponded to minima on the potential energy surfaces, as indicated by the absence of imaginary frequencies. See section SXVII and SXVIII of the Supporting Information for coordinates and energies of optimized geometries. The interaction energies were calculated following Eq. S1 using the sum of the electronic energy ($E$) and zero-point correction to the electronic energy ($E_{ZPE}$). In all cases, the most energetically favorable geometry of the dimer that could be located was used for the calculation of the interaction energy.

$$\text{Interaction Energy} = (E_{dimer} + E_{ZPE,\ dimer}) - 2(E_{monomer} + E_{ZPE,\ monomer}) \quad \text{(Eq. S1)}$$

## Dynamic light scattering experiments

DLS was used as an aggregation detection method for the examination of clioquinol and methylclioquinol. The instrument used in this study was a Nanotemper Prometheus Panta and standard 10 µL capillaries were used. Clioquinol was purchased from TCI chemicals via Fisher Scientific, and methylclioquinol was prepared and purified following standard methylation conditions using iodomethane (see section SXIX of the Supporting Information for NMR spectra).[42] Stock solutions of 1 mM were prepared for each compound in DMSO. The stock solutions underwent serial dilution using a 40 mM sodium phosphate buffer at pH 7.4 to prepare solutions of 100, 75, 50, 25, 10, 8, 5, 5, 3, 1, and 0.5 µM. The concentrations tested were chosen based on the concentrations that standard HTS screens are run at. Each solution was run in triplicate on the instrument, and aggregation was determined based on hydrodynamic radius.

## Data and code availability

Our code for model training and counterfactual analysis can be found on Github https://github.com/aimat-lab/megan_aggregators. In our repository, we also include a persistent representation of the already trained model, which can be used directly. The full dataset can be downloaded at https://bwsyncandshare.kit.edu/s/4r9kgyCFQL6PTcF. The cleaned aggregator and non-aggregator datasets, external validation dataset and $^{1}H$ and $^{13}C$ NMR spectra can be downloaded at https://github.com/DavisGroup/MEGAN-aggregation-data. Furthermore, we provide an interactive web interface for the manual prediction of the aggregation behavior of single molecules at https://megan.aimat.science/predict/megan_aggregator. Given a SMILES representation of a molecule, the interface shows the predicted classification, visualization of local explanations, and the top counterfactuals.

## SII. Model performance comparison

The performance of the MEGAN model was compared against the XGboost model of Yang et al. (ChemAgg), using our balanced test set.[11] The results in **Table S1**, in particular a comparison of Entries 1 and 2, show that the dataset presented here (or at least the balanced test split chosen here) is more difficult to predict than the dataset presented in Yang et al.[11] However, when comparing the performance of the feature based XGBoost model used in Yang et al. (Entry 2) with the MEGAN model (Entry 3), we see a clear advantage of the graph neural network. This can be due to the fact that the model input for the graph neural network is "complete" in a sense that the full molecular structure is modeled, which is not the case for the fingerprint and feature representations, or it can be related to the higher complexity and expressiveness of graph neural networks. Similar trends of superior performance of graph neural networks compared to classical machine learning models can also be seen with many other datasets of molecules and materials of similar size.[43]

**Table S1.** Accuracy and F1 score of aggregation classifiers from literature (Yang et al.[11]), compared to our data and our MEGAN model.

| Entry | Model | Data | Accuracy (test) | F1 score (test) |
|---|---|---|---|---|
| 1 | Yang et al. | Yang et al.[1] | 0.937 | 0.899 |
| 2 | Yang et al.[2] | Ours (balanced test set) | 0.733 | 0.735 |
| 3 | Ours (MEGAN) | Ours (balanced test set) | 0.818 | 0.807 |

[1] The dataset used in Yang et al. is not published, which is why we could not reproduce the experiments with our model on their dataset.

[2] The code used in Yang et al. is not published, so we used a re-implementation based on the methodology described in Yang et al. However, not all 5 sets of descriptors could be re-implemented, which is why we used a subset of 3 sets of descriptors (see Methods in section SI of Supporting Information). The best performance was achieved with a XGBoost model but the performance of random forest models and gradient boosting models was almost identical.

## SIII. Global concept extraction report

Global concept extraction report is located at https://github.com/DavisGroup/MEGAN-aggregation-data

## SIV. Examples of correct model predictions for molecule pairs with small structural changes

Additional examples of aggregation cliffs (i.e., where small structural changes flip the label from aggregator to non-aggregator or vice versa) in the training/test data have been identified and select examples are provided below. However, it should be noted that these represent a small sample of hand-picked aggregation cliffs out of the training and test set and it is highly likely given the oversampling method used for the aggregator class during model training that the model will miss many aggregation cliffs. We were unable to further validate our model's sensitivity on unseen sets of molecules exhibiting aggregation cliffs as no additional examples could be found in the literature.

| **Experimentally Known Non-Aggregators** | **Experimentally Known Aggregators** |
| --- | --- |
| Predicted Non-Aggregator: 0.93 Confidence | Predicted Aggregator: 0.60 Confidence |
| Predicted Non-Aggregator: 0.84 Confidence | Predicted Aggregator: 0.91 Confidence |
| Predicted Non-Aggregator: 0.69 Confidence | Predicted Aggregator: 0.92 Confidence |
| Predicted Non-Aggregator: 0.63 Confidence | Predicted Aggregator: 0.87 Confidence |

**SV. Impact of training set imbalance and test set imbalance on accuracy, precision, and recall**

As the dataset is highly imbalanced, we explicitly used oversampling to address the class imbalance, which is a common practice in such cases. This means that in each training epoch, we reused some of the non-aggregating molecules multiple times in order to have the same number of training instances with both label types.
For the evaluation of our model, we specifically constructed a balanced test set consisting of 500 randomly sampled aggregators and 500 randomly sampled non-aggregators that were not used during training. On this balanced test set, we report an accuracy of approximately 82%. However, we can still observe a slight imbalance in the prediction performance. The confusion matrix below shows that the model more often mispredicts true aggregators as non-aggregators than vice versa. This indicates that it is generally more biased to predict non-aggregators, which is most likely a consequence of the (training) dataset imbalance.

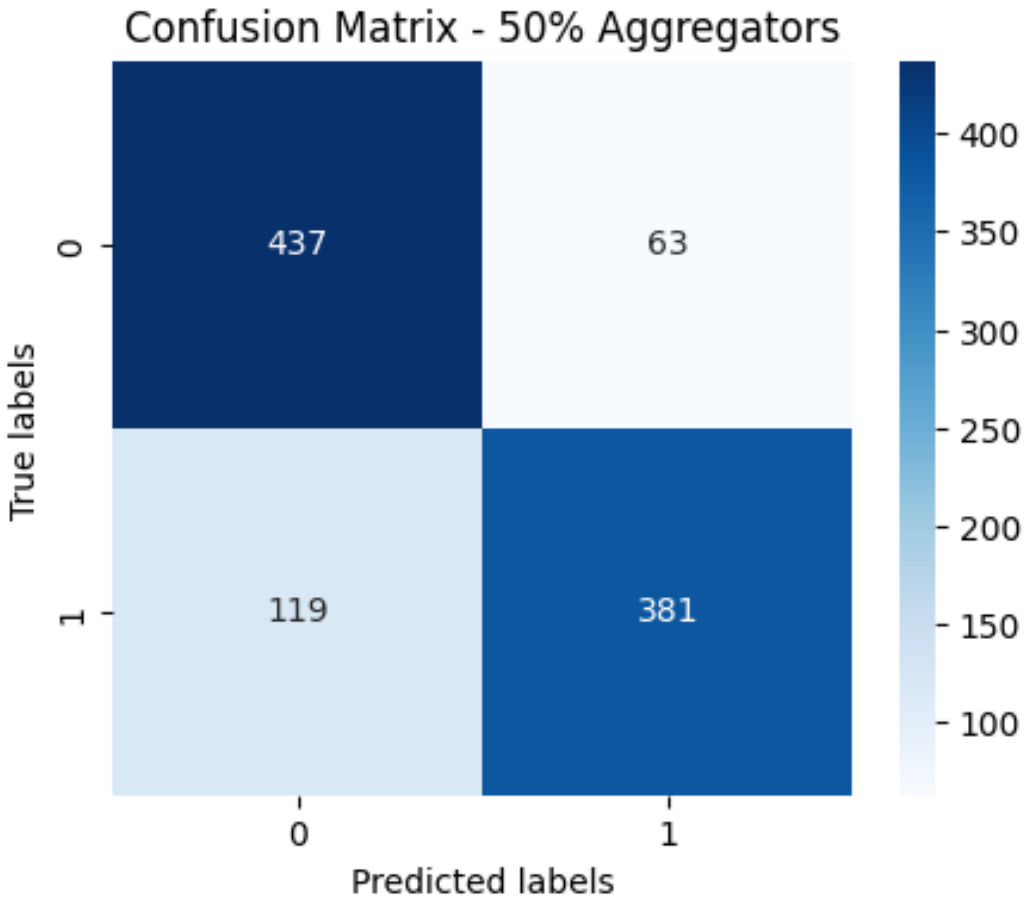

The plot below shows the confusion matrix when artificially reducing the ratio of aggregators by subsampling, creating a test set with high label imbalance (5 aggregators and 500 non-aggregators).

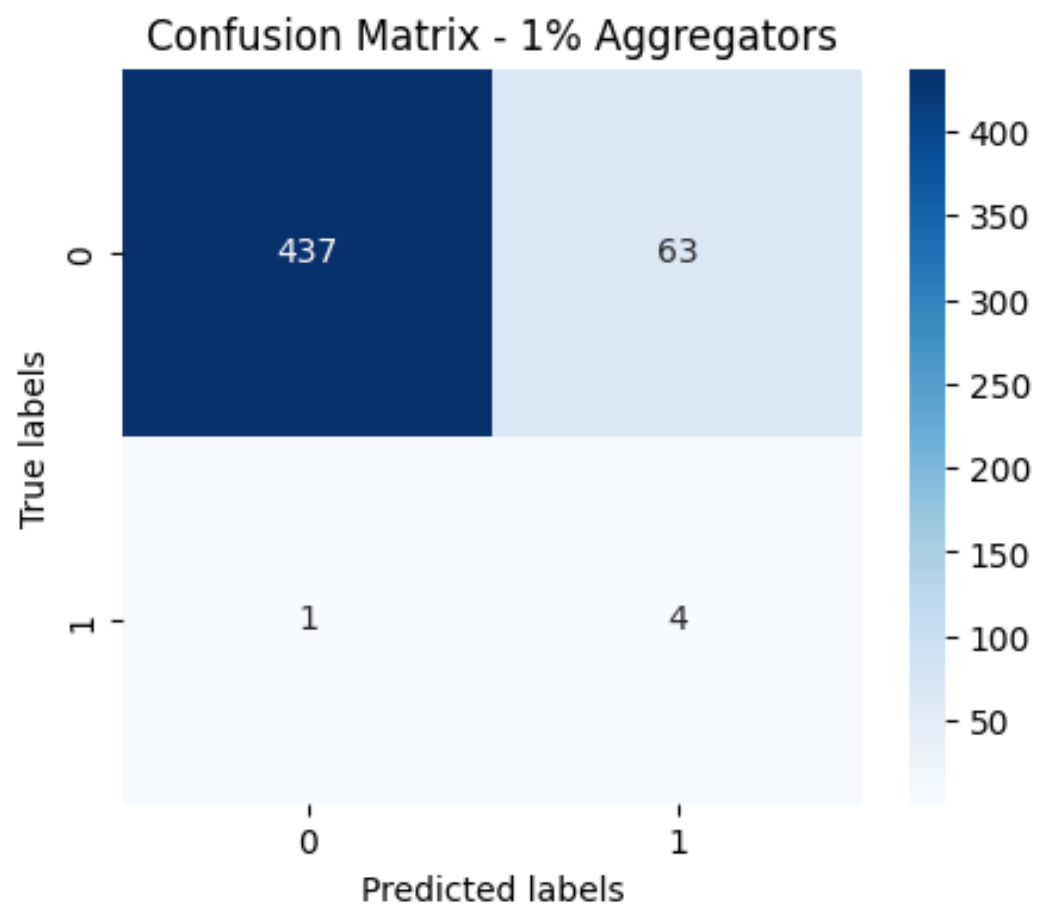

This case illustrates that (with the given threshold) the model has high recall, meaning a high capability of detecting (true) positives. Out of 5 true aggregators in the imbalanced test set, 4 are correctly identified. However, the model has a rather low precision, so there are many false positives as well, which is to be expected even for a well-trained model when tested on a highly imbalanced dataset. The Bayesian probability that a molecule actually is an aggregator, given that the train model predicts that it is an aggregator, i.e. the precision of the model, is dominated by the low overall probability of the aggregator label, so even a very well-trained model with high accuracy and recall have a low precision due to the high intrinsic ground truth label imbalance resulting in a low value of P(agg):

$$precision = P(agg \mid model\ predicts\ agg) = \frac{P(model\ predicts\ agg \mid agg) * P(agg)}{P(model\ predicts\ agg)} \propto recall * P(agg)$$

In agreement with the equation above, the explicit dependence of accuracy, precision, and recall as a function of test set (im)balance can be found below. While the accuracy slightly increases with increasing test set imbalance, the precision linearly drops (at a nearly constant recall).

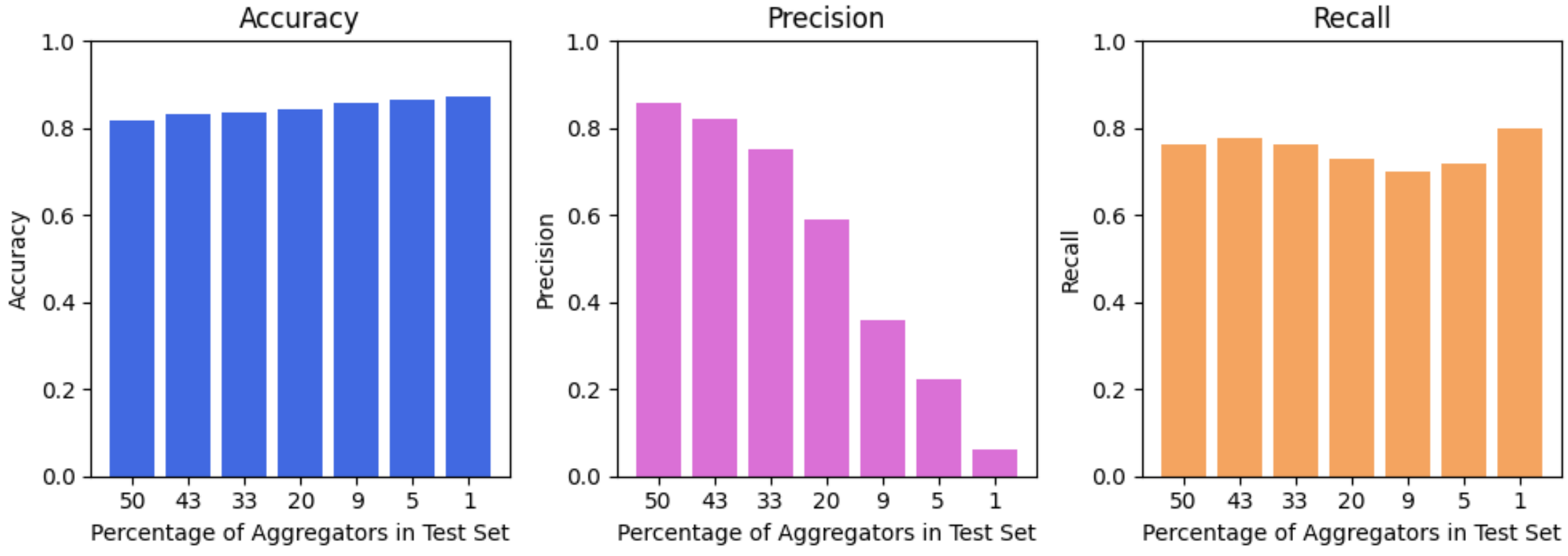

## SVI. Energy decomposition analysis of dimers of E-P

For pyridine derivatives **E – P**, energy decomposition analysis (EDA) was completed to further evaluate the trend between aggregation prediction confidence and the dimer interaction energies (computational methods below). For the EDA, SAPT0 calculations were performed using Psi4 to decompose the dimers interaction energies ($E_{int}$) into the sum of four physically relevant energy terms: electrostatics ($E_{elec}$), exchange ($E_{exch}$), induction ($E_{ind}$), and dispersion ($E_{disp}$) (Eq. S2).

$$E_{int} = E_{elec} + E_{exch} + E_{ind} + E_{disp} \qquad \text{(Eq. S2)}$$

Each of these energy terms and the non-covalent interactions they represent are described in greater detail in the literature.[23, 24, 44] However, the terms can be briefly summarized as follows: the electrostatic interaction term describes the energy of interactions between permanent multipoles on each molecule. The exchange interaction term (sometimes called exchange repulsion) describes the energy needed to be overcome for tight molecular packing, and is often destabilizing. The induction term describes the energy of the induced electrostatic moments of one molecule interacting with the permanent electrostatic moments of another. Finally, the dispersion term describes the energy of interactions between induced dipoles.

The results of the EDA analysis for the pyridine derivative dimers are reported in Figure S1 and Table S2. These results show the total SAPT0 energy (i.e. the interaction energy of the two molecules forming the dimer ($E_{int}$) decomposed into its four contributing energy terms. For all dimers, $E_{exch}$ was found to be destabilizing, whereas $E_{elec}$, $E_{ind}$ and $E_{disp}$ were stabilizing.

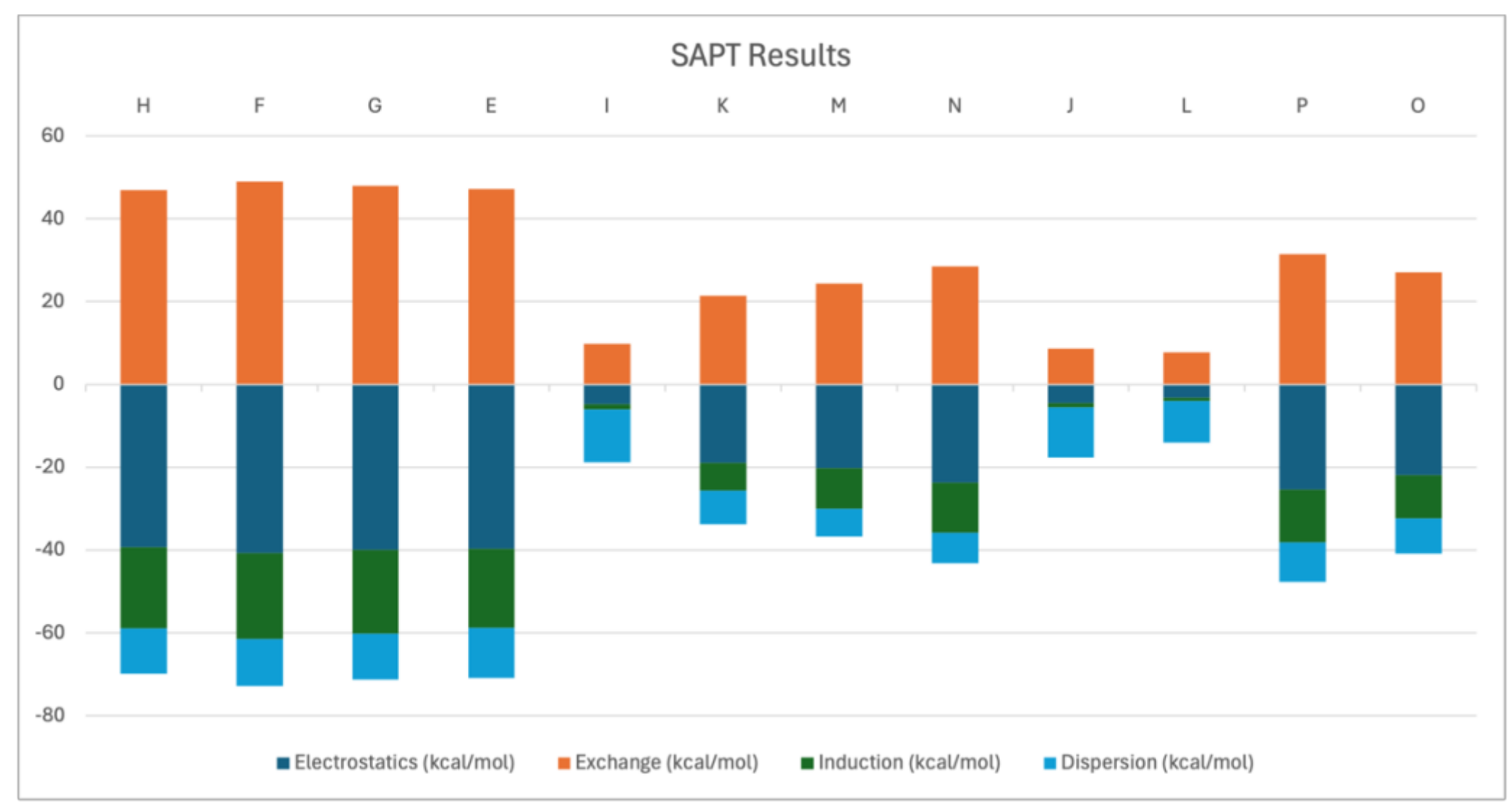

**Figure S1.** SAPT0 energy decomposition analysis results for dimers of **E** - **P**. Sorted from left to right based on highest to lowest aggregator prediction confidence.

**Table S2.** SAPT0 energy decomposition analysis results for dimers of **E** - **P**, sorted by aggregation prediction confidence. All energies are reported in kcal/mol.

| Dimer | Aggregation Prediction | $E_{elec}$ | $E_{exch}$ | $E_{ind}$ | $E_{disp}$ | $E_{int}$ |
|---|---|---|---|---|---|---|
| H | 0.94 | -39.29 | 46.87 | -19.56 | -10.99 | -22.96 |
| F | 0.92 | -40.72 | 48.99 | -20.75 | -11.36 | -23.84 |
| G | 0.88 | -40.00 | 48.02 | -20.23 | -11.03 | -23.23 |
| E | 0.86 | -39.67 | 47.17 | -19.10 | -12.07 | -23.67 |
| I | 0.12 | -4.82 | 9.76 | -1.29 | -12.67 | -9.02 |
| K | 0.09 | -18.89 | 21.38 | -6.75 | -8.11 | -12.36 |
| M | 0.07 | -20.25 | 24.36 | -9.78 | -6.75 | -12.41 |
| N | 0.07 | -23.64 | 28.44 | -12.16 | -7.32 | -14.69 |
| J | 0.06 | -4.52 | 8.61 | -1.00 | -12.07 | -8.97 |
| L | 0.06 | -3.15 | 7.72 | -0.81 | -10.09 | -6.32 |
| P | 0.05 | -25.35 | 31.47 | -12.80 | -9.47 | -16.15 |
| O | 0.05 | -21.84 | 27.08 | -10.43 | -8.55 | -13.74 |

Analysis of the magnitude of $E_{elec}$, $E_{exch}$, $E_{ind}$ and $E_{disp}$ for each dimer revealed three distinct groups. Group 1 consists of the dimers of all of the predicted aggregators (**E**, **F**, **G** and **H**), group 2 consists of the dimers of predicted non-aggregators **K**, **M**, **N**, **O** and **P**, and group 3 consists of the dimers of predicted non-aggregators **I**, **J** and **L**. Interestingly, the molecules of group 1 each form two hydrogen bonds in their dimers, the molecules of group 2 each form one hydrogen bond in their dimer with the exception of the dimer of **K** which forms 2 hydrogen bonds, and molecules of group 3 are those exhibiting pi-stacking interactions.

For group 1 (dimers of the predicted aggregators **E**, **F**, **G** and **H**) the destabilizing exchange energy ($E_{exch}$) had the largest magnitude out of the four energy terms. Of the three stabilizing interactions, $E_{elec}$ had the largest magnitude. This term would encompass the two favourable hydrogen bonding interactions present in each of the dimers of this group. The second most stabilizing energy term for dimers of molecules of this group was $E_{ind}$, followed by $E_{disp}$.

Group 2 (dimers of the predicted non-aggregators **K**, **M**, **N**, **O** and **P**) demonstrated the same energy trends as dimers in group 1. $E_{exch}$ was destabilizing and had the largest magnitude while the three stabilizing interactions (from most to least favourable) were $E_{elec}$, $E_{ind}$ and $E_{disp}$. Interestingly, for group 2, each of the four energy terms were roughly half the magnitude of the energies for dimers of group 1. This fits with the observation that dimers of both groups interact through hydrogen bonding, and that dimers of group 1 form two hydrogen bonds whereas dimers of group 2 form only one hydrogen bond (with the exception of the dimer of **K**). While **K** forms two hydrogen bonds in its dimer, the weaker nature of the N-H-N hydrogen bonding interactions in this dimer, relative to the more polarized O-H-N hydrogen bonding interactions present in dimers of **E-H,** matches the observed trend. The decreased strength of the N-H-N hydrogen bonding can be seen in the smaller electrostatic and induction terms in **K** relative to **E-H**.

Group 3 (dimers of the predicted non-aggregators **I**, **J**, and **L**) was found to have the smallest SAPT0 interaction energies and showed a different trend compared to groups 1 and 2. In group 3, $E_{disp}$ was found to have the greatest magnitude and for all three dimers of this group was larger than the destabilizing exchange energy term. For this group, $E_{ind}$ was the least stabilizing energy term preceded by $E_{elec}$. This was an interesting, yet unsurprising trend as the dimers of **I**, **J**, and **L** were the only dimers in our study that did not participate in hydrogen bonding and instead participate in pi-stacking interactions, explaining why $E_{disp}$ had the largest magnitude of the energy terms. Overall, the energetic decomposition of the interaction energies of the dimers of the pyridine derivatives showed interesting trends and were consistent with the trends provided by the DFT calculations reported in the main text.

<u>EDA Computational Methods:</u> For SAPT0 calculation setup with Psi4, the default flags were used in all cases, except for the basis flag which was set to def2-TZVP and the freeze_core flag which was set to TRUE. The coordinates of the DFT optimized pyridine dimer geometries were used for the EDA (see section SI of the Supporting Information for DFT methods).

## SVII. Determination of relative frequency of thiourea fragments for all compounds in the aggregator and non-aggregator datasets from SMARTS pattern

In order to determine the number of thiourea fragments present in each compound in the aggregator and non-aggregator datasets, the SMILES strings for the molecules in each dataset were searched for the thiourea substructure using RDKit. The thiourea substructure was defined using the SMARTS pattern [NX3][CX3](=[SX1])[NX3] and the search was performed on the SMILES strings for the aggregator and non-aggregator datasets which had been protonated to protonated to the most energetically favourable ionization state at physiological pH (pH=7.4) using the FixpKa functionality of OpenEye's QUACPAC 2.1.2.1.[45]

In the non-aggregator dataset, 4,268 molecules were found to contain one thiourea group and 37 molecules were found to contain between two and four thiourea groups. In the aggregator dataset, 334 molecules contained one thiourea group and 1 molecule contained two thiourea groups. The relative frequencies of the thiourea substructure in each dataset is illustrated below. As the relative frequency of molecules containing more than one thiourea group is negligible, it has been excluded from the image below.

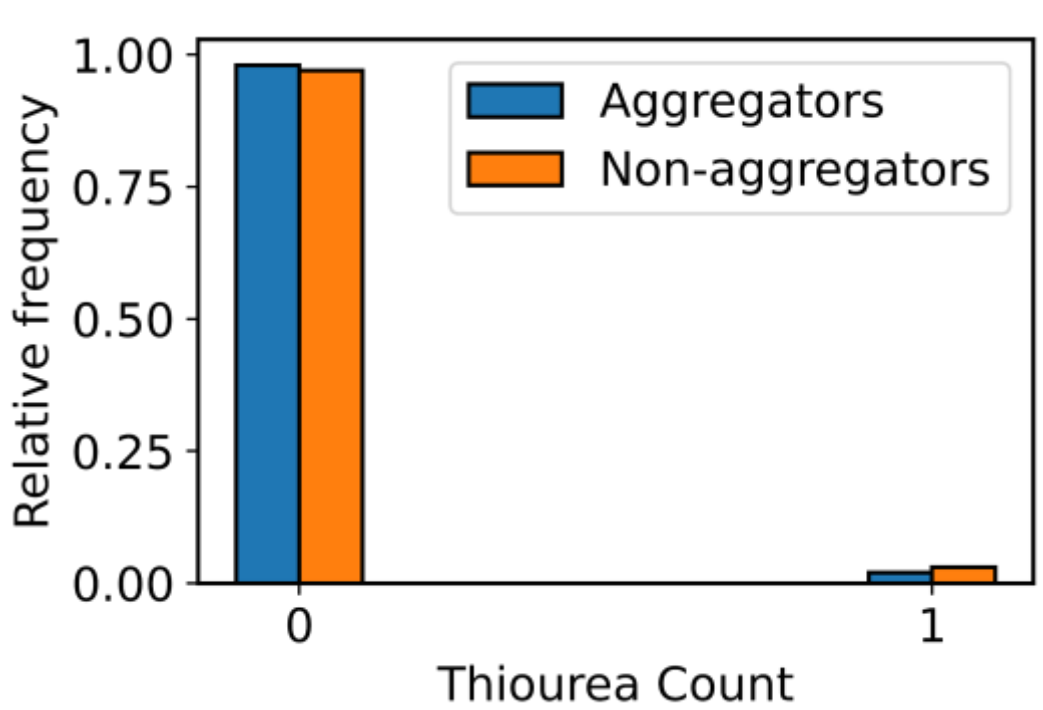

## SVIII. Correlation of explanation masks and atomic contributions of simple molecular descriptors

We conducted a feature importance analysis for the given aggregation classification task and find that several classic molecular descriptors show a weak separability between the aggregator and non-aggregators classes, i.e. they have slightly different (shifted) distributions when comparing their values for the aggregating and the non-aggregating samples. Notably, we find MolLogP to be the most important feature, but also find MolMR (Wildman-Crippen molar refractivity value - accounting for molecular size and polarizability) and LabuteASA (Labute's Approximate Surface Area) to be important as well[1].

These findings are reinforced by Figure S2, which show the slightly differing distributions for the three previously mentioned properties between the two ground truth classes in the test dataset. However, neither of these simple descriptors alone is capable of sufficiently separating the classes on their own, as indicated by the high distributional overlap coefficients of >0.75.

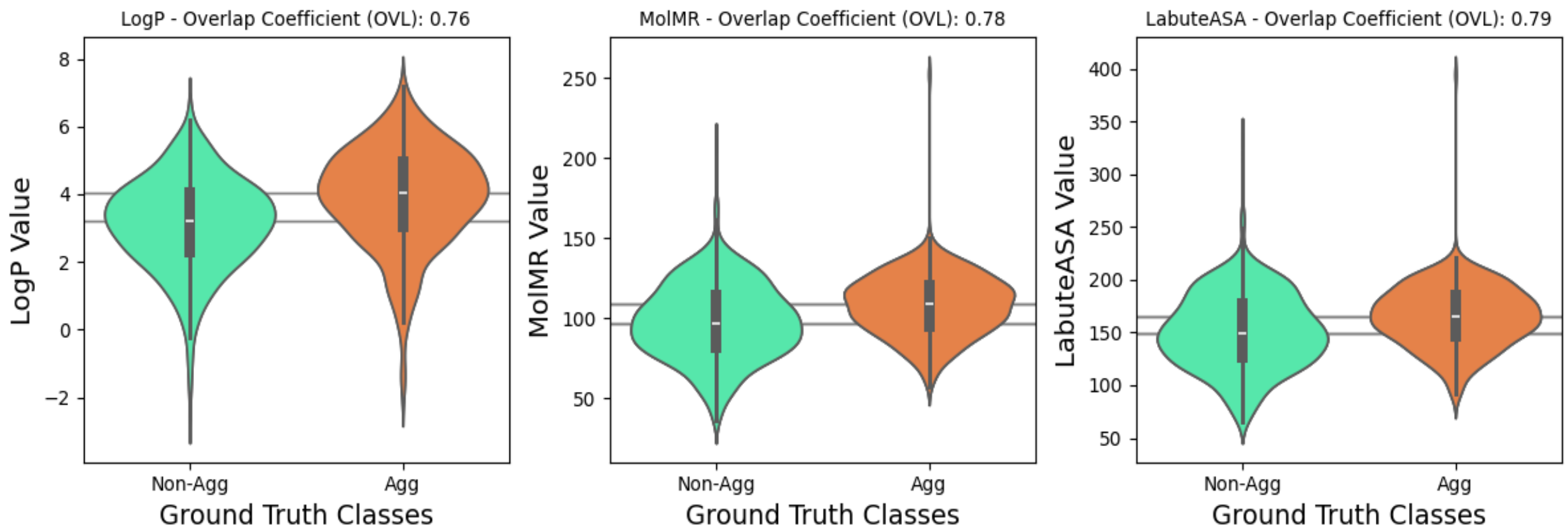

**Figure S2.** Class-separated distributions of three influential molecular descriptors for the 1000 elements of the test set. Non-Aggregator distributions are shown in green and Aggregator in orange. The three molecular descriptors LogP, MolMR and LabuteASA (left to right) have previously been identified as especially influential during a feature importance analysis. For each molecular descriptor, the overlap coefficient provides a measure of overlap between the two classe's distributions where higher values indicate lower separability.

In this context, it is important to emphasize that while simple molecular descriptors certainly provide some foundation for the model's explanations, they only contribute a small fraction to the overall class separability.

In contrast, as illustrated in Figure S3, the distribution of the MEGAN model's output logits shows a much higher capability of separating the two ground truth classes with a distributional overlap coefficient of only 0.34.

[1] We find TPSA to have little capability in distinguishing between aggregators and non-aggregators, hence we did not include it in this discussion.

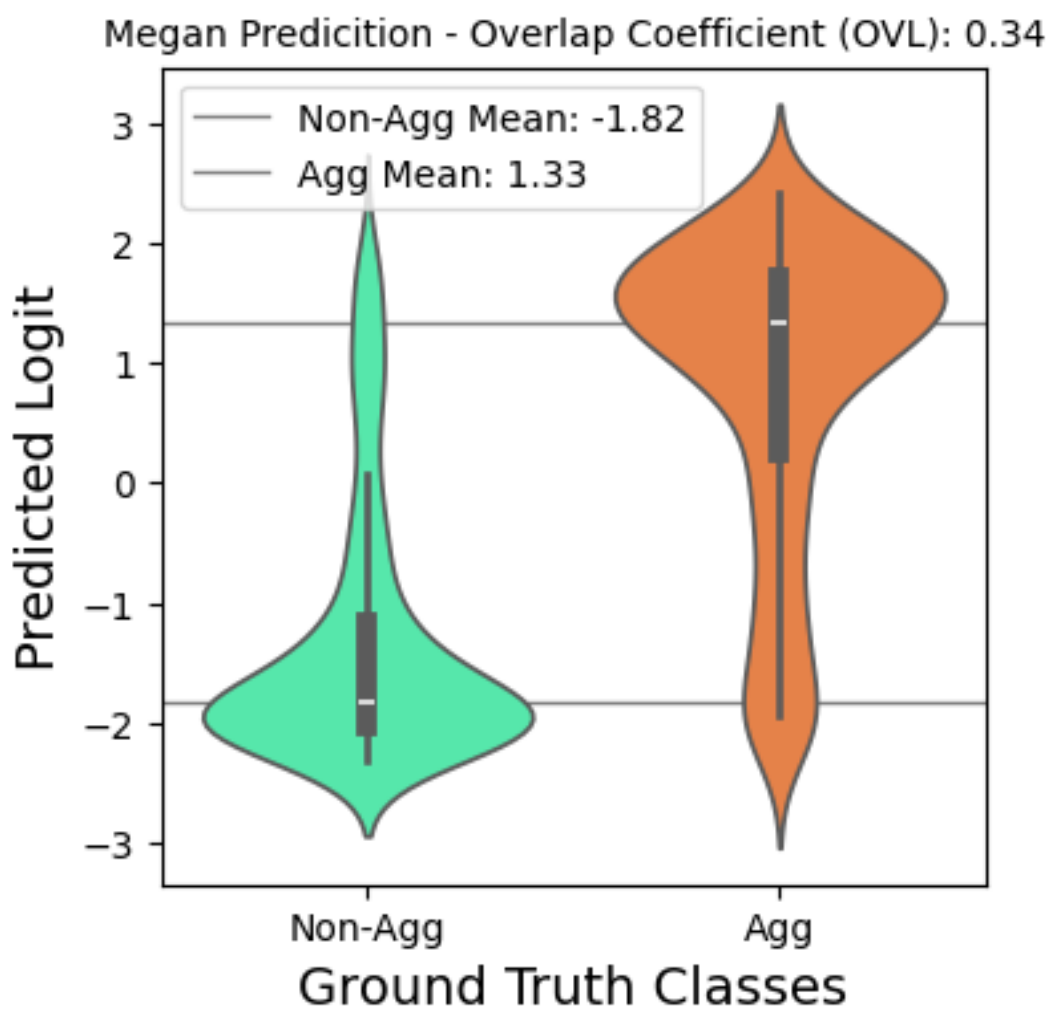

**Figure S3.** Class-separated distributions of the of MEGAN's prediction logits for the 1000 elements of the test set. Non-Aggregator distributions are shown in green and Aggregator in orange. The overlap coefficient provides a measure of overlap between the two classe's distributions where higher values indicate lower separability.

In light of this presumption, we have conducted a quantitative analysis to compare the MEGAN model's atom-contribution-based explanation masks with the decomposed atom contributions available for the LogP, MolMR, and LabuteASA properties. For each molecule in the test set, we compute the Spearman correlation coefficient between the explanation mask and the atom contributions. Figure S4 shows the average over these individual correlation coefficients for both of the model's explanation channels, respectively. We find that the "non-aggregator" explanations are anticorrelated with atomic logP contributions and that the "aggregator" explanations are correlated with the MolMR contributions. This means that the parts of molecules indicated by the model as non-aggregating have, on average, lower logP scores and thus higher water-solubilities, which makes intuitive sense. It also means that the parts of molecules indicated by the model as aggregating contribute more to the molecular polarizability, potentially related to aromatic systems, which have high polarizability values and at the same time promote stacking and aggregation. In terms of the LabuteASA contributions, we find both "aggregator" and "non-aggregator" explanations to be inversely correlated, albeit on different populations of molecules.

Ultimately, for this analysis, it is again important to emphasize that these are small correlation values with rather large standard deviations, indicative of relatively small effects. However, in continuation of the previous argument, no individual simple descriptor can be expected to fully explain aggregation considering their relatively small individual contributions to class separability.

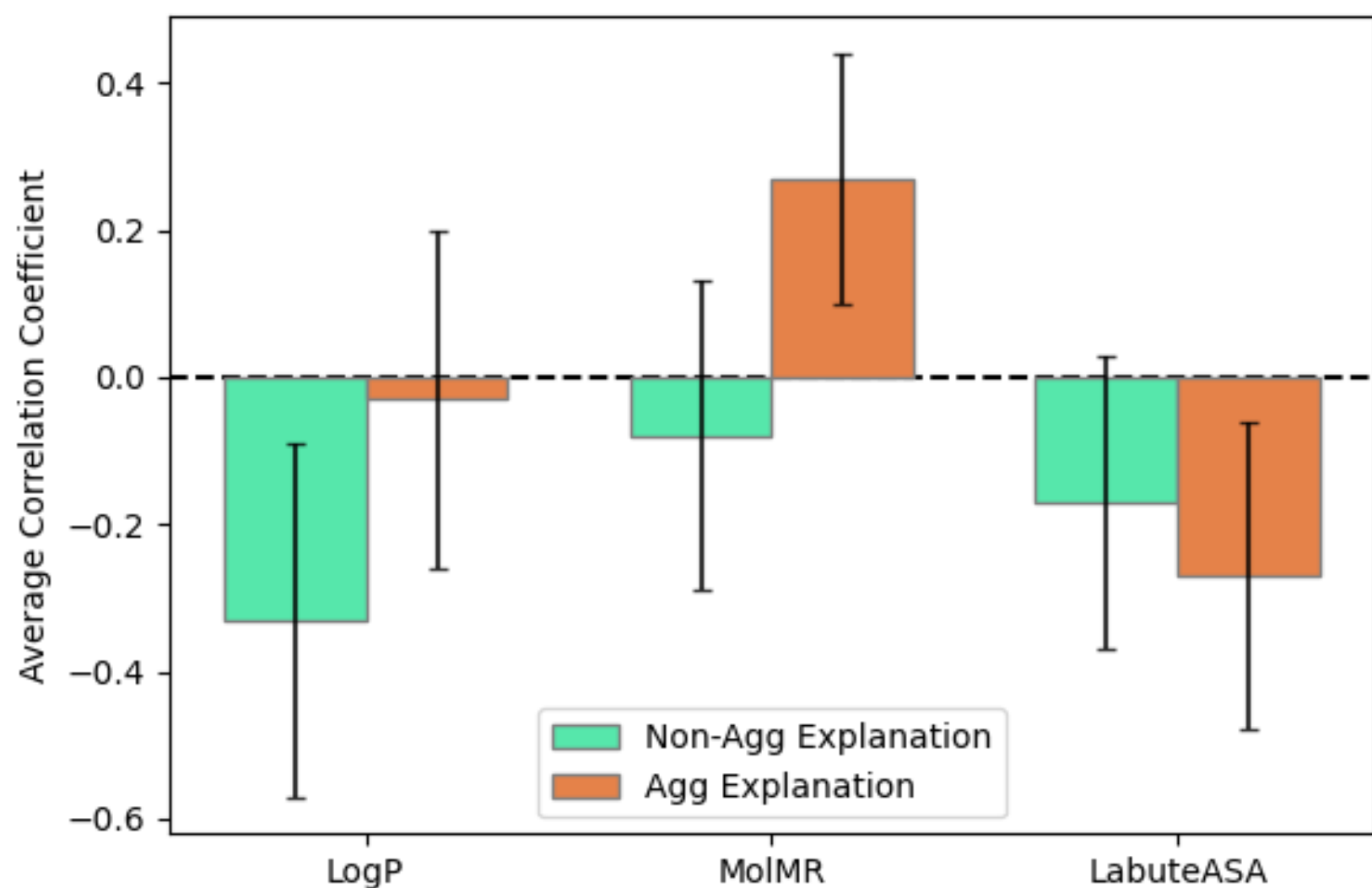

**Figure S4.** Distribution of spearman correlation coefficient of MEGAN's explanation masks and the atom contributions of the three molecular descriptors LogP, MolMR and LabuteASA. Non-Aggregator distributions are shown in green and Aggregator in orange. The extent of bar plots indicates the average spearman coefficient over the 1000 elements of the test set and error bars indicate the corresponding standard deviation.

**SIX. Automated generation of explanations using large language models**

To analyze whether the explanations generated based on the MEGAN model can be interpreted in terms of structural and physicochemical properties in an automated way, we experimented with large language models, specifically GPT-4o. The goal was to see whether the graph explanations can be automatically related to more broad chemical concepts and human-understandable explanations, in order to stimulate further ideas for more detailed analysis by experts and to potentially derive design rules in an automated way.

Prompts were created with the following template, which was adapted from OpenAI's examples:

> "[SYSTEM PROMPT]: You are a chemistry expert with the task of proposing possible hypotheses about the underlying structure-property relationships of molecular properties. You will be presented with some empirical evidence, which links a molecular fragment to a certain statistical impact on a given molecular property. You will create a hypothesis about the underlying physical and chemical mechanism that can explain why the given structure may have the observed effect.
>
> The property in question is the following: Molecular Aggregation
>
> Your answer should follow the structure below:
>
> Detailed Explanation: [Elaboration of the causal reasoning for the suggested substructure-property relationship]
>
> Hypothesis: [One sentence describing the structure and the linked property. Two sentences about the hypothesized causal explanation.]
>
> [USER PROMPT]: The structure is represented by the following scaffold in SMILES representation: {{ SMILES }}. This structure has been linked to {{ non-aggregating | aggregating }} behavior."

In contrast to previous large language models such as GPT-3.5, the GPT-4o model in most cases correctly interpreted the SMILES code and translated it to a description of the chemical structures as well as substructures contained in the explanatory motif. However, in most cases the connection to physicochemical properties did not reveal informative insights. In most cases where non-aggregating motifs were queried, the answer of the GPT-4o model referred to functional groups that enhanced hydrogen bonding with water molecules, and steric hindrance of π-stacking, which are potentially correct but also rather obvious explanations. In case of aggregating motifs, the answers frequently included functional groups that enhanced hydrogen bonding to form molecular networks, as well as planar structures that promote π-stacking. Few explanations provided further insight that might be validated and quantified in further experiments, e.g. electron-donating effects which can lead to increased electron density in the π system of the benzene ring, potentially facilitating better solvent interactions over intermolecular π-π stacking interactions.

In further tests, we used GPT-4o to analyse and summarize common structural and physicochemical characteristics in all explanation concepts, with and without revealing that the explained property is molecular aggregation. The detailed results including the prompts can be found below. One of the obtained summaries was the following: "Positive influence [on aggregation]: High conjugation, electron-donating groups, structural rigidity, hydrophobicity, and bulky groups (especially with sulfur and nitrogen heterocycles). Negative influence [on aggregation]: Electron-withdrawing groups, polar substituents (carbonyl, amides), smaller or flexible structures, and reduced conjugation." While being highly aggregated and thus not very specific, those outputs can be used for fully automated feature engineering to train simpler subsymbolic models or even derive analytical models using methods such as symbolic regression.

Overall, we observe a drastic improvement of the usefulness of large language models in automatically interpreting results of explainable AI methods (from GPT-3.5 to GPT-4o), revealing basic insights into structure-property relations governing molecular aggregation. However, more specific insights that inspire more detailed analysis approaches or even lead to immediate understanding are still lacking.[26]

**Detailed prompts:**

> We trained an explainable AI methods based on graph neural networks to predict a molecular property, called PROP. We then analysed the model's explanations to identify relevant molecular motifs that increase and decrease property PROP.
>
> The following motifs increase PROP: [LIST OF SMILES CODES OF ALL MOTIFS]
> The following motifs increase PROP: [LIST OF SMILES CODES OF ALL MOTIFS]

## SX. Dataset structural feature and property analysis

Numerous molecular descriptors were explored to evaluate the differences in structural features and properties of the compounds in the aggregating and non-aggregating datasets. Prior to the calculation of molecular descriptors, each compound from the cleaned datasets of aggregating and non-aggregating molecules was protonated to its most energetically favourable ionization state at physiological pH (pH=7.4) using the FixpKa functionality of OpenEye's QUACPAC 2.1.2.1.[45] Using RDKit, 111 molecular descriptors were then calculated for both the aggregator and non-aggregator compounds (full list of descriptors provided in section SXI of the Supporting Information). Of the 111 descriptors, 85 were fragment descriptors which indicate the number of occurrences of specific fragments (e.g., halogens, carboxylic acids, nitro groups) in a given molecule. The remaining 26 descriptors were for general features (e.g., molecular weight, logP, ring count, number of hydrogen bond donors) of the molecules.

Histograms of each feature were plotted for both the aggregator and non-aggregator molecules to assess the distribution of the data (all histograms presented in sections SXII-SXIV of the Supporting Information). Nearly all 85 fragment descriptors, as well as 10 of the 26 general descriptors, showed a non-normal distribution. As such, the fragment descriptors were analyzed separately from the general descriptors, and only the 16 general descriptors following a normal distribution were considered for further analysis.

To assess to statistical significance of each of the 16 general descriptors for the aggregating and non-aggregating molecules, p-values were calculated for each descriptor from a standard independent two sample t-test assuming equal population variances (calculated using scipy.stats.ttest_ind).[46] With the exception of the NumRotatableBonds descriptor, all p-values were determined to be less than 0.005, indicating a high statistical significance for these descriptors. As such, the effect size was computed for the 15 statistically significant descriptors as the magnitude of the Cohen's d value.

$$Cohen's\ d = \frac{\mu_1 - \mu_2}{\sqrt{\frac{1}{2}((\sigma_1)^2 + (\sigma_2)^2)}} \qquad \text{(Eq. S3)}$$

Where $\mu_1$ and $\mu_2$ are the means of the descriptor values for aggregators and non-aggregators, and $\sigma_1$ and $\sigma_2$ are the standard deviations. The p-values, means, standard deviations and Cohen's d values for the 15 statistically significant general molecular descriptors are presented in section SXV of the Supporting Information.

Out of the 16 normally distributed general molecular descriptors, 15 were calculated to have p-values of less than 0.005, indicating that they are highly statistically significant. For these 15 descriptors, the effect size was calculated in order to look for meaningful relationships between the properties of the molecules in the aggregating and non-aggregating datasets. As a measure of the difference in the mean of each descriptor of the two data sets, Cohen's d values were calculated (Figure S5). The generally accepted interpretation of Cohen's d values is that a value of less than 0.2 is a small effect size, values near 0.5 are a medium effect size, and values greater than 0.8 are a large effect size.[47] Cohen's d values indicate that the logP has a medium to large effect size. Further analysis of the histogram for this feature (Figure S6a) indicates that the molecules in the aggregator data set have a higher logP, on average, than those in the non-aggregator data set. The number of aromatic rings (both NumAromaticCarbocycles and NumAromaticRings) and the fraction of sp3 hybridized carbon atoms (FractionCSP3) are found to have medium effect sizes, with the aggregators having higher values on average for the number of aromatic rings and lower values on average for the fraction of sp3 hybridized carbons. The observation that the aggregators have a smaller fraction of sp3 hybridized carbons is consistent with the greater aromaticity observed in the aggregators dataset. Consistent with many other reports on aggregation features, our data indicates that both logP and aromaticity are linked to aggregation. Alternatively, other features proposed in the literature to be important to small molecule aggregation (i.e. number of sulfur atoms and number of hydroxyl groups) were not found to vary significantly between the molecules in the aggregating and non-aggregating datasets (Figure S6d).[11]

Analysis of the fragment descriptors through comparison of the relative frequency of each of the fragments between the aggregator and non-aggregator molecules revealed that the fragments provide little insight into the difference in the chemical nature of the aggregator and non-aggregator molecules. Plots of the relative frequency of each of the fragment descriptors are presented in section SXVI of the Supporting Information.

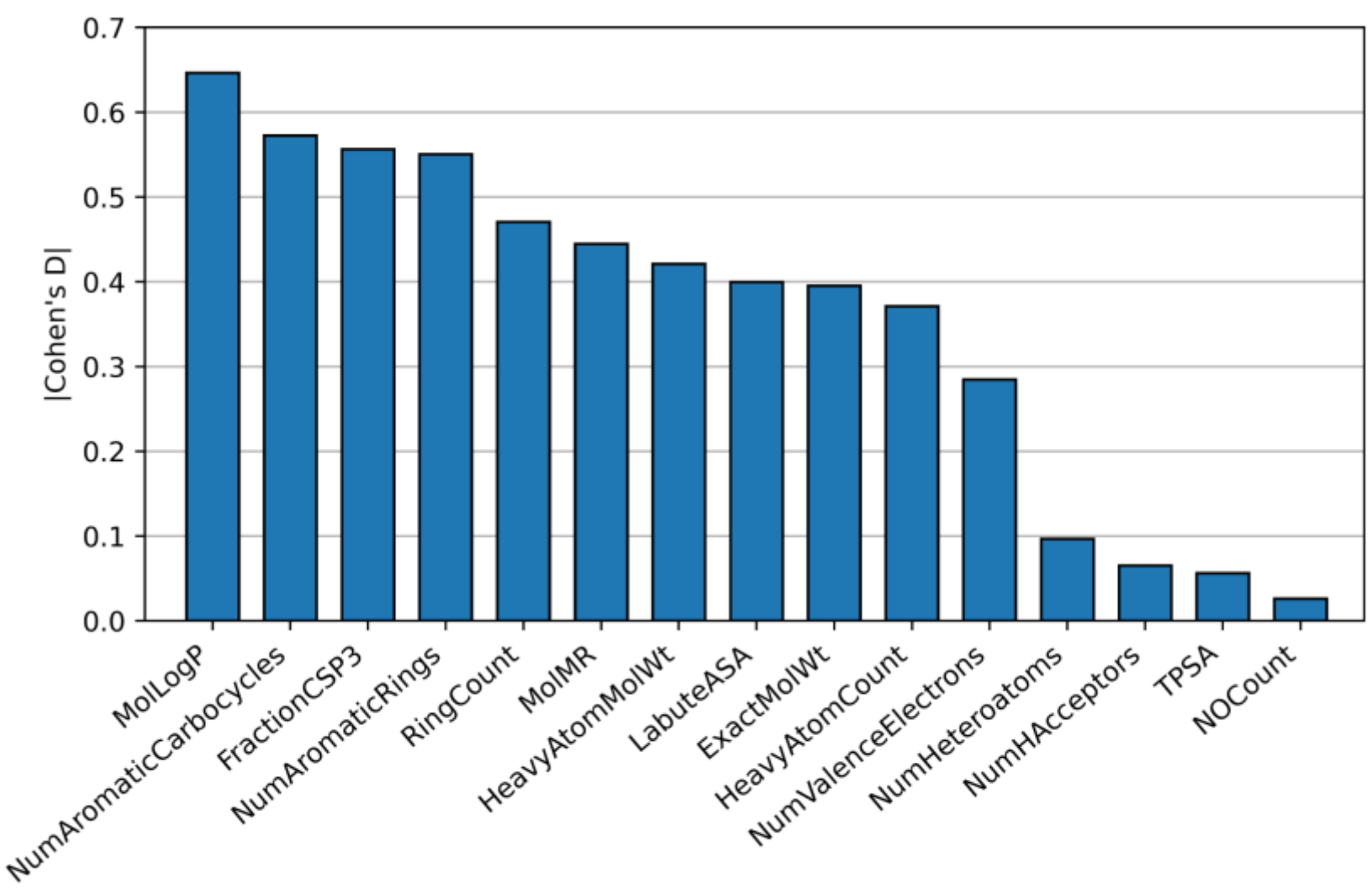

**Figure S5.** The absolute value of the Cohen's d for the 15 statistically significant general molecular descriptors.

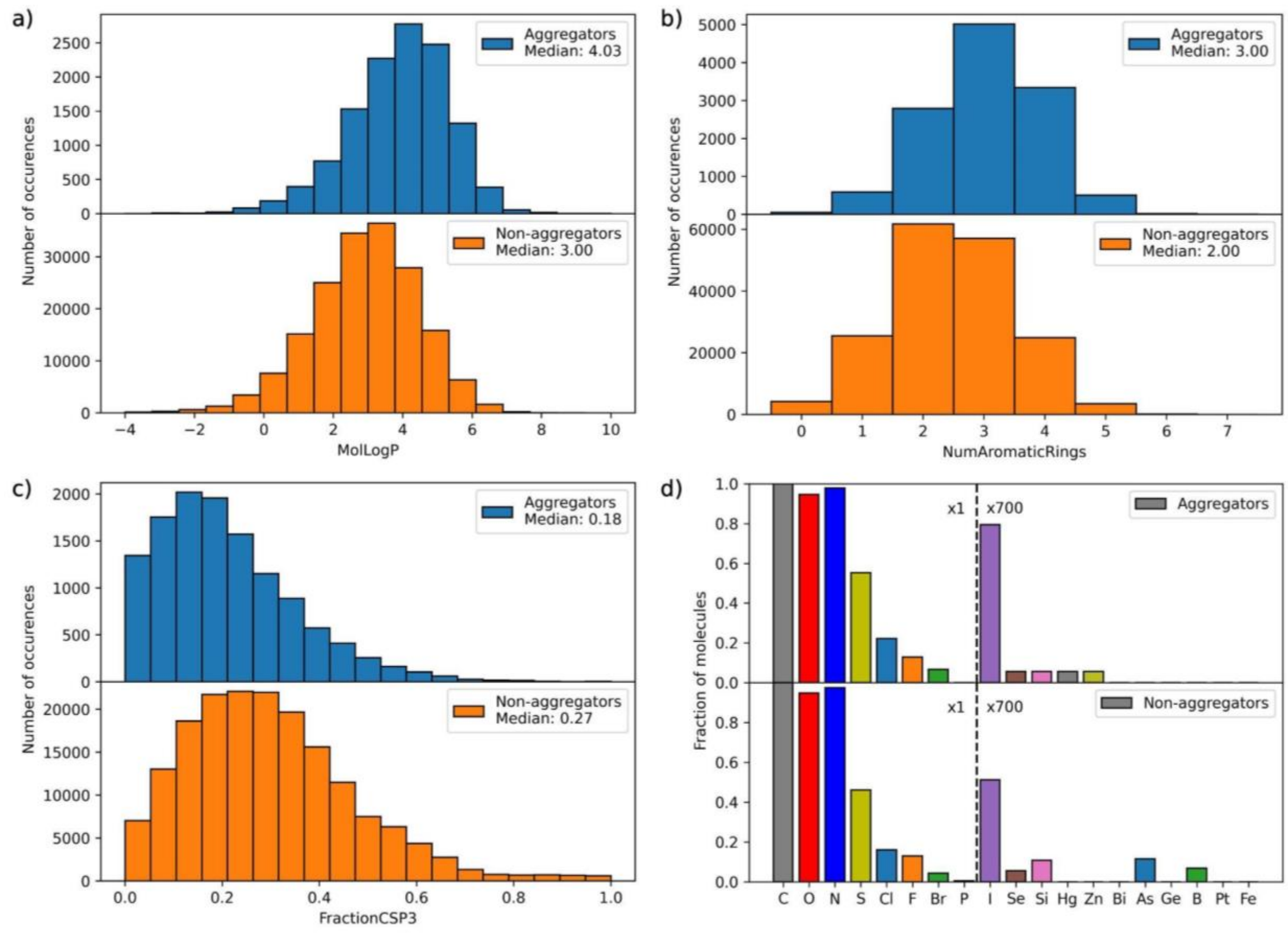

**Figure S6.** Histograms of a) MolLogP, b) Number of aromatic rings per molecule, c) Fraction of SP3 hybridized carbon, and d) element frequencies for the aggregator and non-aggregator datasets.

**SXI. List of all RDKit molecular descriptors calculated for all compounds in the aggregator and non-aggregator datasets**

ExactMolWt, FractionCSP3, HeavyAtomCount, HeavyAtomMolWt, LabuteASA, MolLogP, MolMR, NHOHCount, NOCount, NumAliphaticCarbocycles, NumAliphaticHeterocycles, NumAliphaticRings, NumAromaticCarbocycles, NumAromaticHeterocycles, NumAromaticRings, NumHAcceptors, NumHDonors, NumHeteroatoms, NumRadicalElectrons, NumRotatableBonds, NumSaturatedCarbocycles, NumSaturatedHeterocycles, NumSaturatedRings, NumValenceElectrons, RingCount, TPSA, fr_Al_COO, fr_Al_OH, fr_Al_OH_noTert, fr_ArN, fr_Ar_COO, fr_Ar_N, fr_Ar_NH, fr_Ar_OH, fr_COO, fr_COO2, fr_C_O, fr_C_O_noCOO, fr_C_S, fr_HOCCN, fr_Imine, fr_NH0, fr_NH1, fr_NH2, fr_N_O, fr_Ndealkylation1, fr_Ndealkylation2, fr_Nhpyrrole, fr_SH, fr_aldehyde, fr_alkyl_carbamate, fr_alkyl_halide, fr_allylic_oxid, fr_amide, fr_amidine, fr_aniline, fr_aryl_methyl, fr_azide, fr_azo, fr_barbitur, fr_benzene, fr_benzodiazepine, fr_bicyclic, fr_diazo, fr_dihydropyridine, fr_epoxide, fr_ester, fr_ether, fr_furan, fr_guanido, fr_halogen, fr_hdrzine, fr_hdrzone, fr_imidazole, fr_imide, fr_isocyan, fr_isothiocyan, fr_ketone, fr_ketone_Topliss, fr_lactam, fr_lactone, fr_methoxy, fr_morpholine, fr_nitrile, fr_nitro, fr_nitro_arom, fr_nitro_arom_nonortho, fr_nitroso, fr_oxazole, fr_oxime, fr_para_hydroxylation, fr_phenol, fr_phenol_noOrthoHbond, fr_phos_acid, fr_phos_ester, fr_piperdine, fr_piperzine, fr_priamide, fr_prisulfonamd, fr_pyridine, fr_quatN, fr_sulfide, fr_sulfonamd, fr_sulfone, fr_term_acetylene, fr_tetrazole, fr_thiazole, fr_thiocyan, fr_thiophene, fr_unbrch_alkane, fr_urea

**SXII. Histograms of normally distributed general molecular descriptors for the aggregator and non-aggregator datasets**

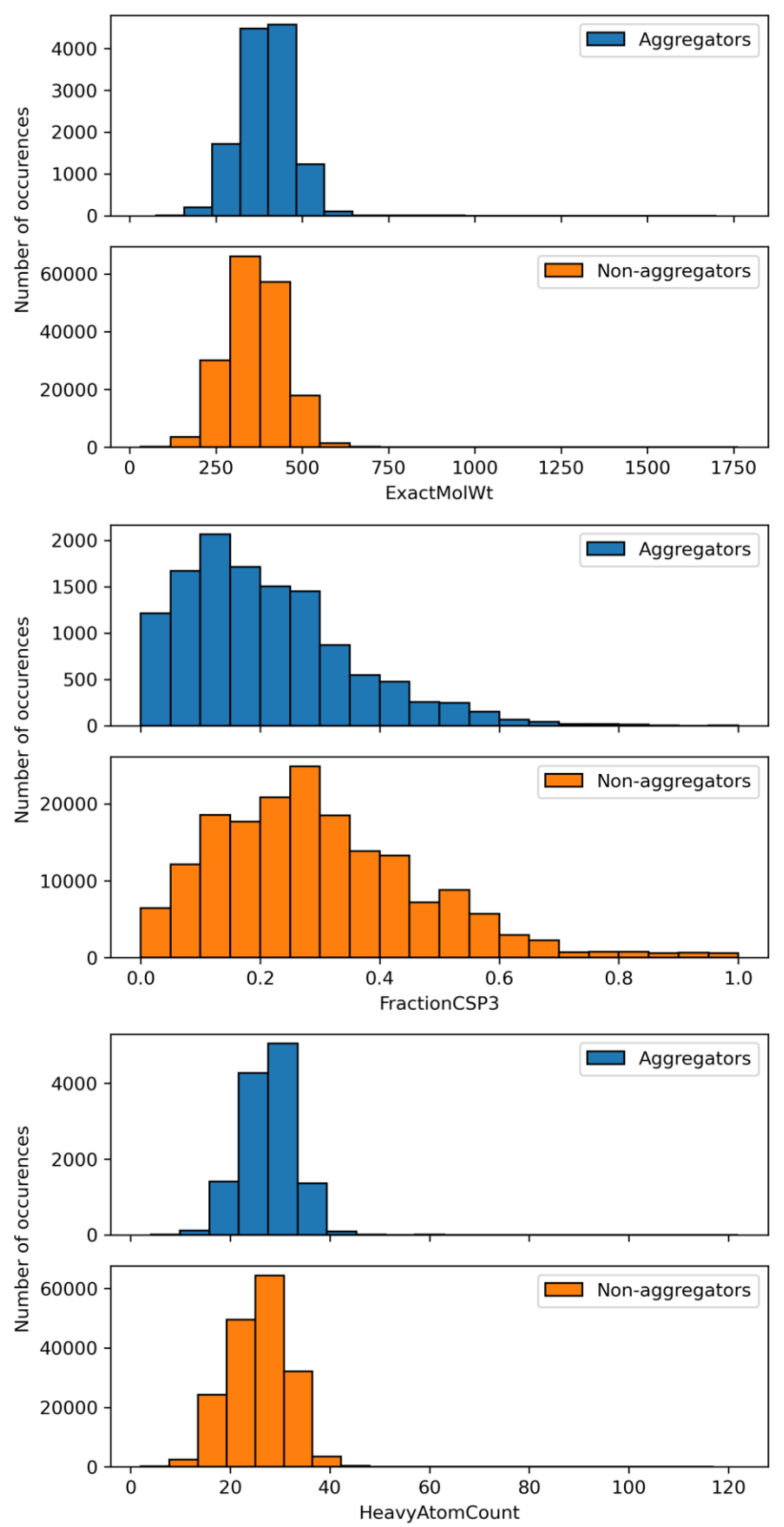

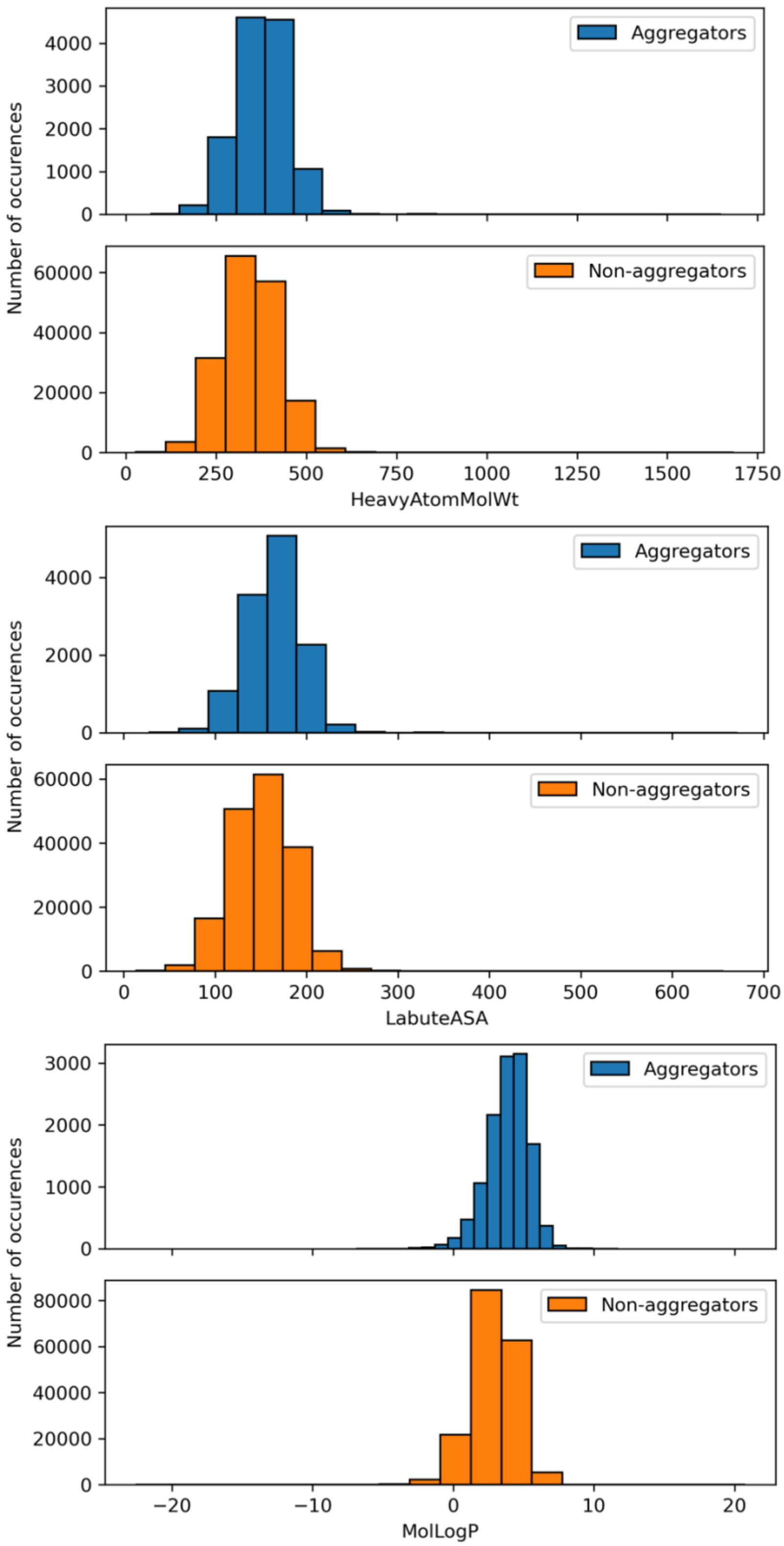

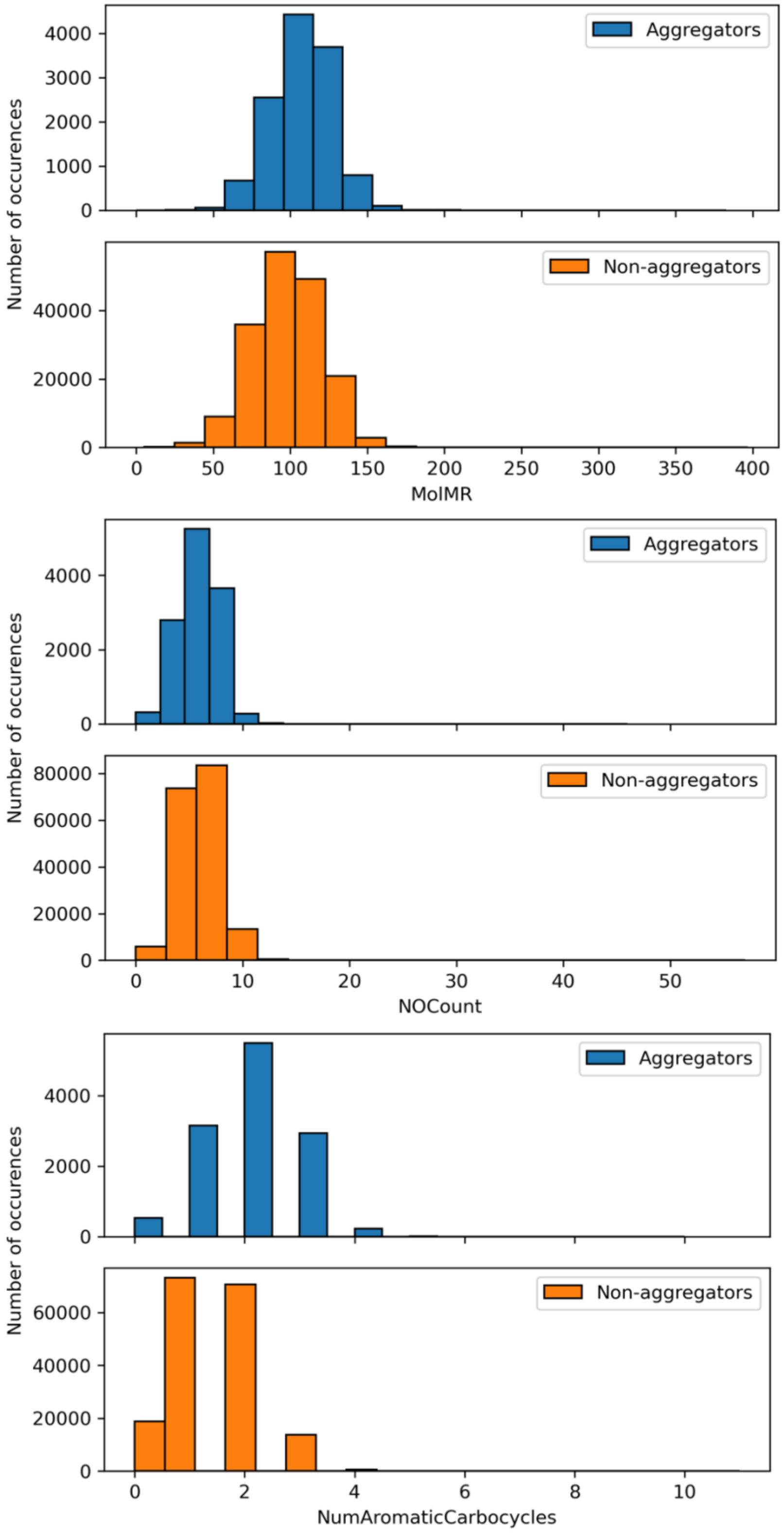

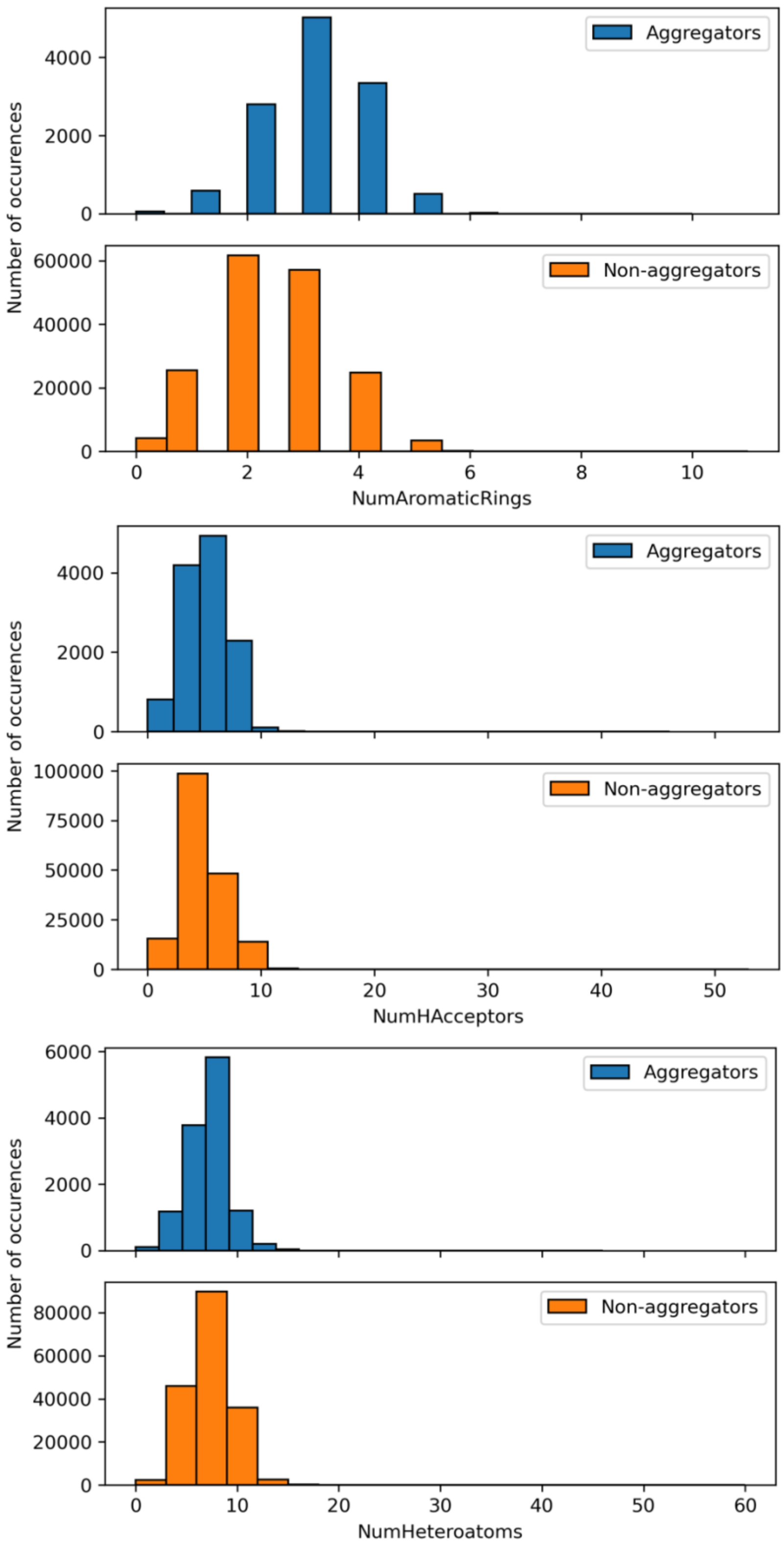

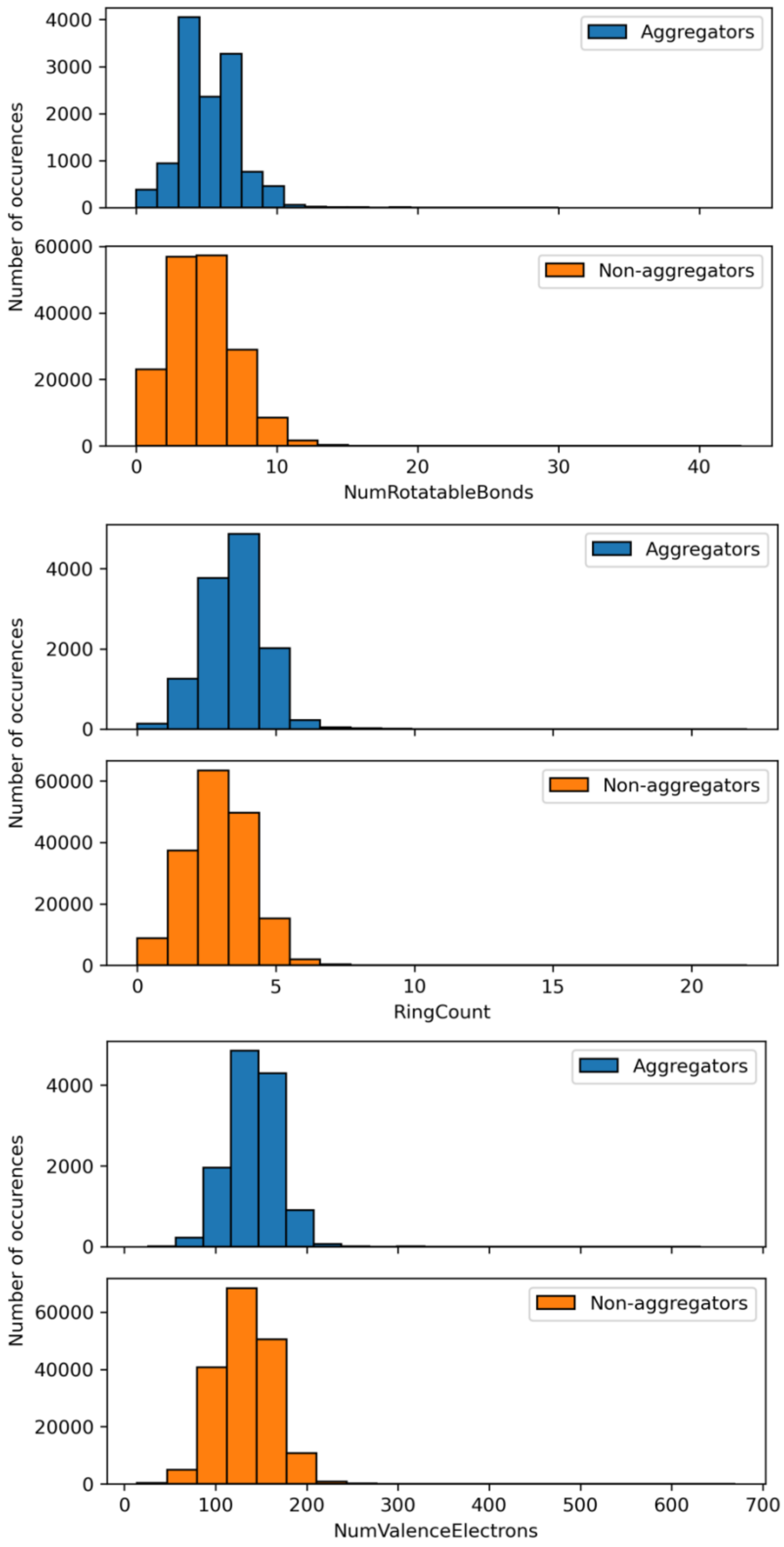

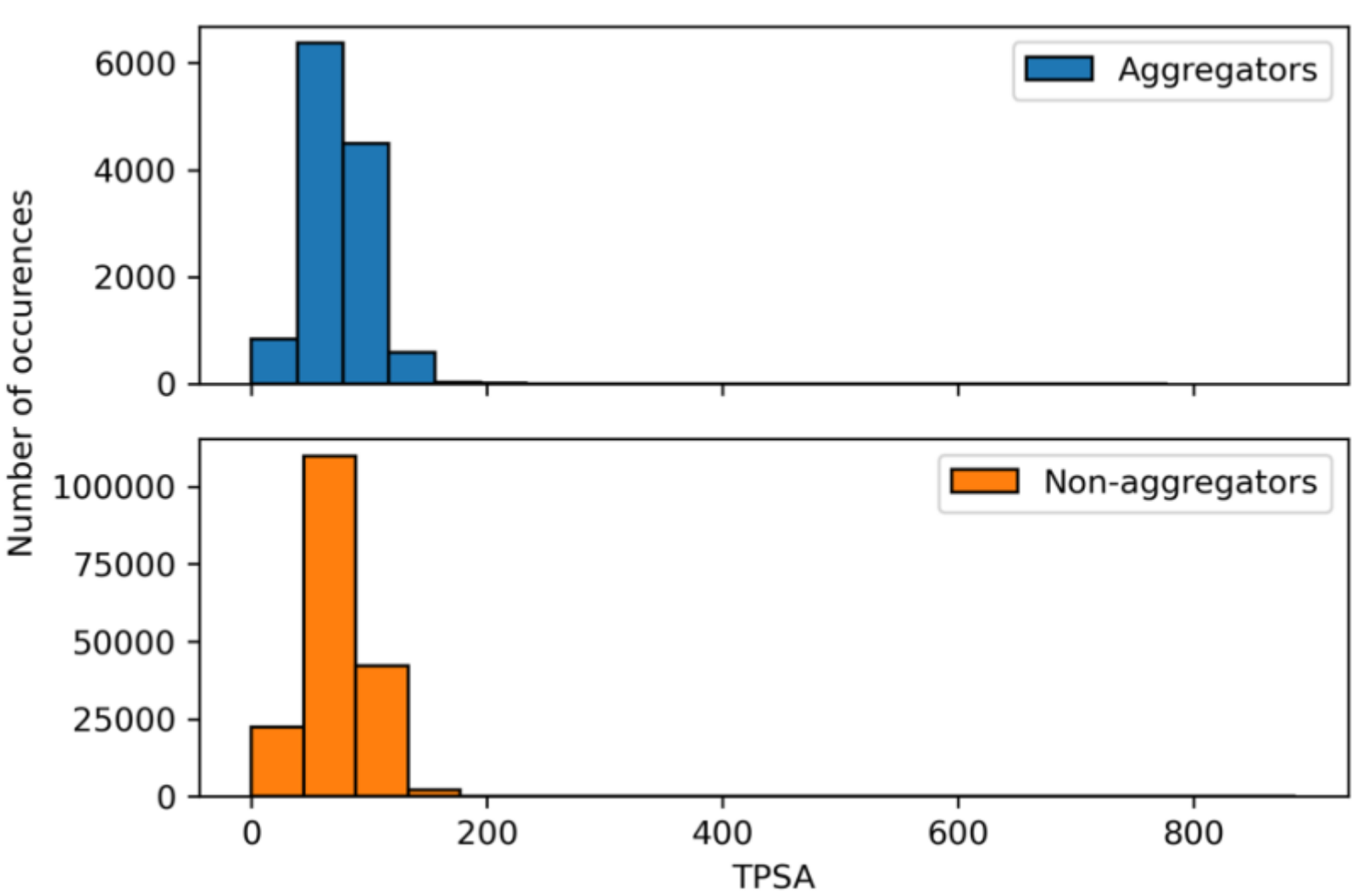

**SXIII. Histograms of non-normally distributed general molecular descriptors for the aggregator and non-aggregator datasets**

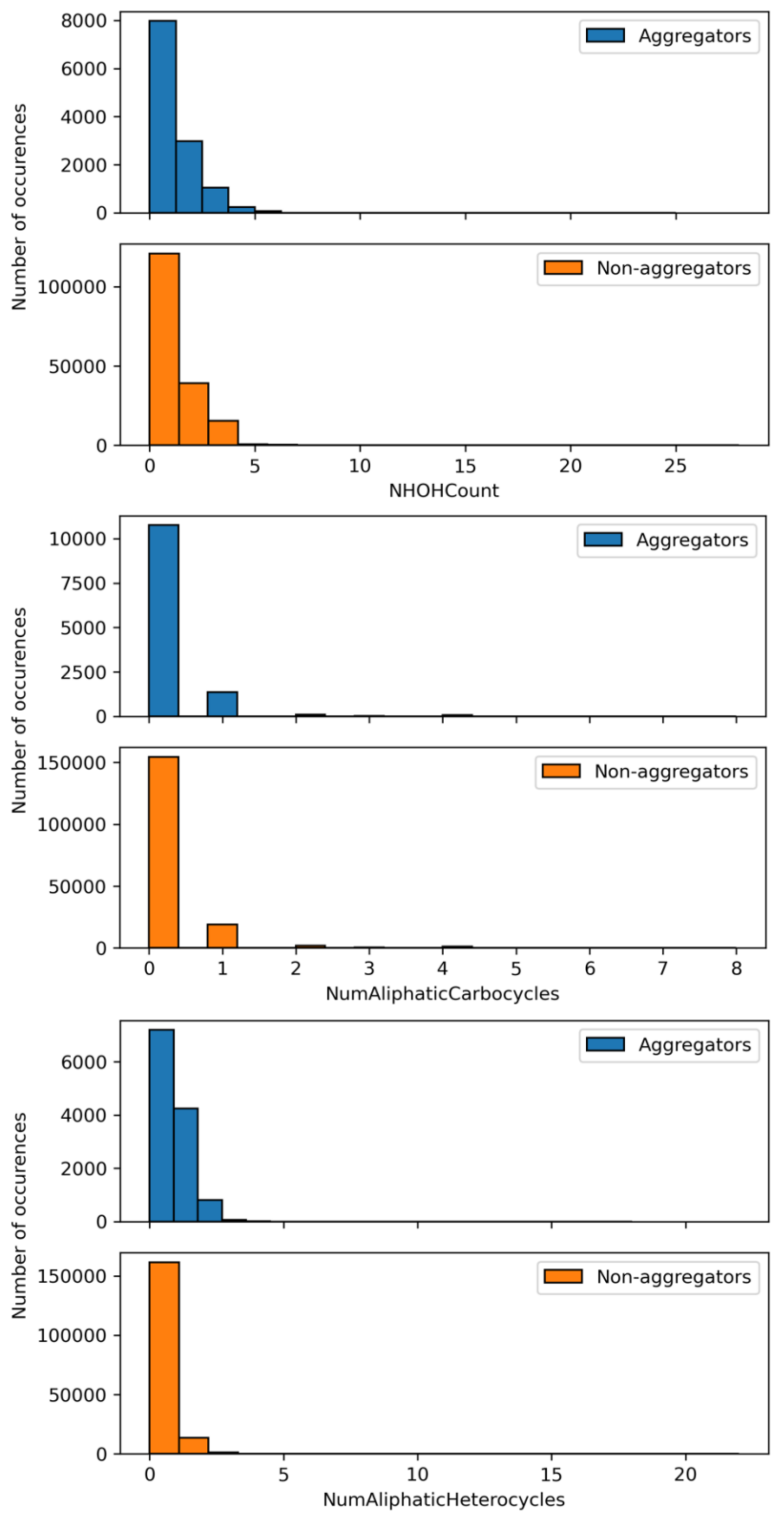

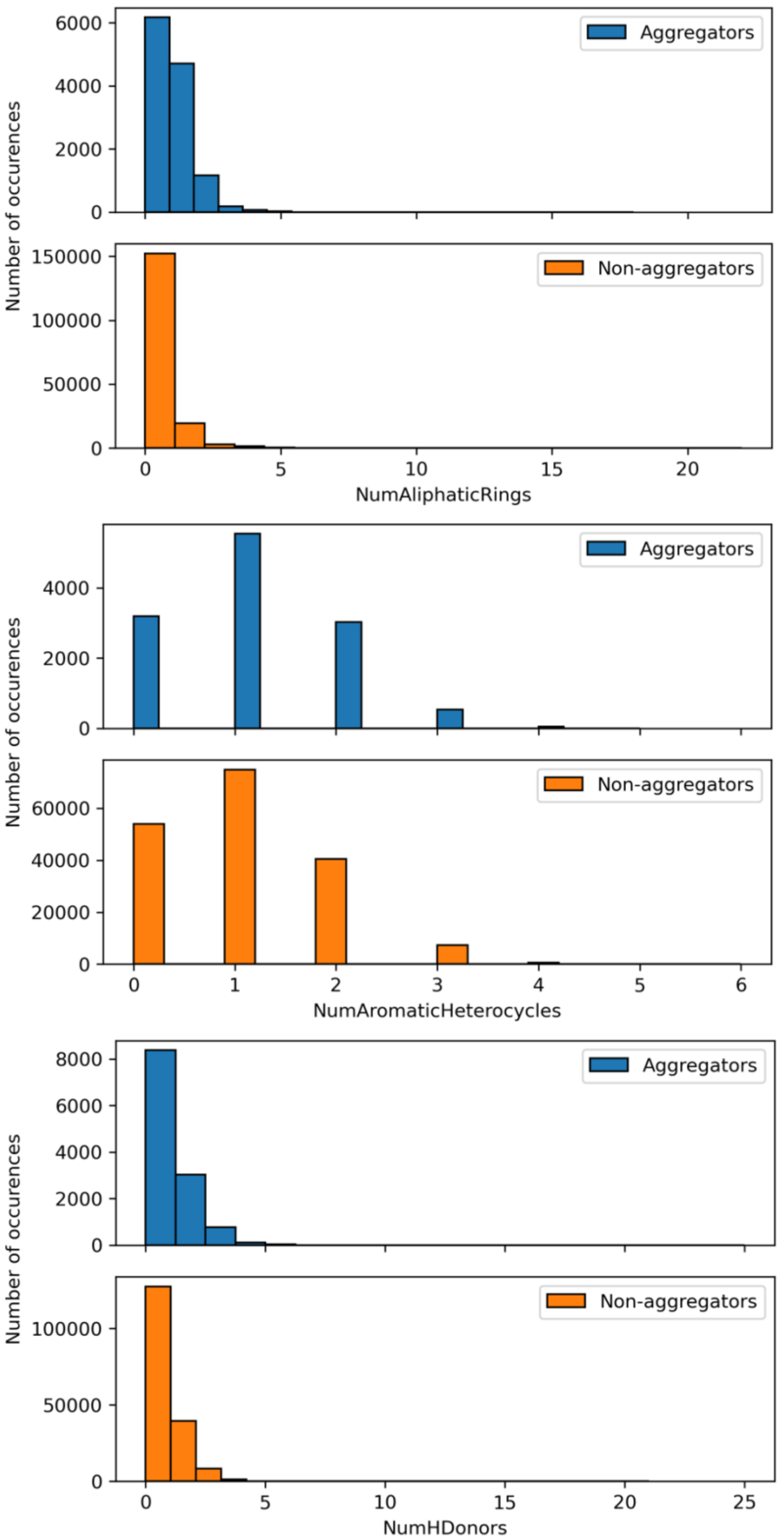

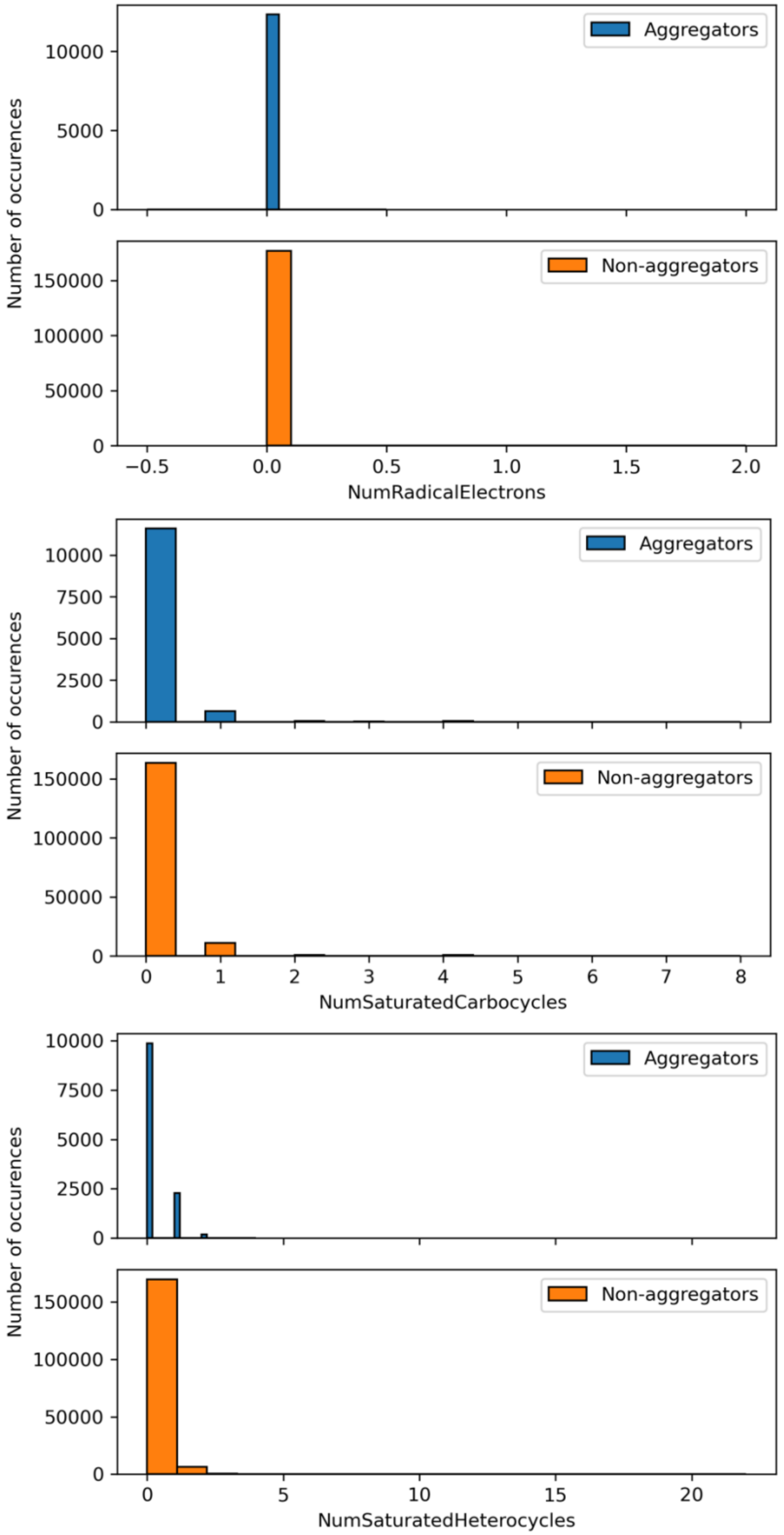

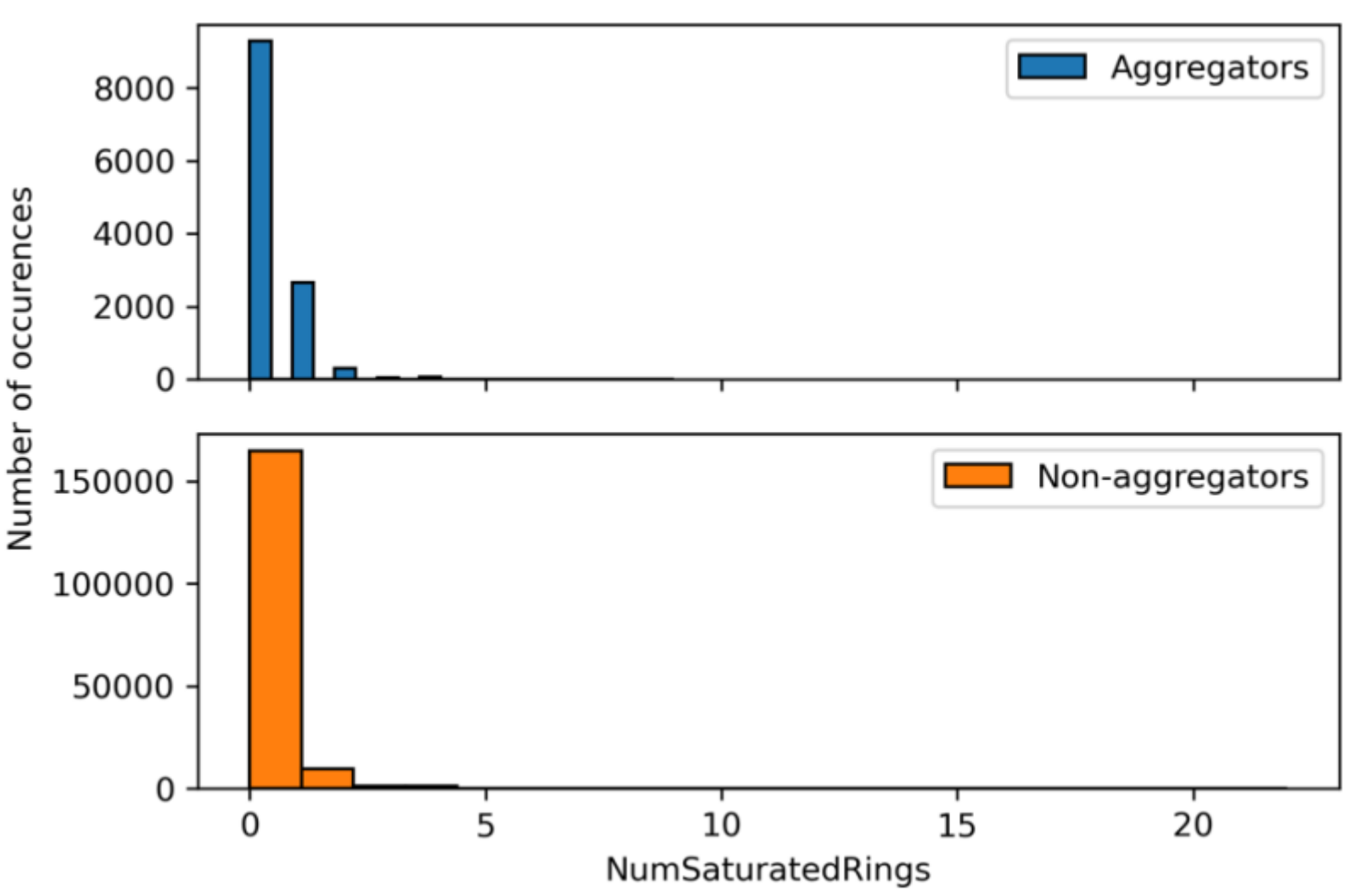

**SXIV. Histograms of fragment molecular descriptors for the aggregator and non-aggregator datasets**

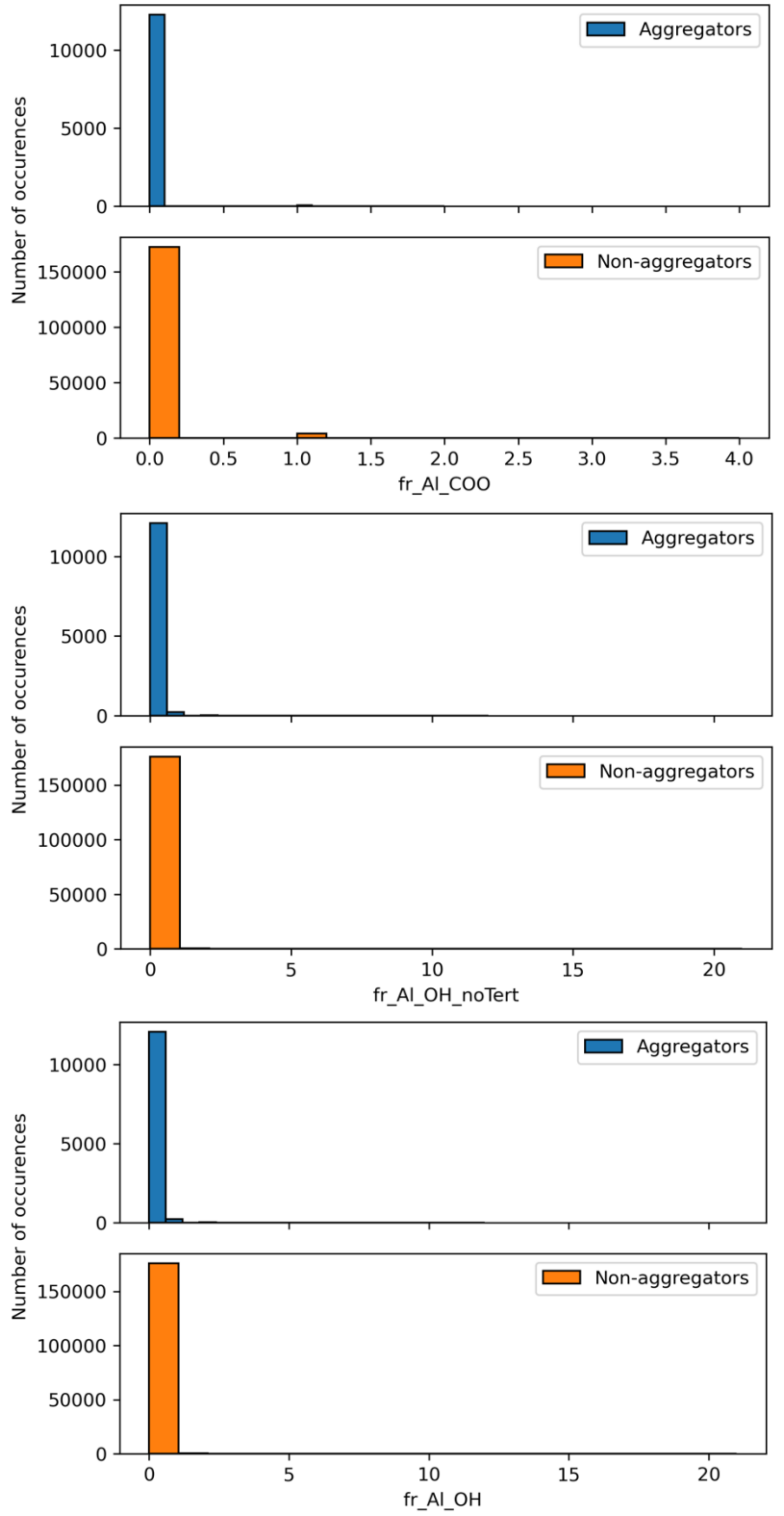

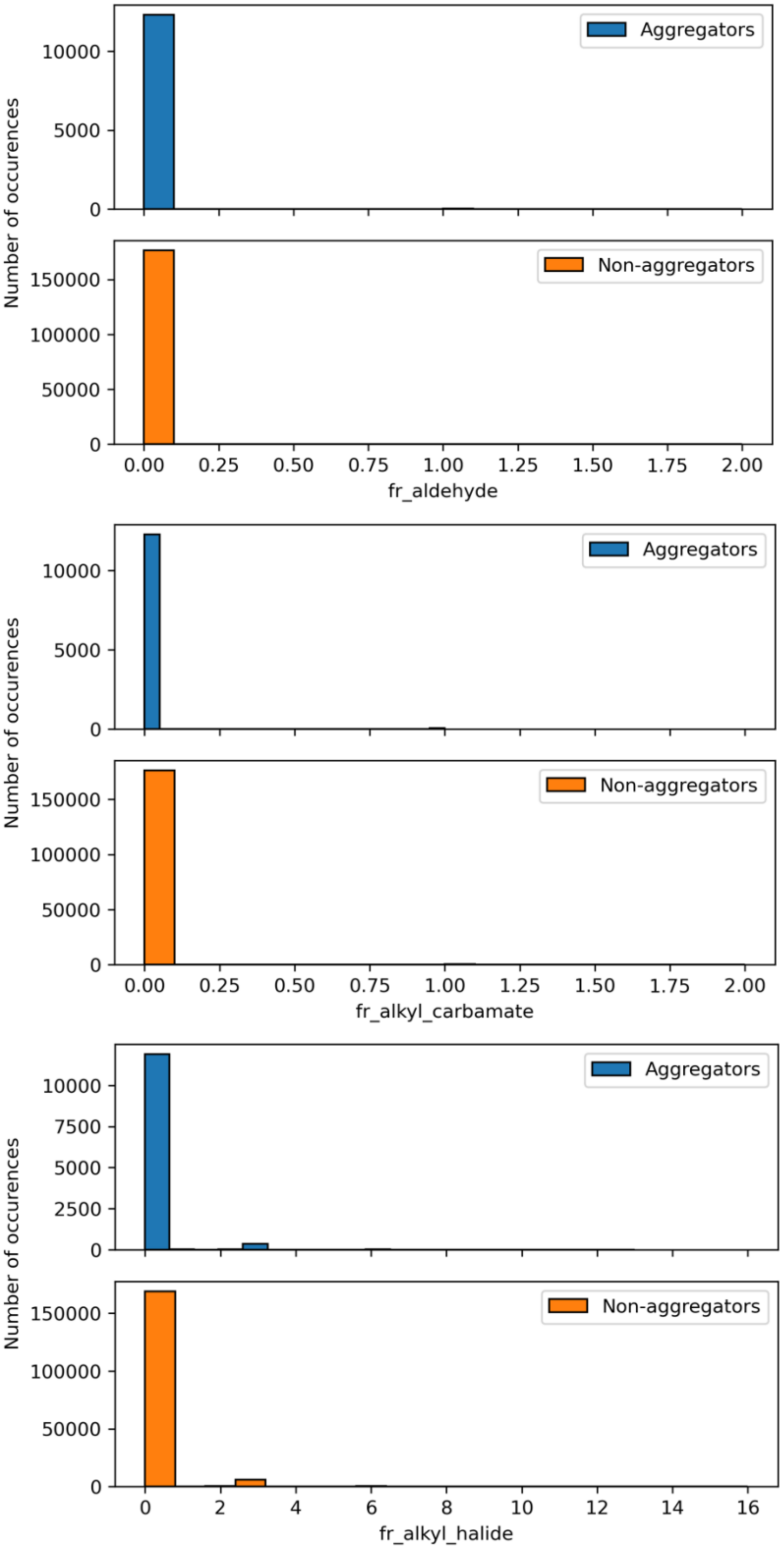

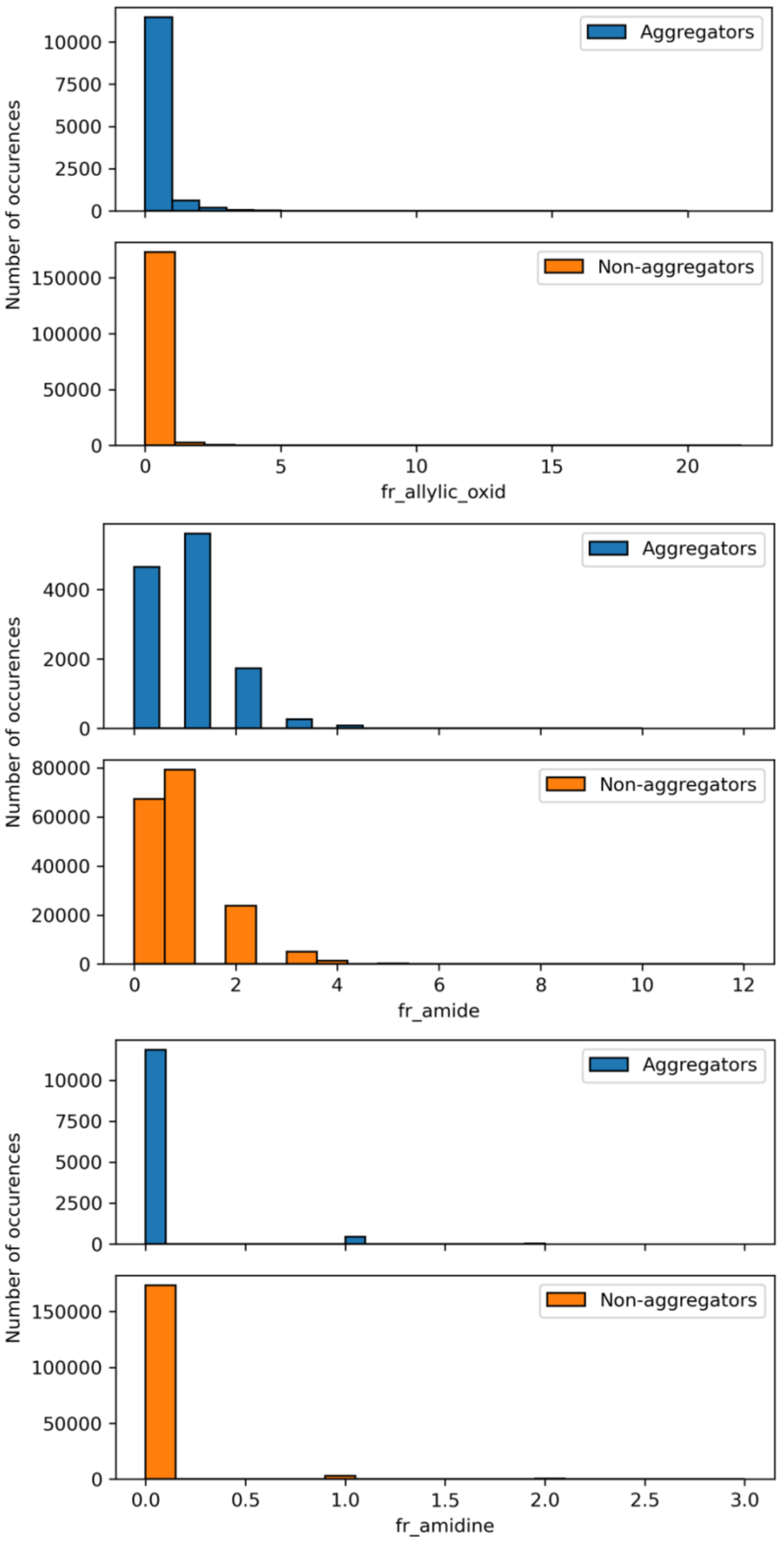

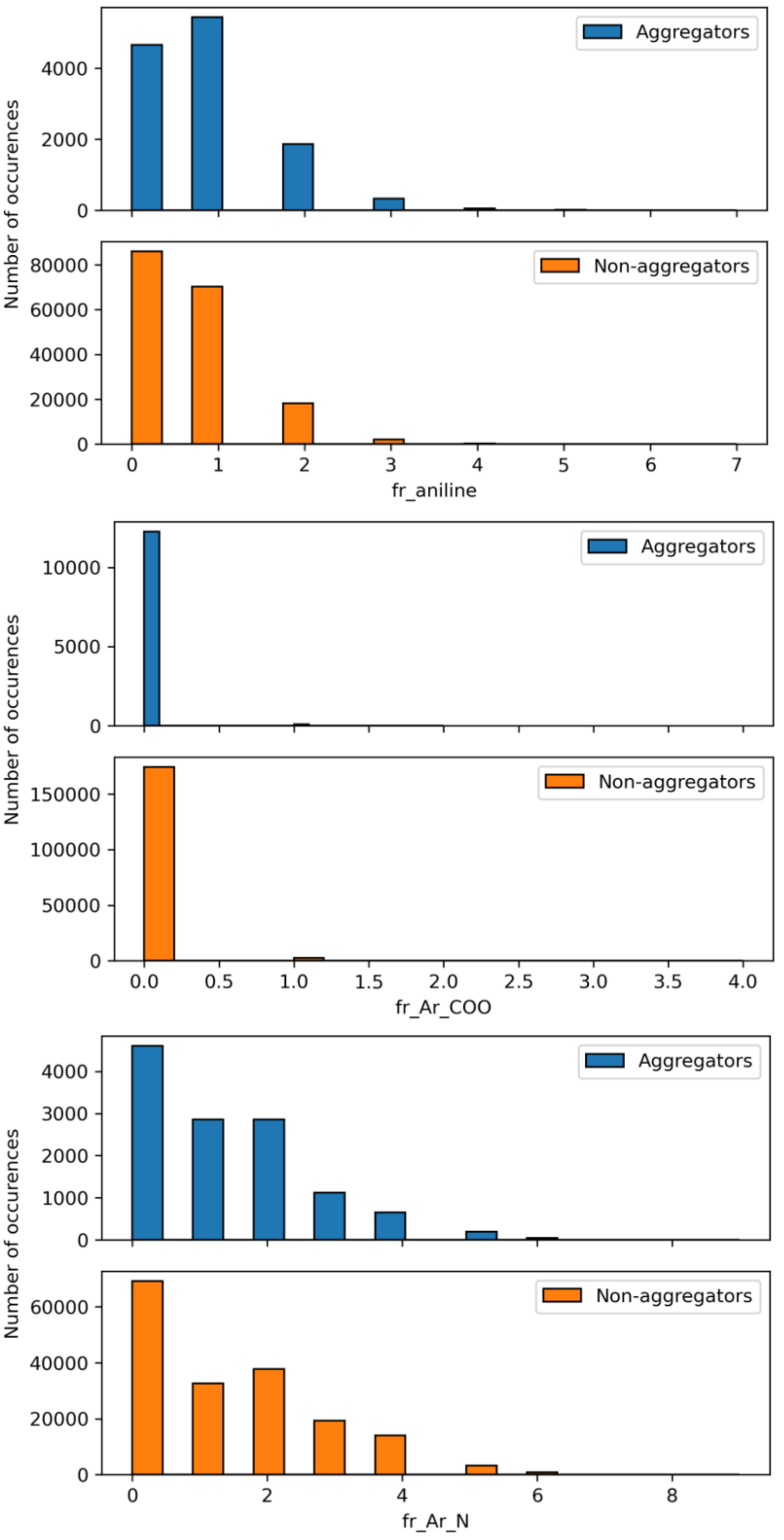

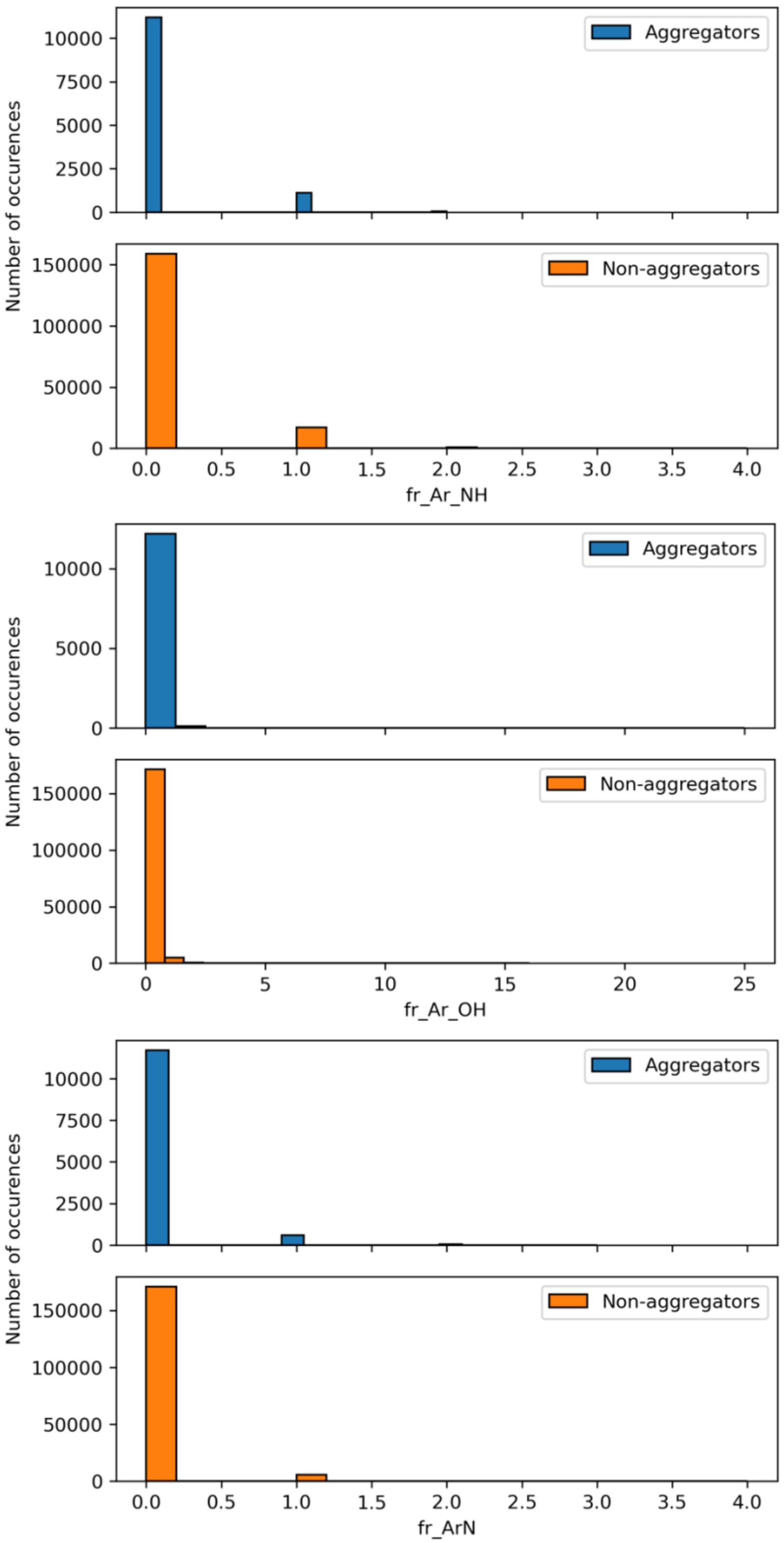

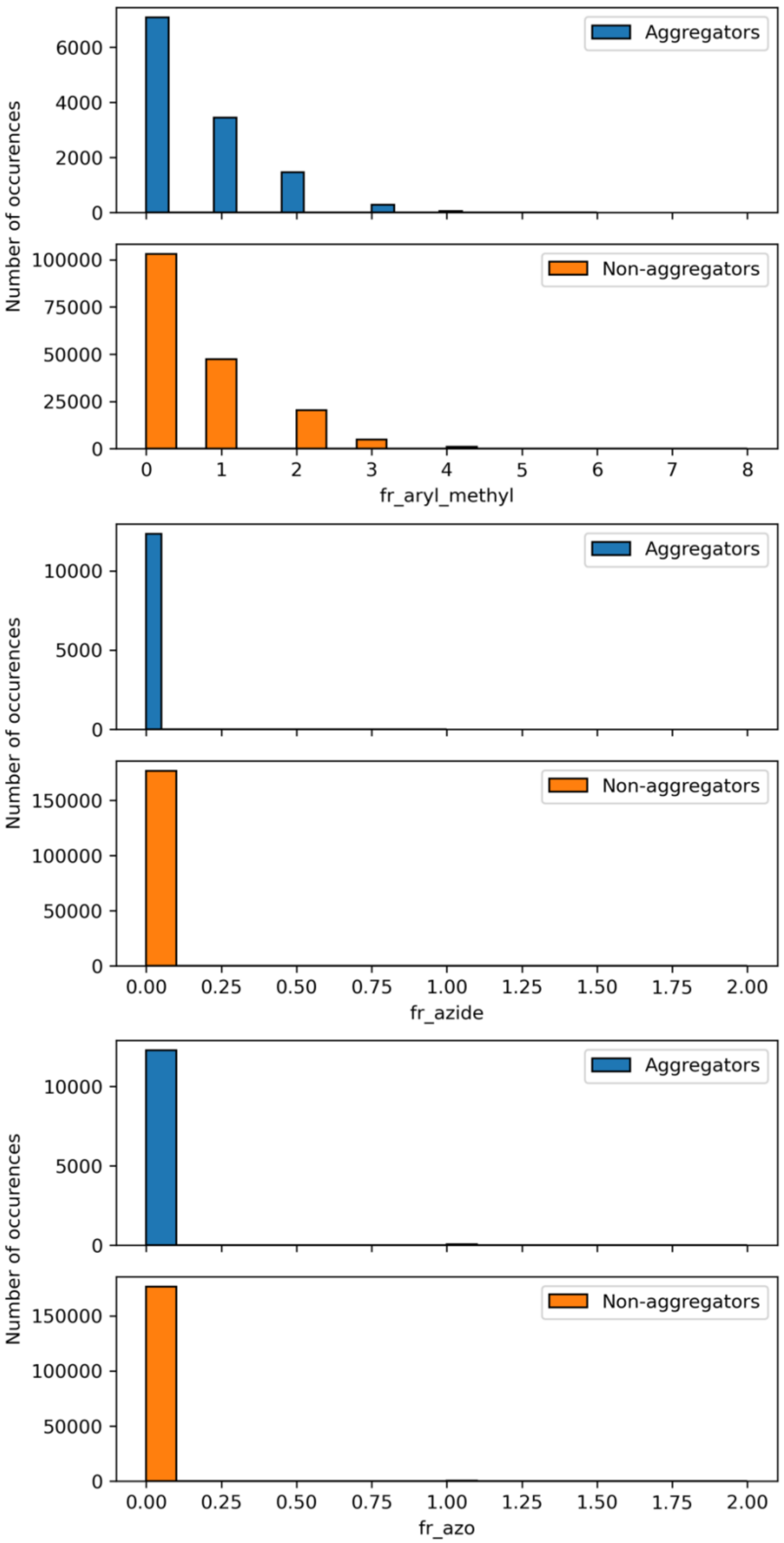

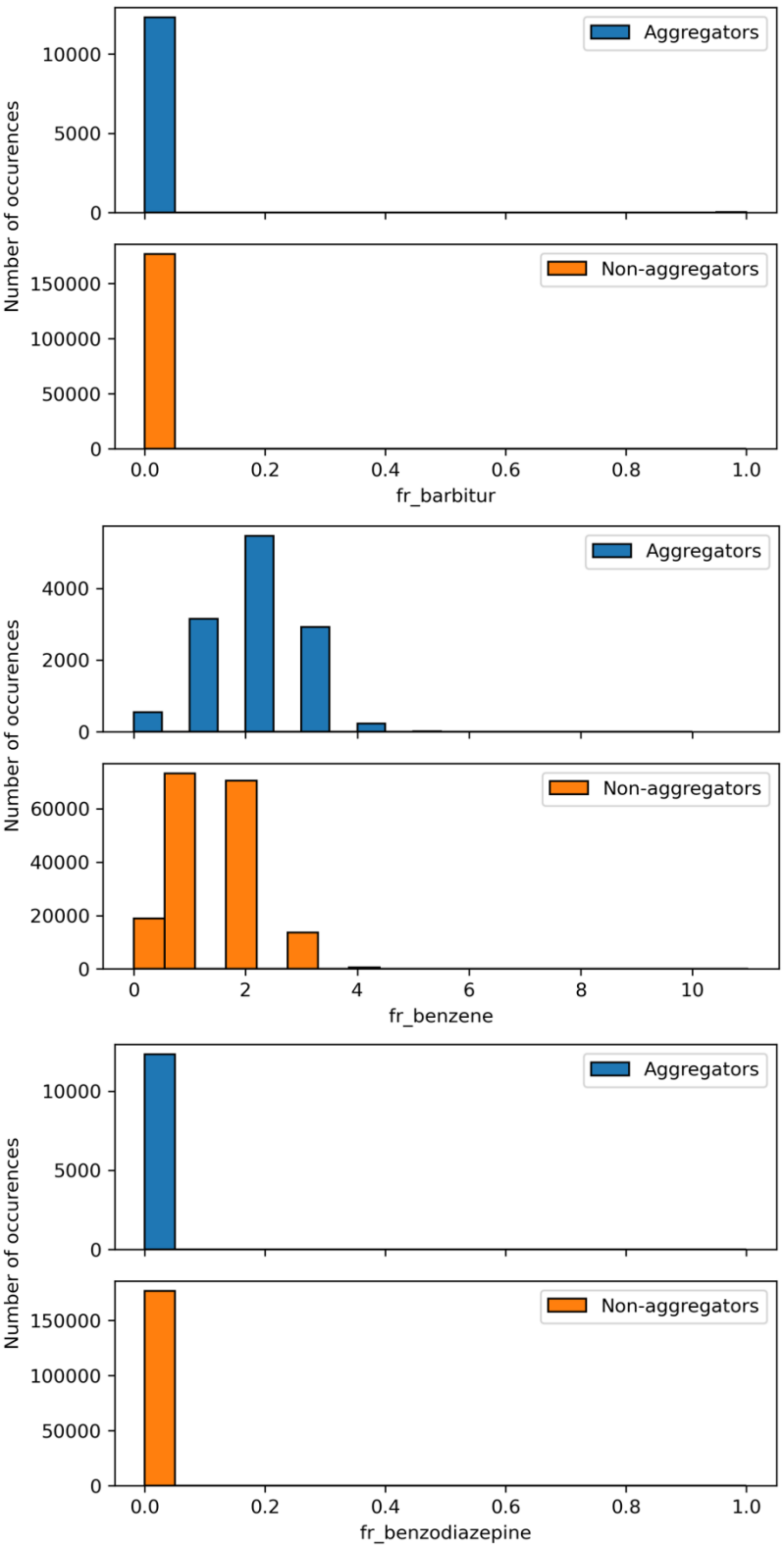

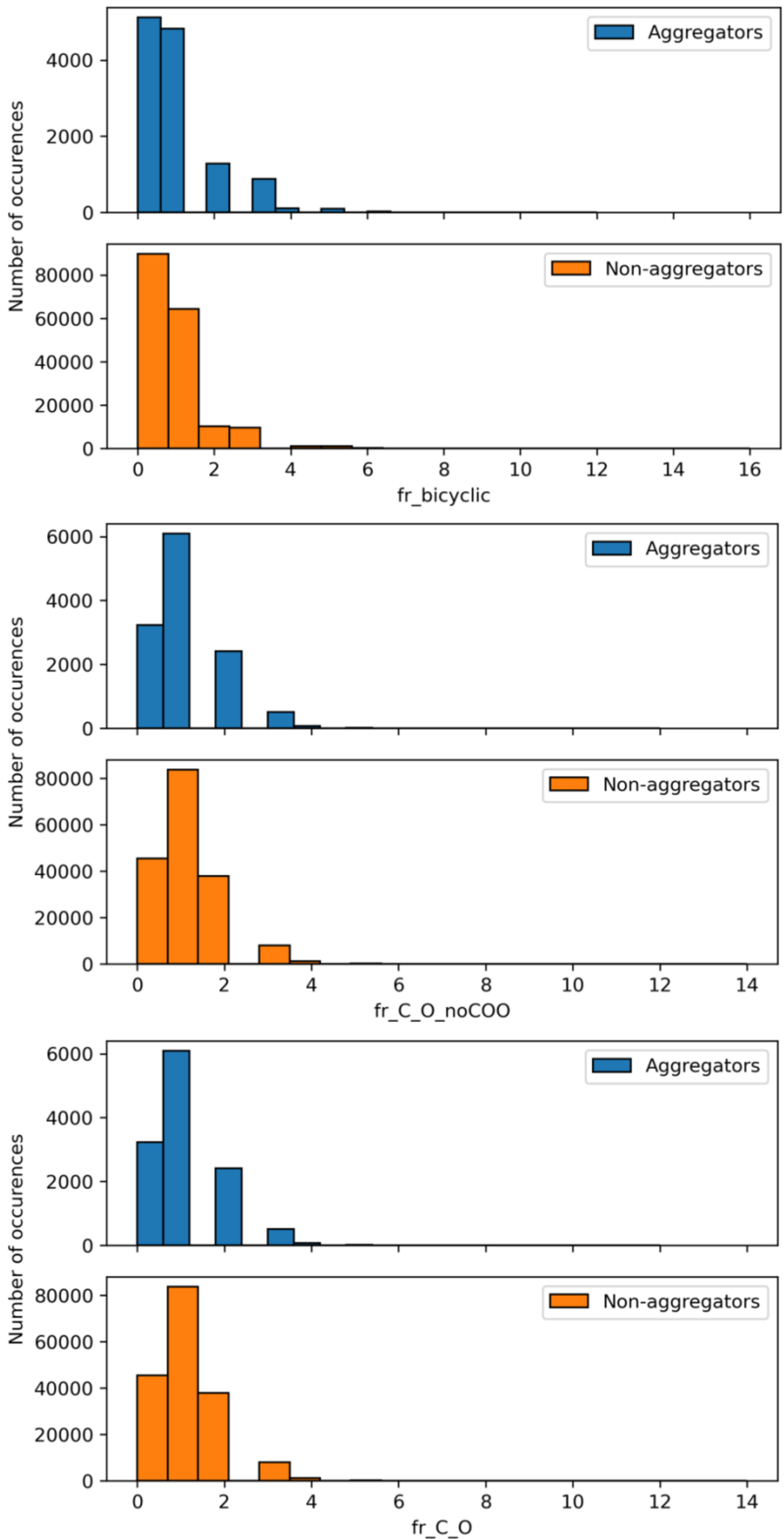

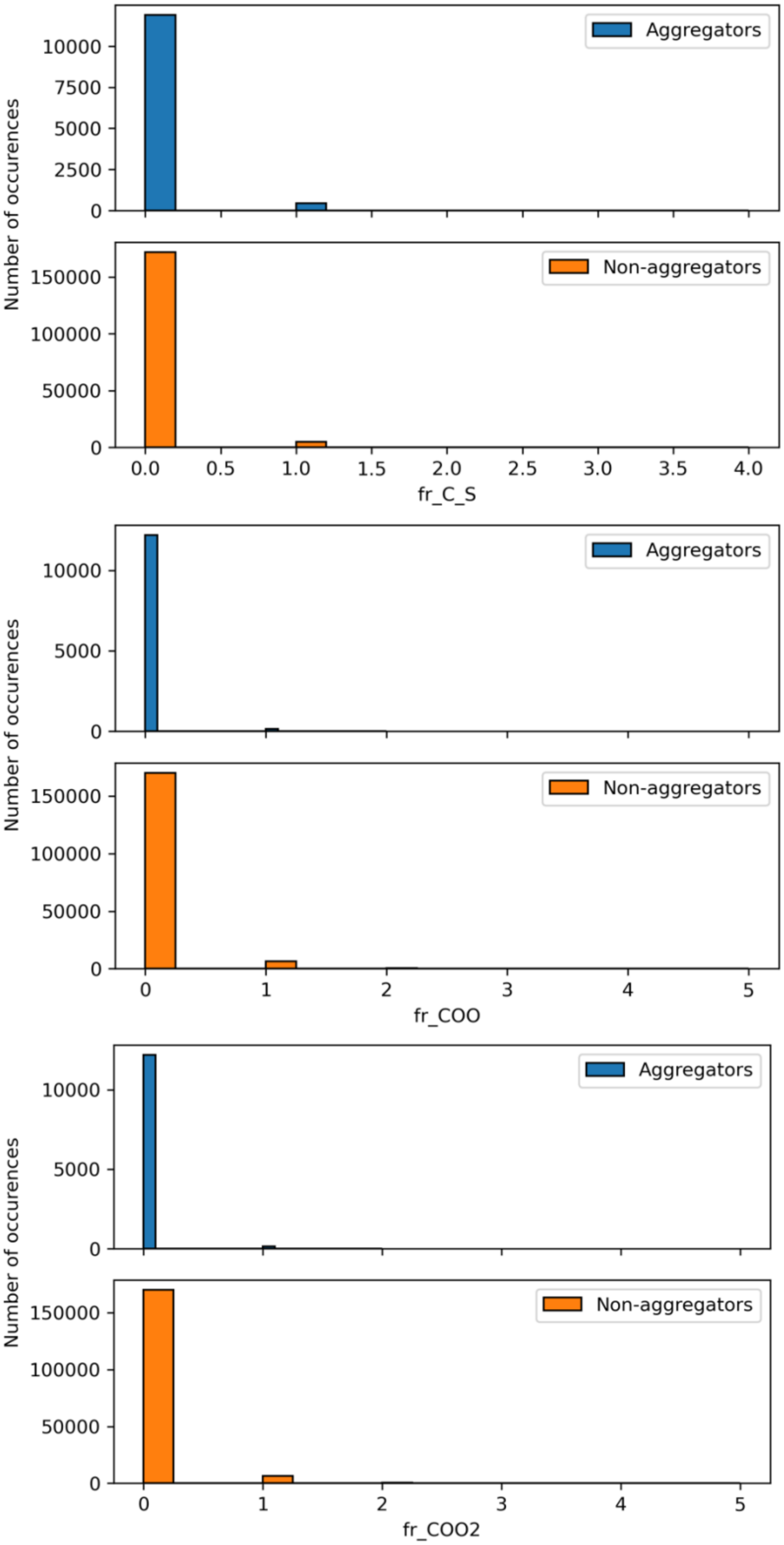

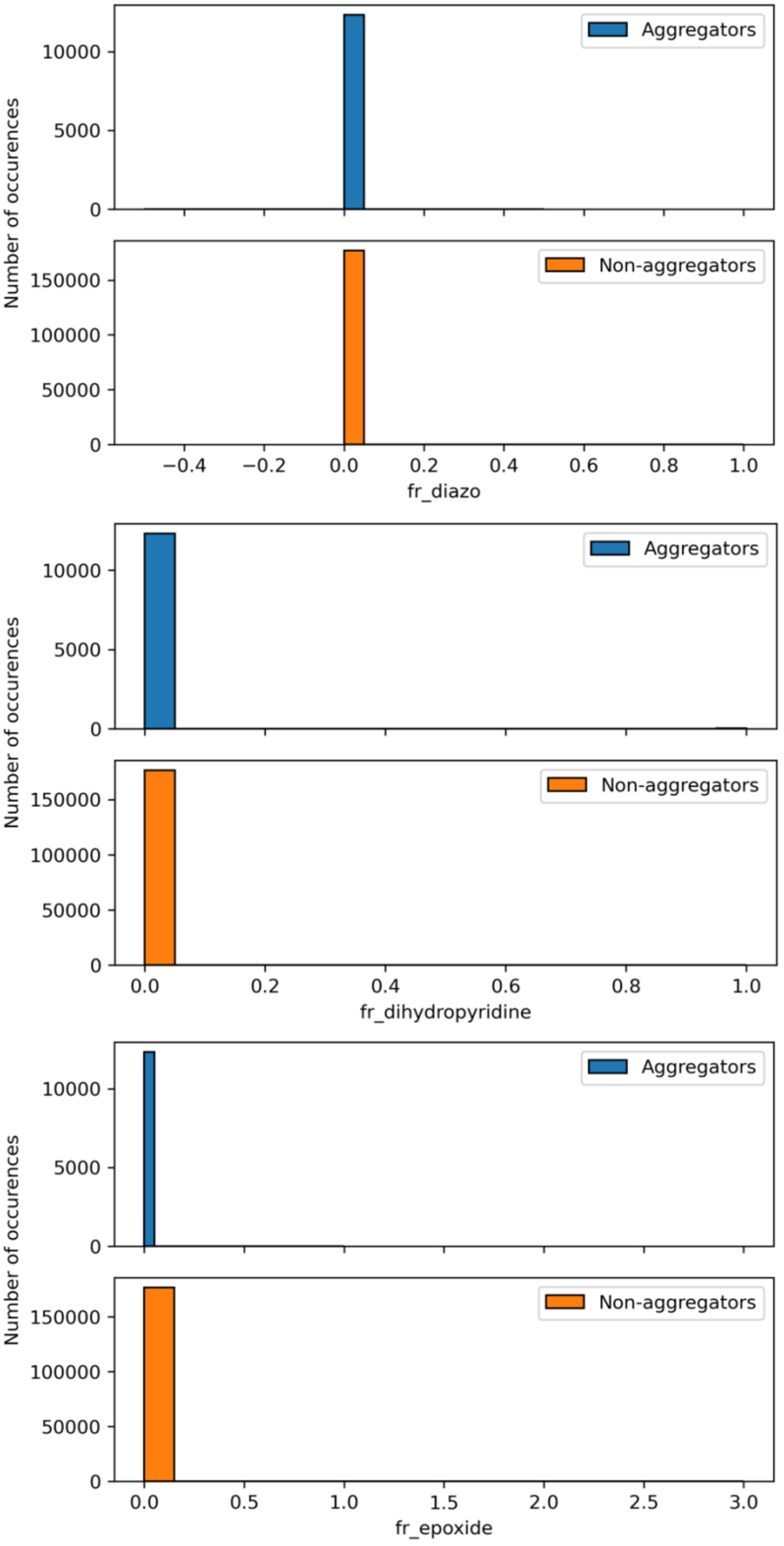

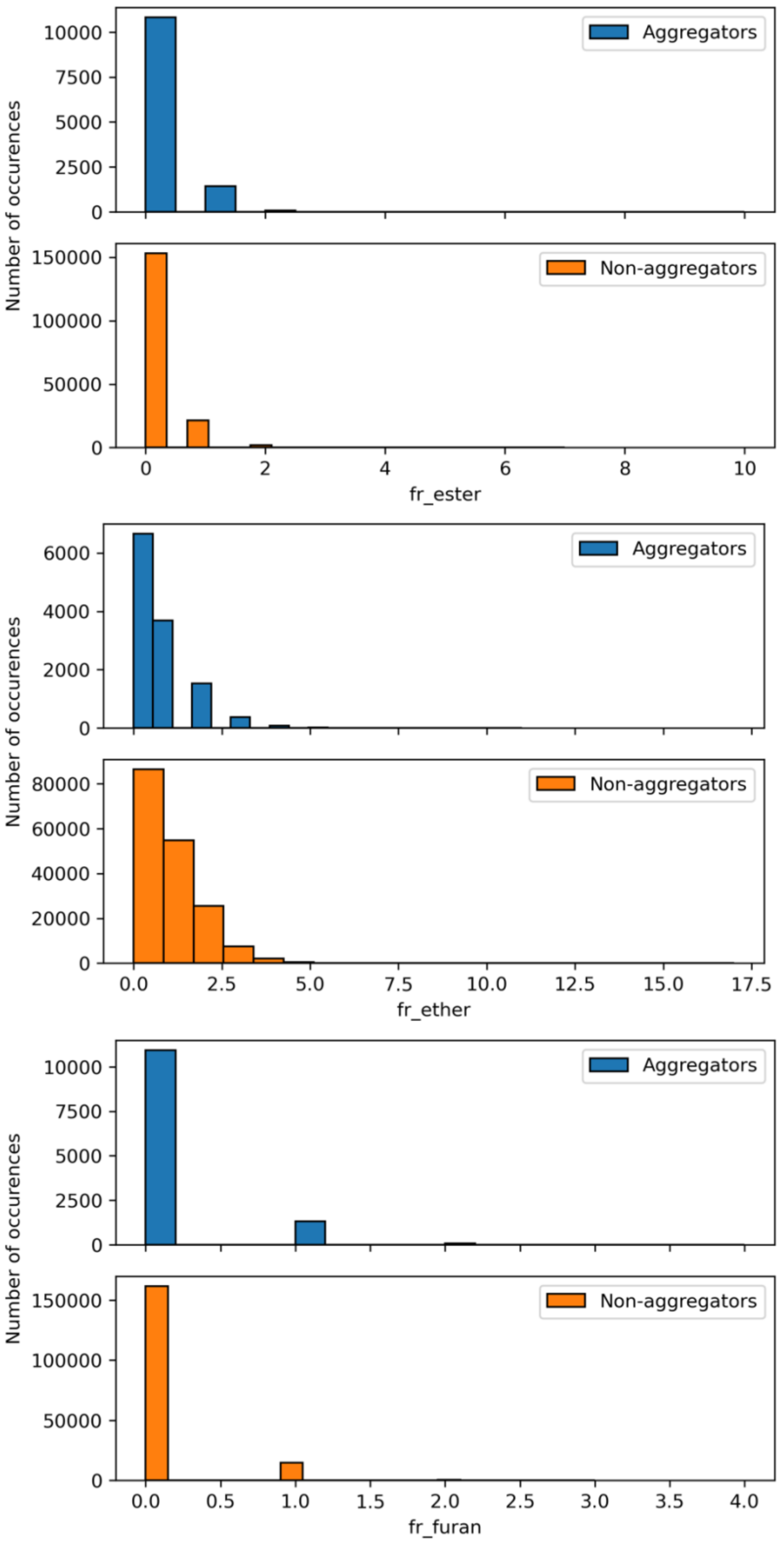

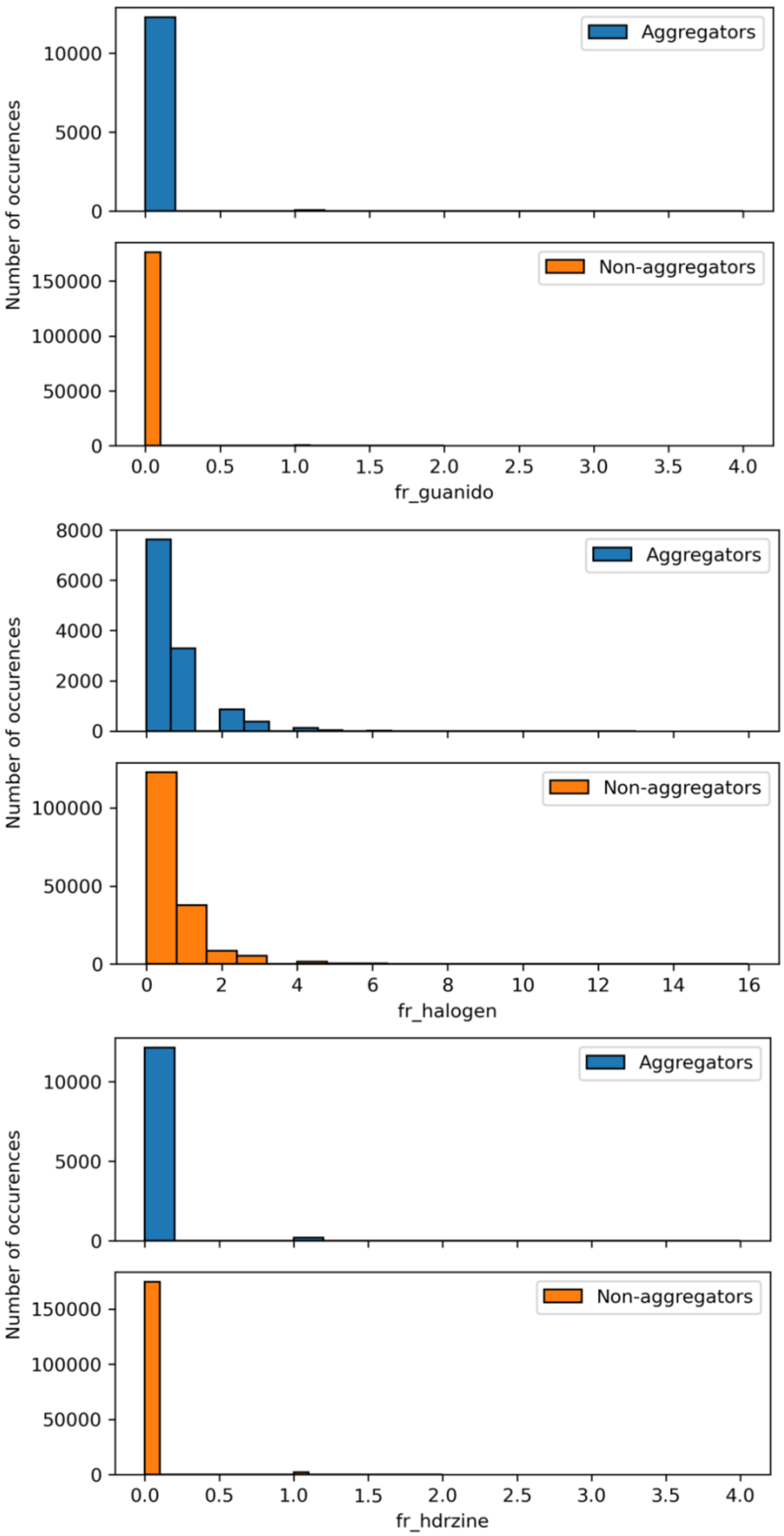

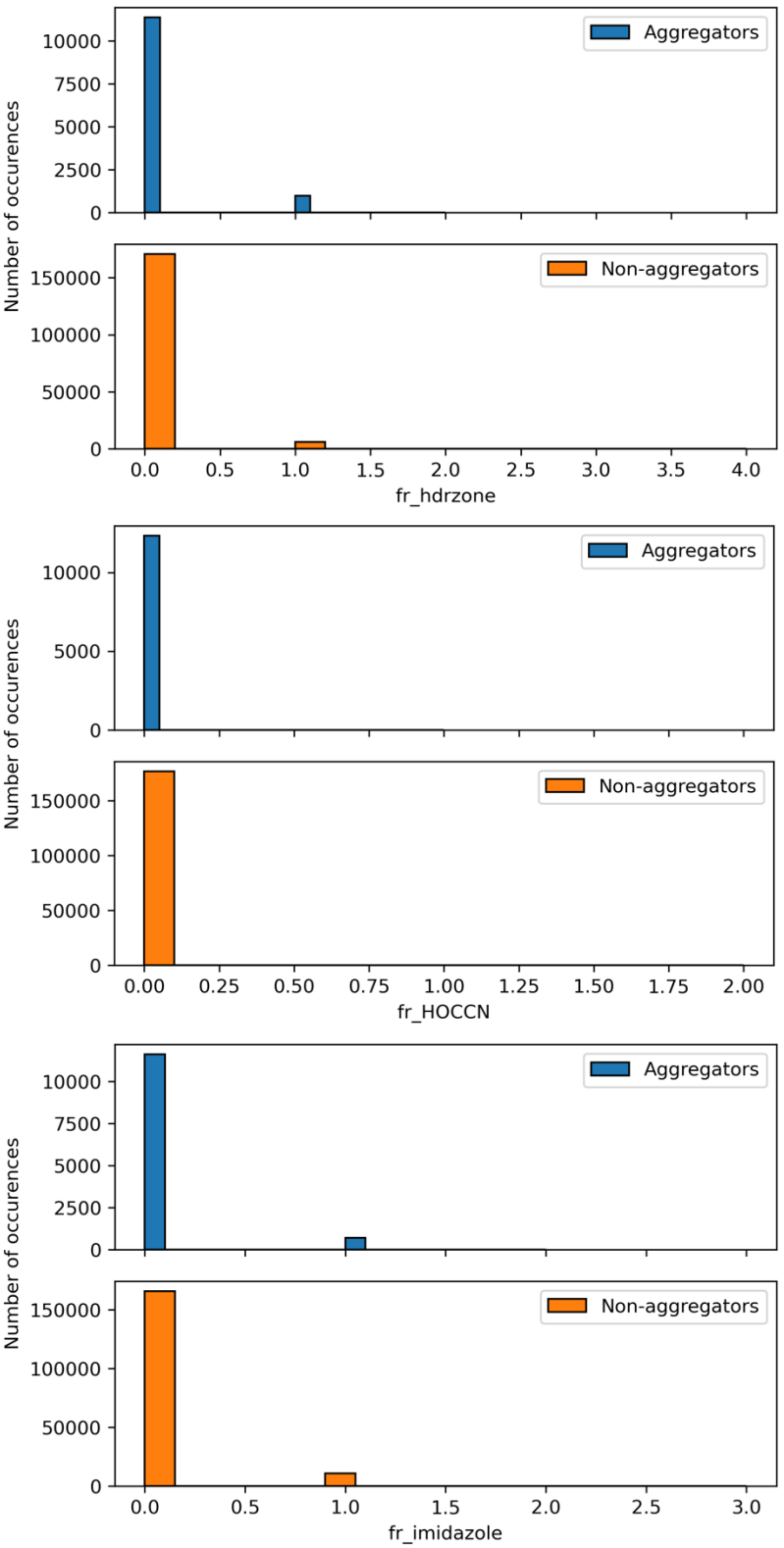

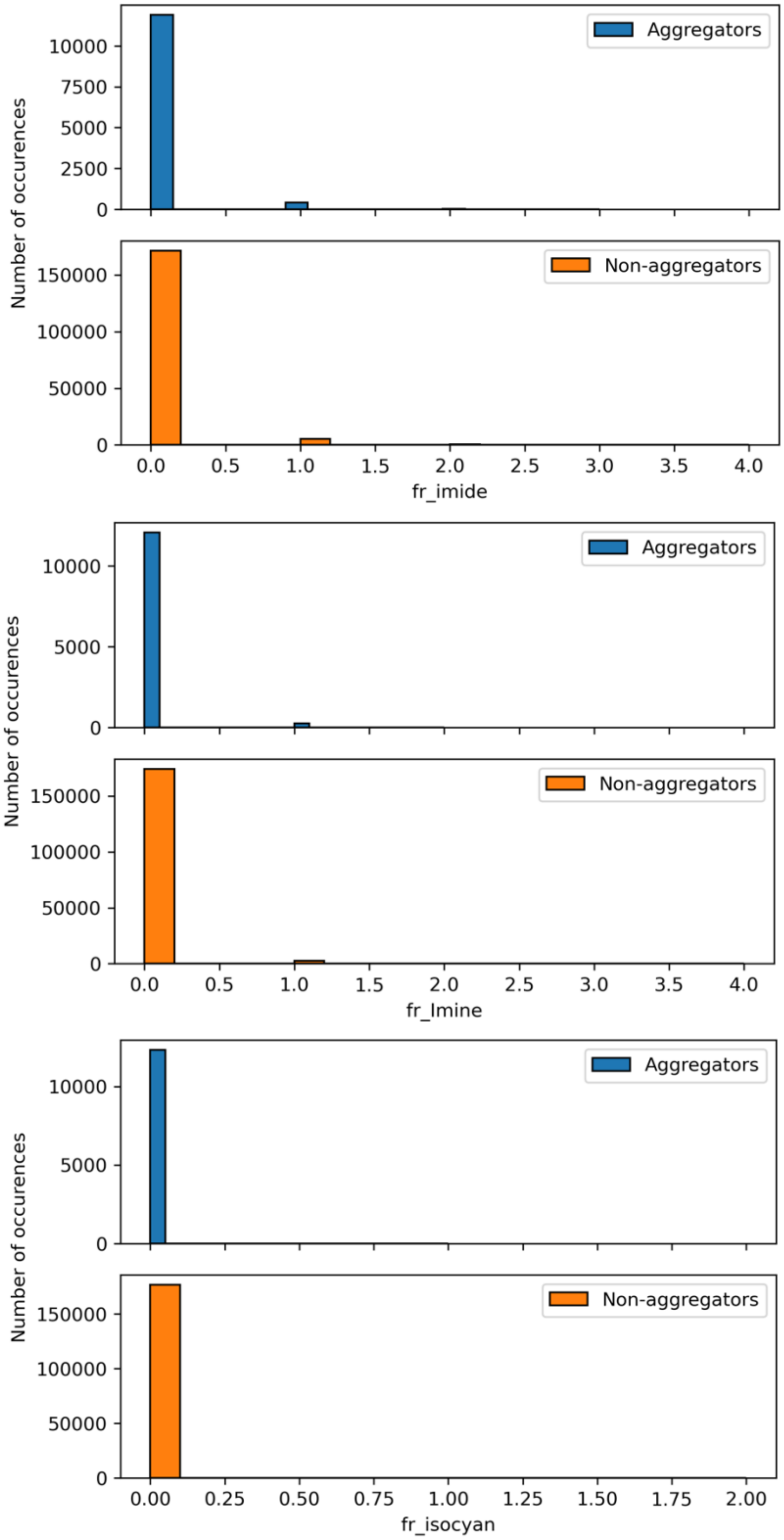

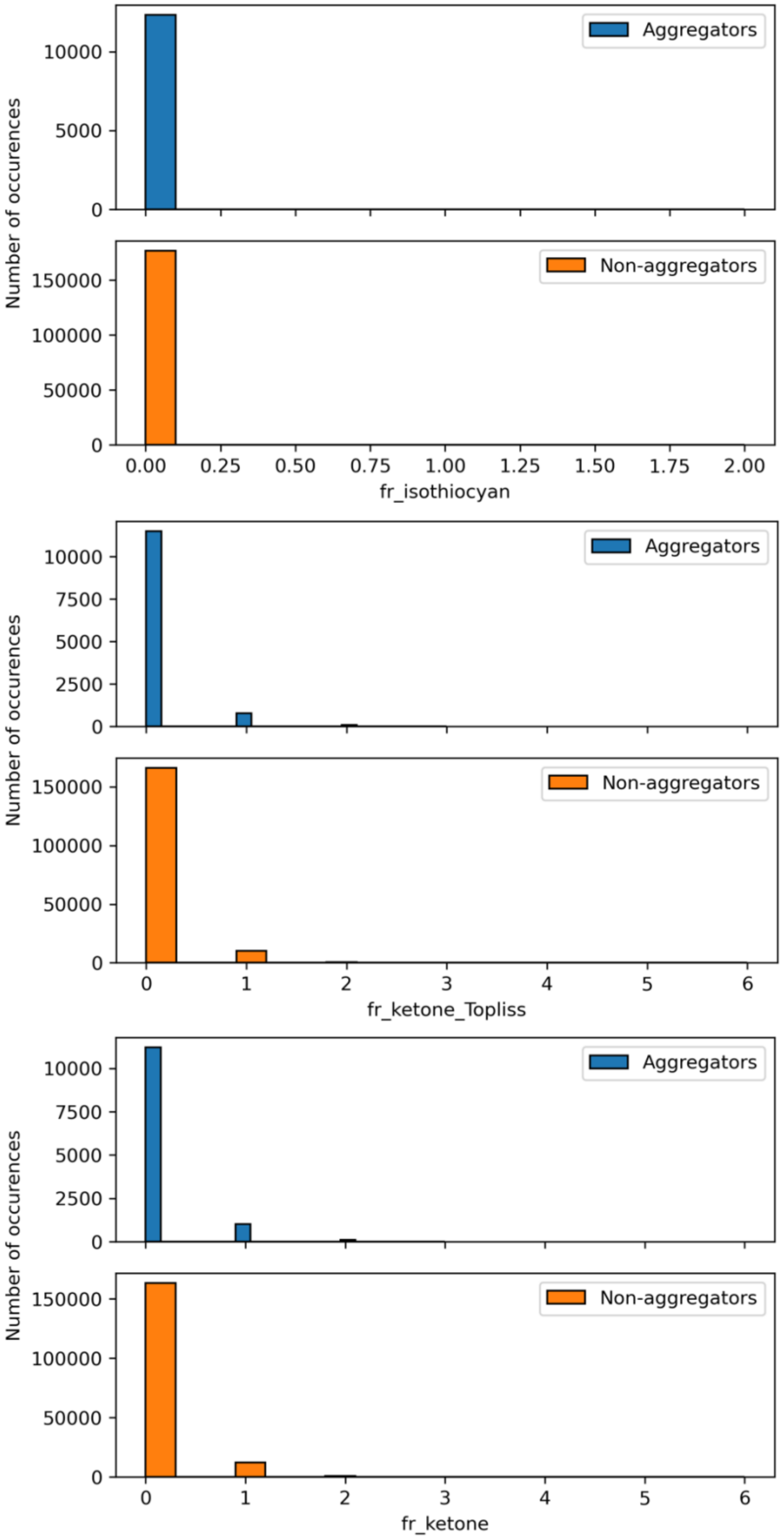

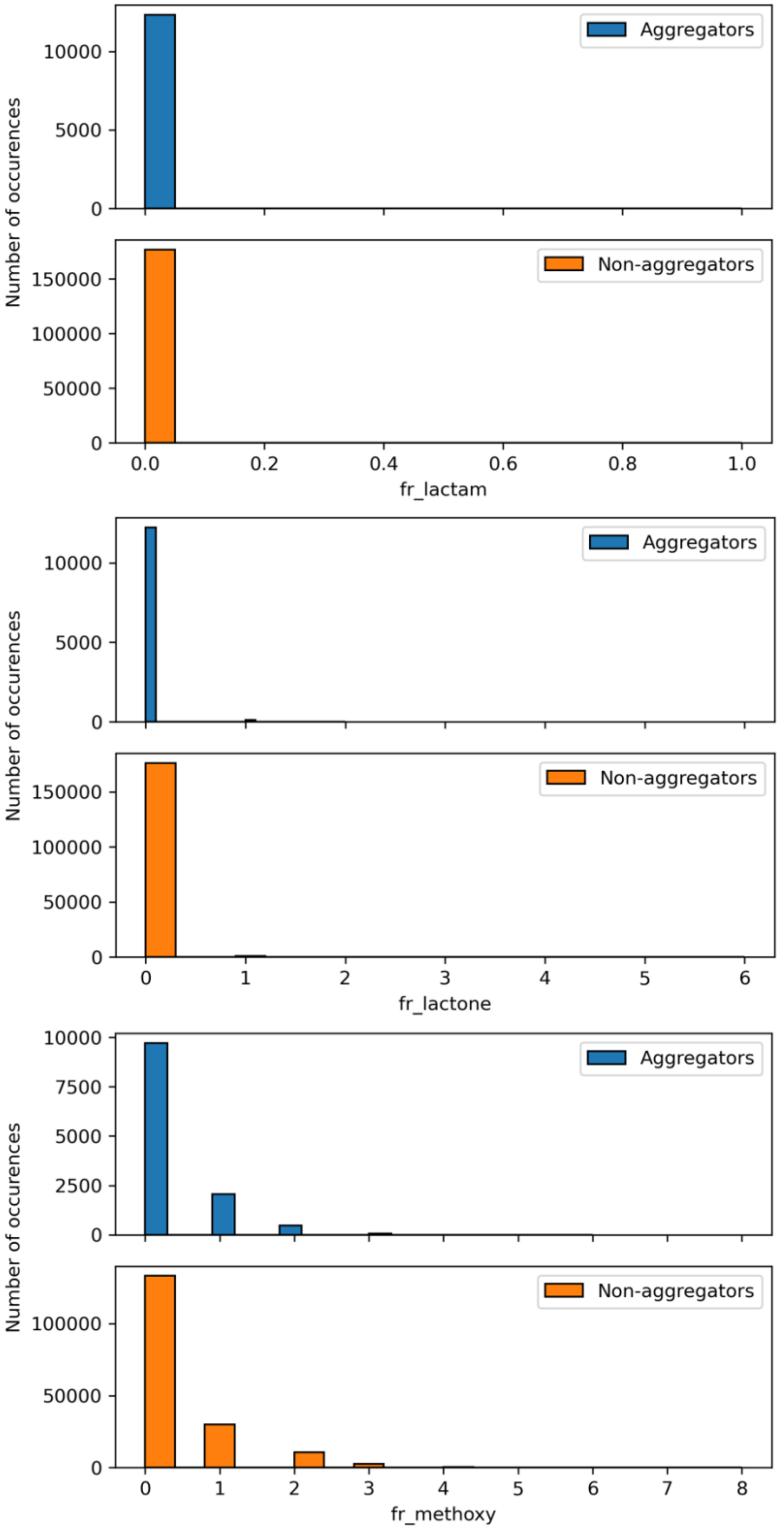

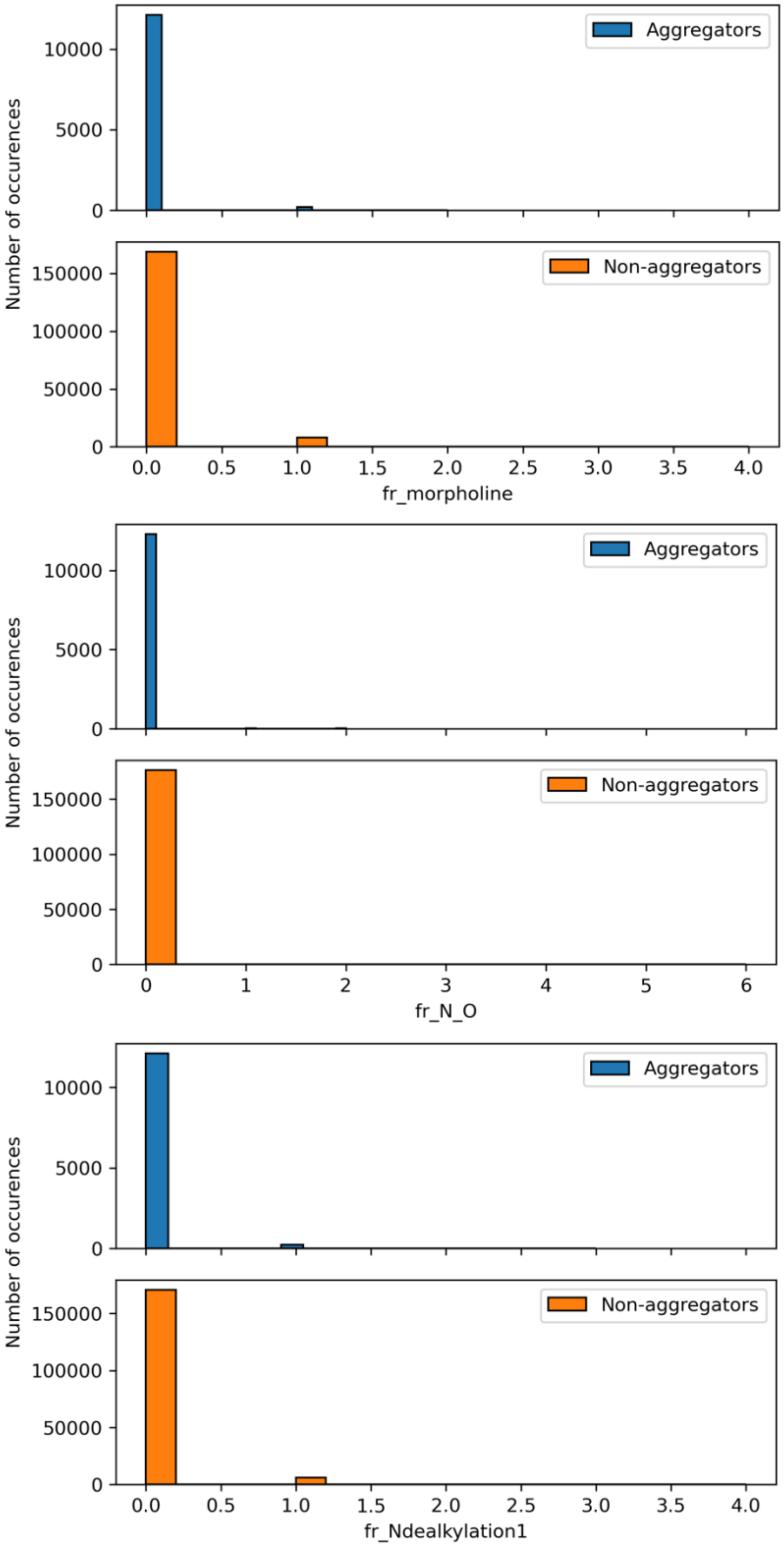

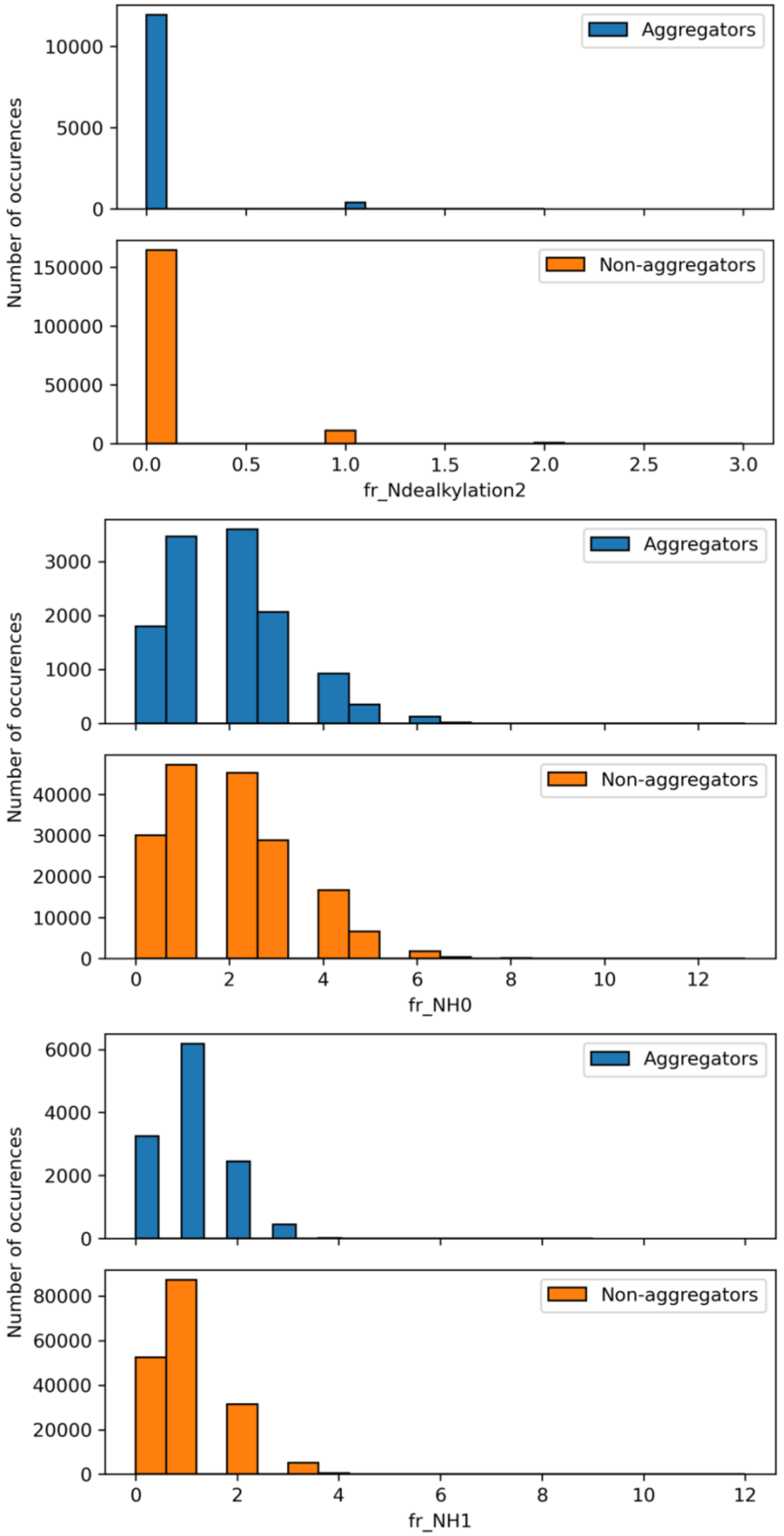

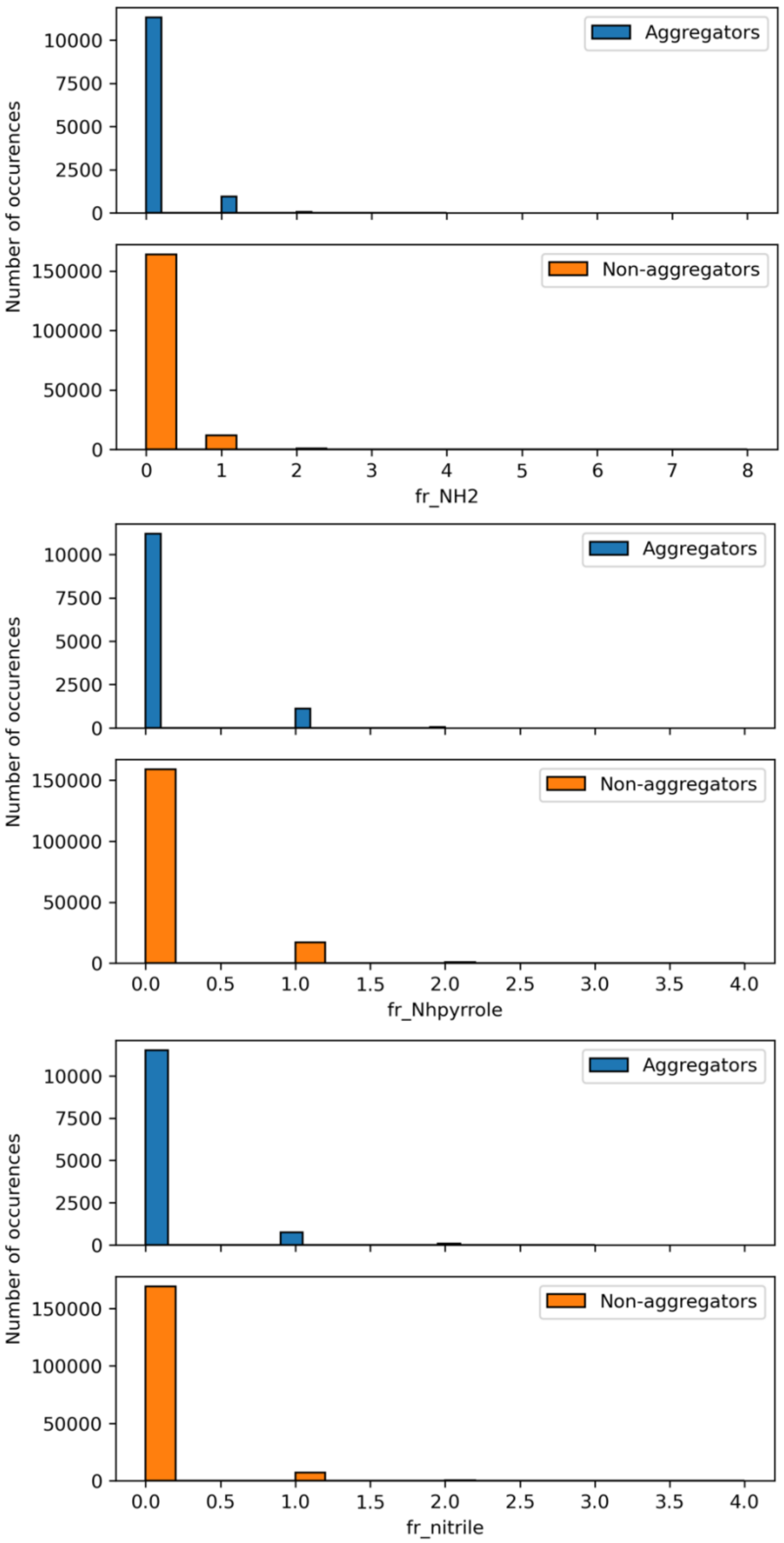

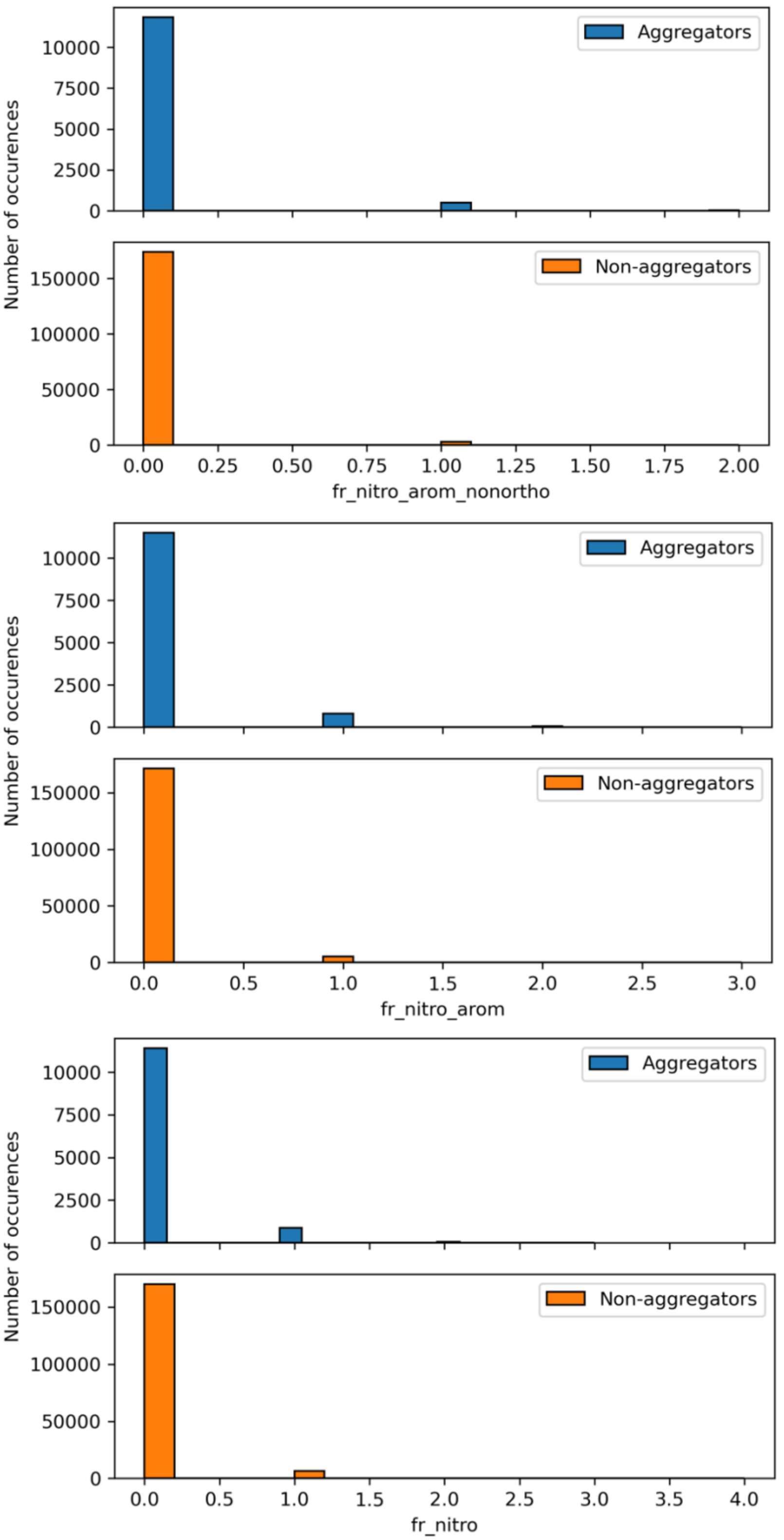

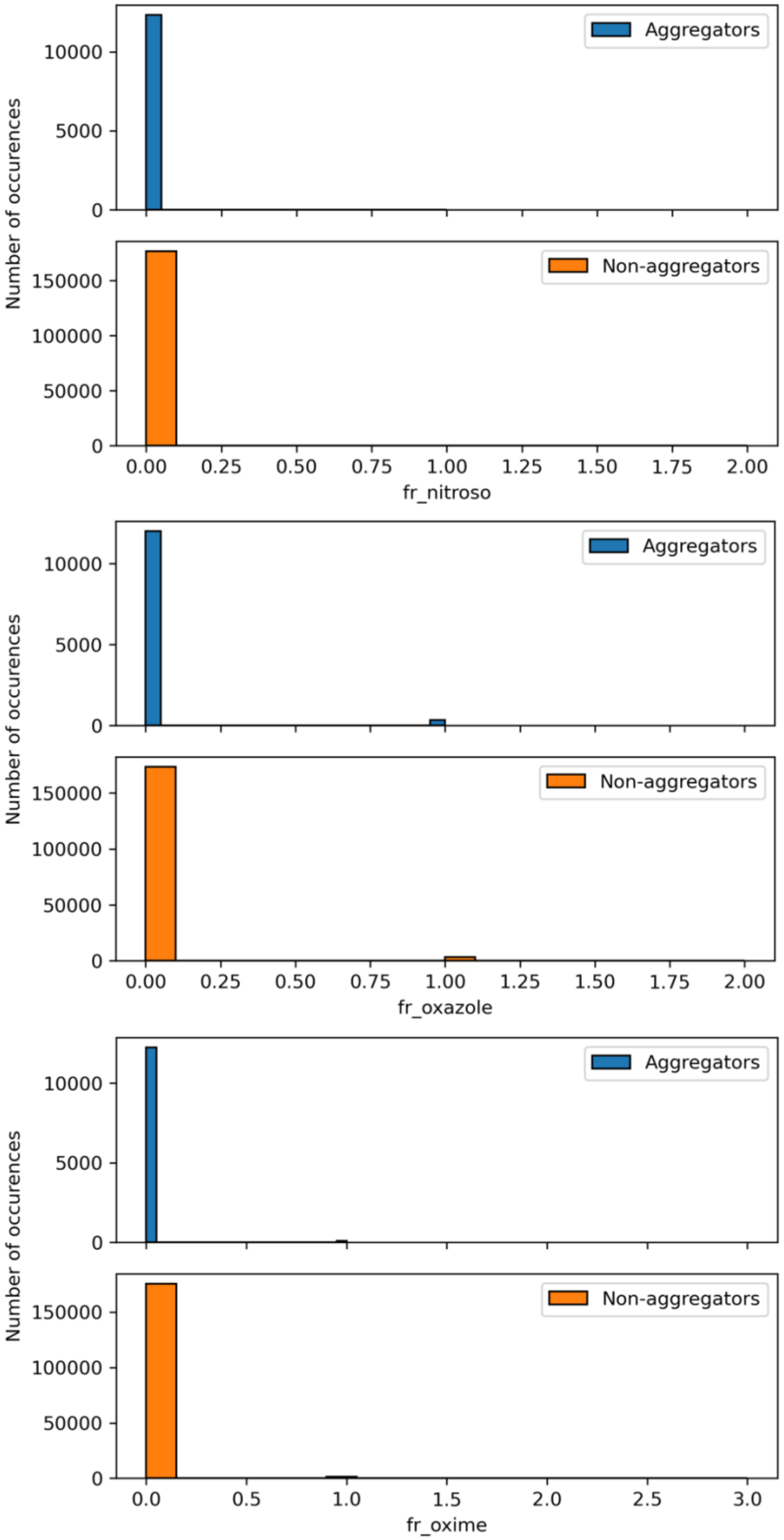

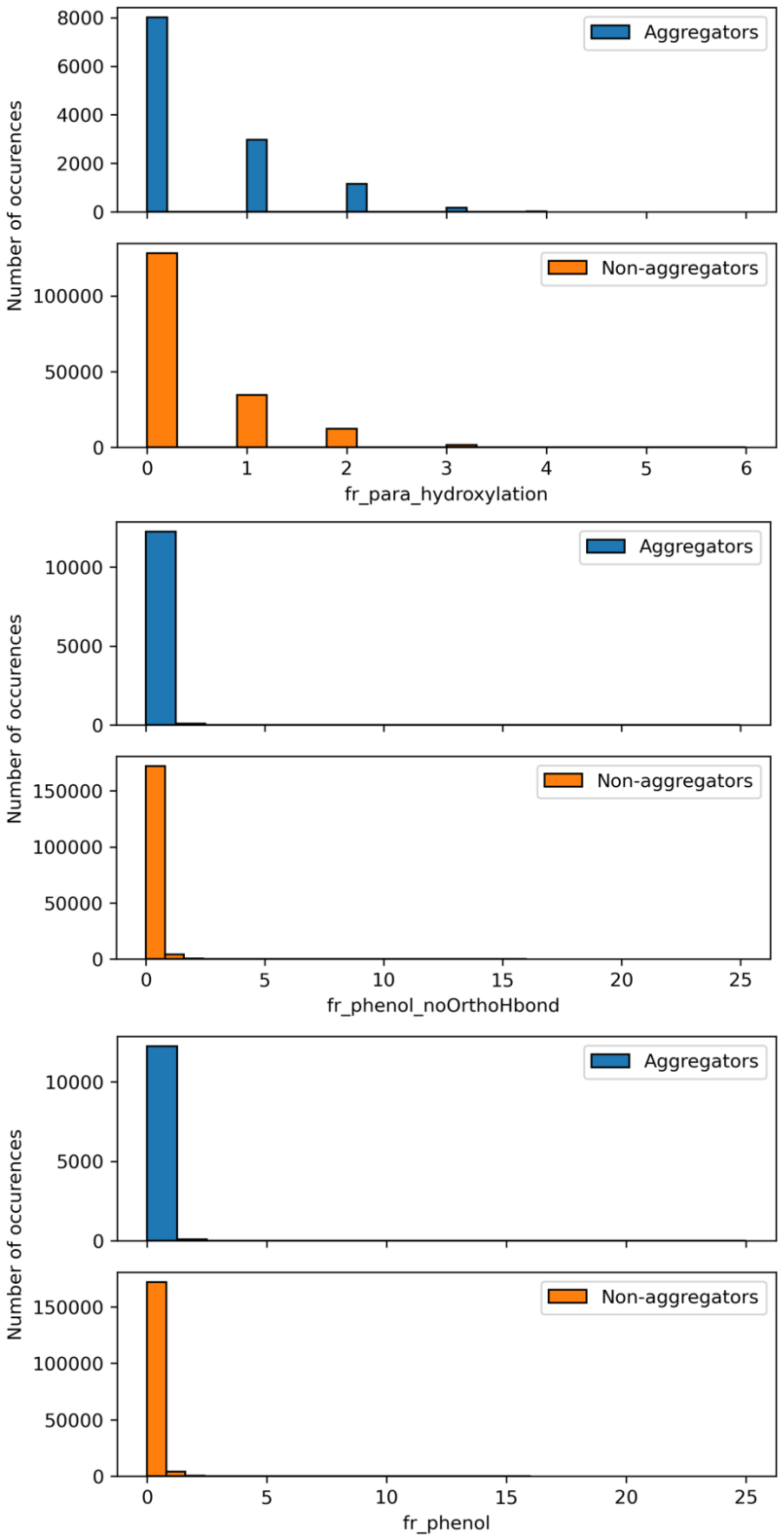

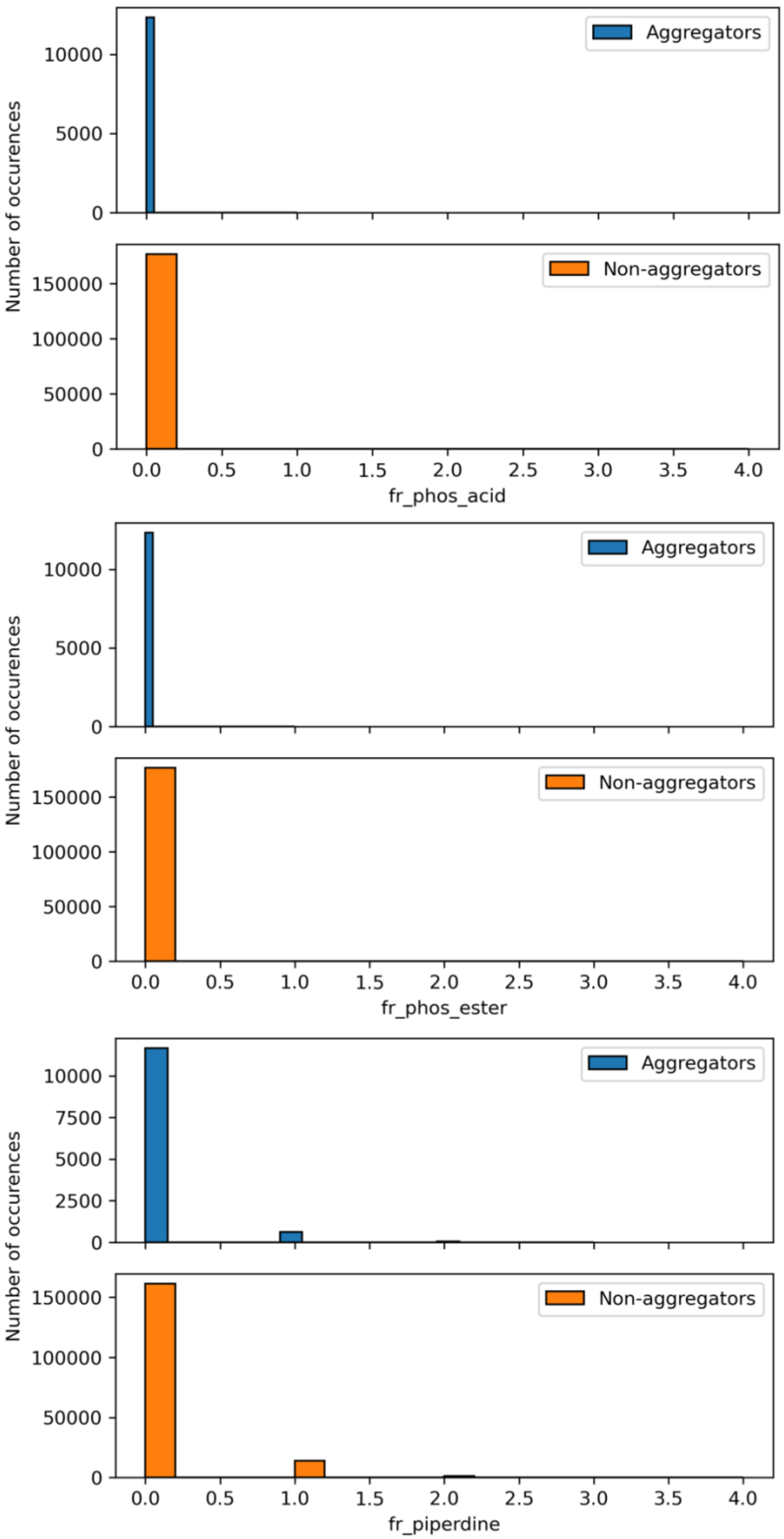

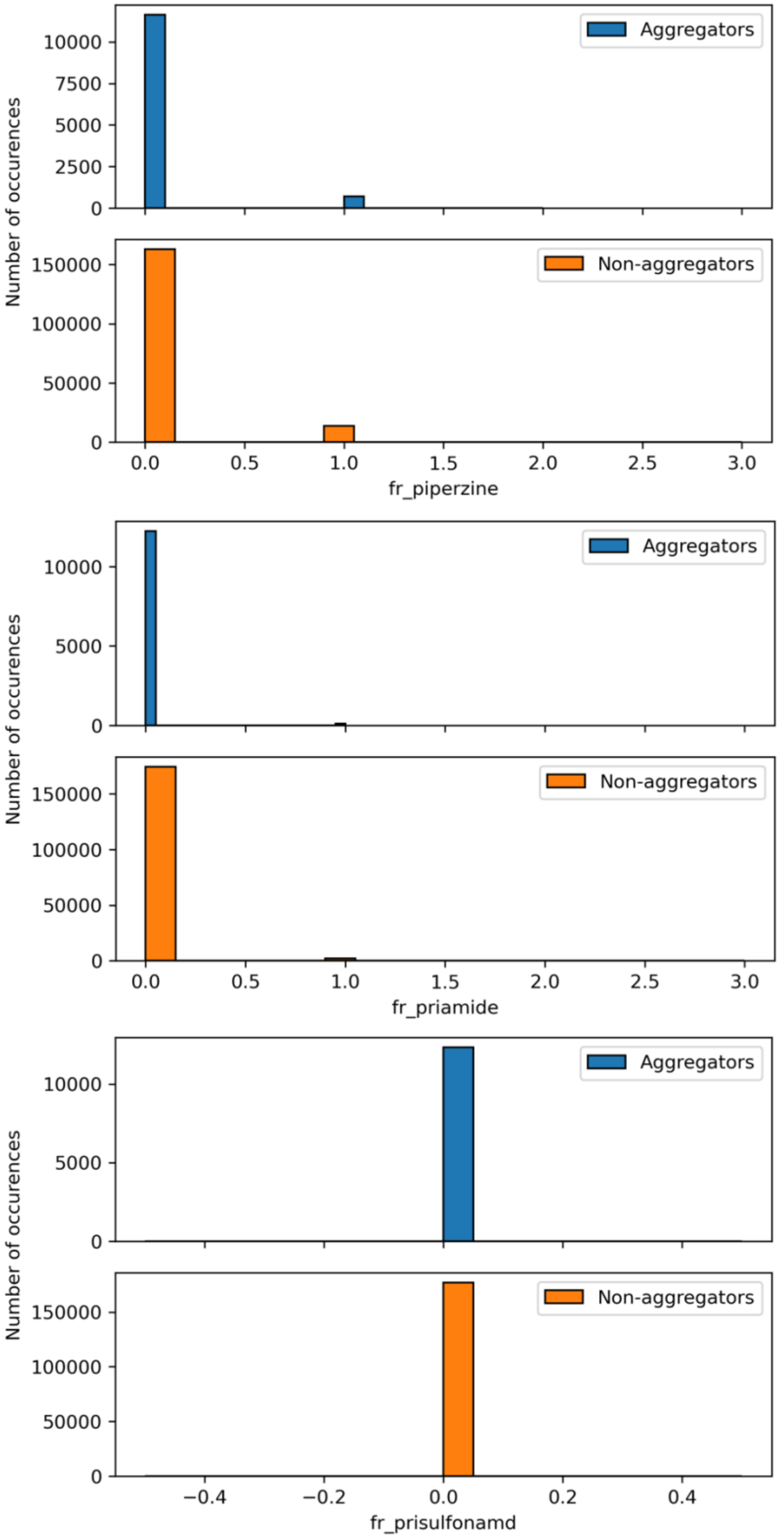

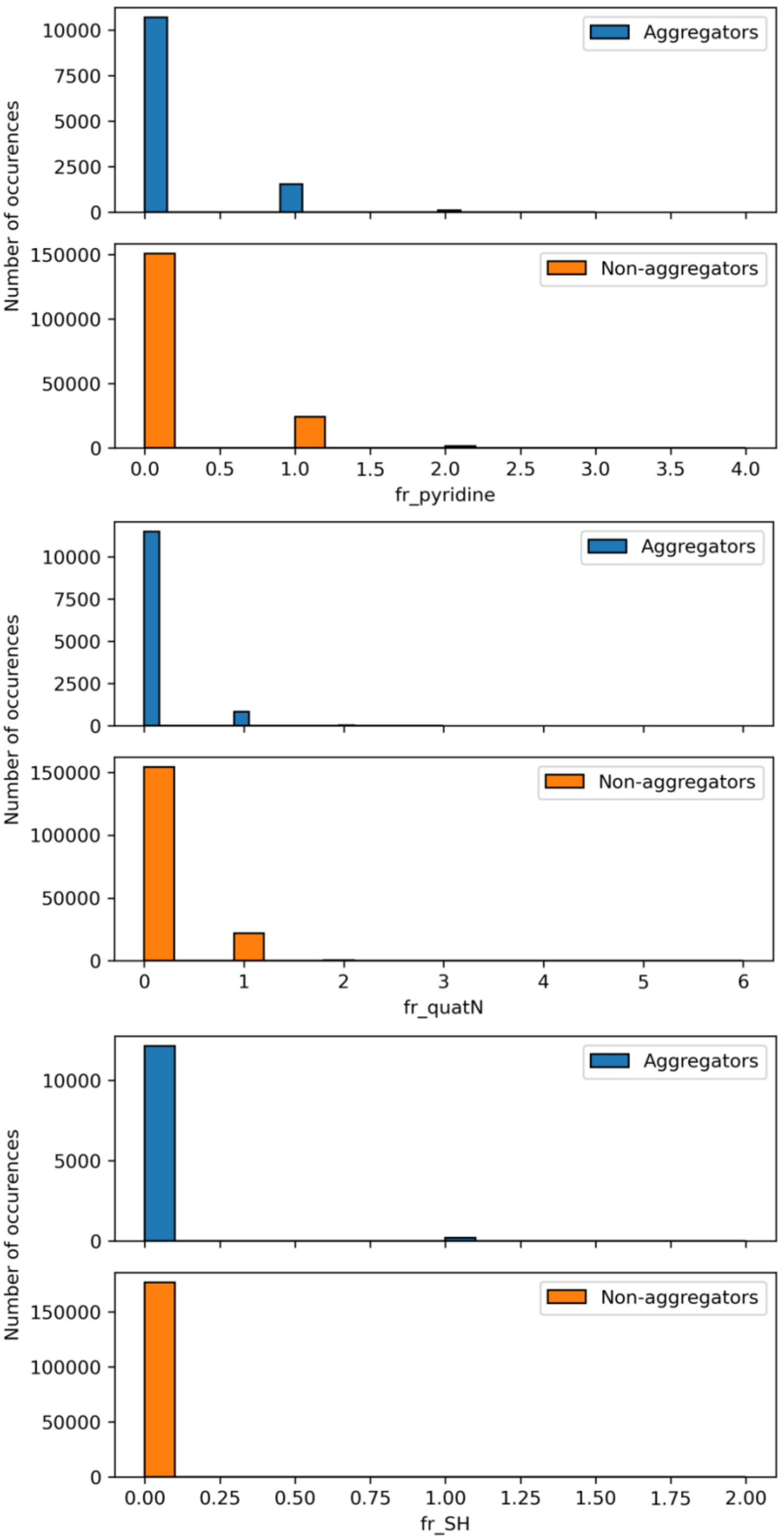

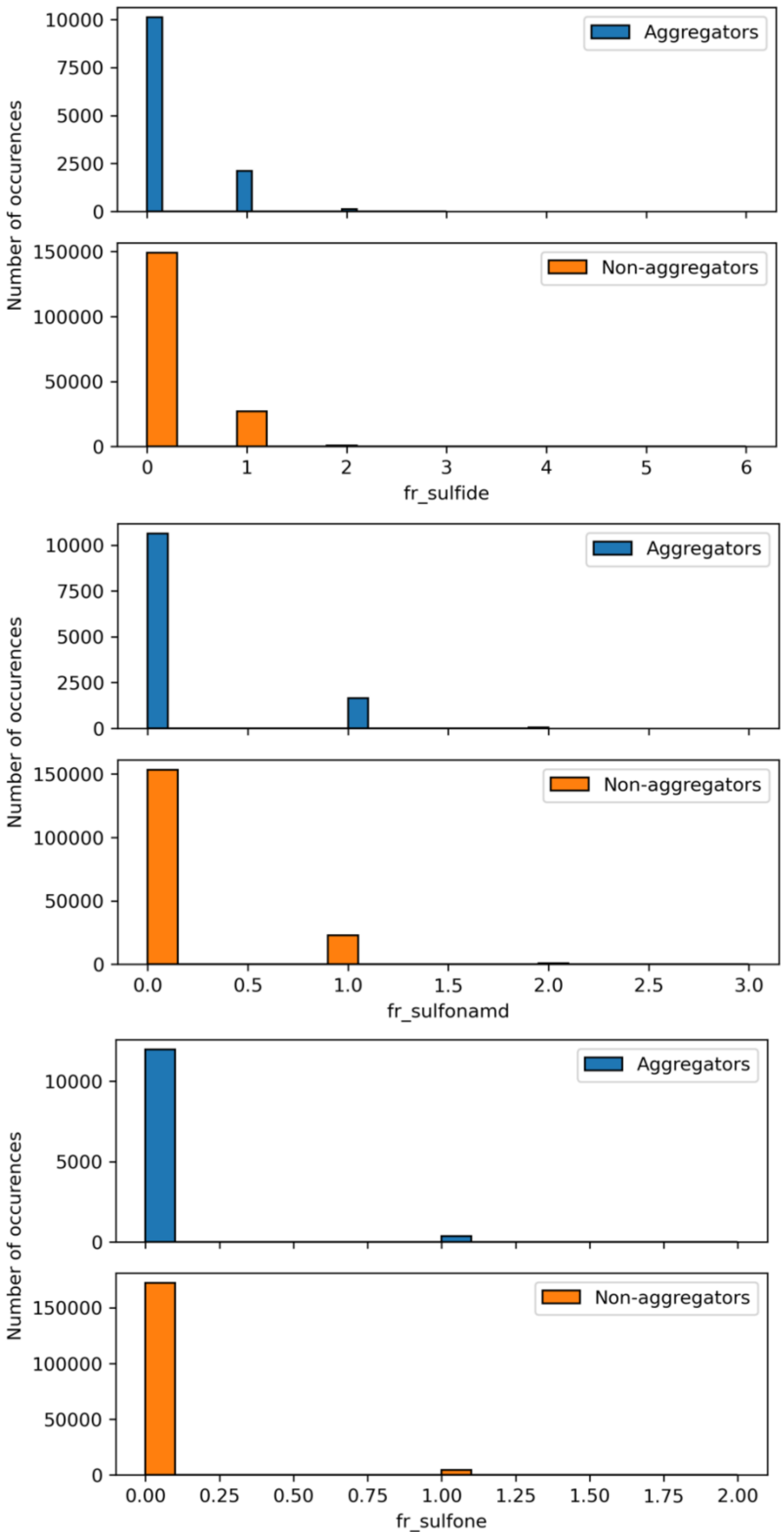

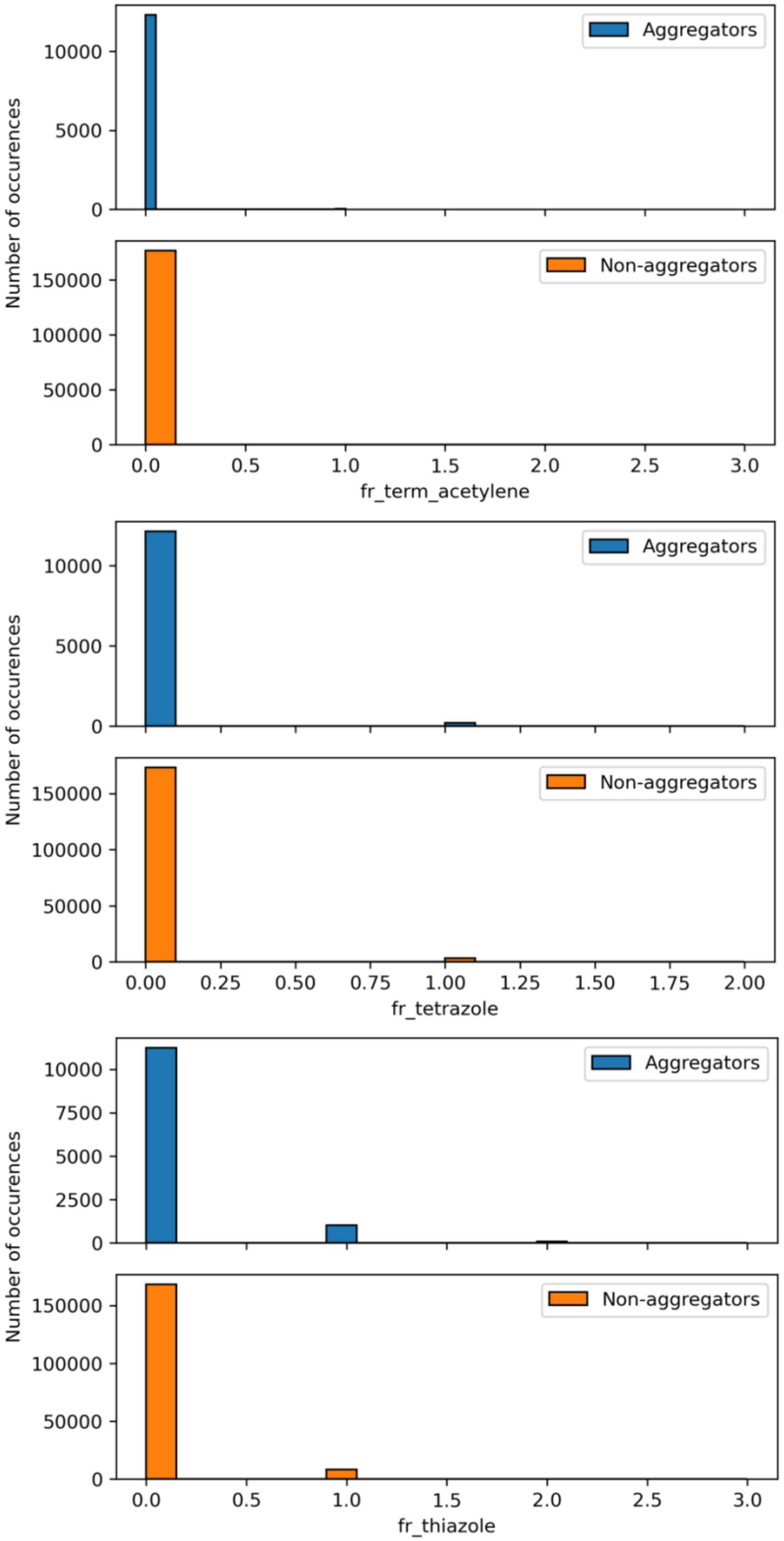

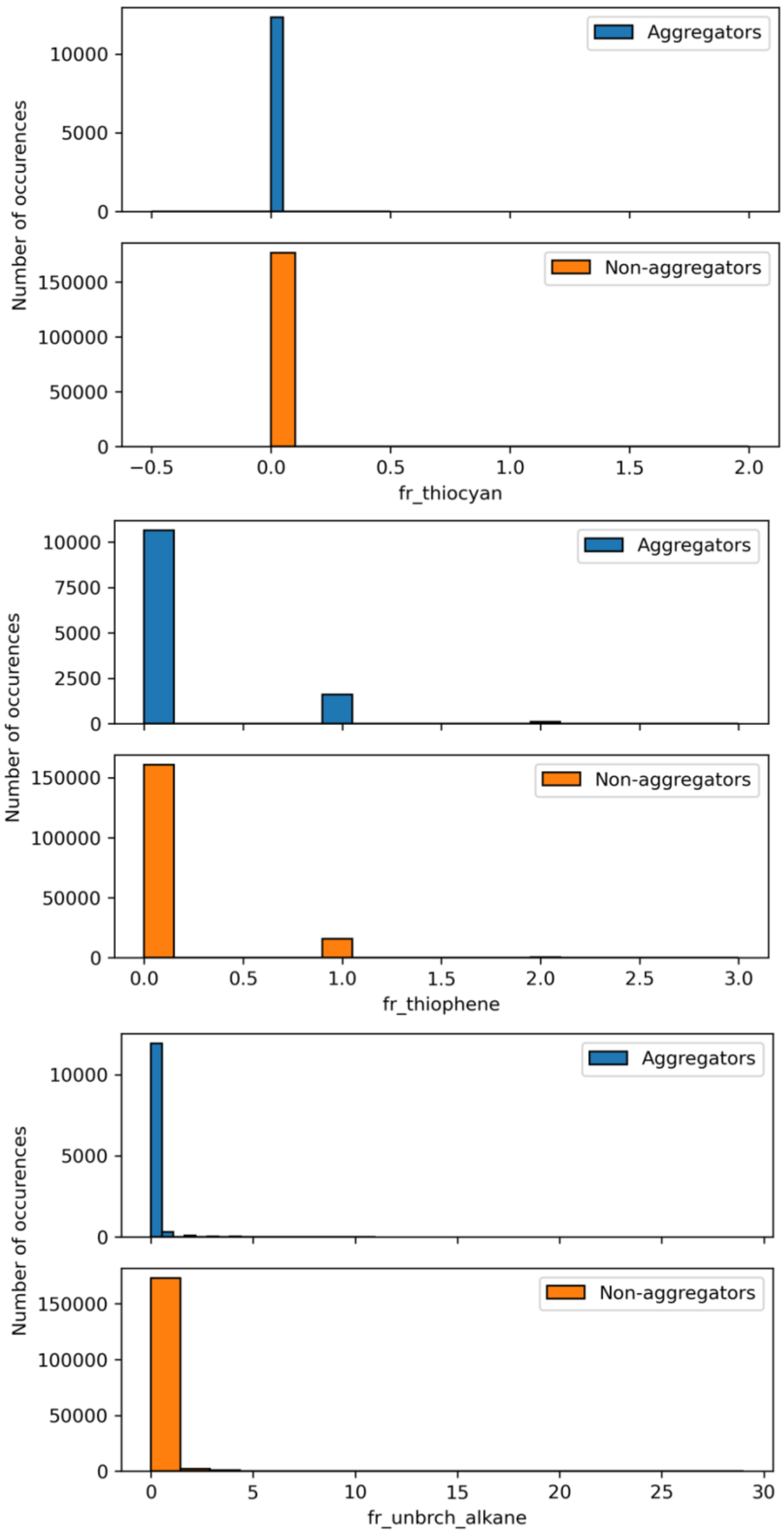

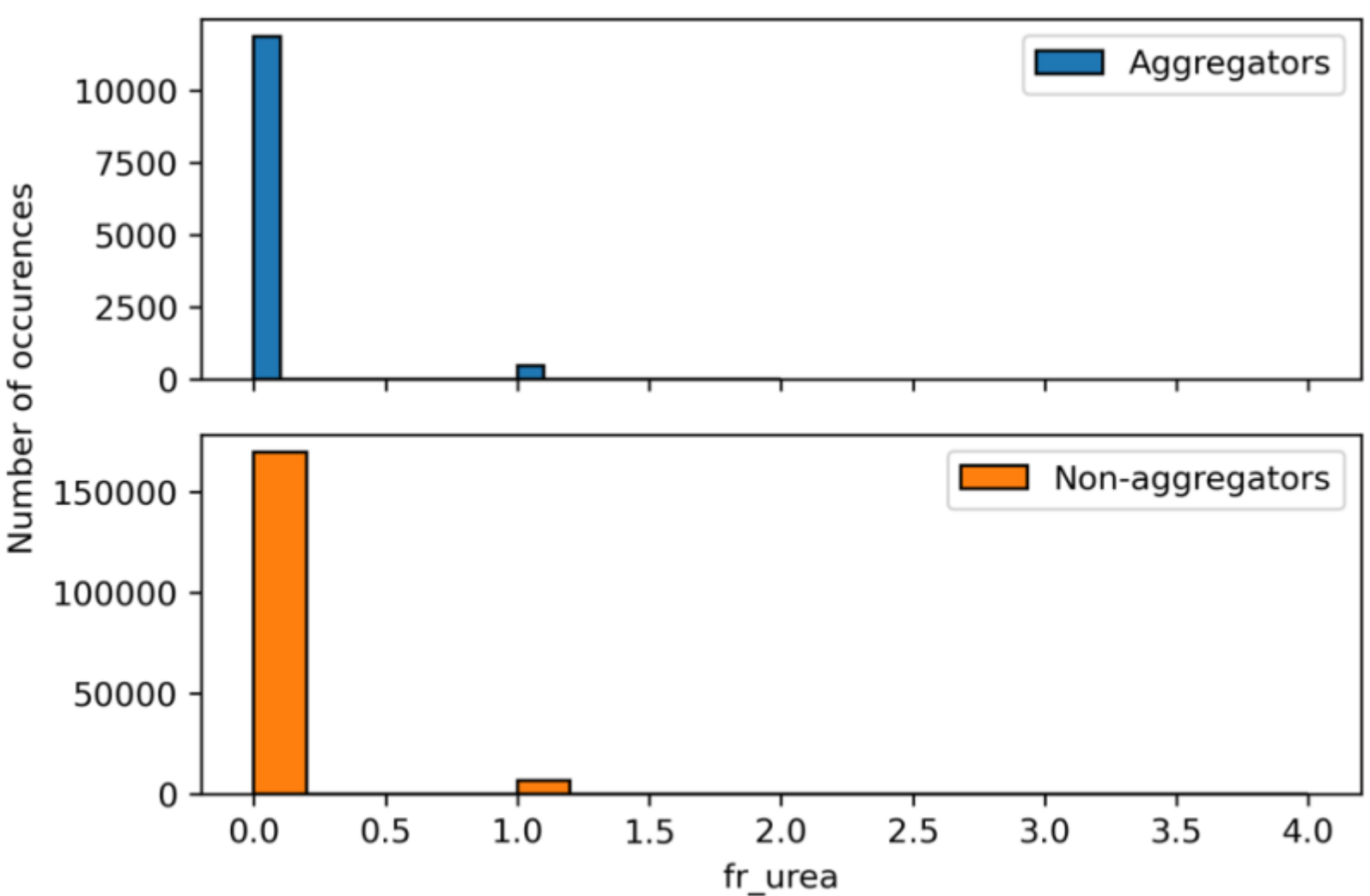

## SXV. p-values, means, standard deviations and Cohen’s d values for the 15 statistically significant general molecular descriptors calculated for the aggregator and non-aggregator datasets

**Table S3. Statistics for the 15 statistically significant general molecular descriptors calculated for the aggregator and non-aggregator datasets.**

| Molecular Descriptor | p-Value | Aggregators Mean | Aggregators Std. Dev. | Non-aggregators Mean | Non-aggregators Std. Dev. | Cohen's D | \|Cohen's_D\| |
|---|---|---|---|---|---|---|---|
| MolLogP | 0.000 | 3.888 | 1.455 | 2.906 | 1.581 | 0.647 | 0.647 |
| NumAromaticCarbocycles | 0.000 | 1.940 | 0.873 | 1.459 | 0.802 | 0.573 | 0.573 |
| FractionCSP3 | 0.000 | 0.208 | 0.143 | 0.296 | 0.174 | -0.557 | 0.557 |
| NumAromaticRings | 0.000 | 3.027 | 0.958 | 2.475 | 1.044 | 0.551 | 0.551 |
| RingCount | 0.000 | 3.681 | 1.029 | 3.182 | 1.092 | 0.471 | 0.471 |
| MolMR | 0.000 | 107.895 | 20.149 | 98.343 | 22.730 | 0.445 | 0.445 |
| HeavyAtomMolWt | 0.000 | 377.782 | 74.078 | 345.340 | 79.833 | 0.421 | 0.421 |
| LabuteASA | 0.000 | 164.944 | 30.714 | 151.925 | 34.314 | 0.400 | 0.400 |
| ExactMolWt | 0.000 | 396.802 | 76.906 | 365.012 | 83.778 | 0.395 | 0.395 |
| HeavyAtomCount | 0.000 | 27.697 | 5.375 | 25.592 | 5.952 | 0.371 | 0.371 |
| NumValenceElectrons | 0.000 | 141.820 | 27.812 | 133.437 | 31.001 | 0.285 | 0.285 |
| NumHeteroatoms | 0.000 | 7.074 | 2.168 | 6.863 | 2.206 | 0.097 | 0.097 |
| NumHAcceptors | 0.000 | 5.015 | 1.868 | 4.892 | 1.887 | 0.065 | 0.065 |
| TPSA | 0.000 | 74.603 | 27.422 | 73.071 | 27.139 | 0.056 | 0.056 |
| NOCount | 0.006 | 5.771 | 1.920 | 5.822 | 1.972 | -0.026 | 0.026 |

**SXVI. Comparison of the relative frequency of each of the fragments described by the fragment descriptors for the aggregator and non-aggregator molecules**

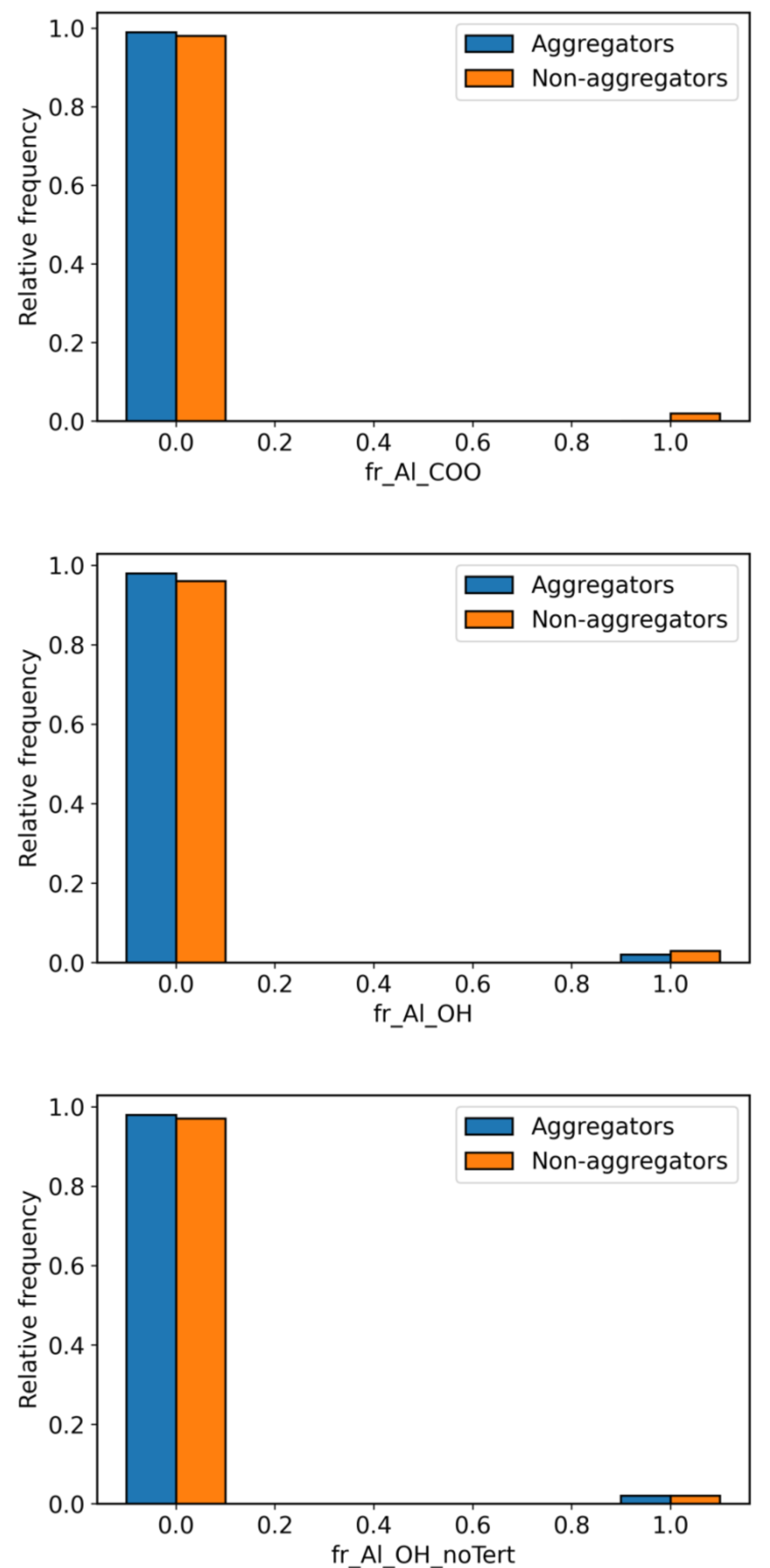

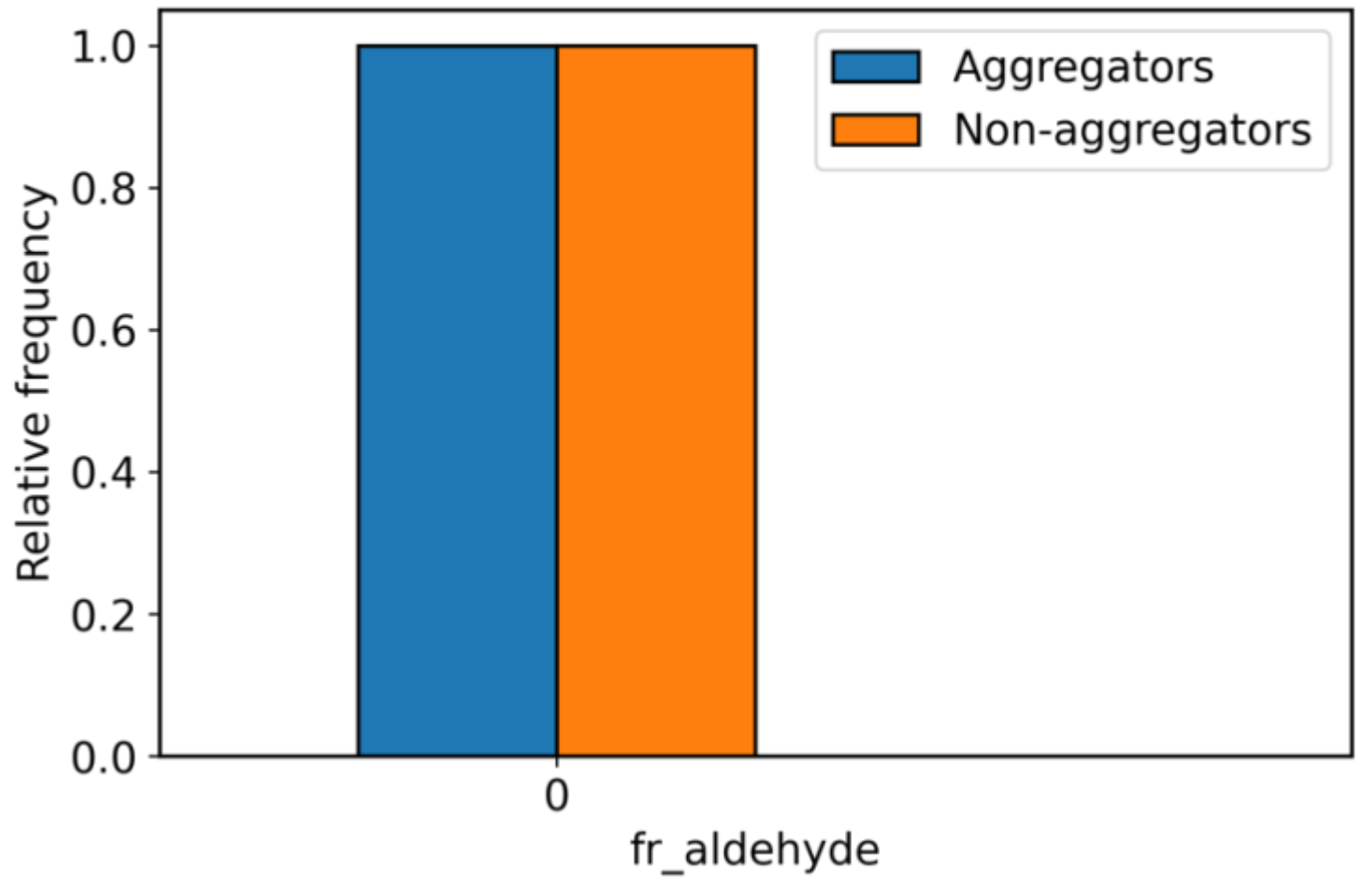

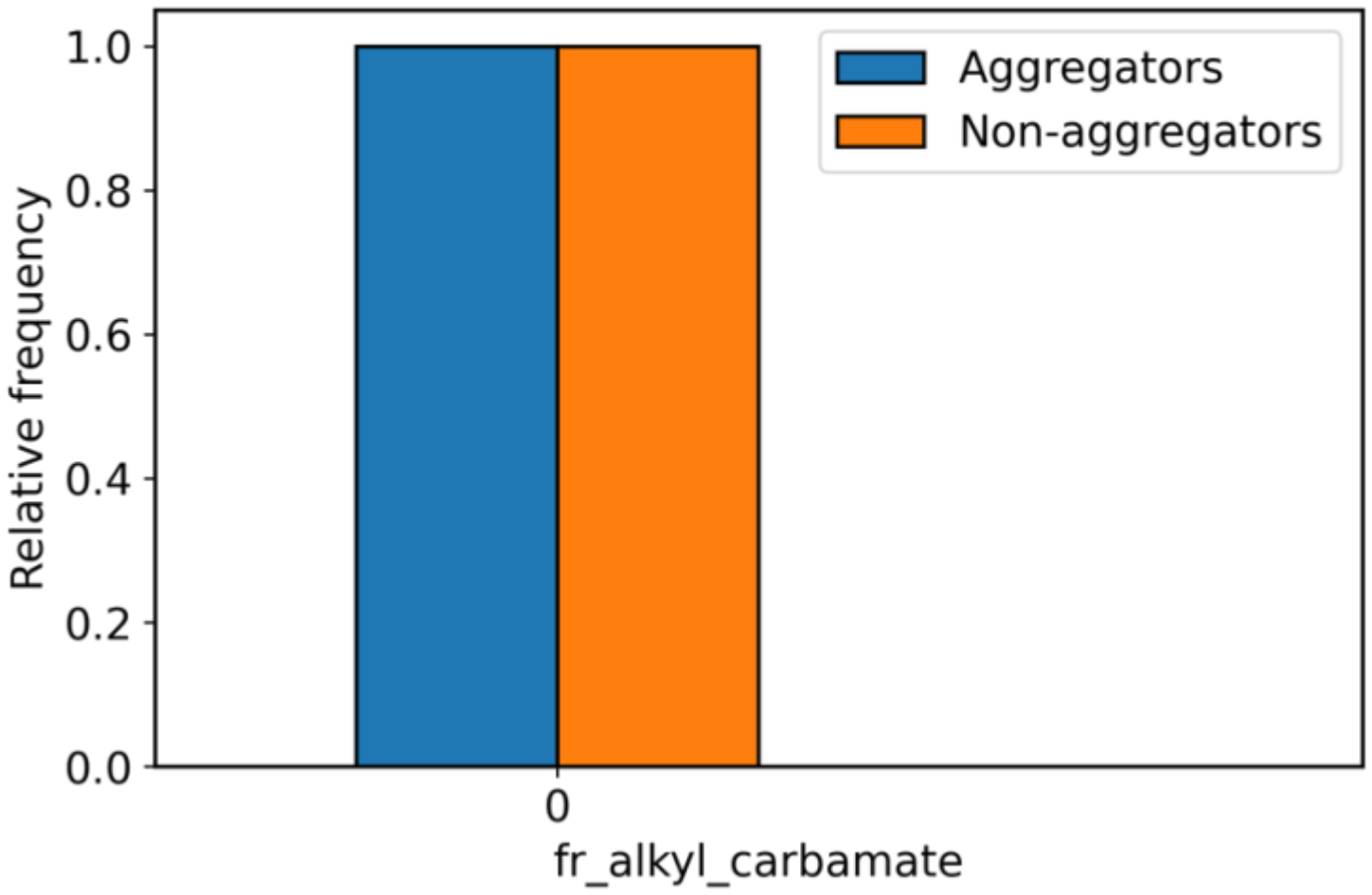

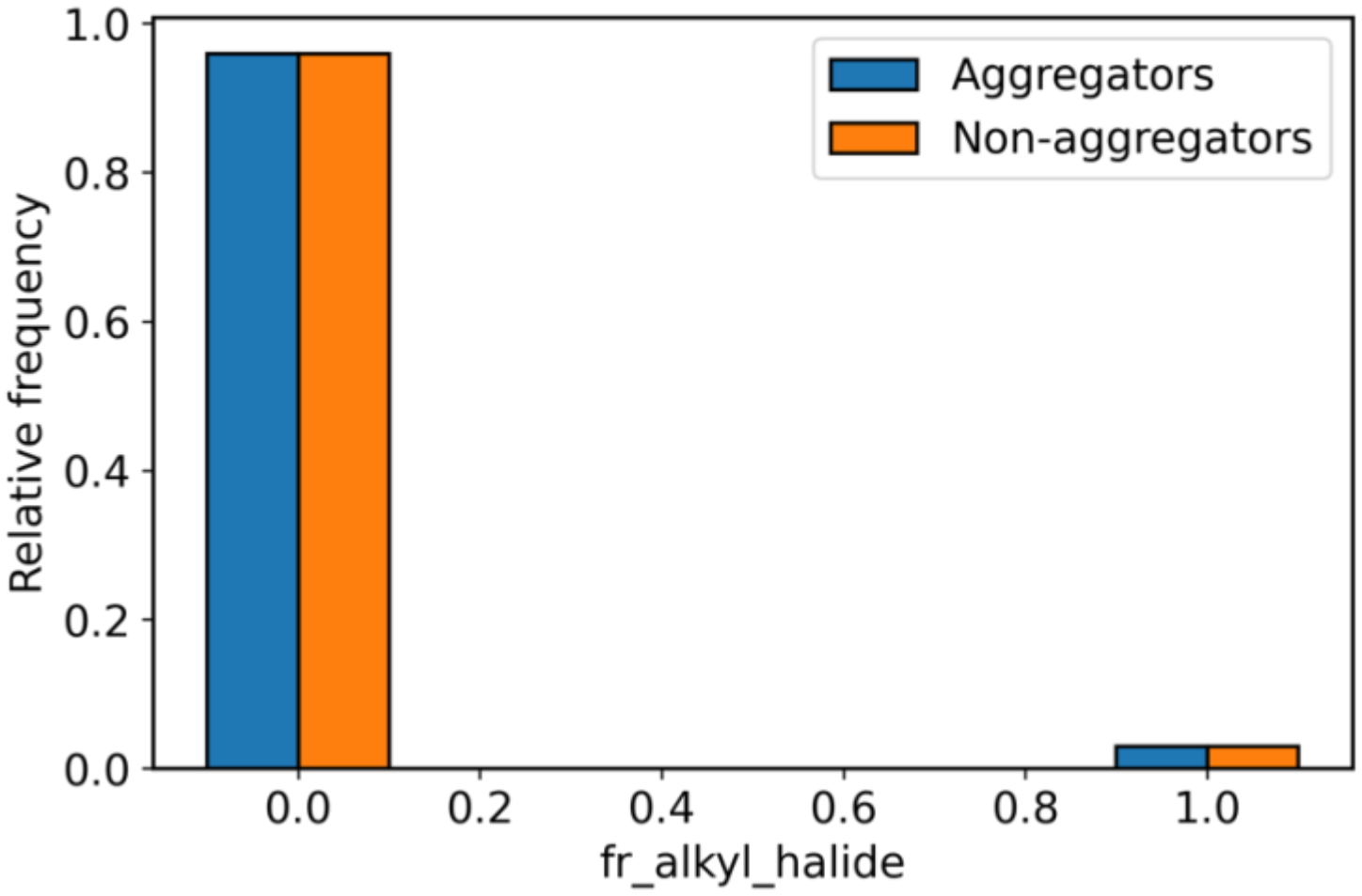

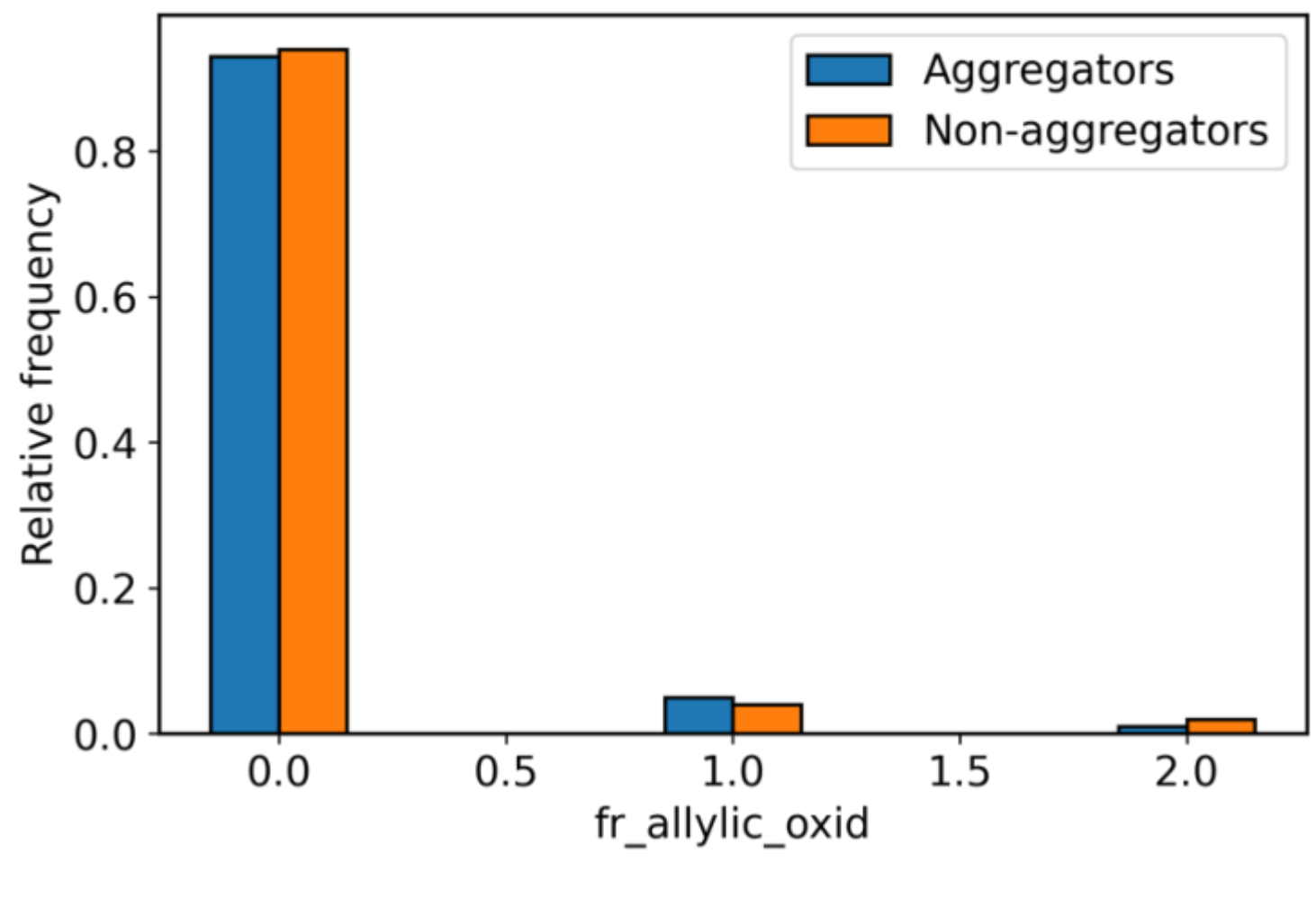

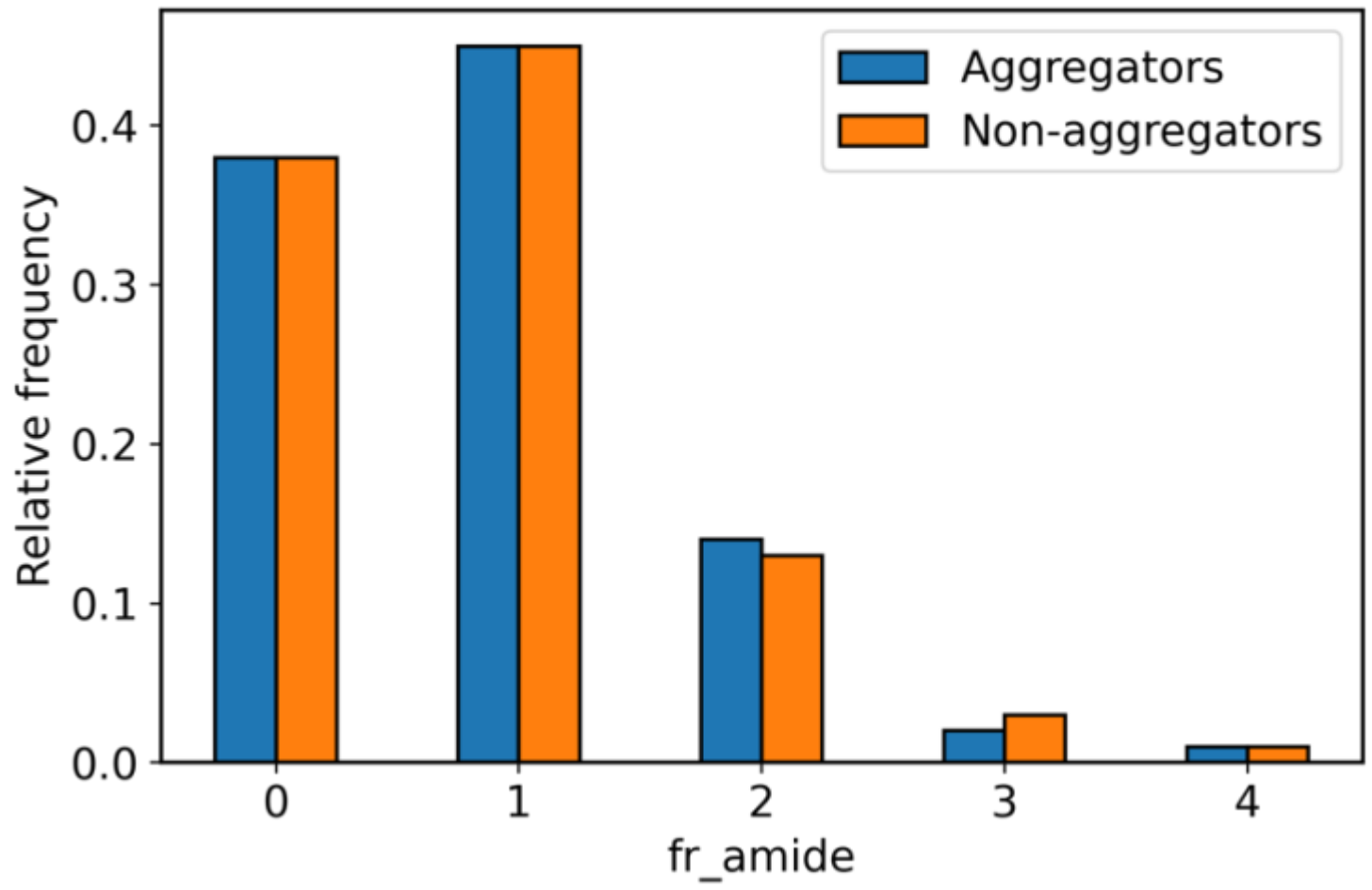

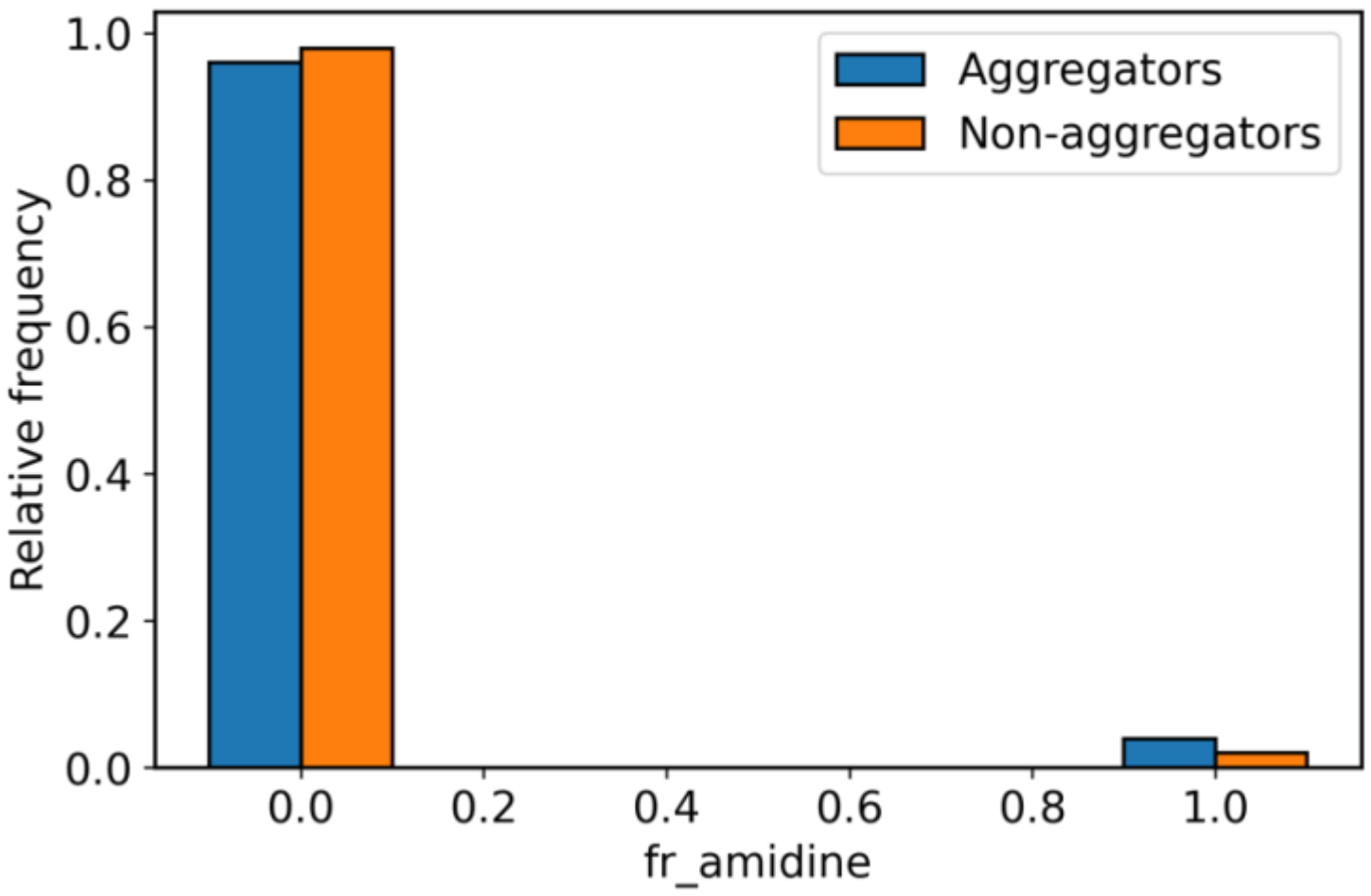

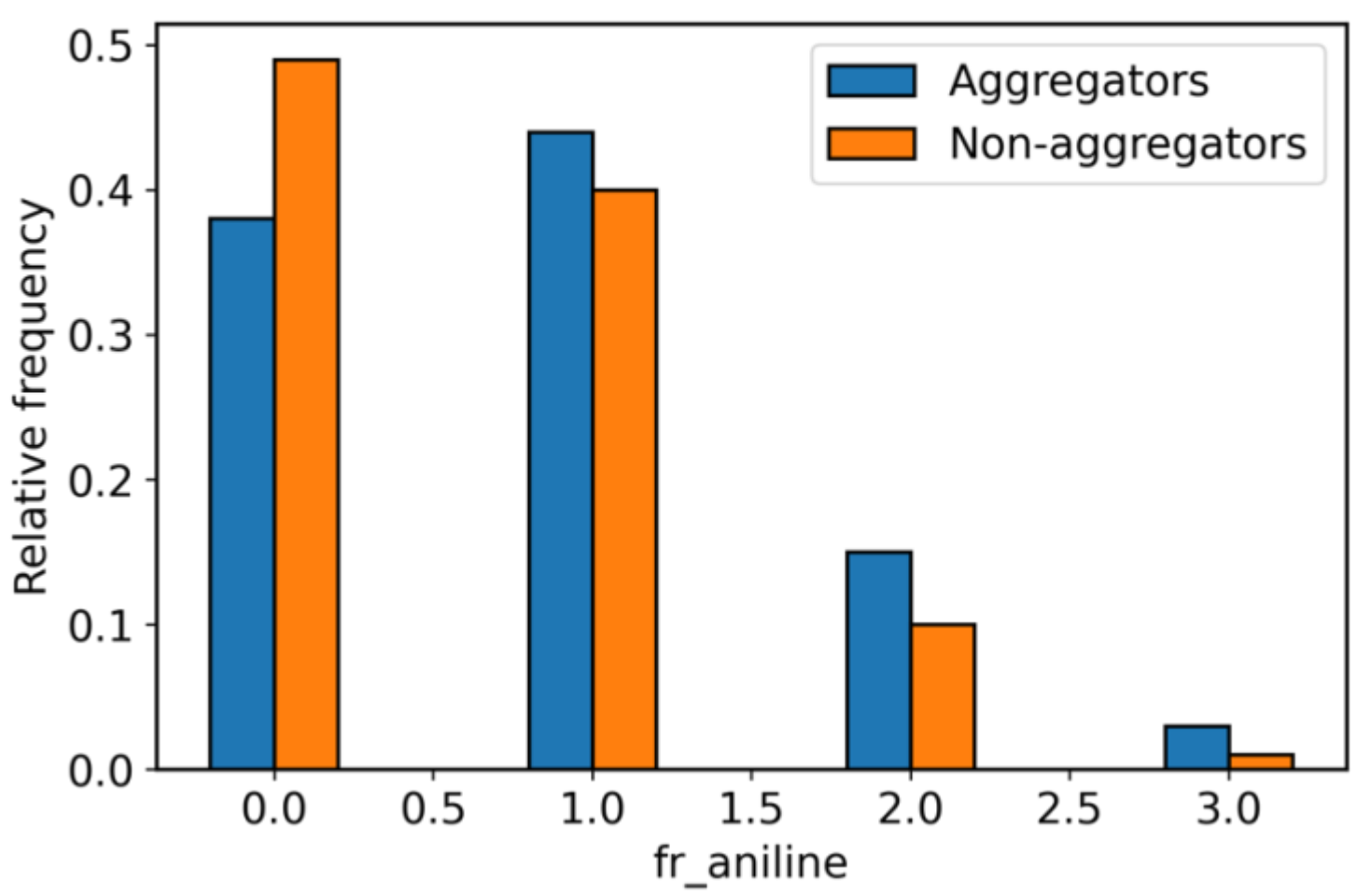

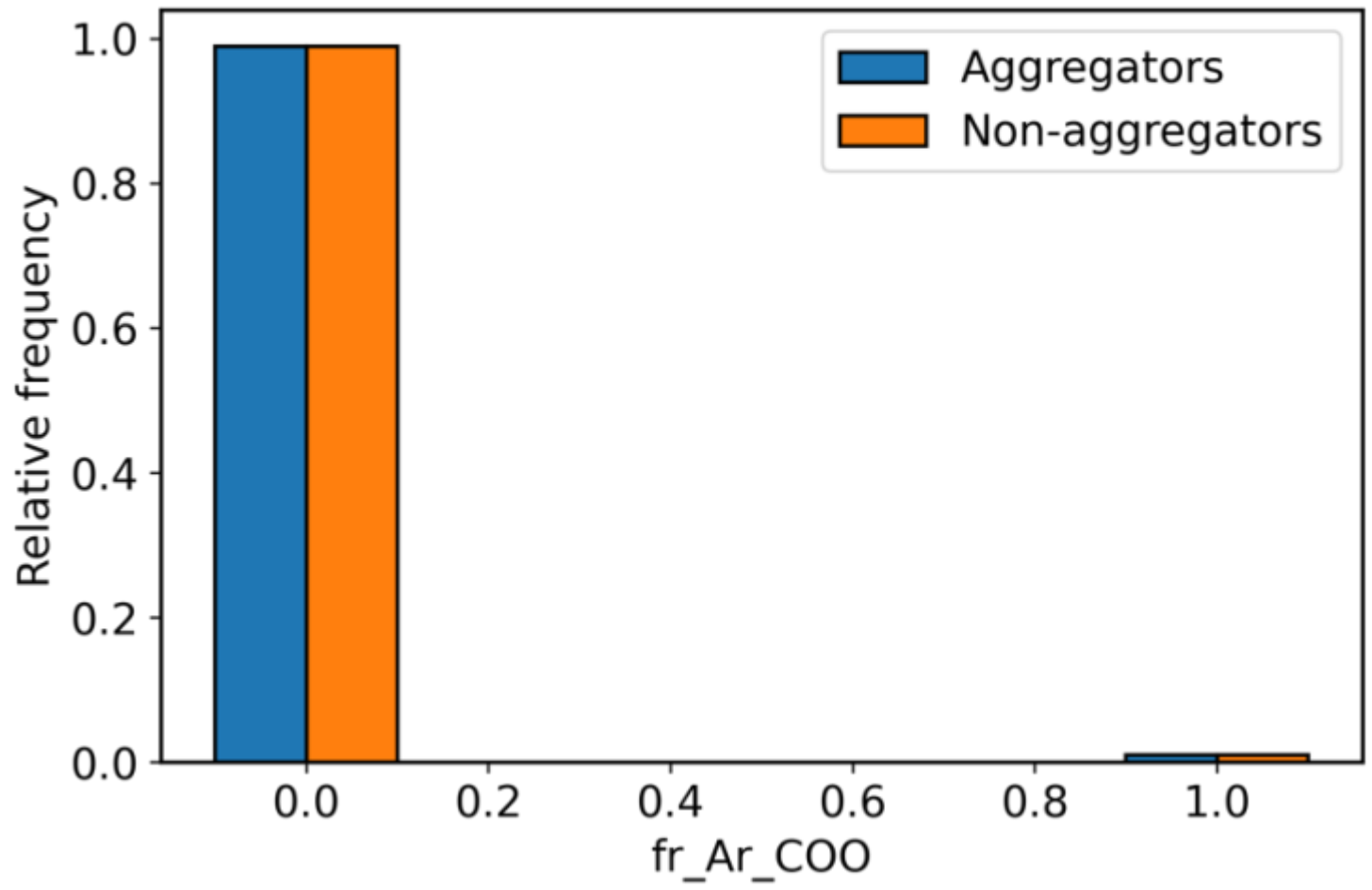

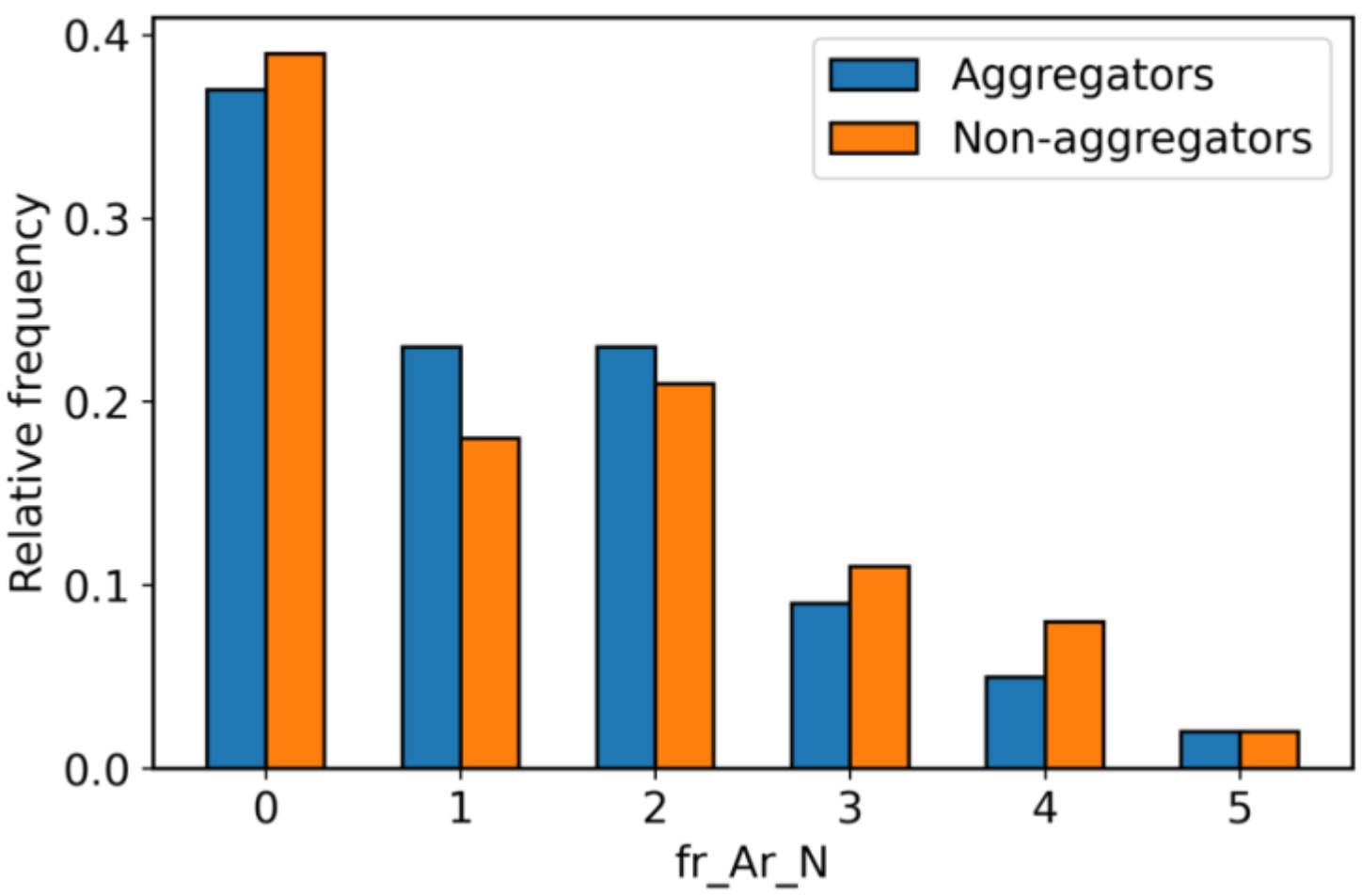

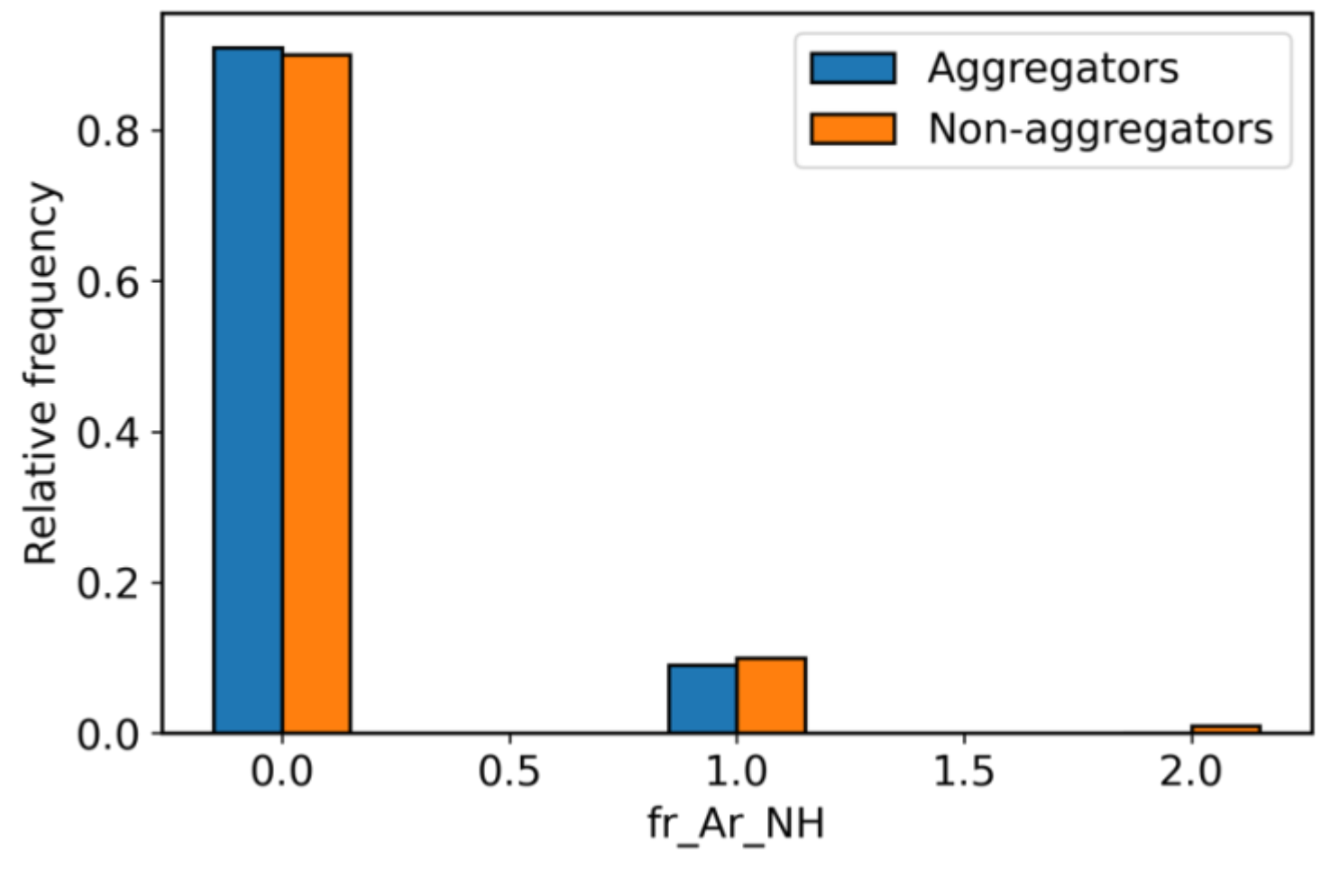

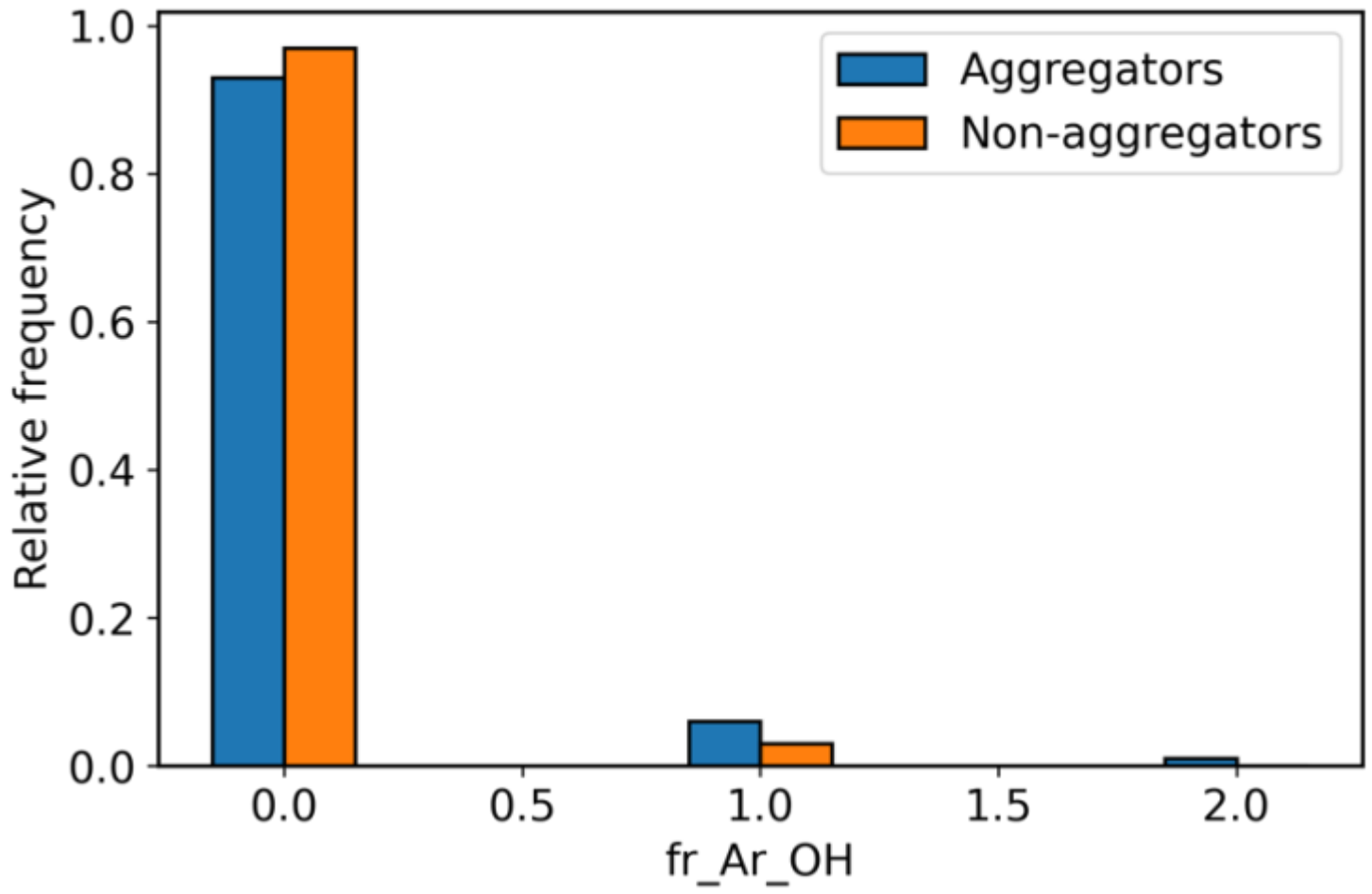

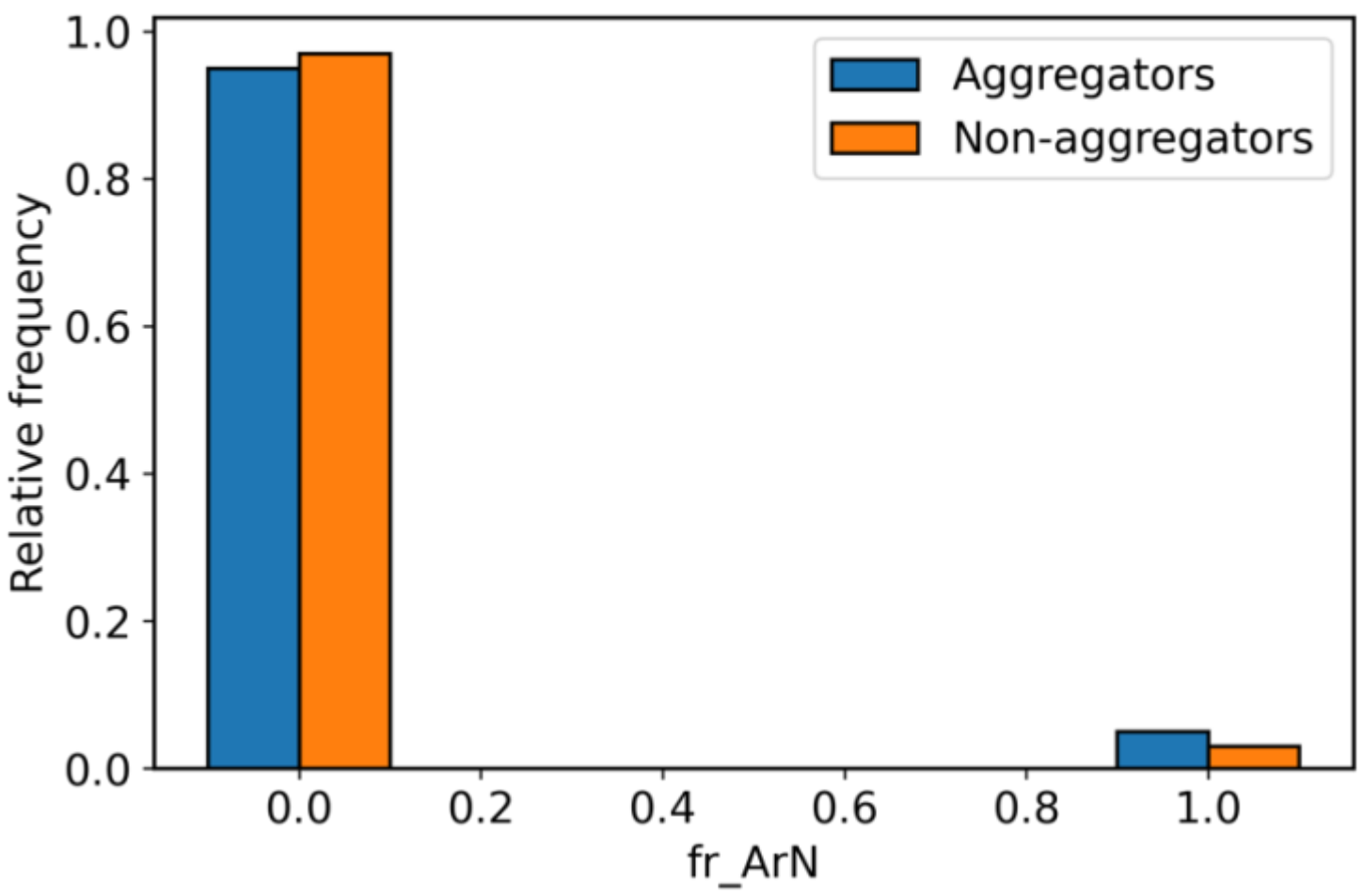

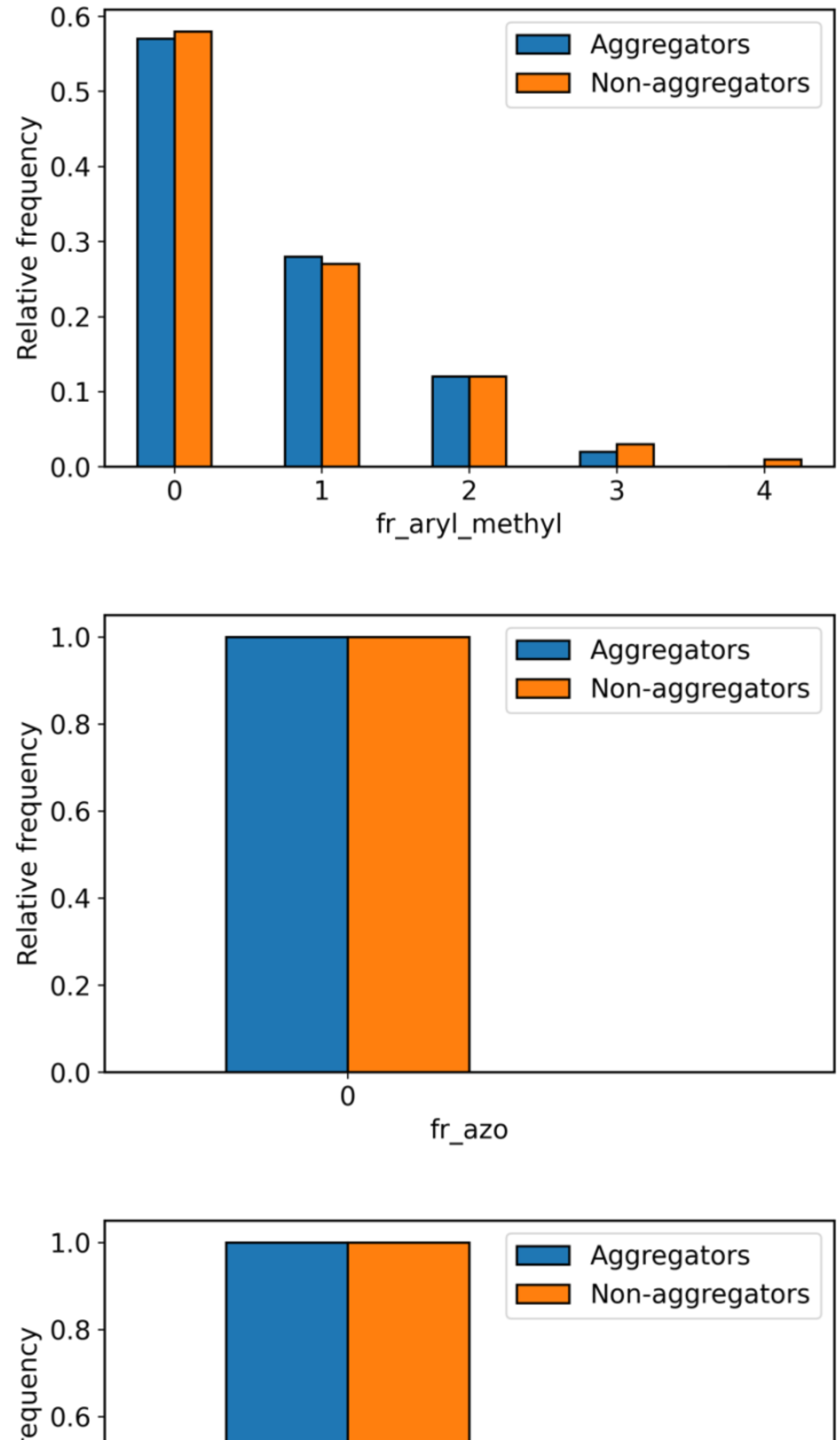

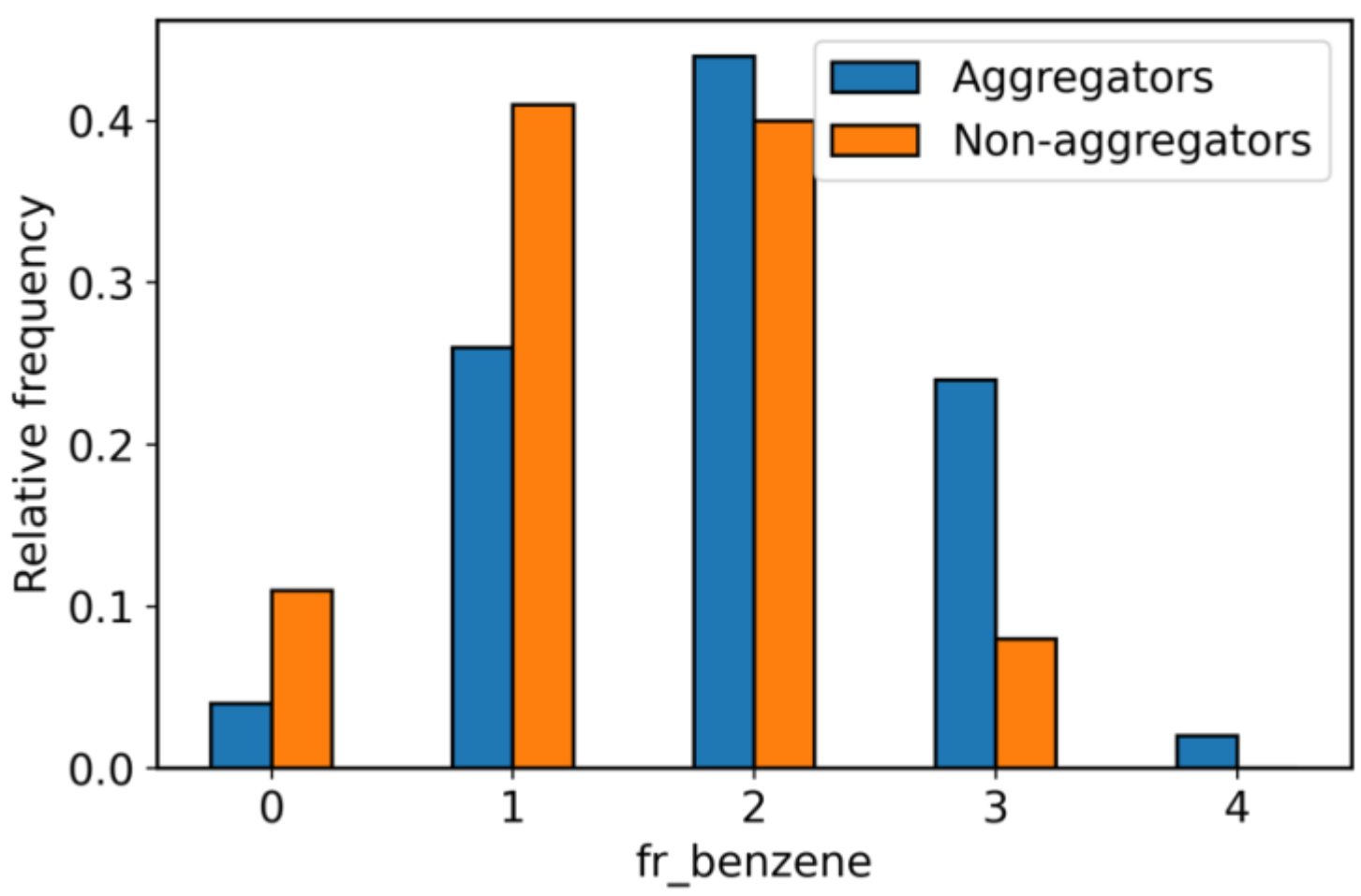

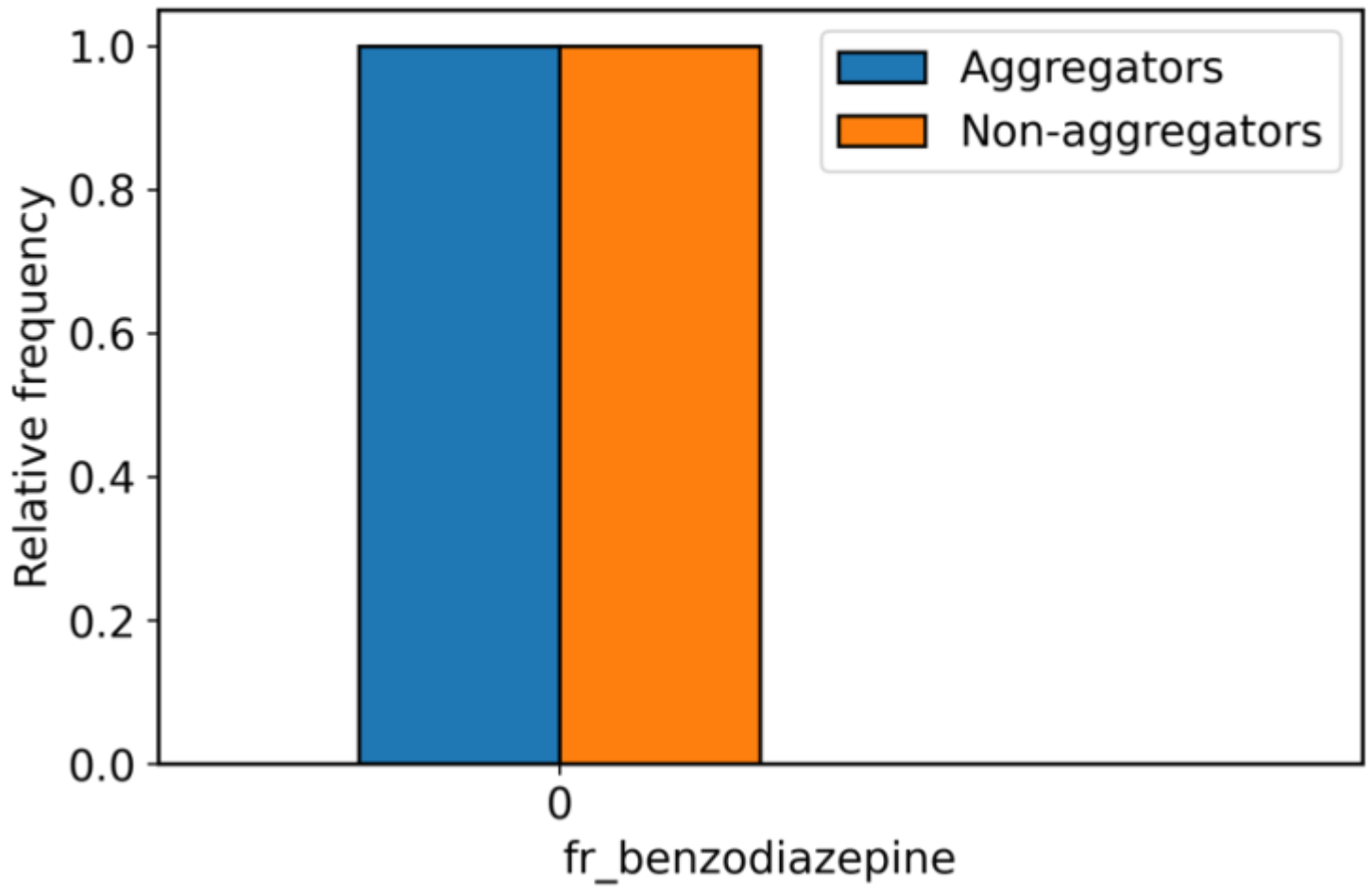

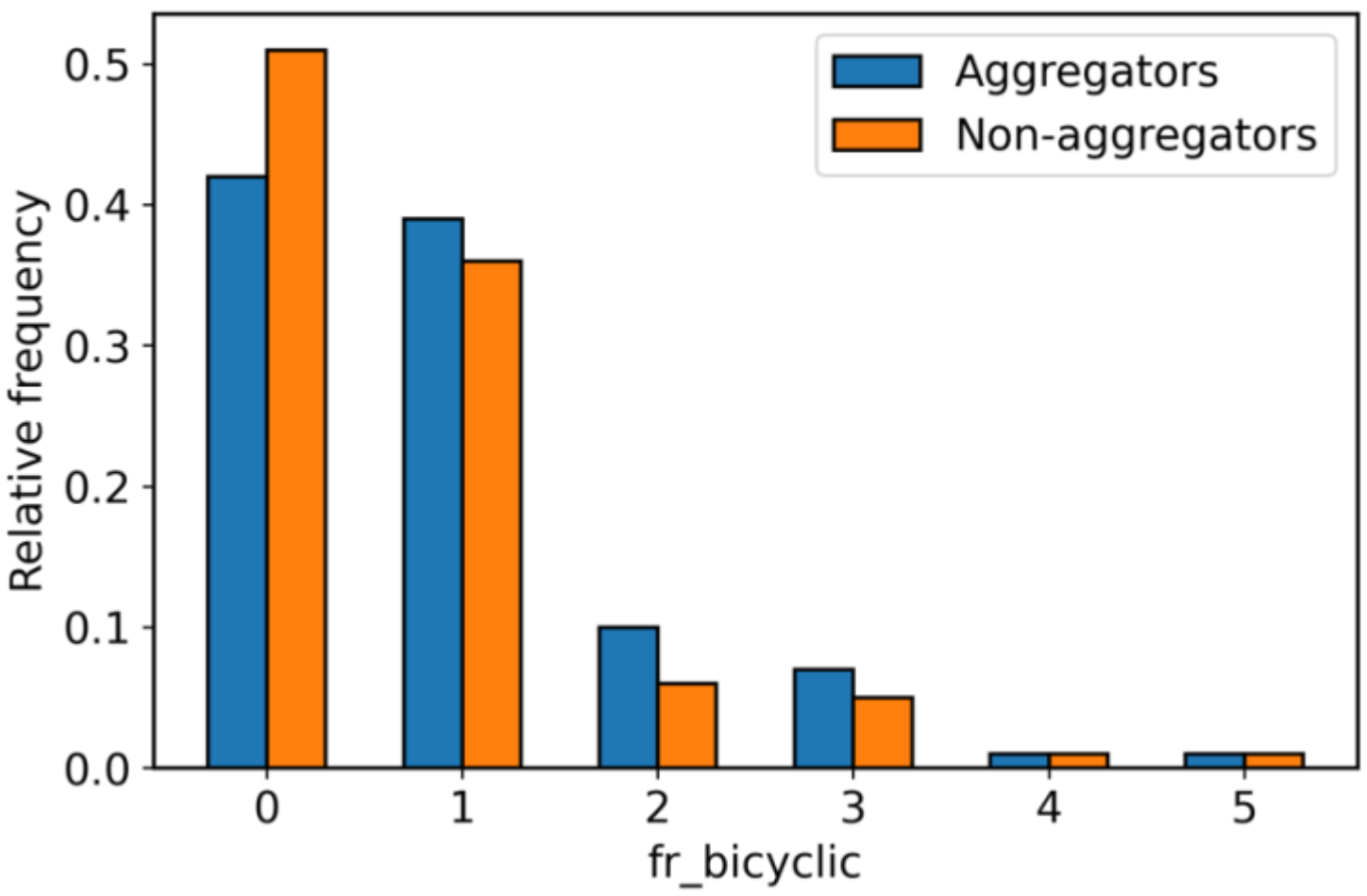

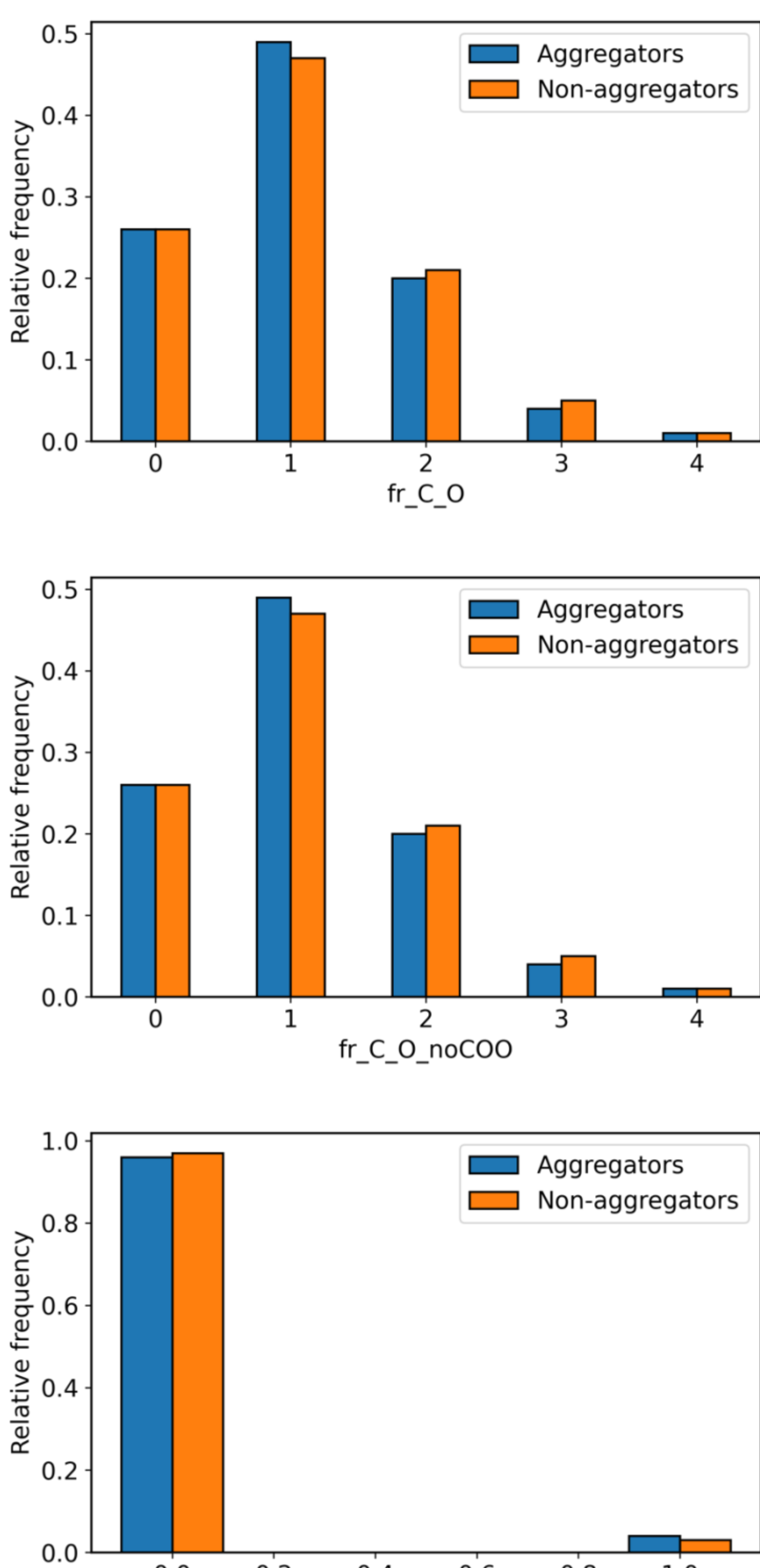

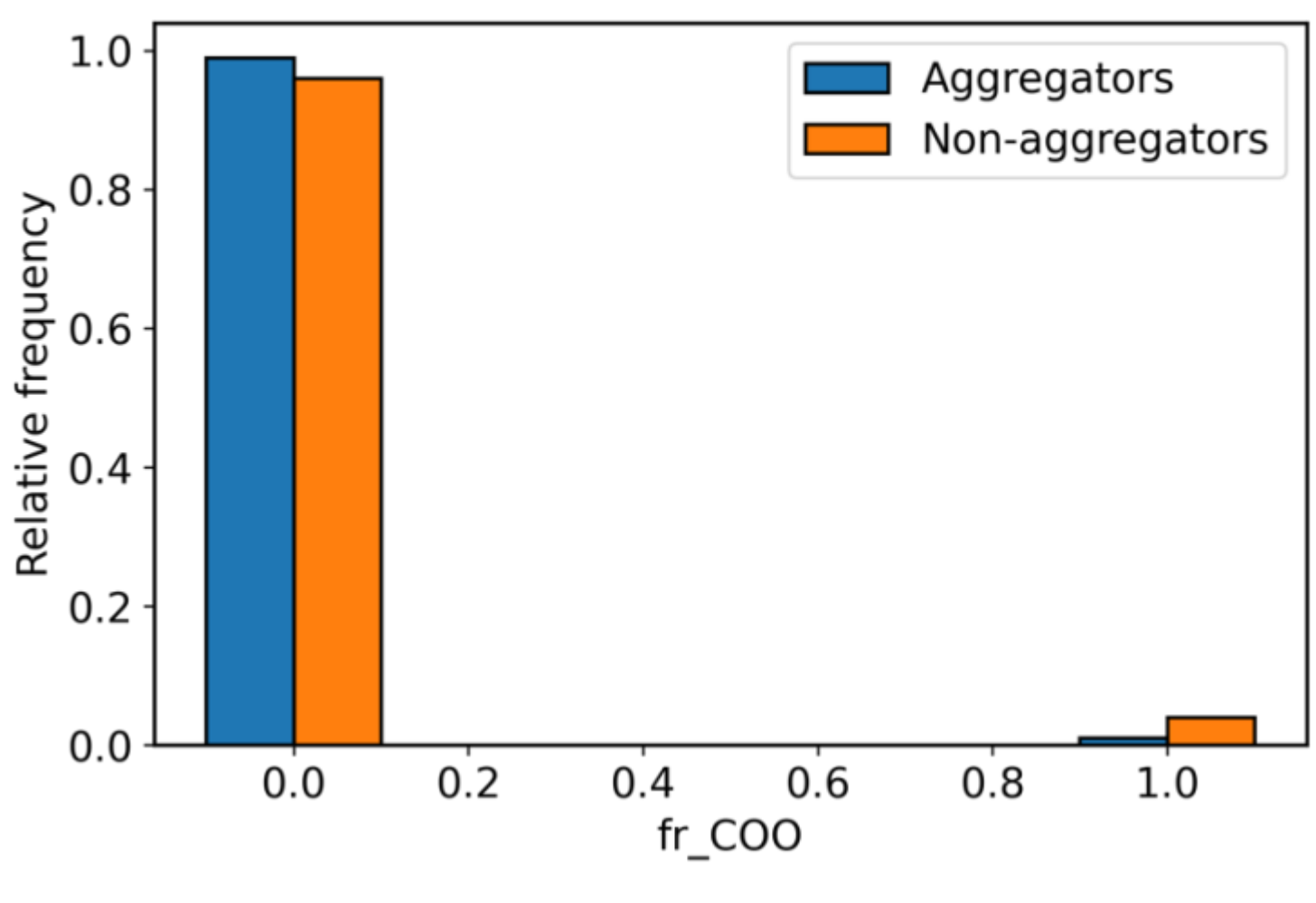

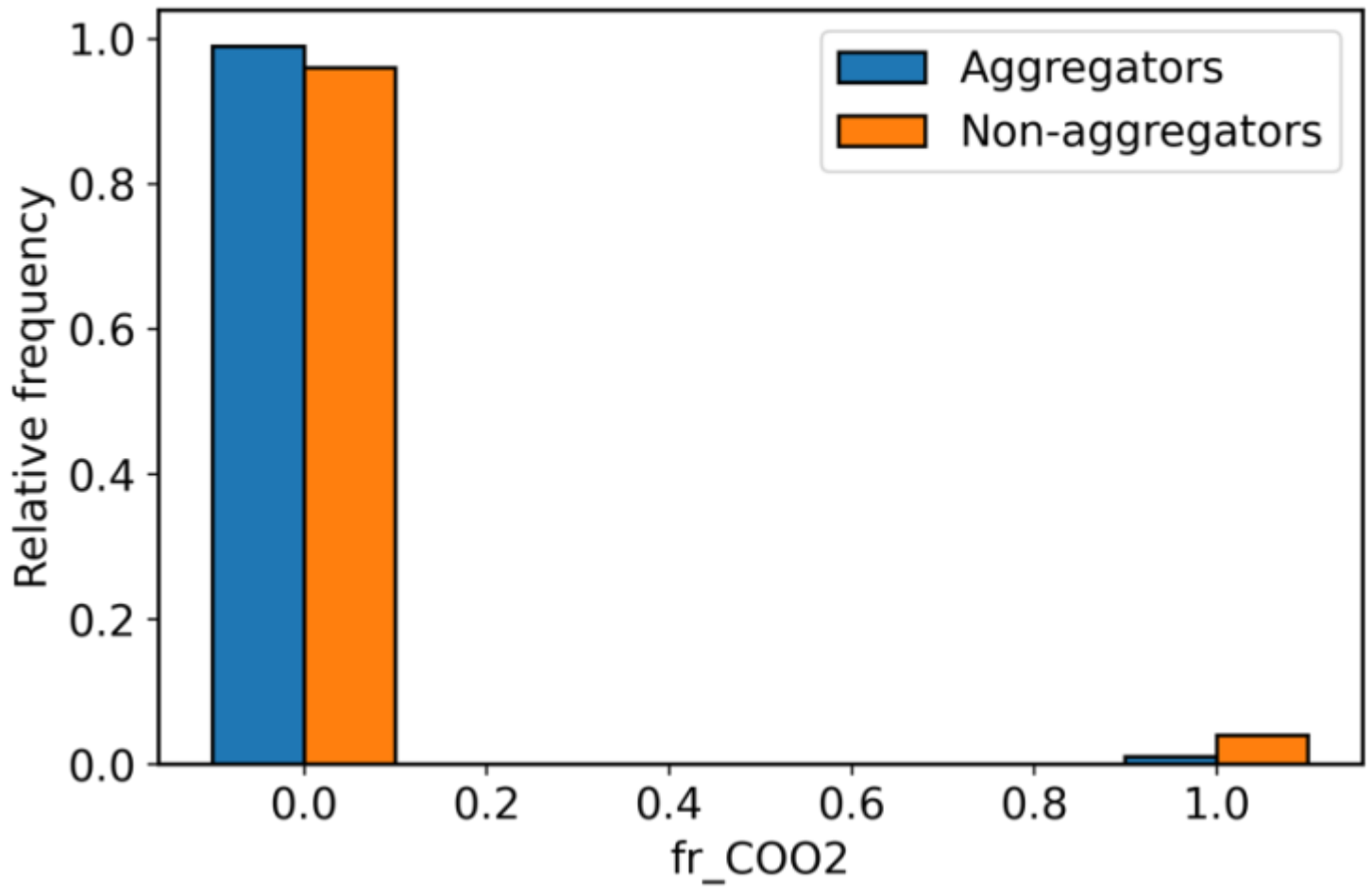

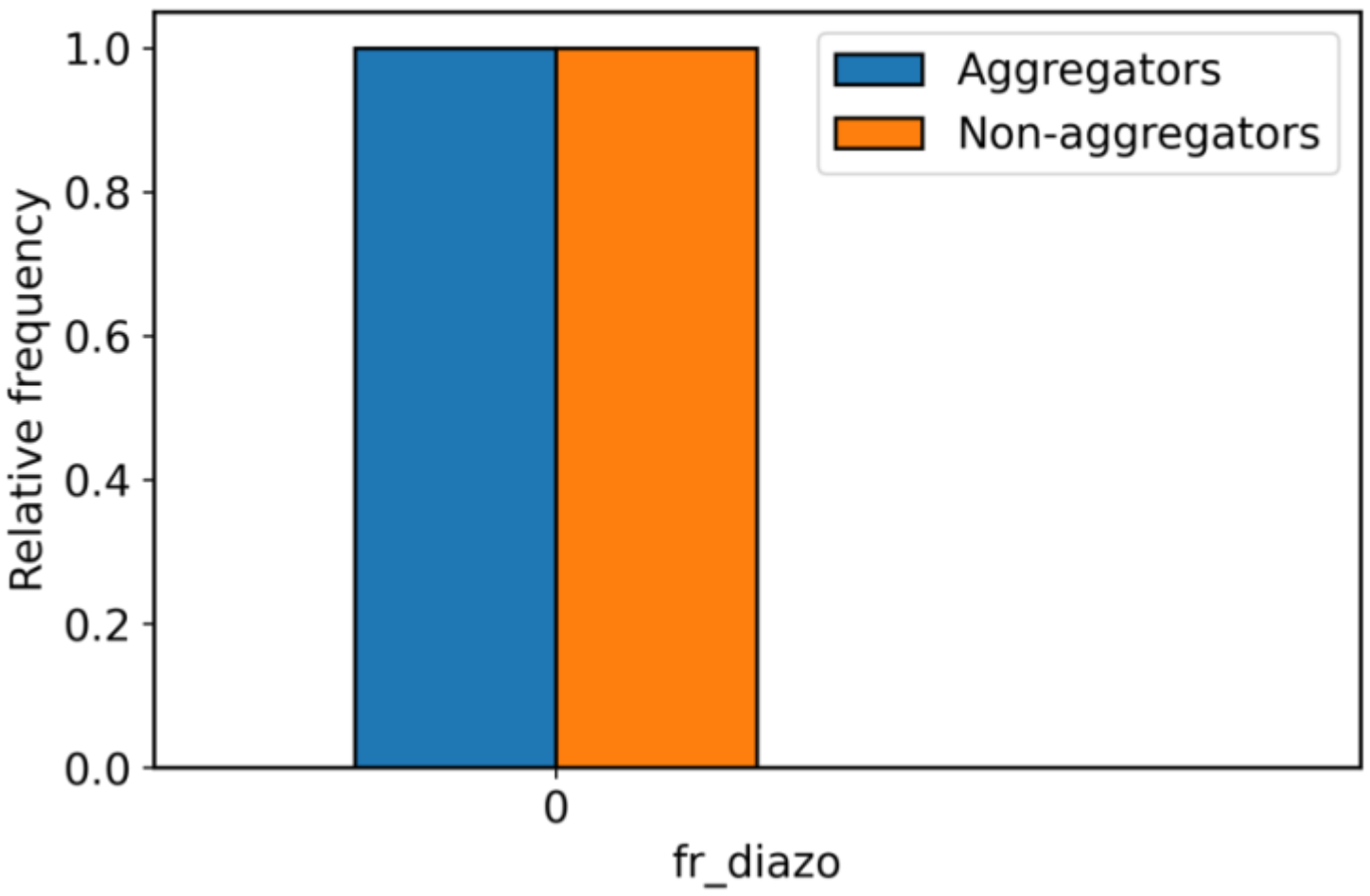

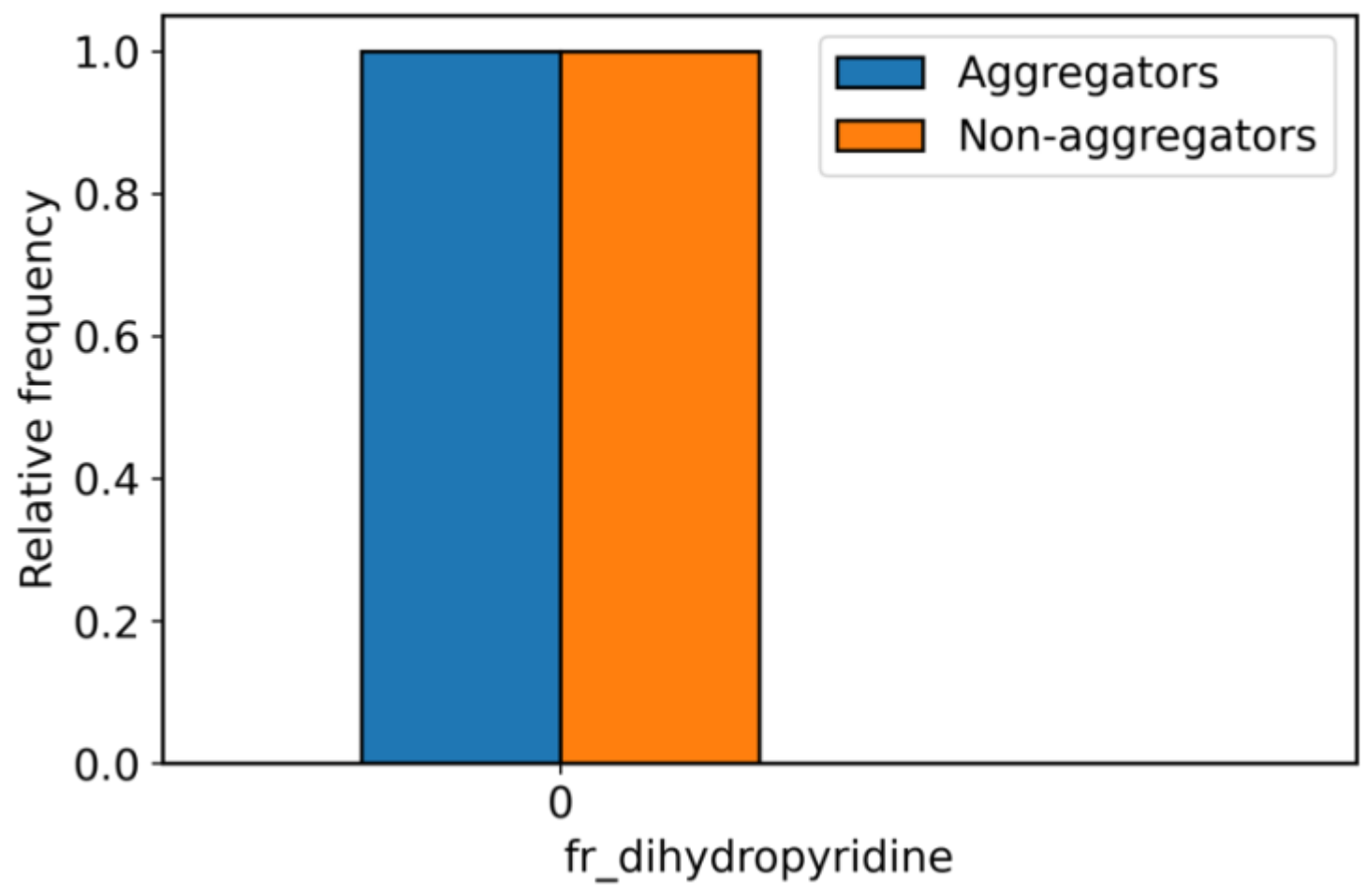

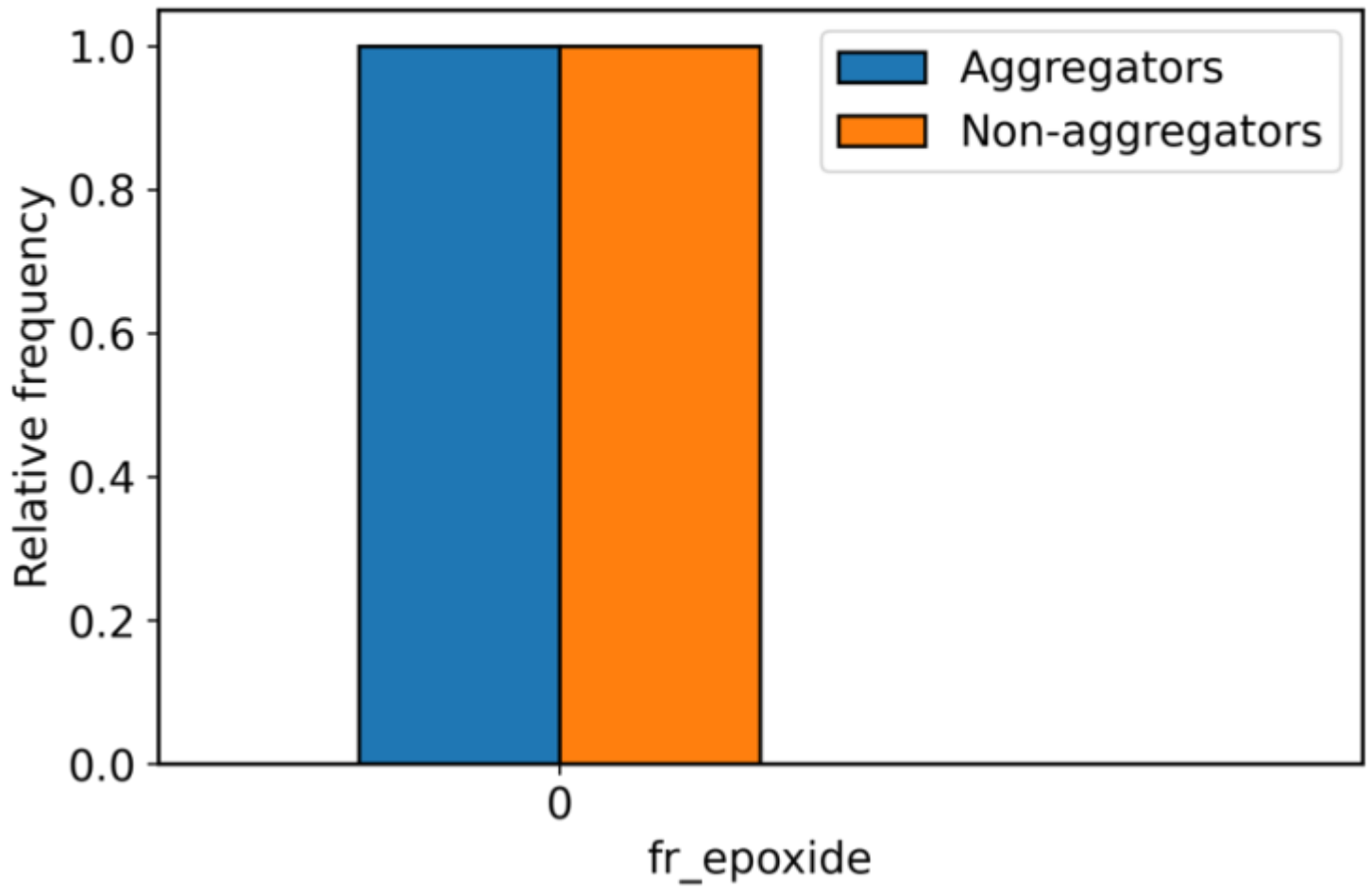

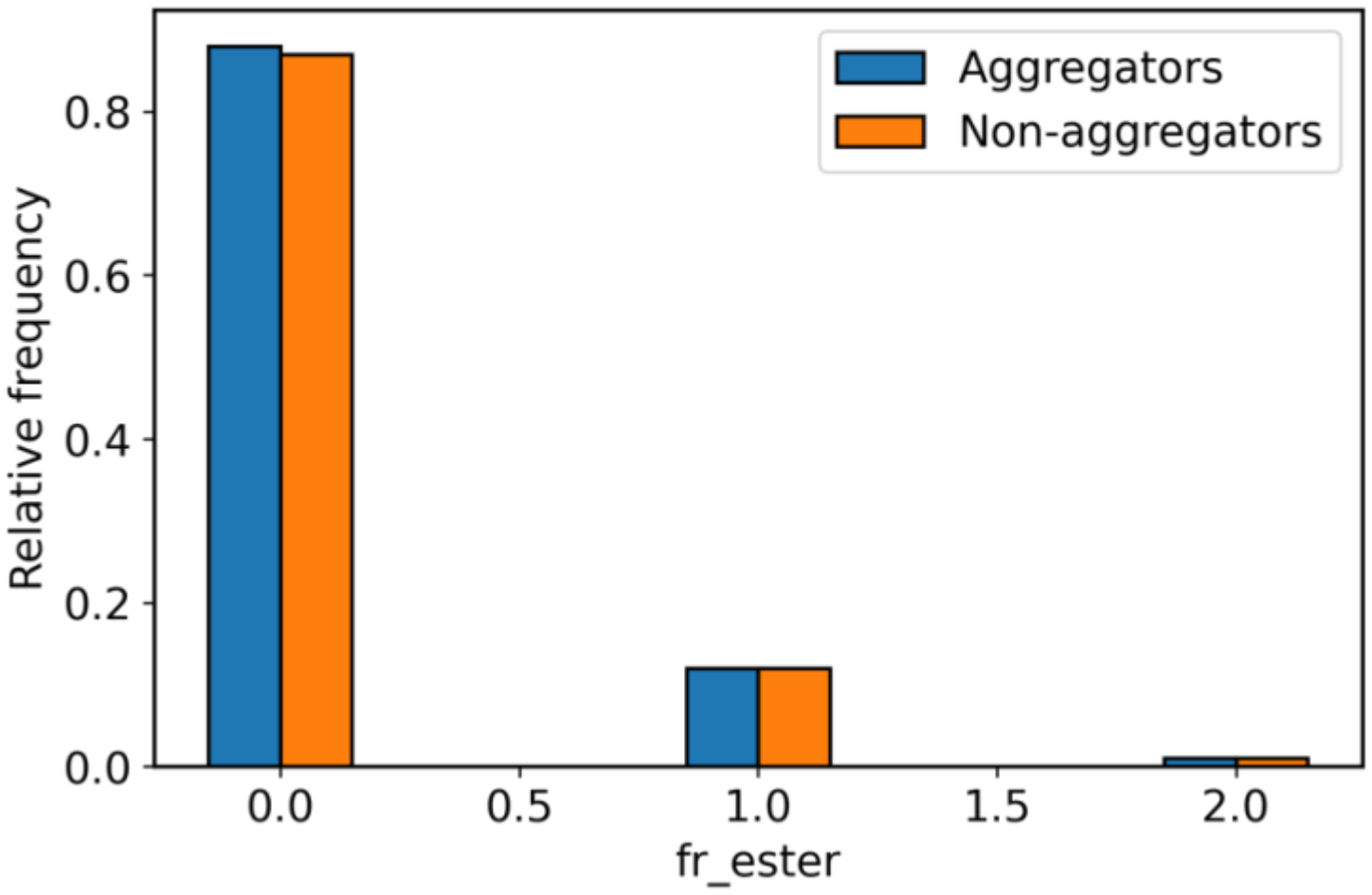

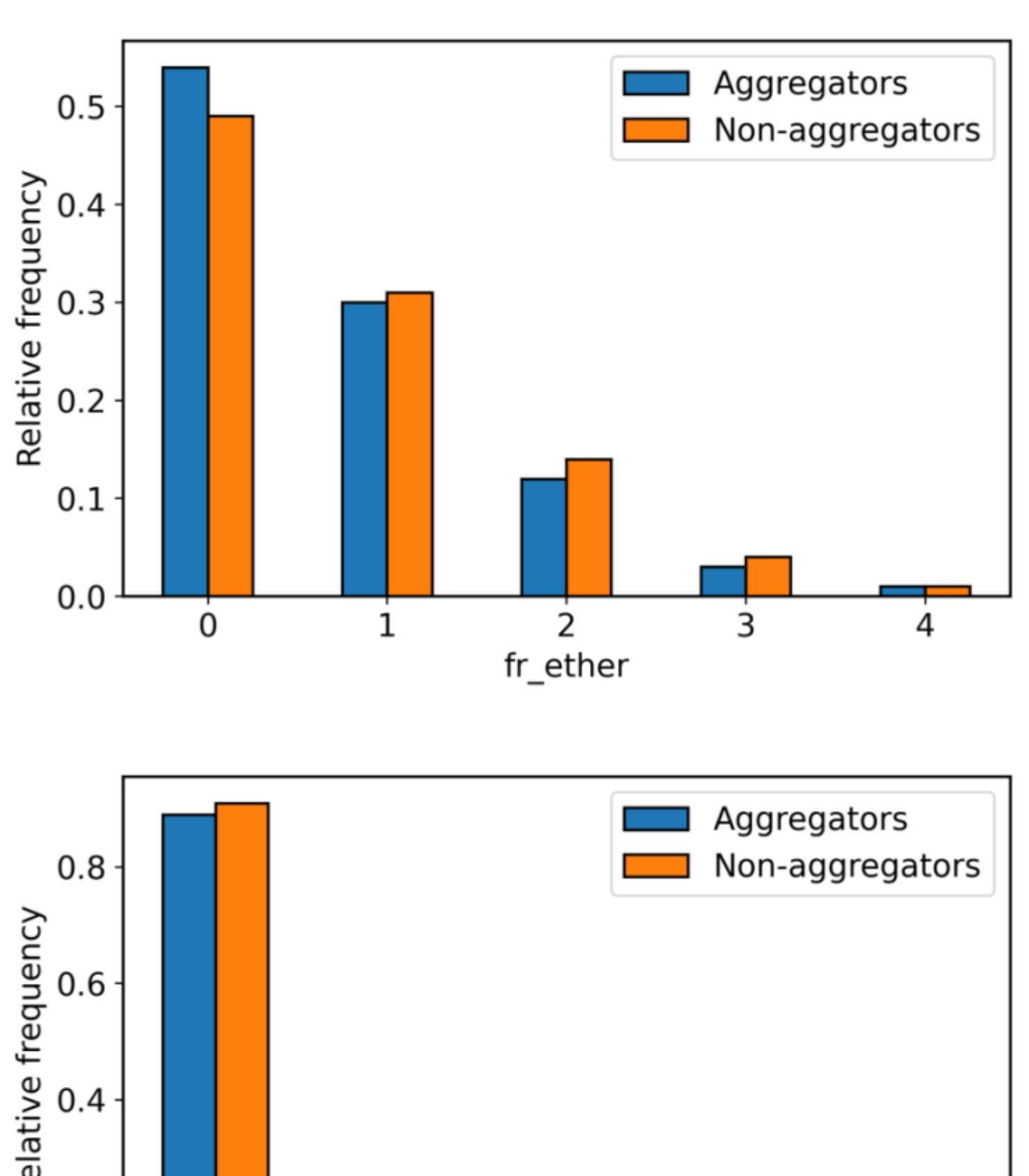

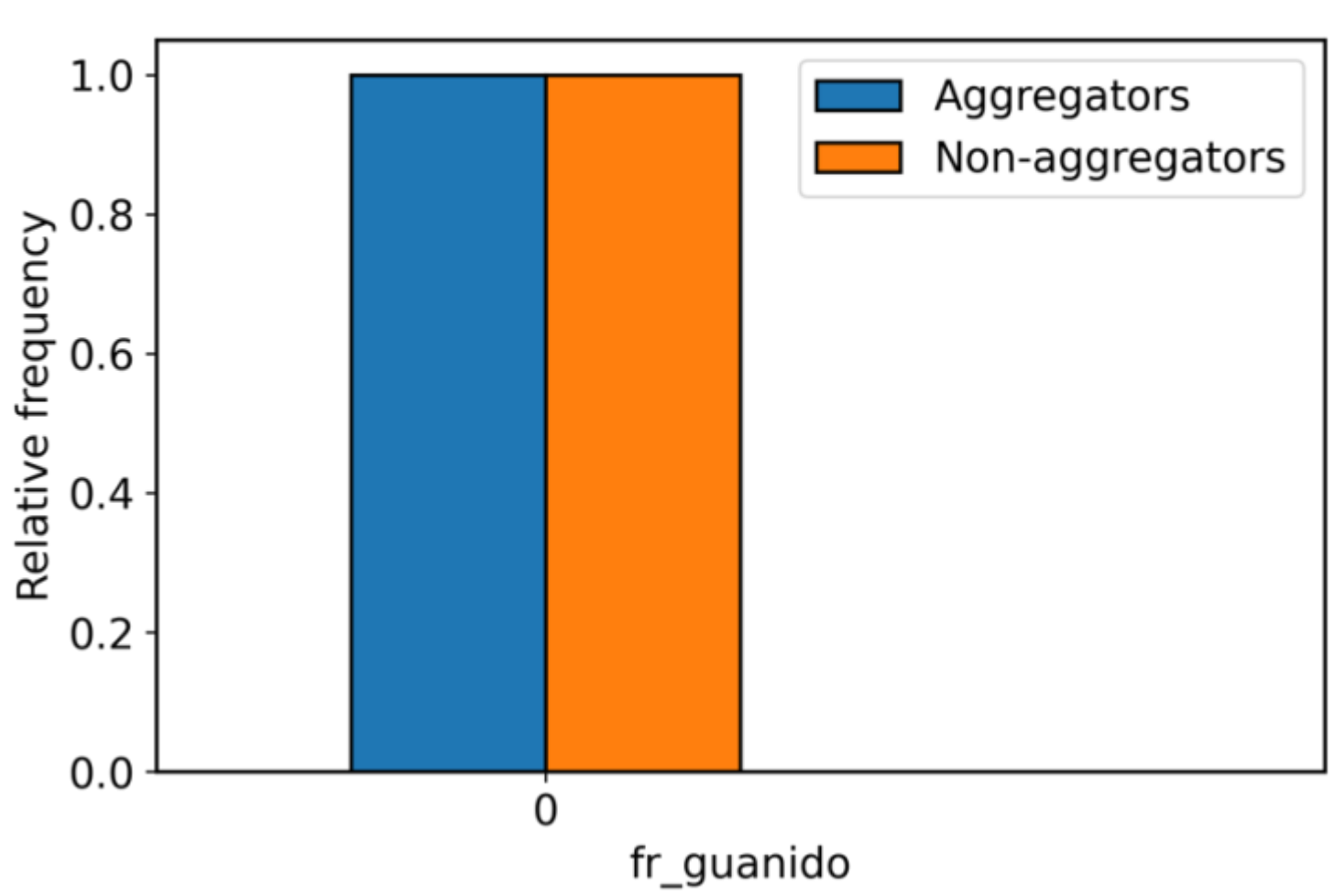

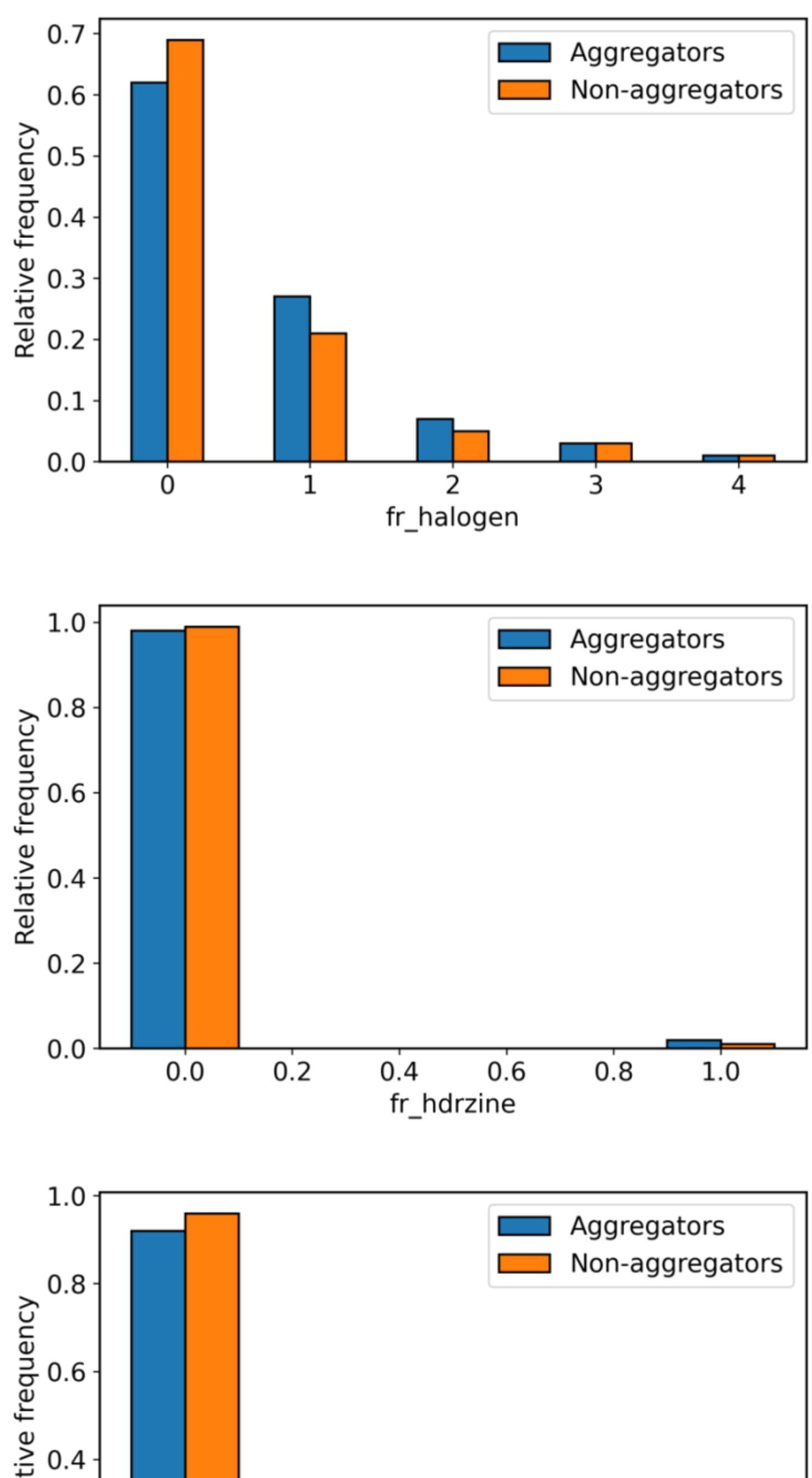

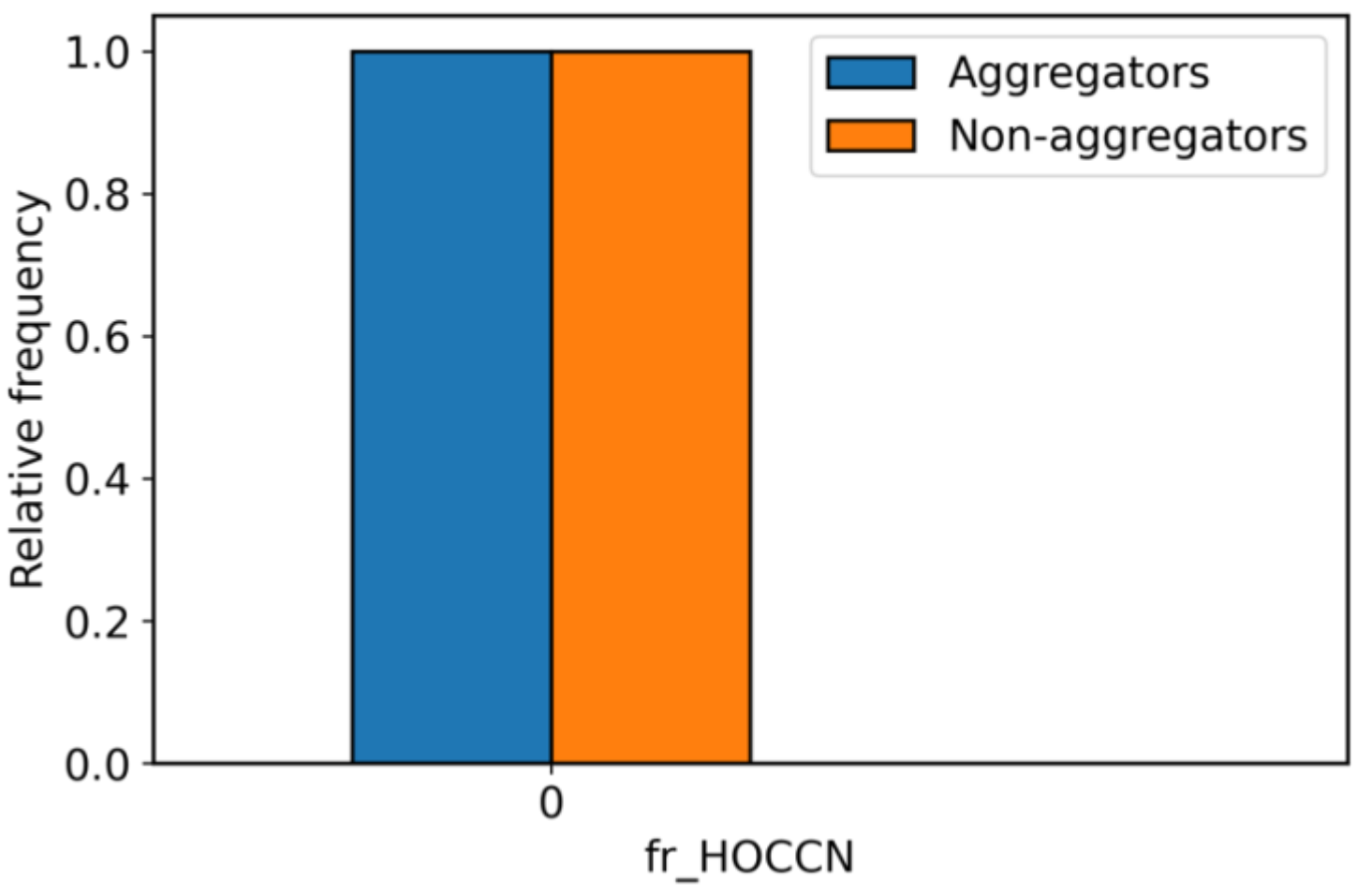

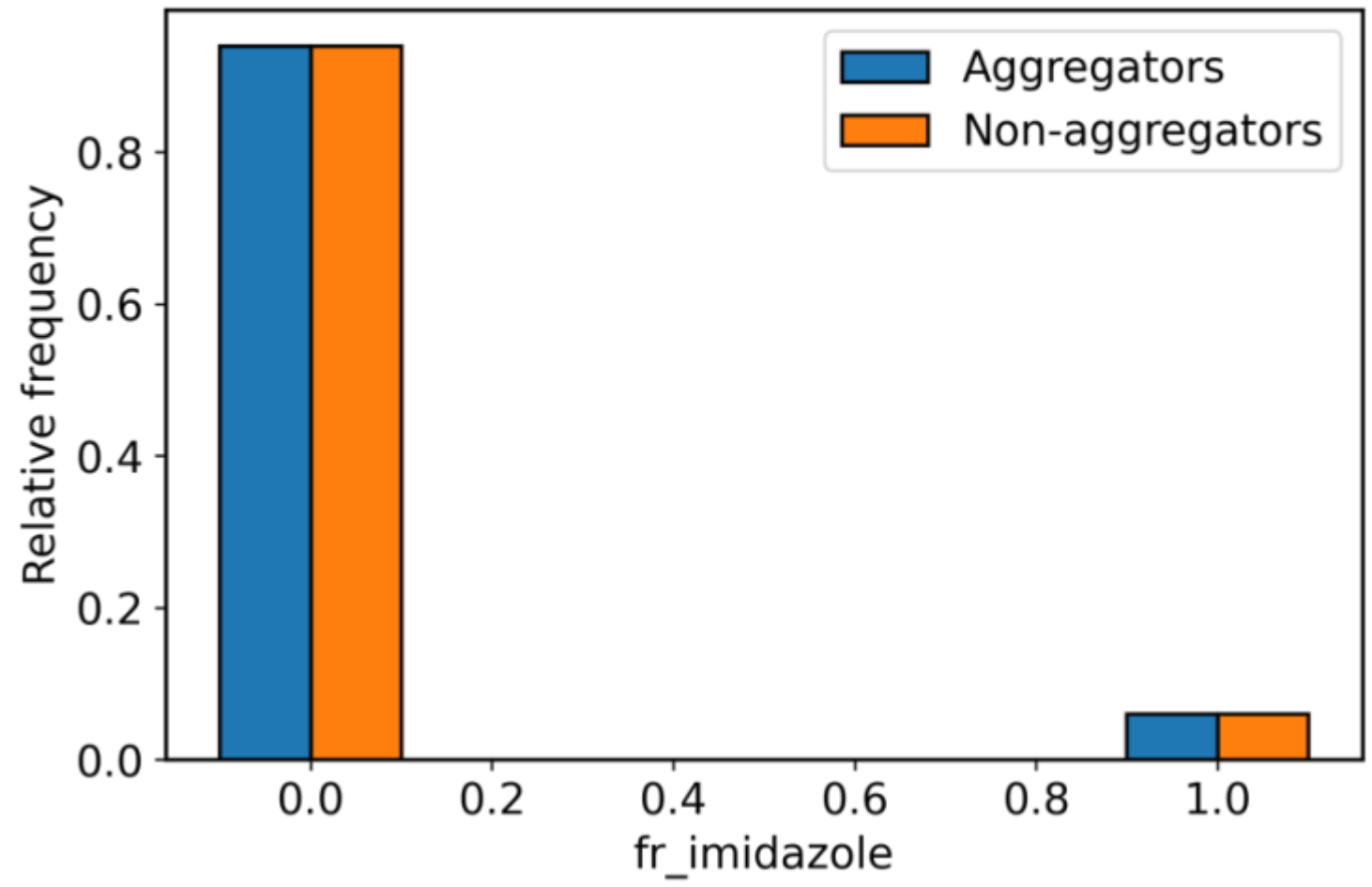

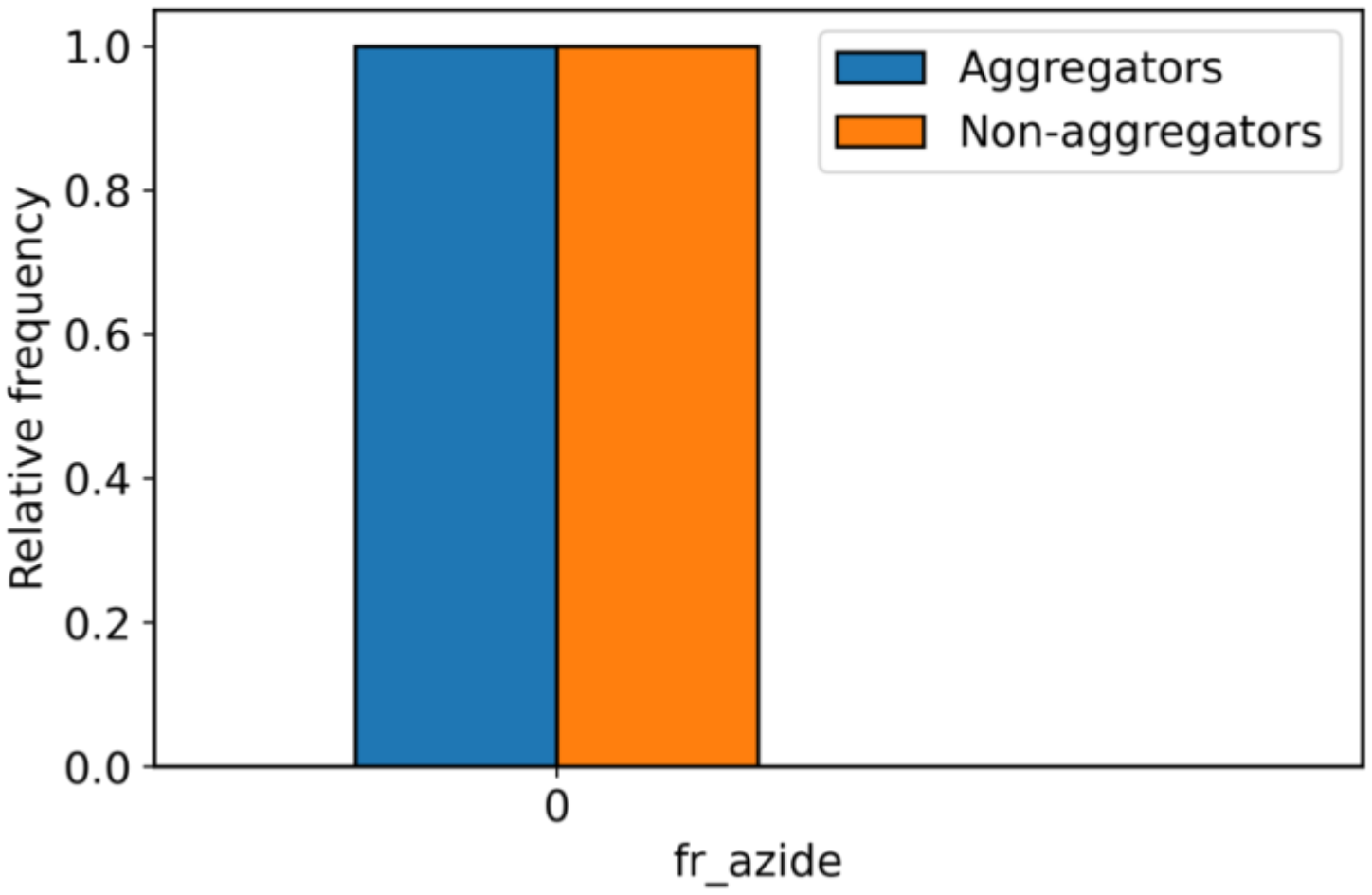

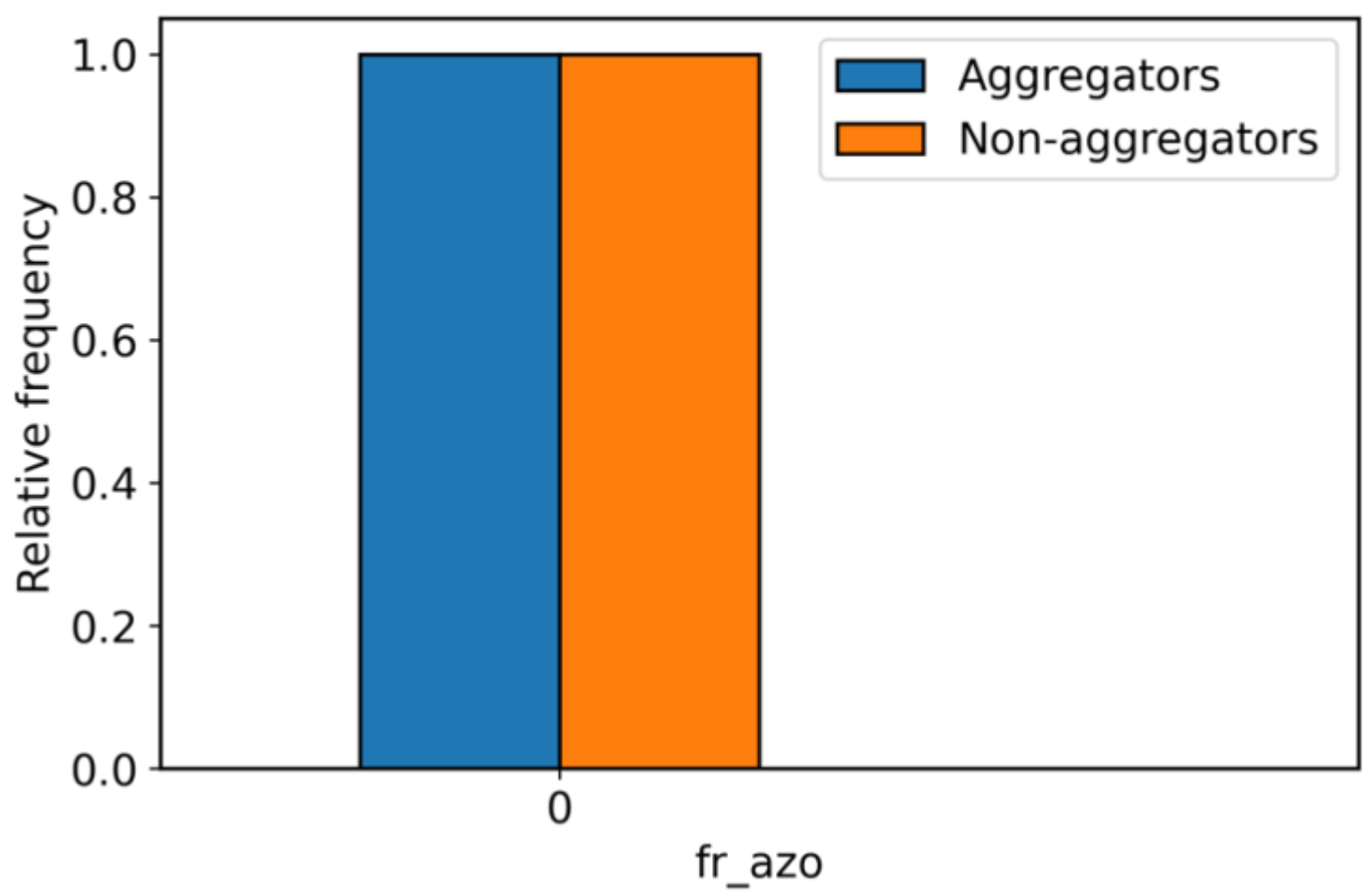

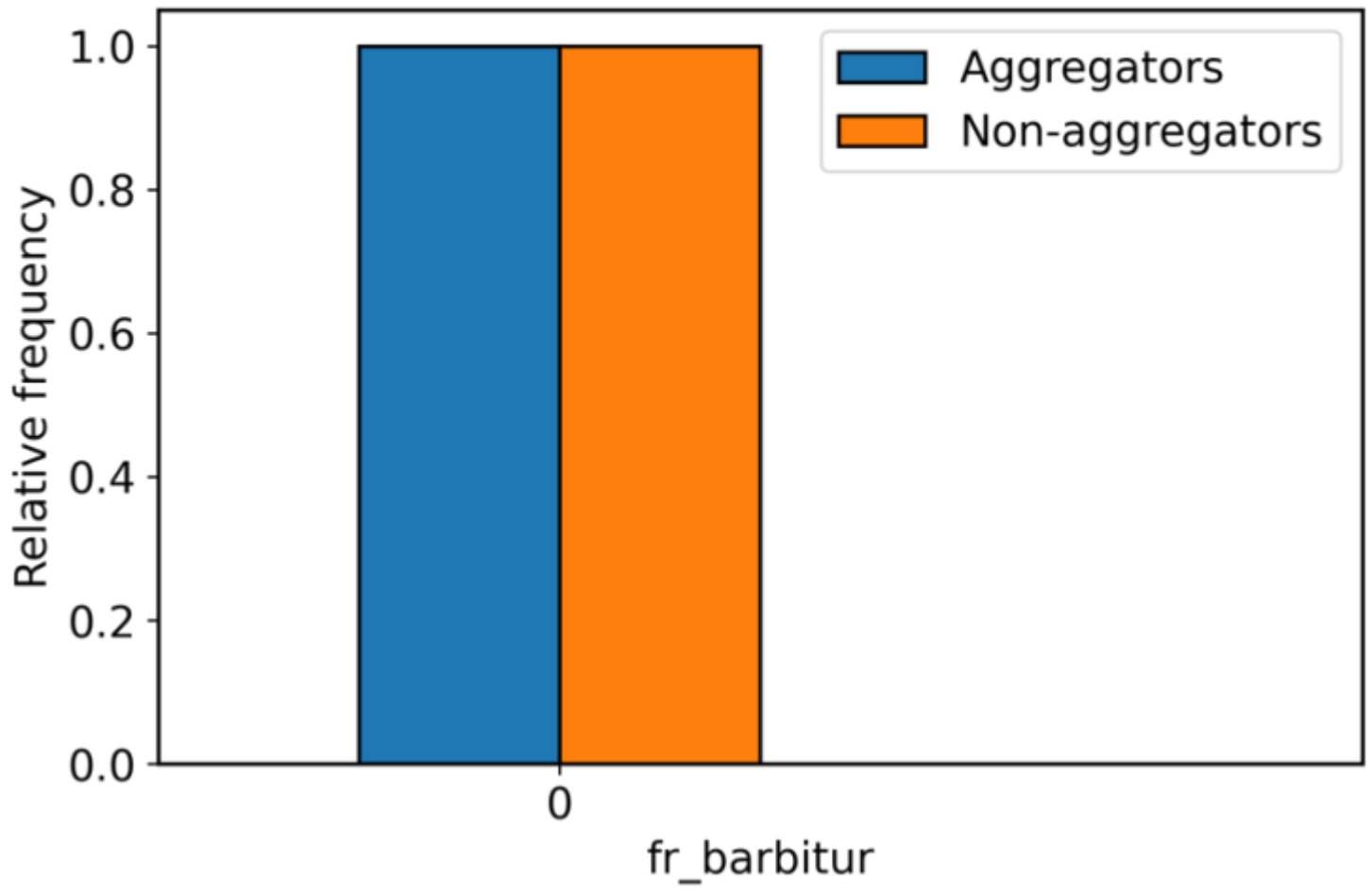

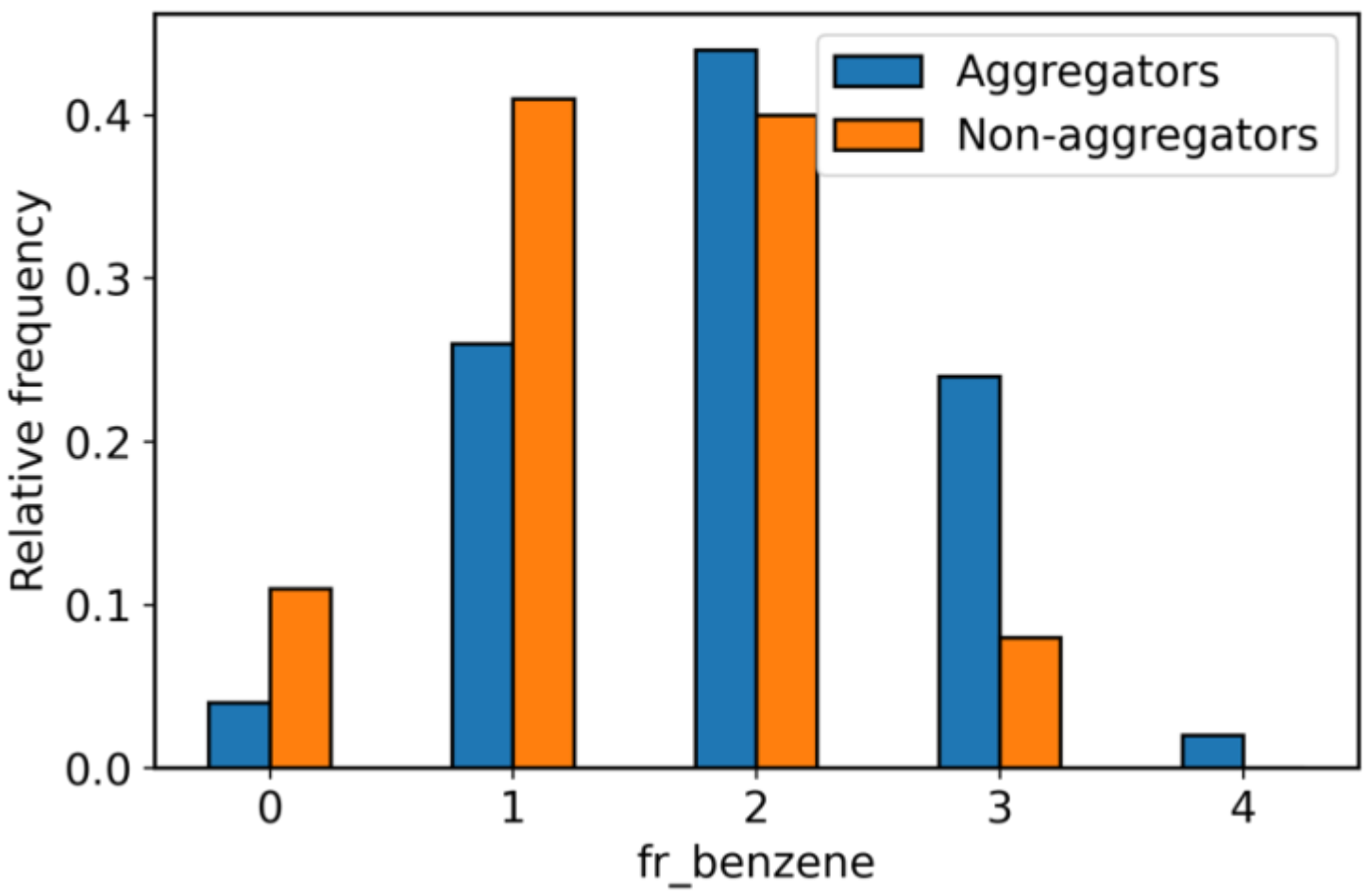

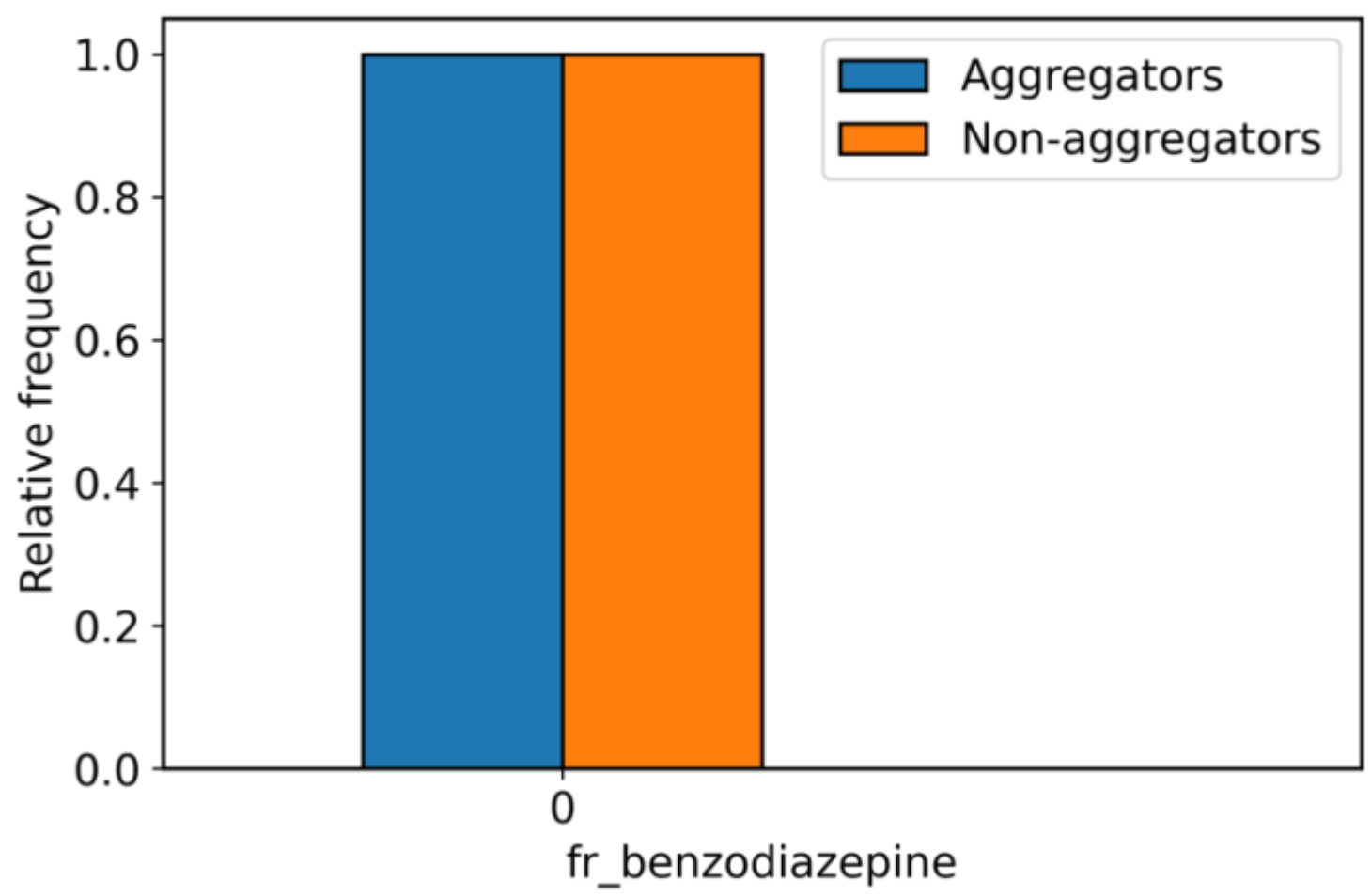

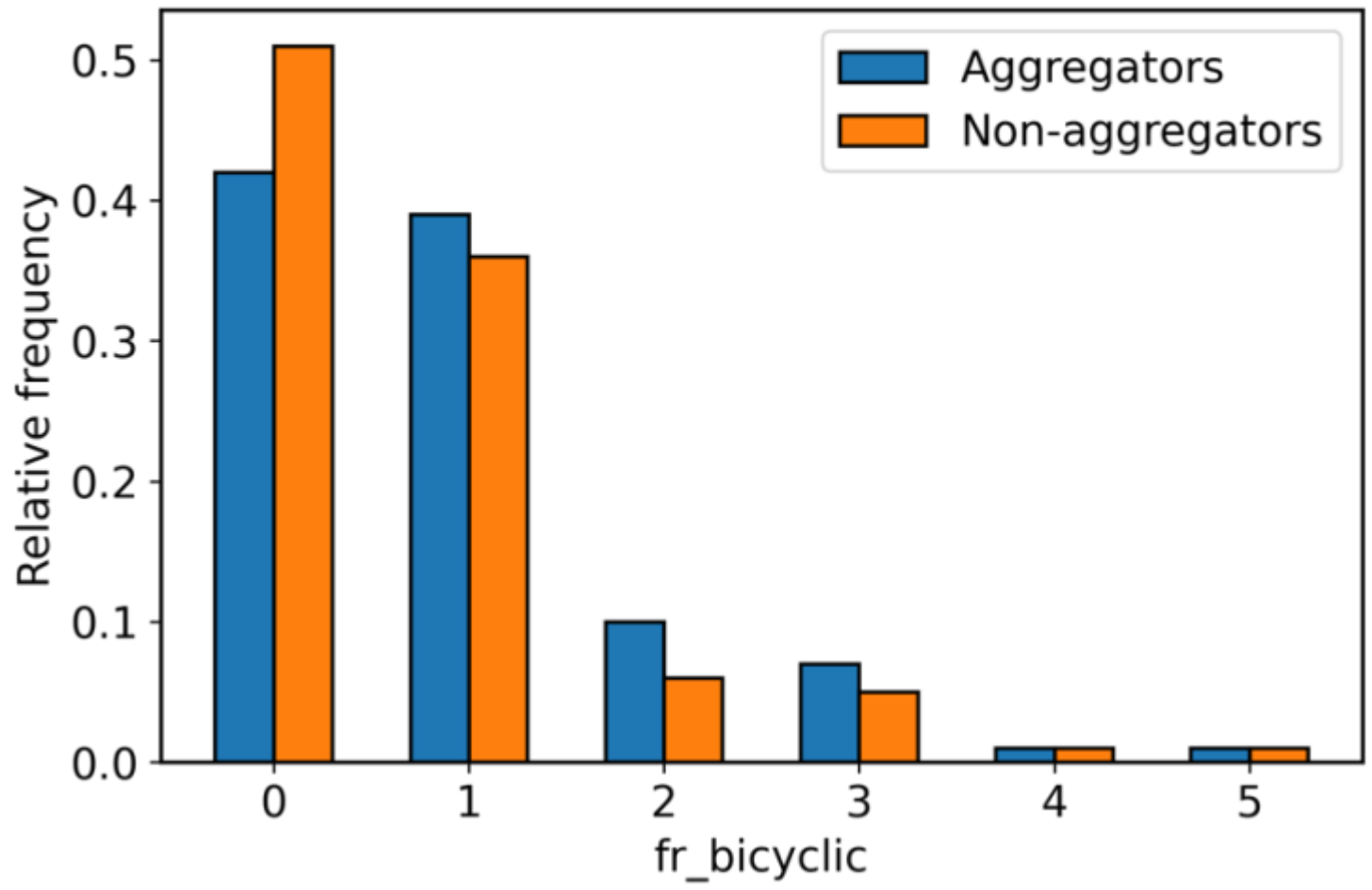

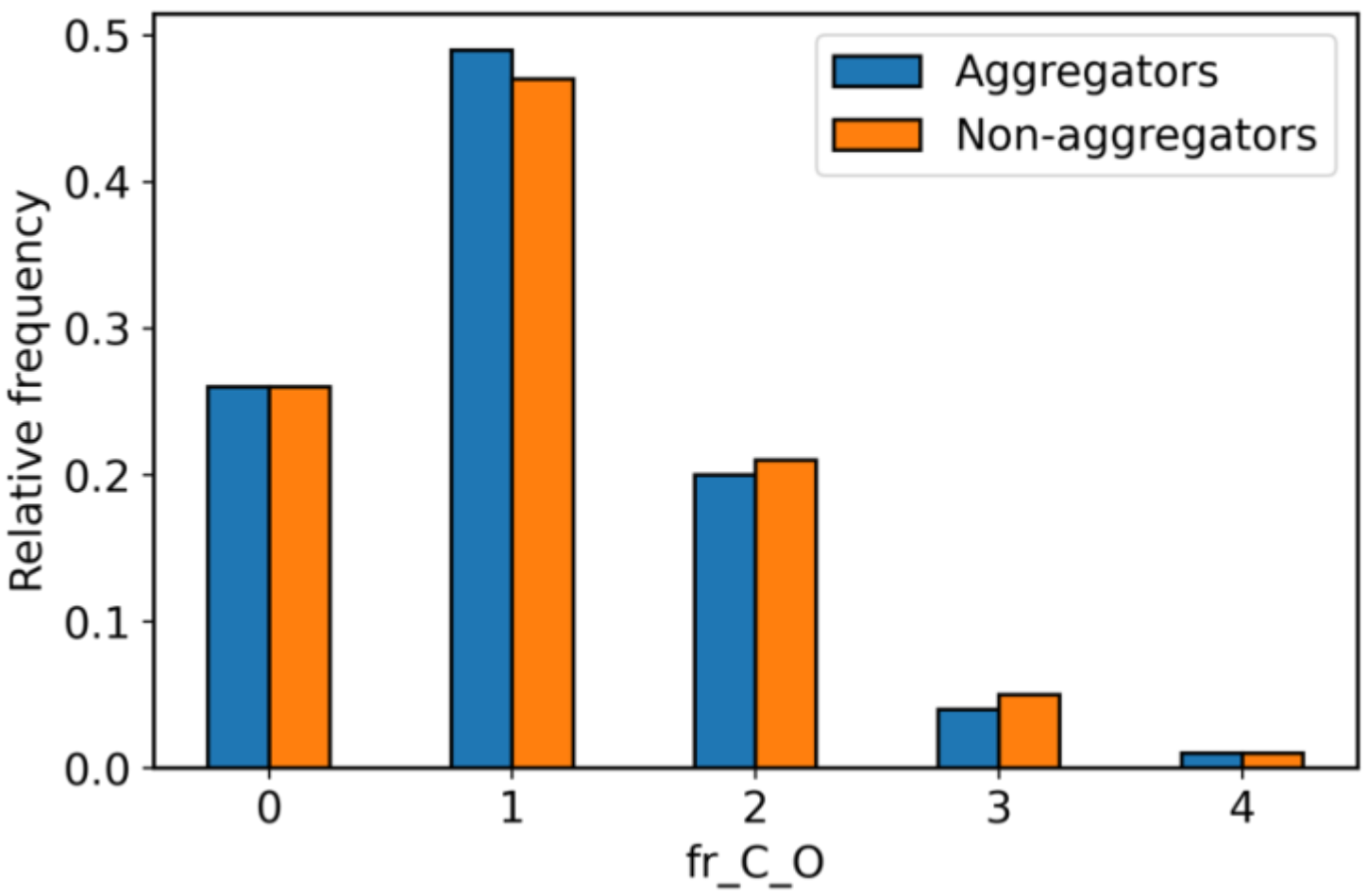

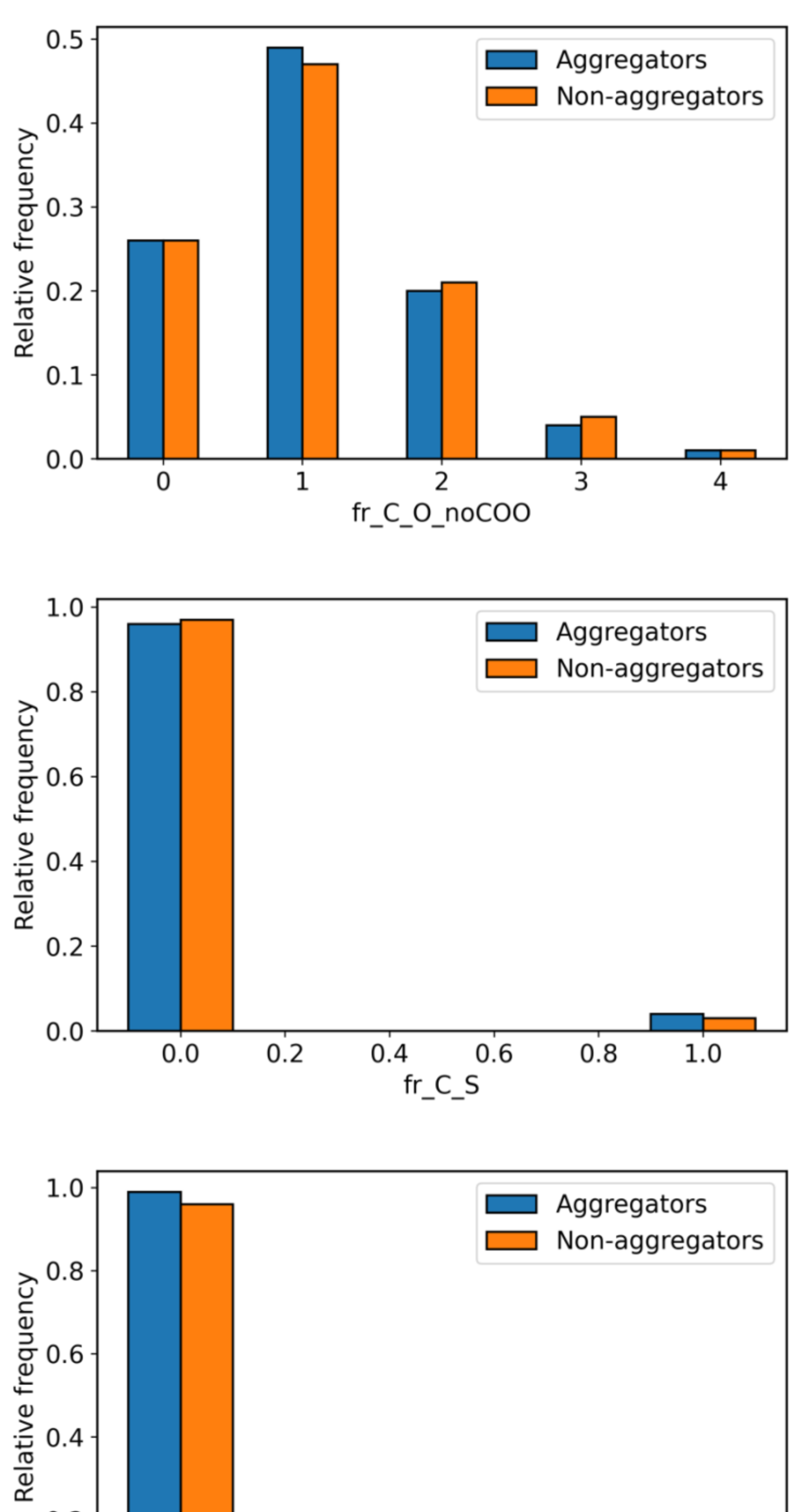

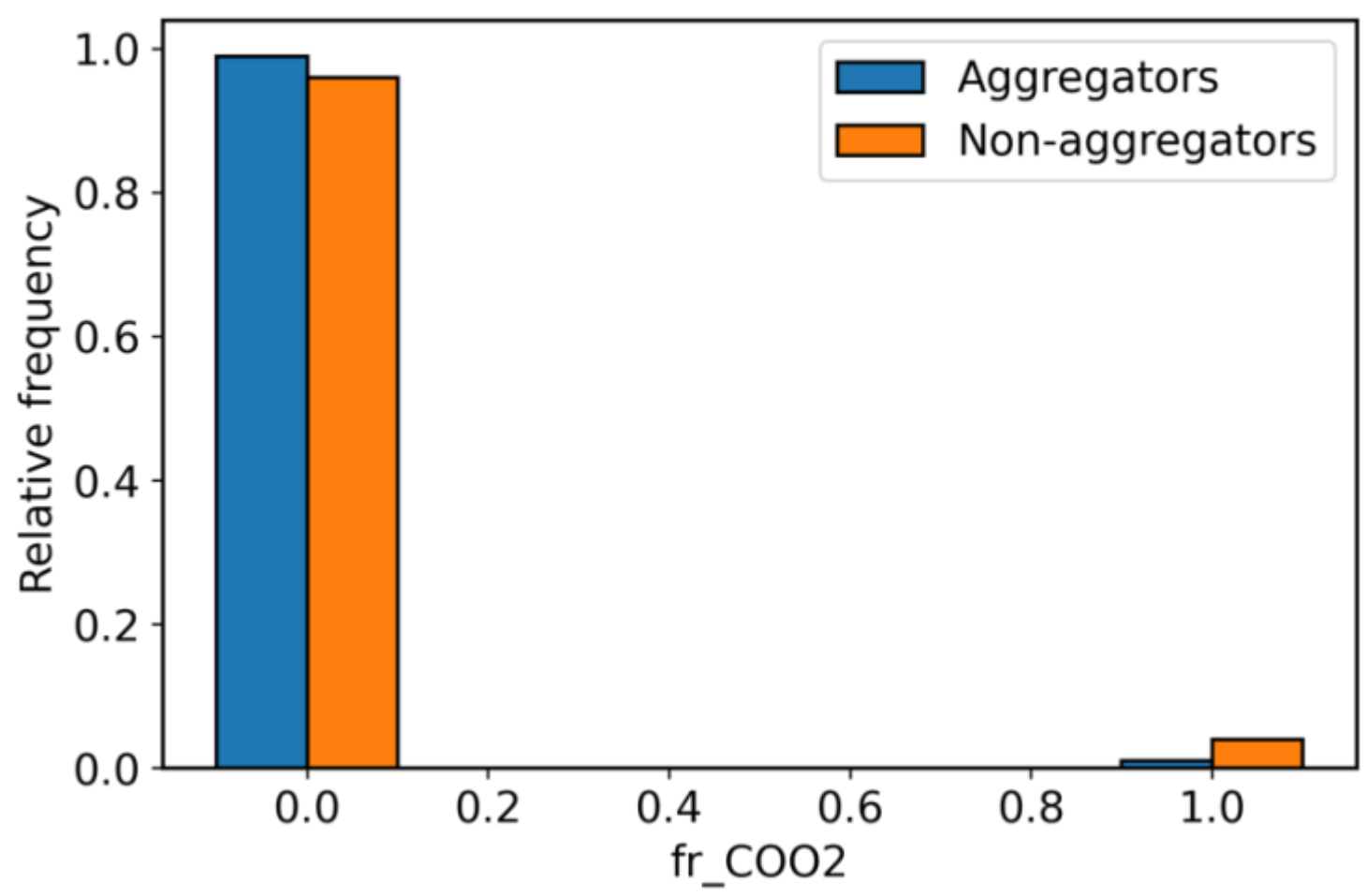

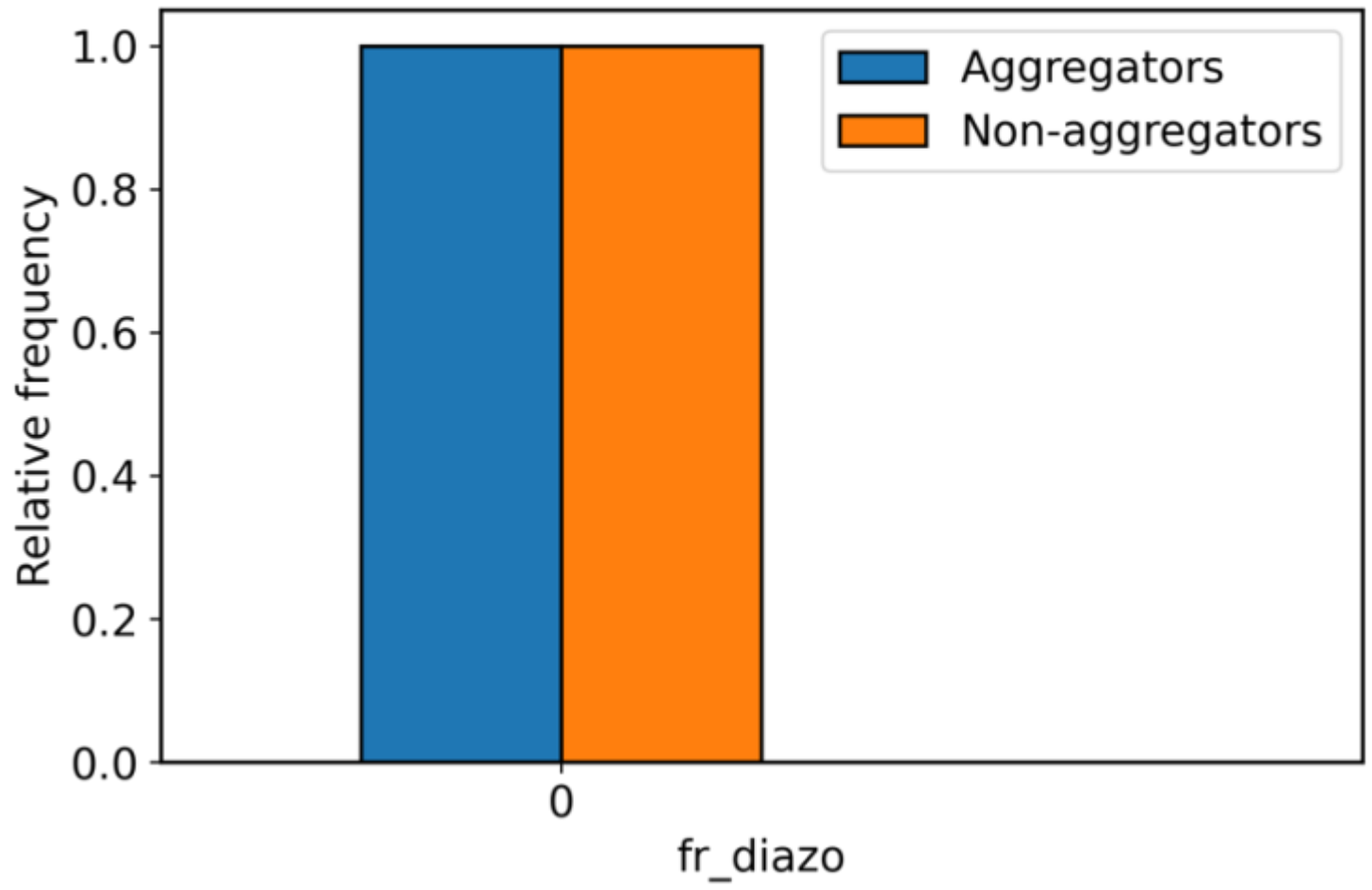

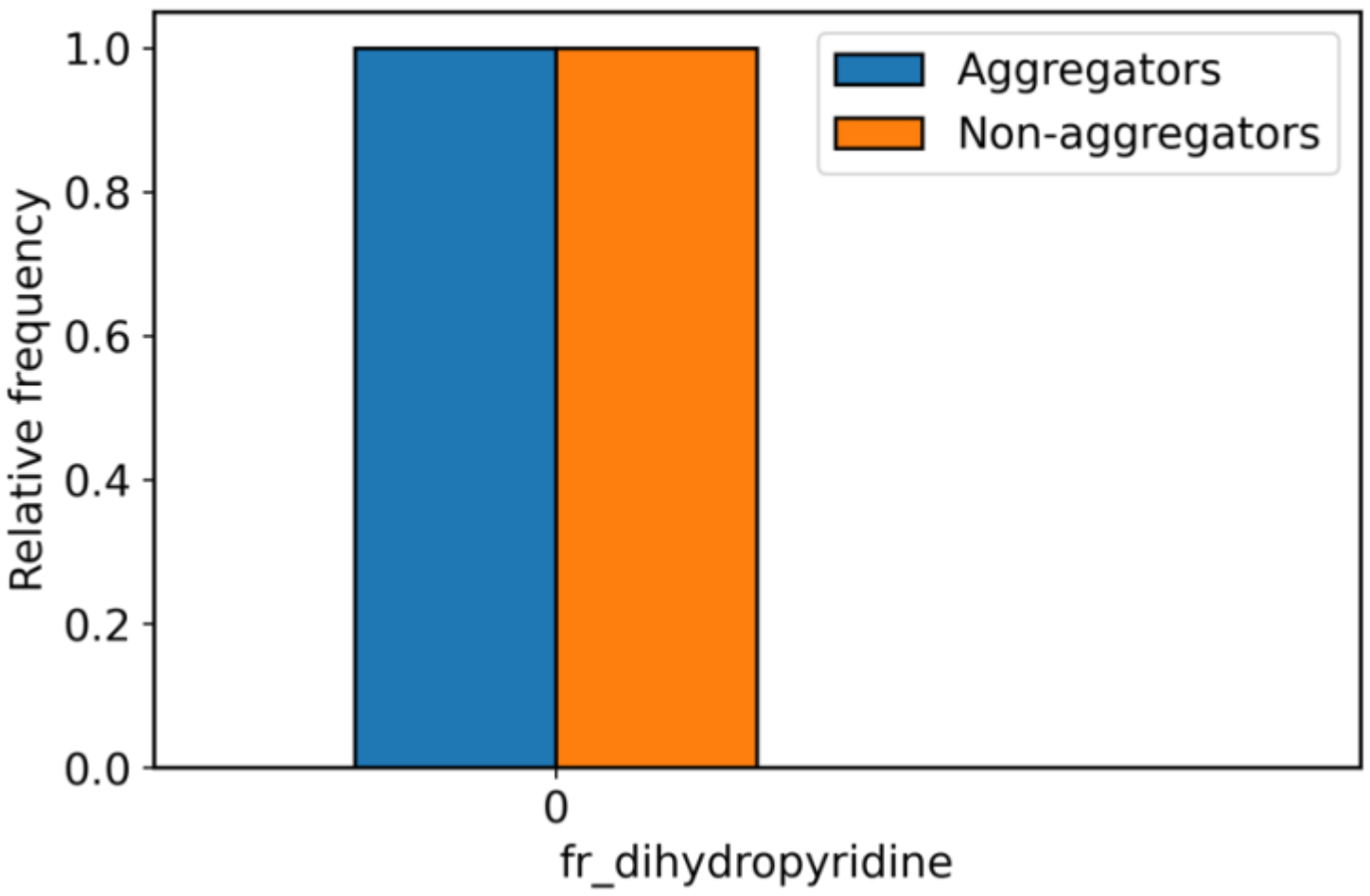

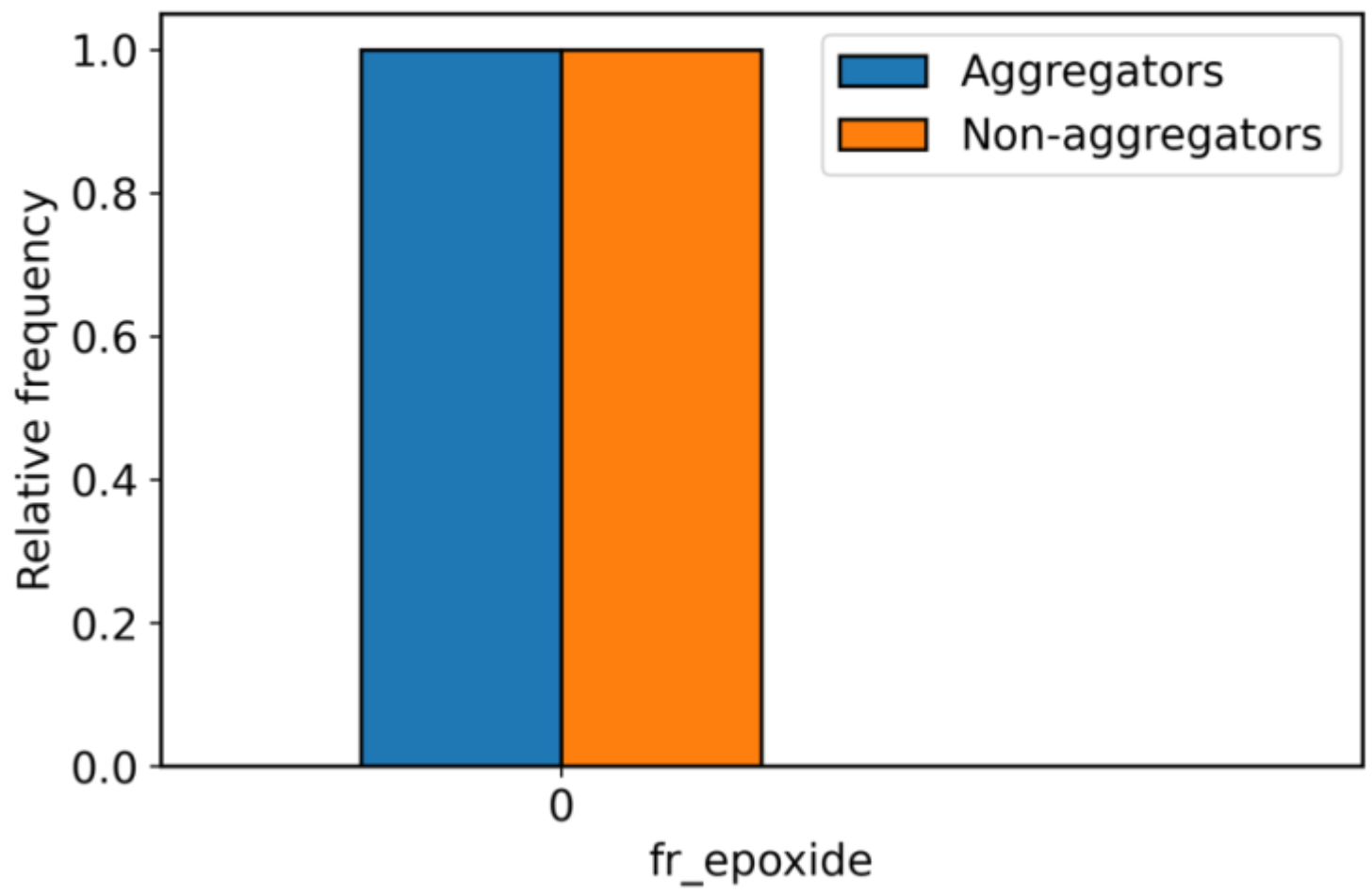

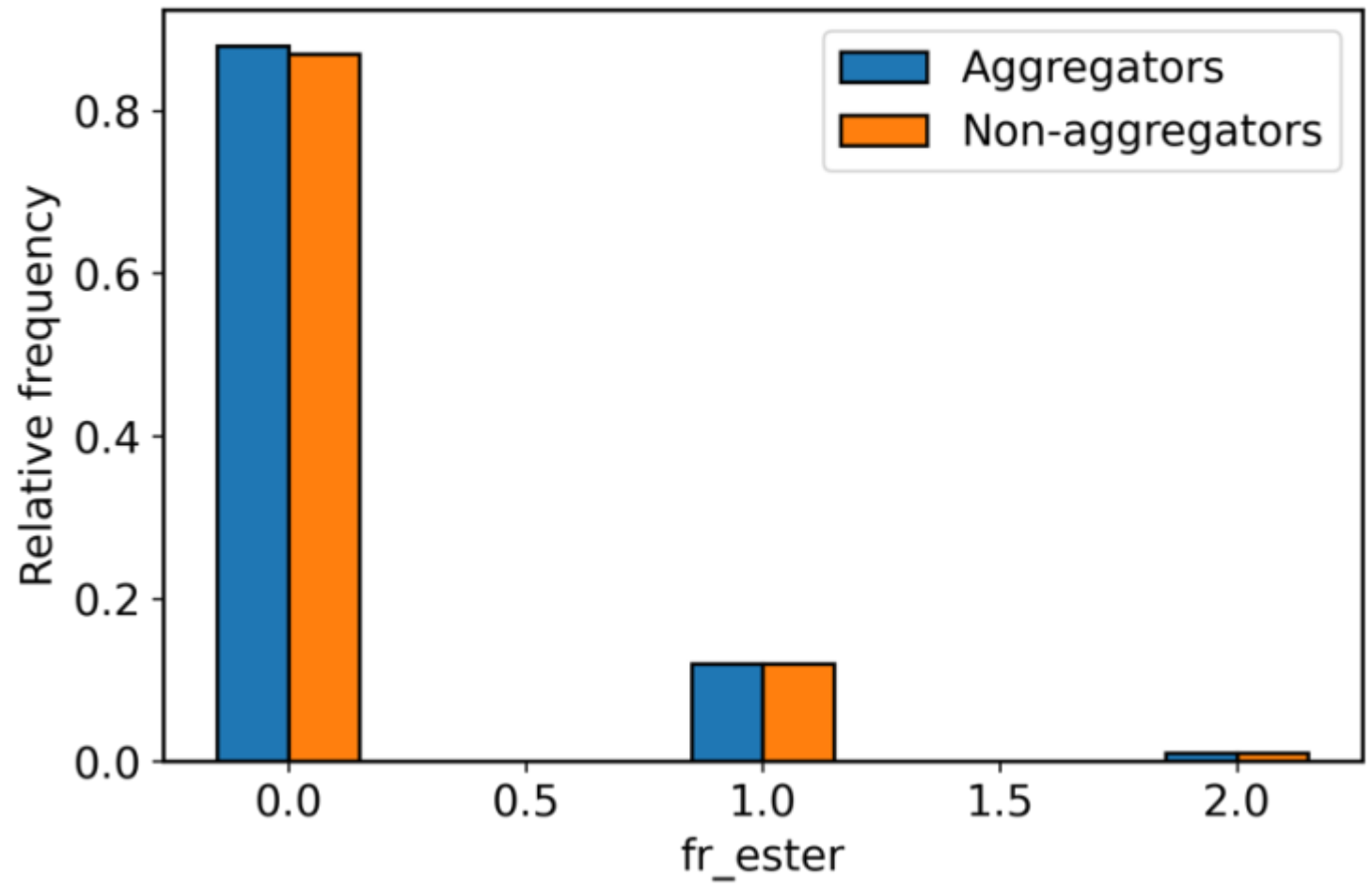

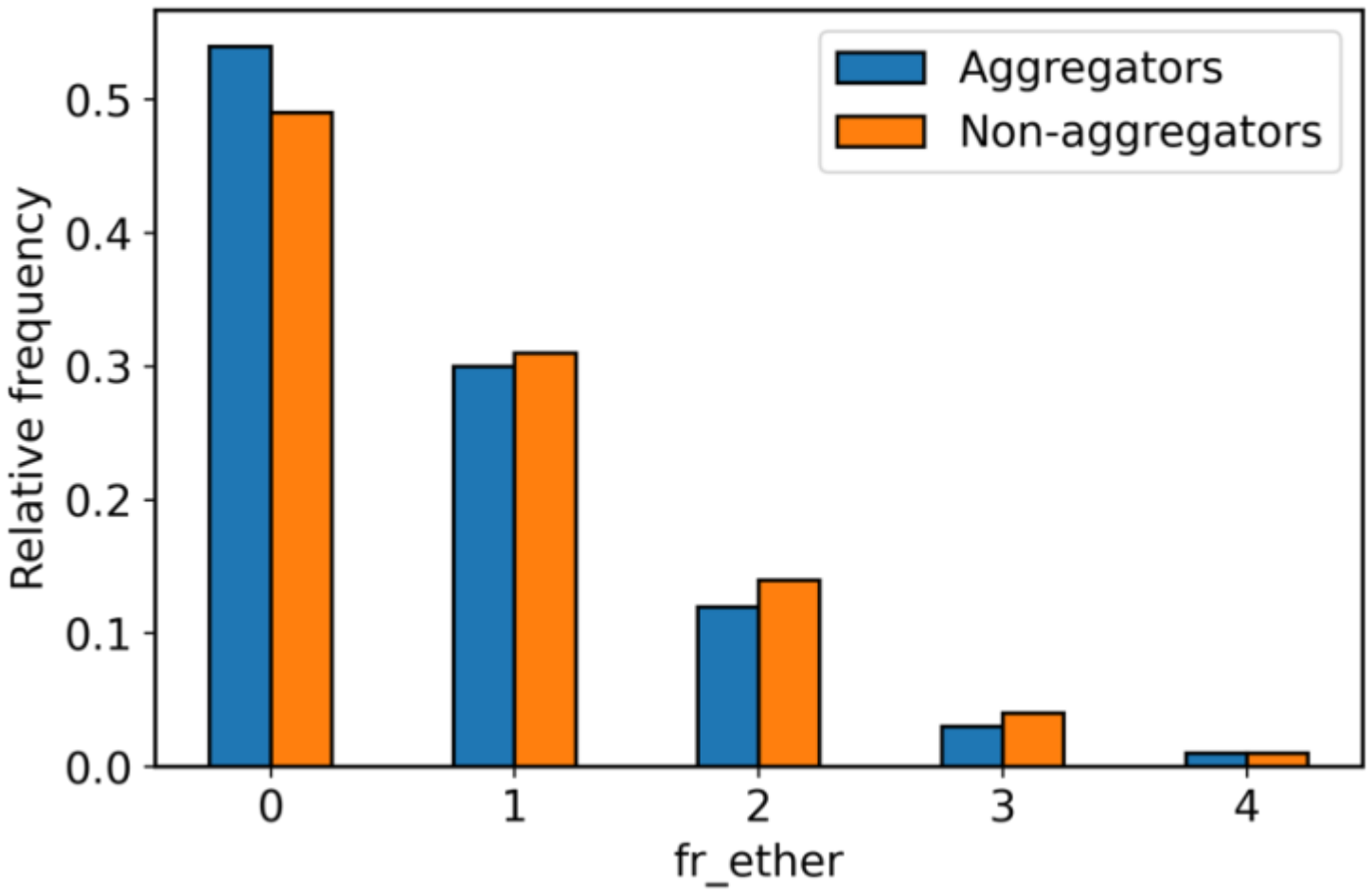

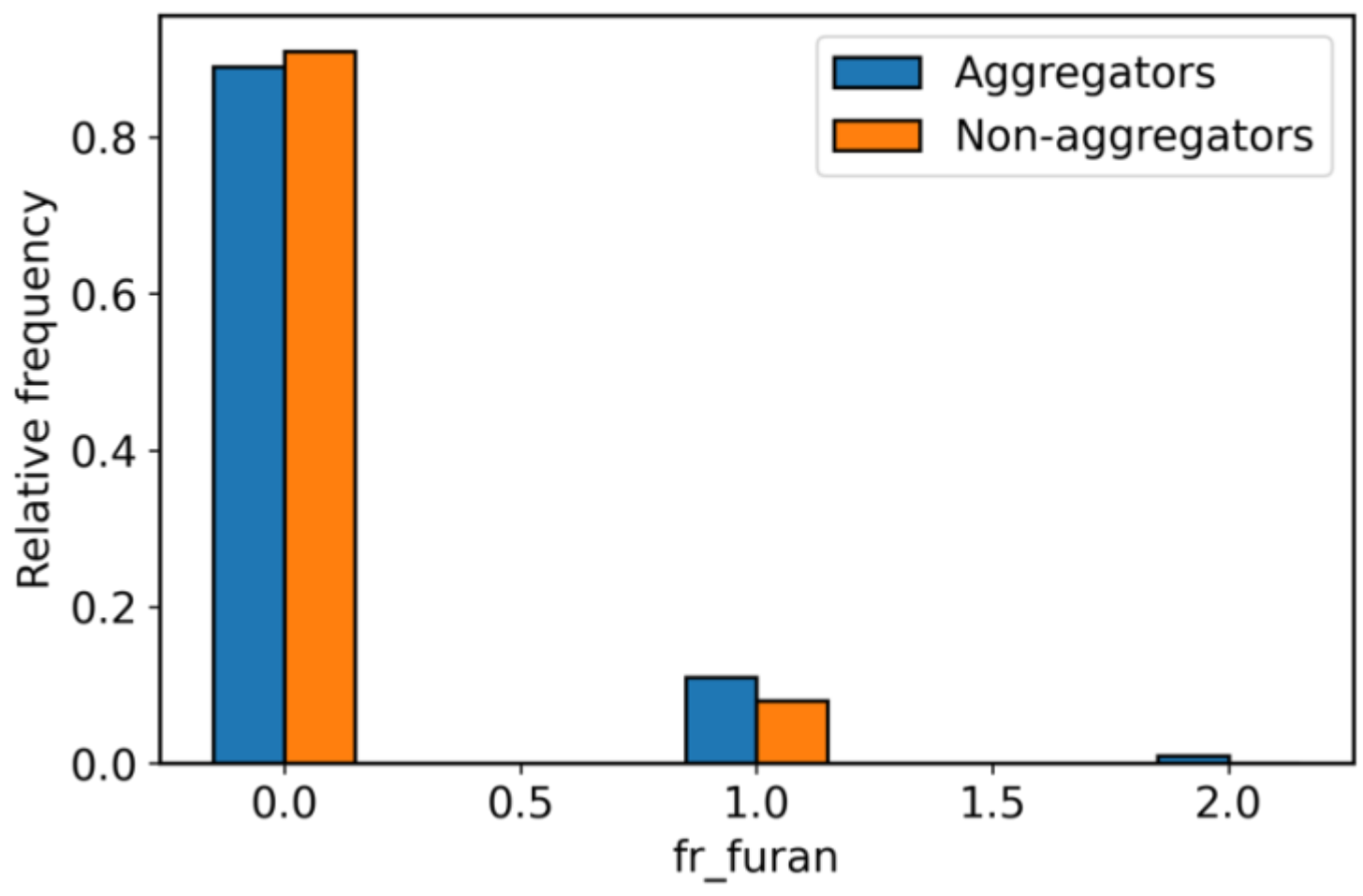

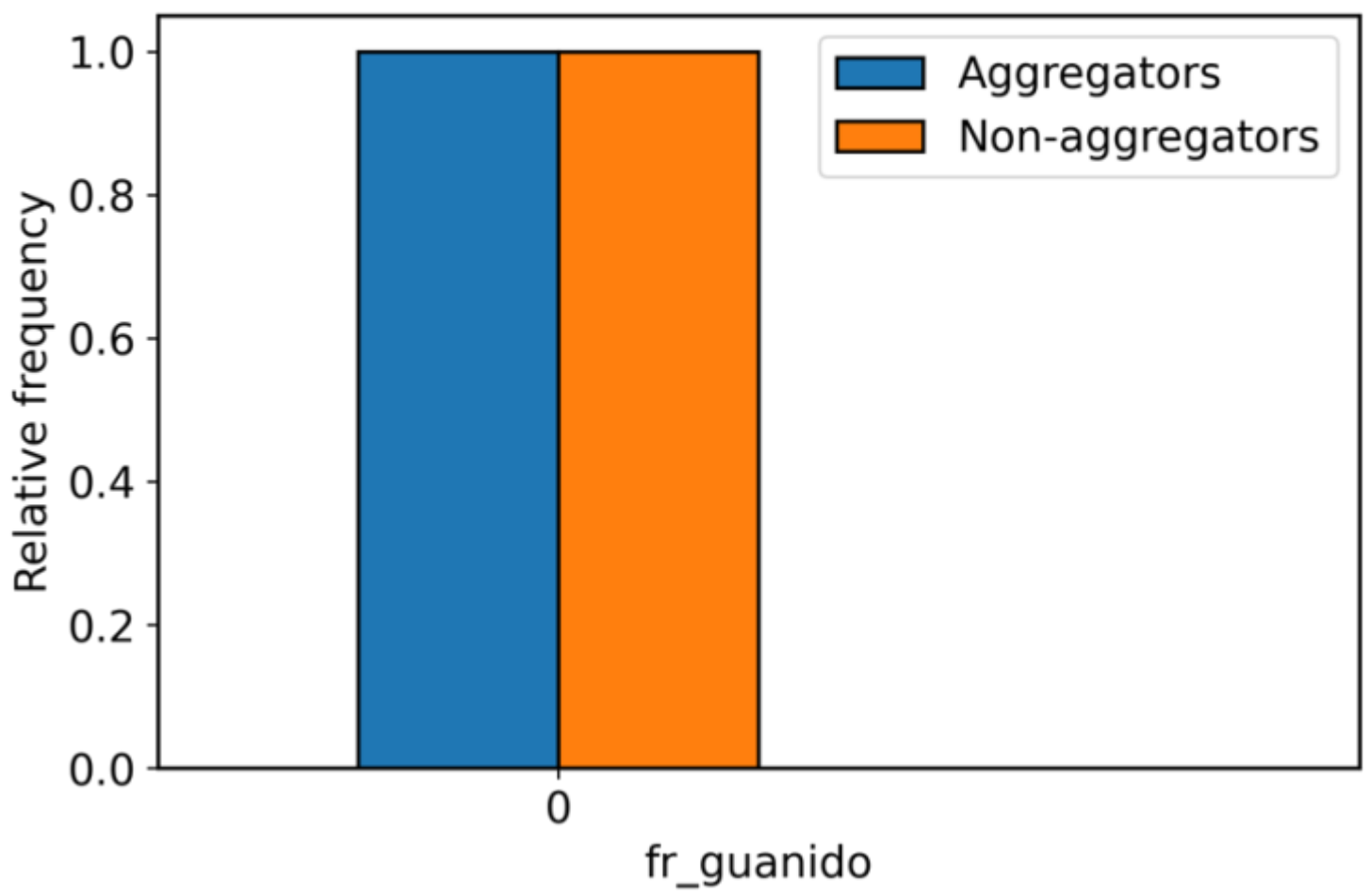

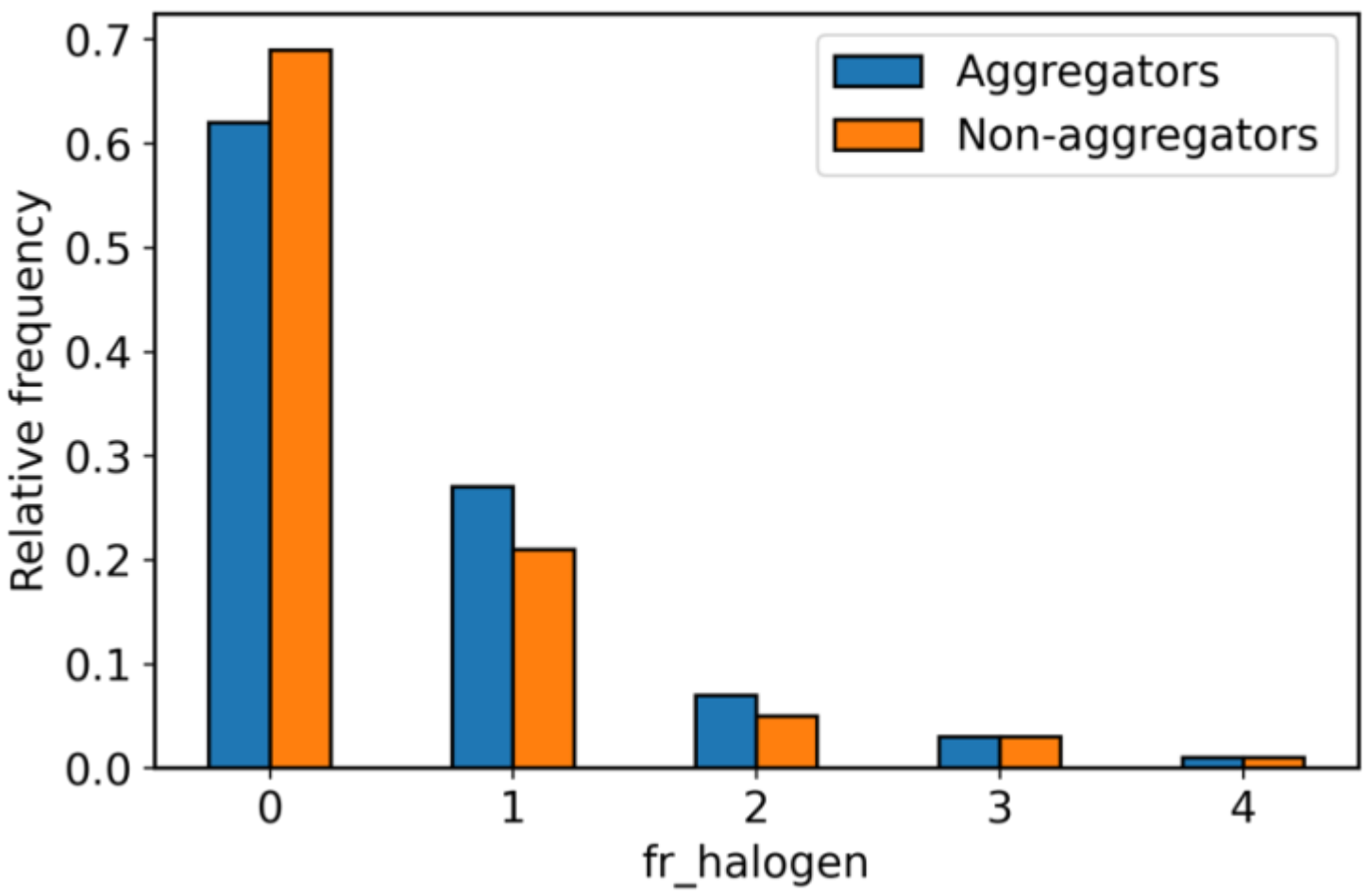

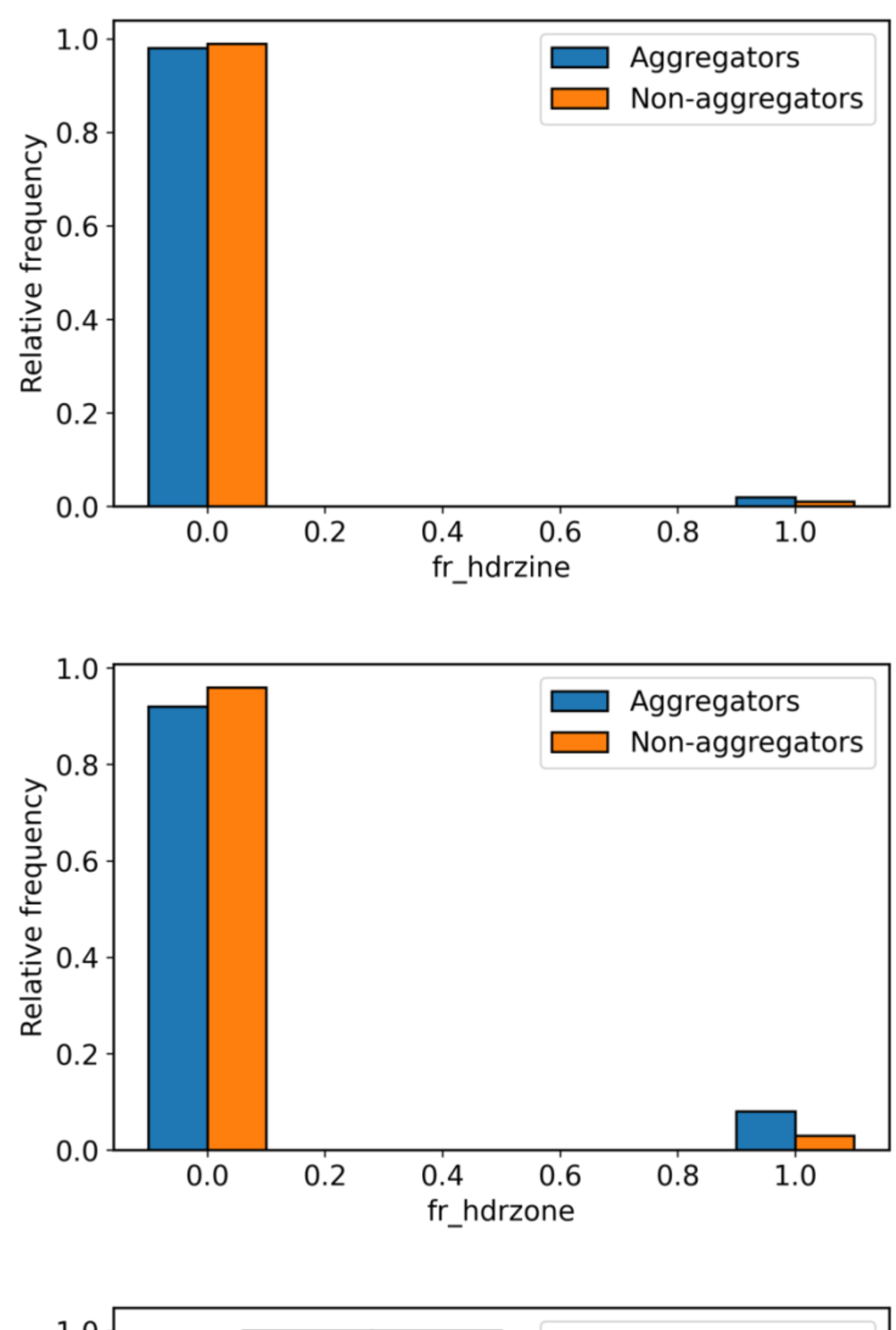

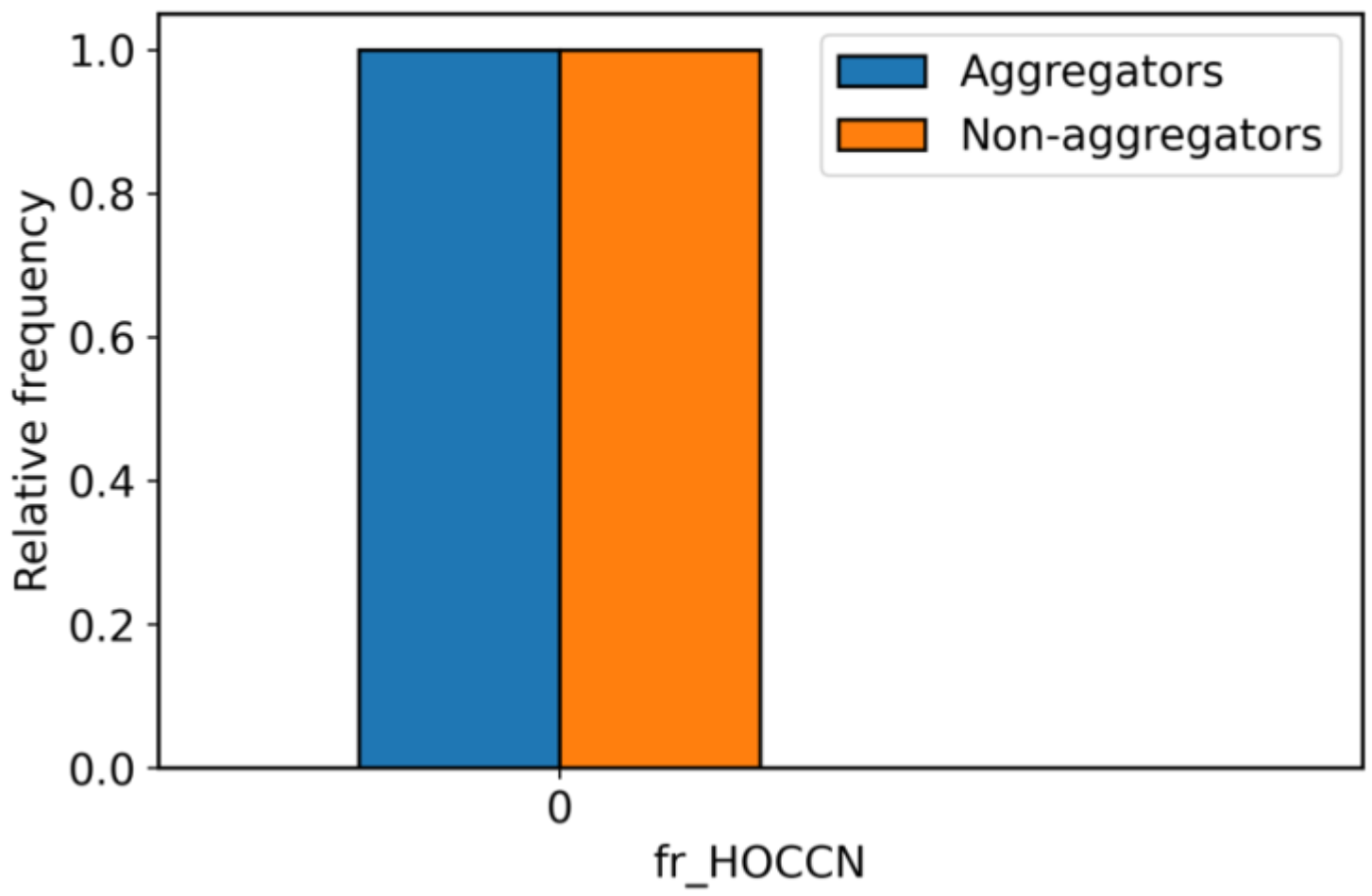

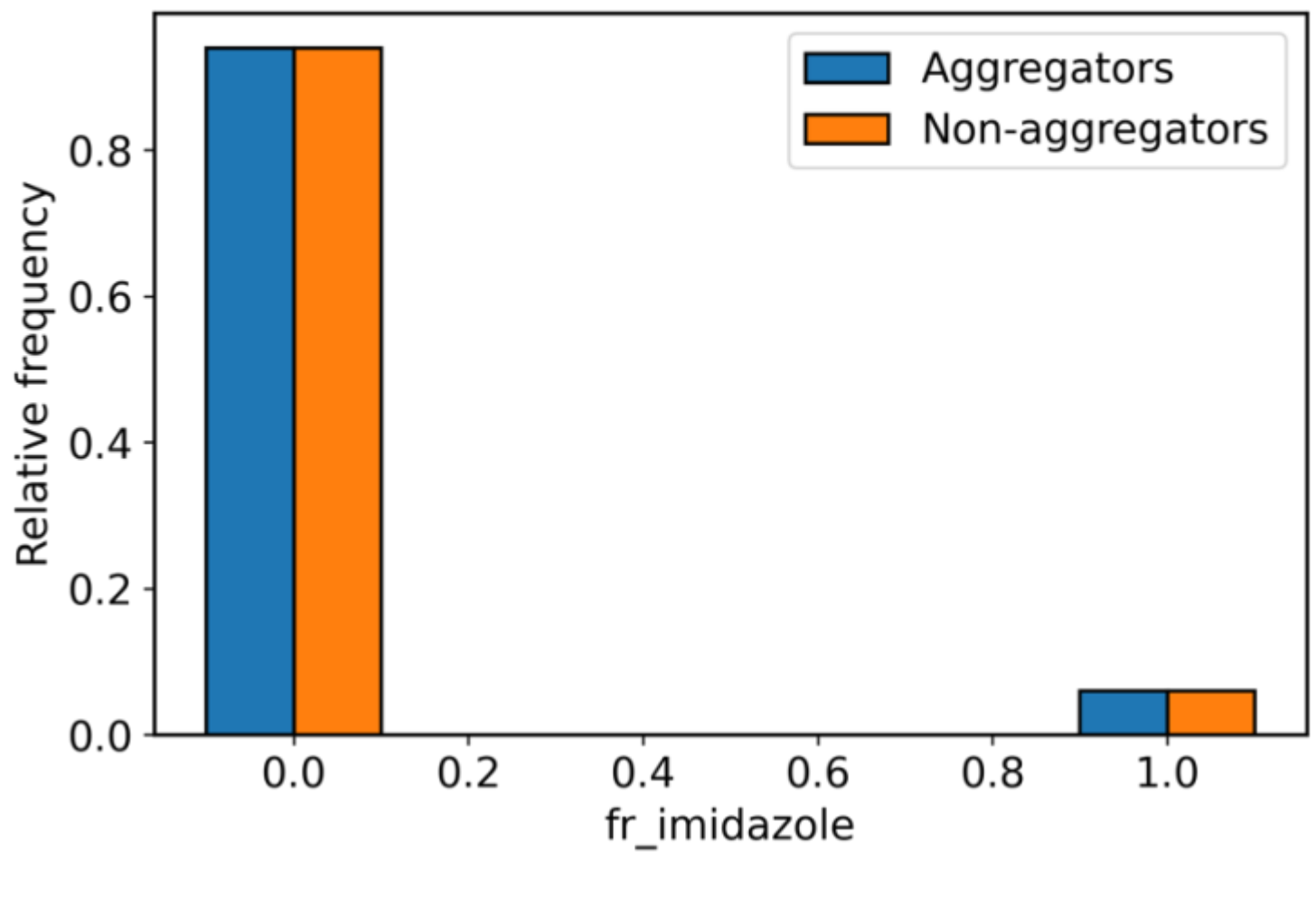

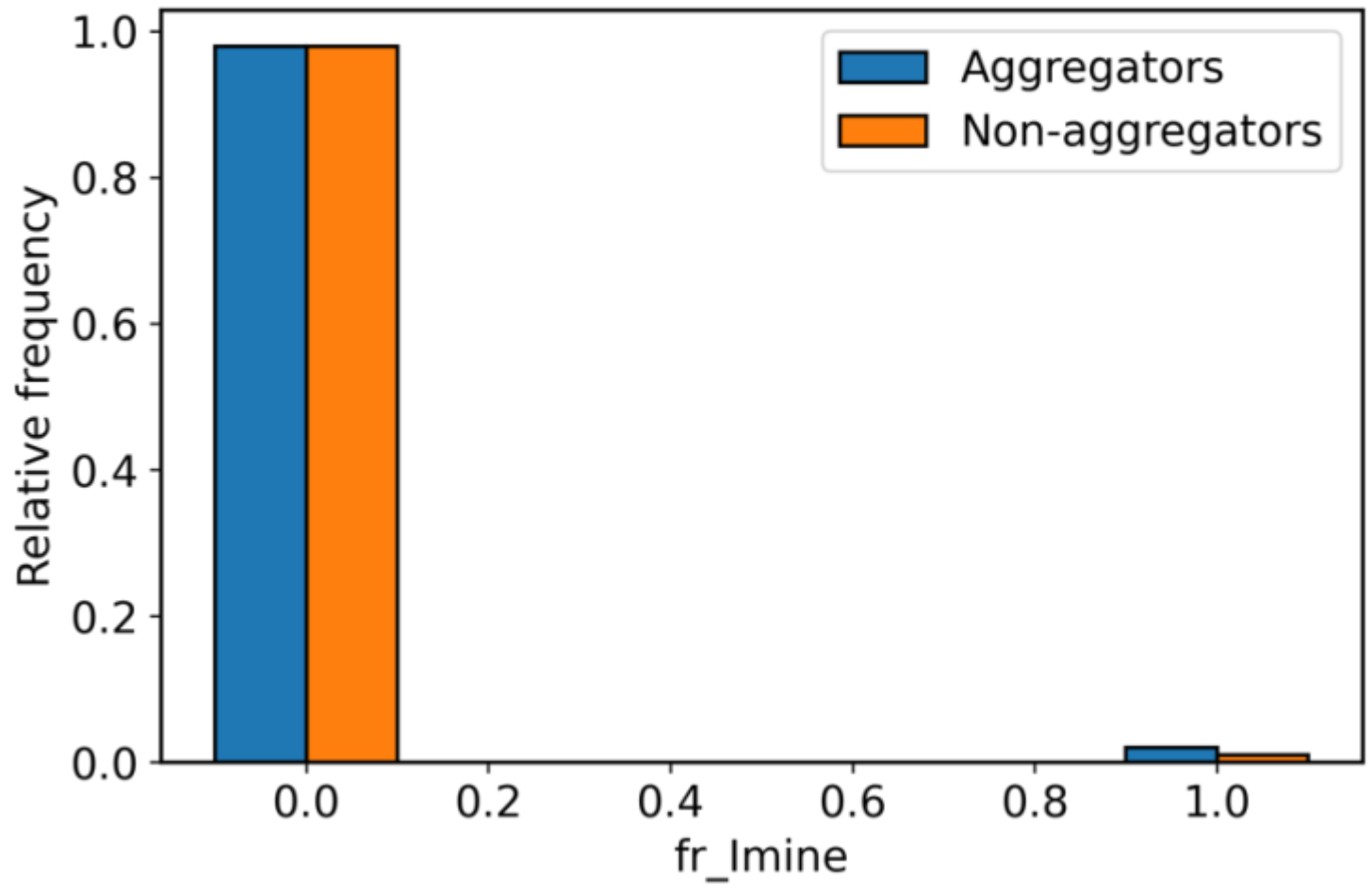

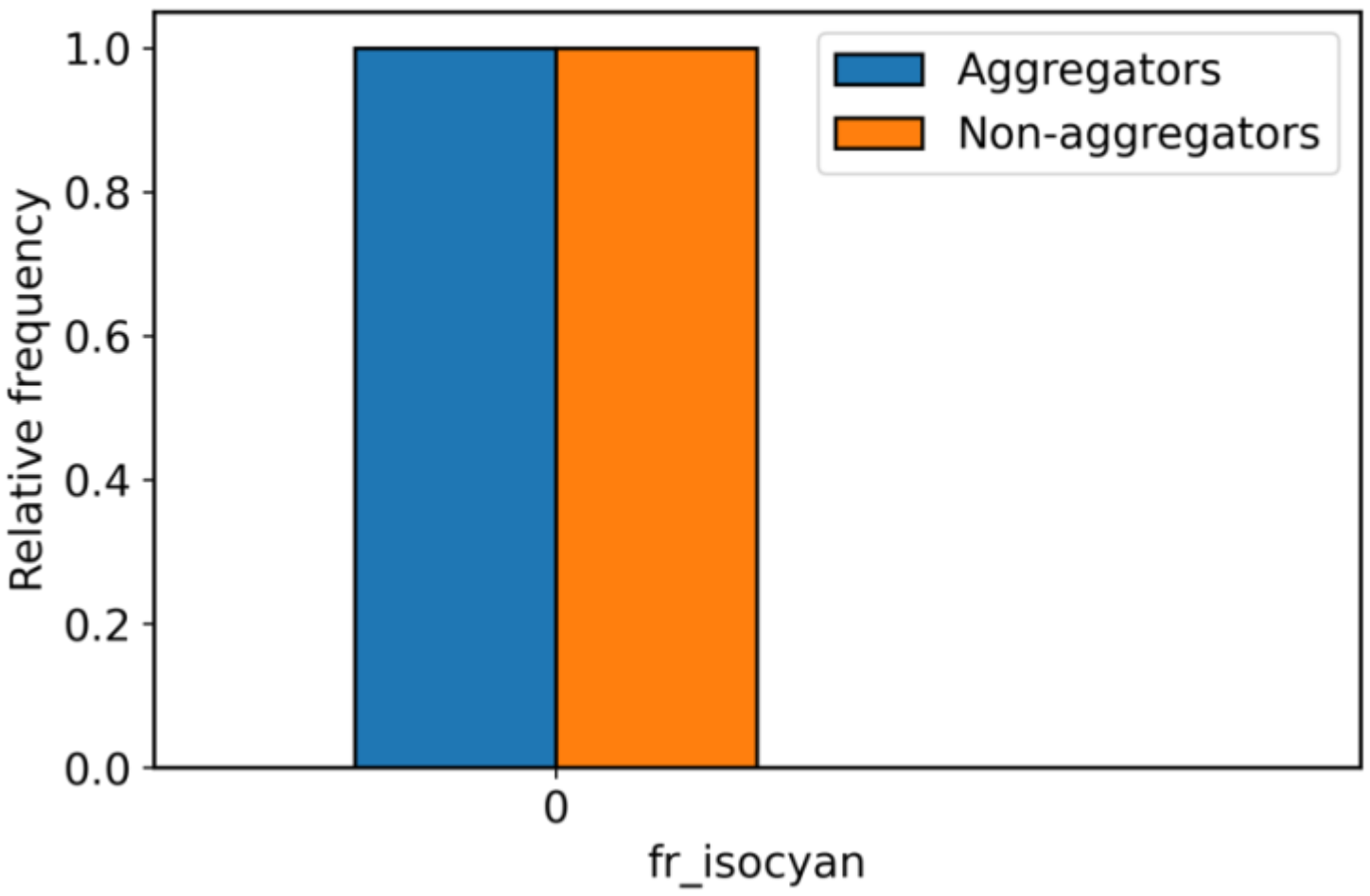

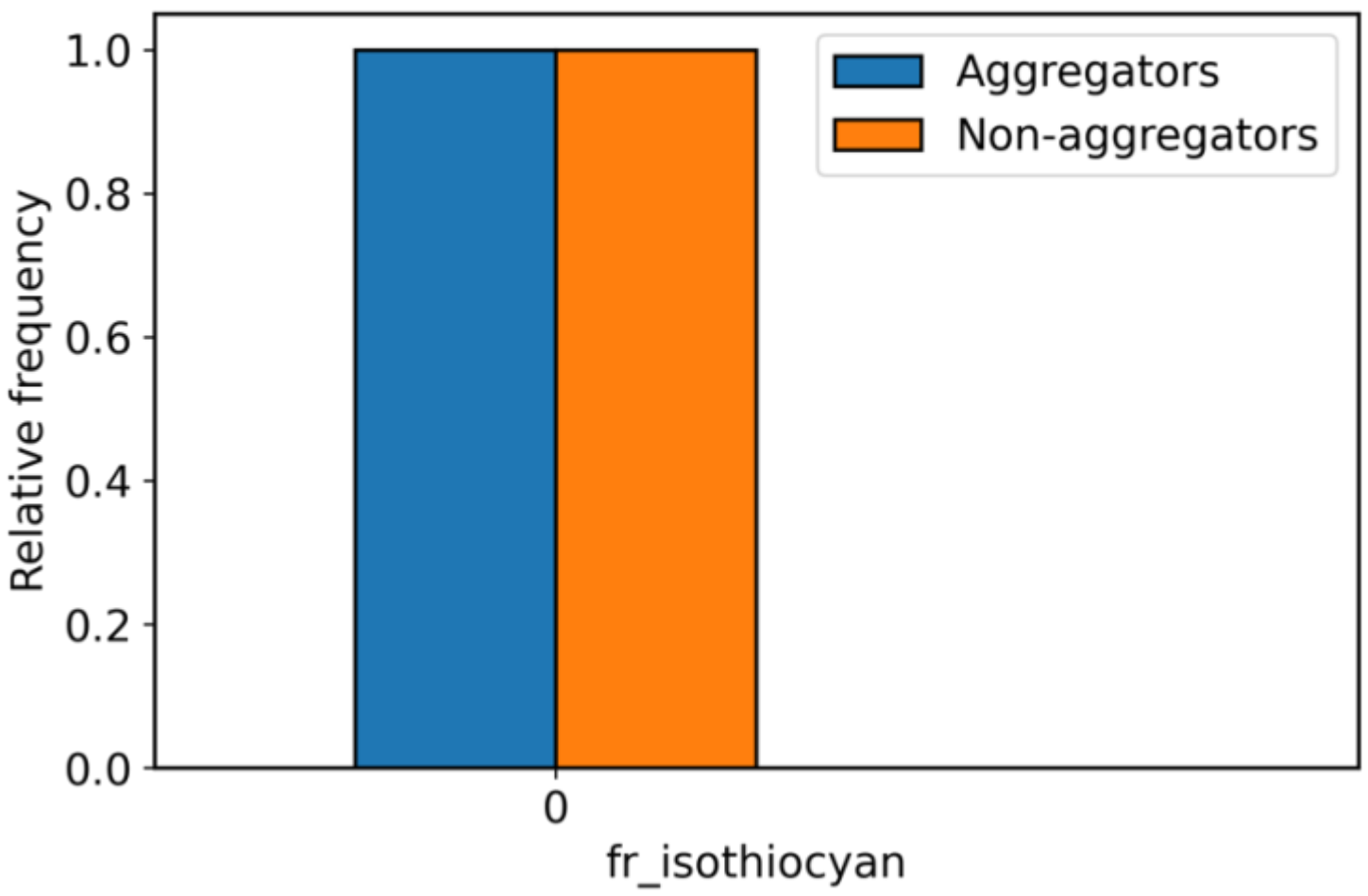

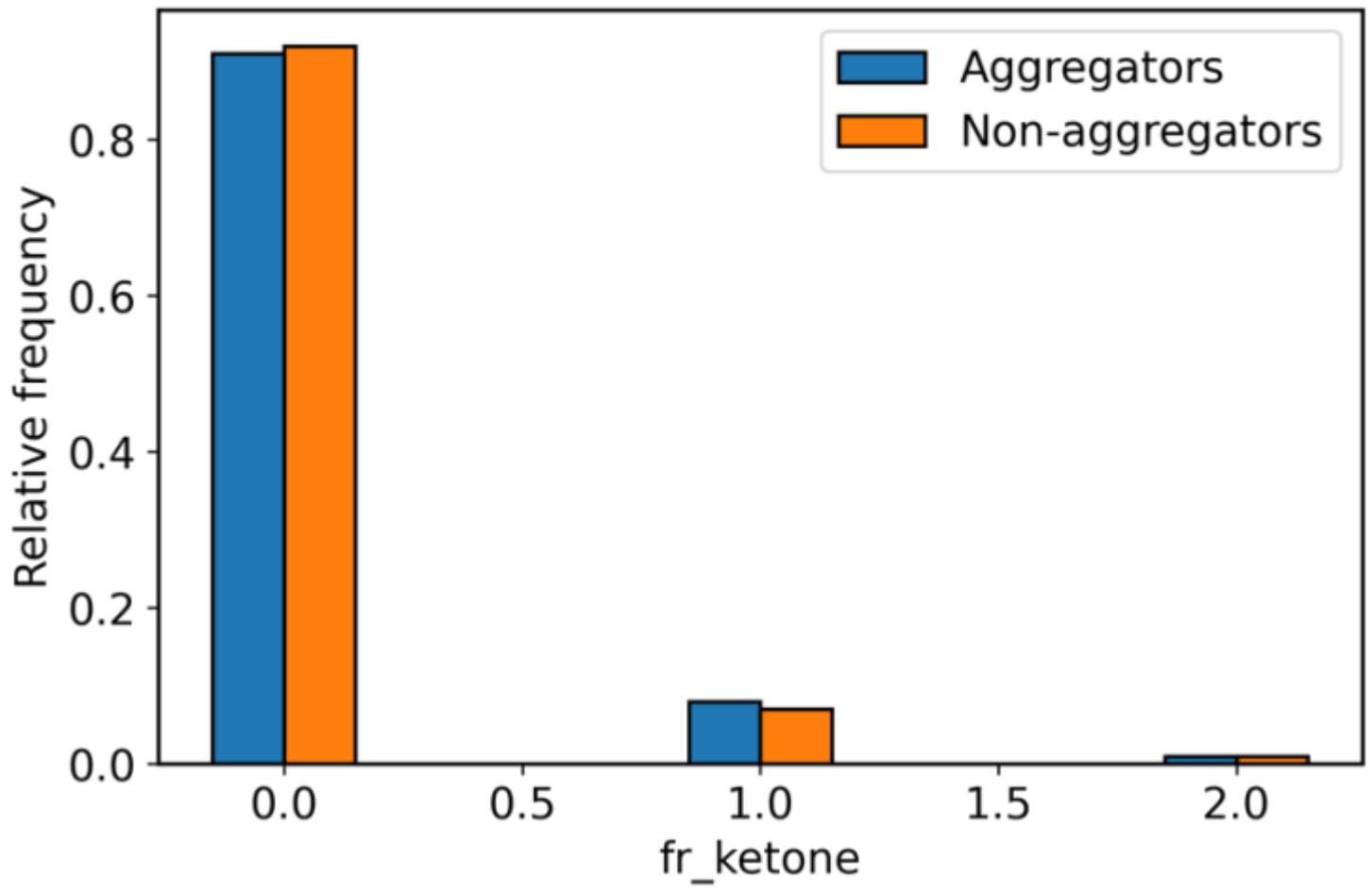

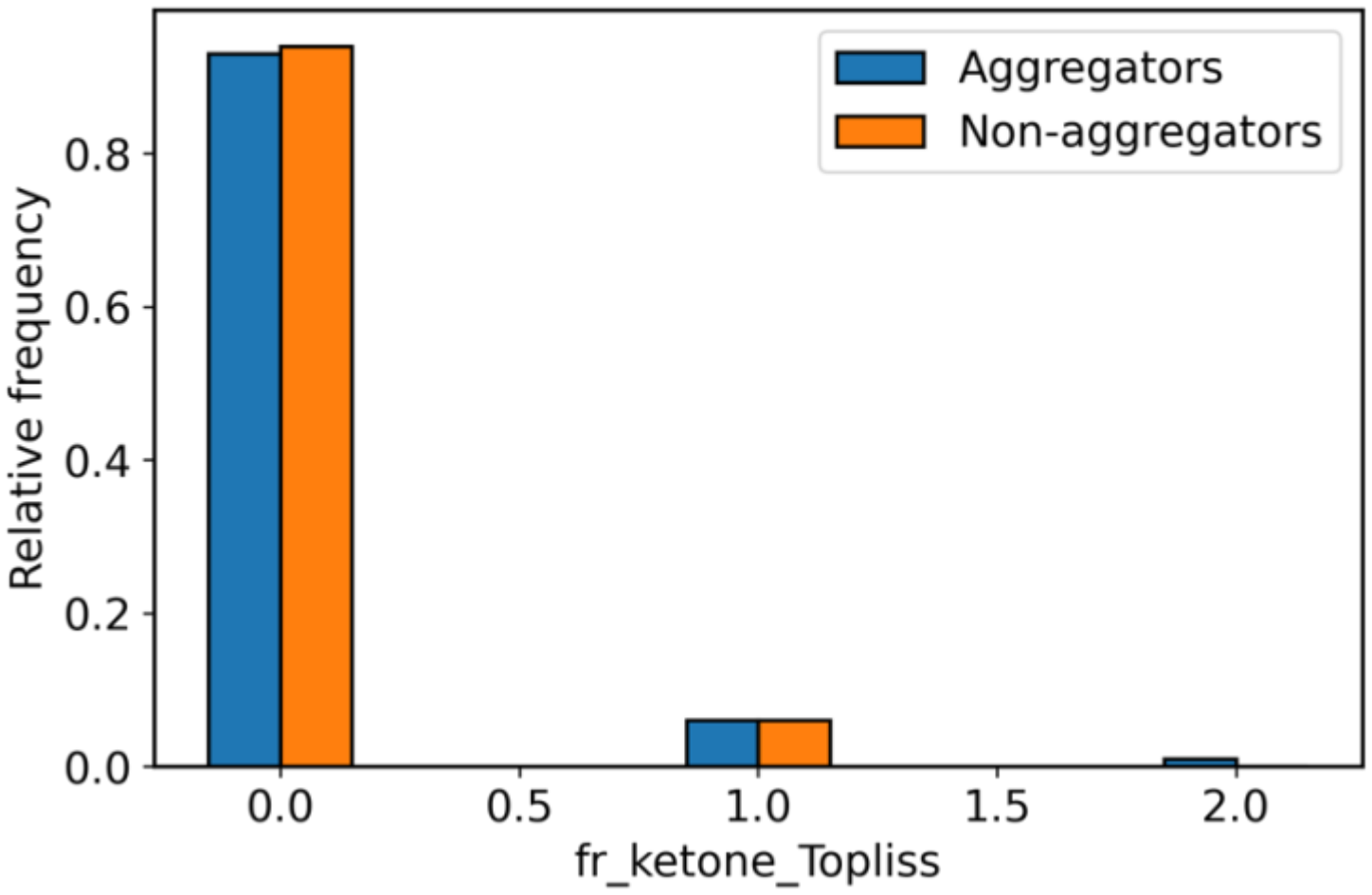

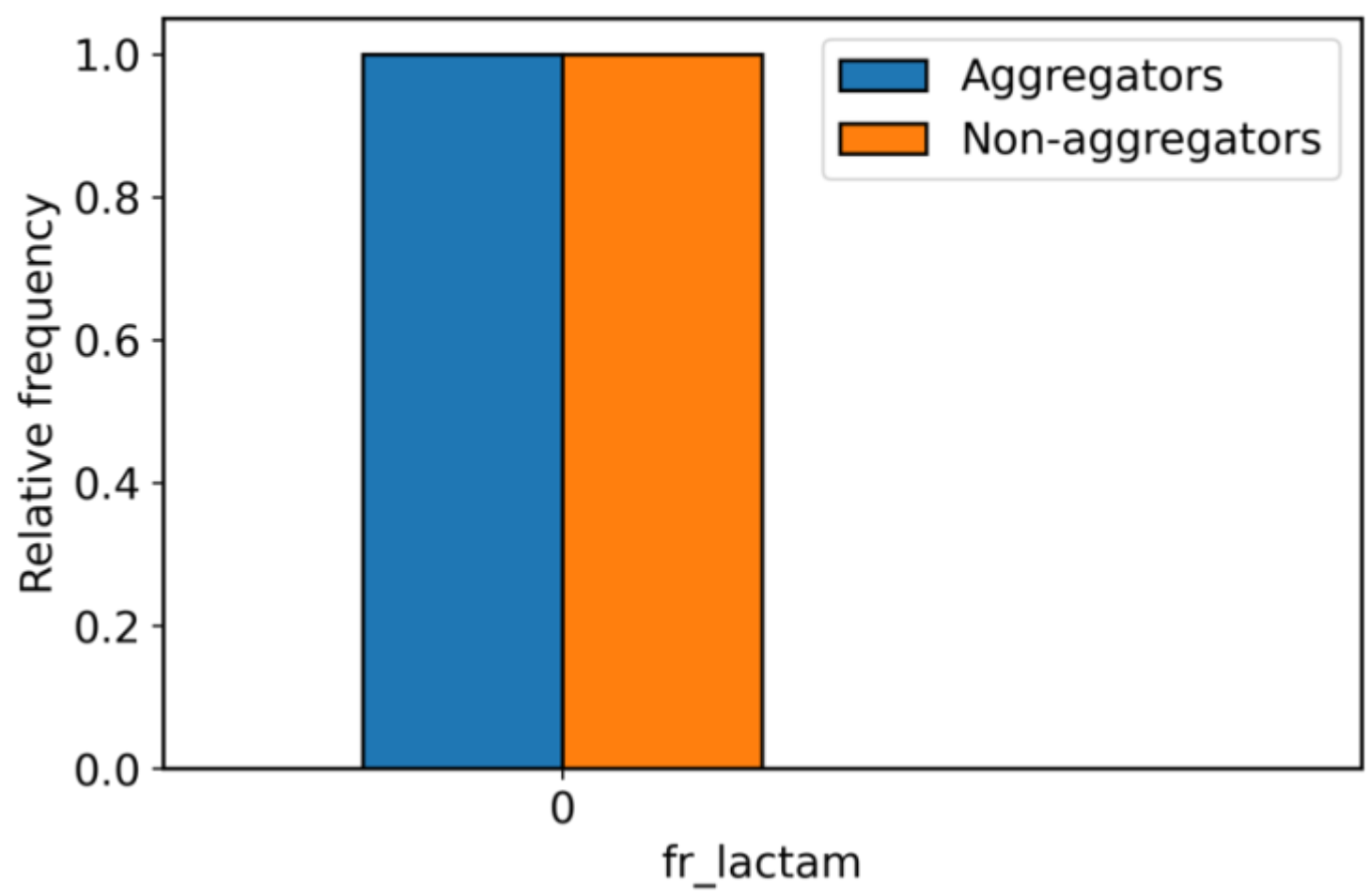

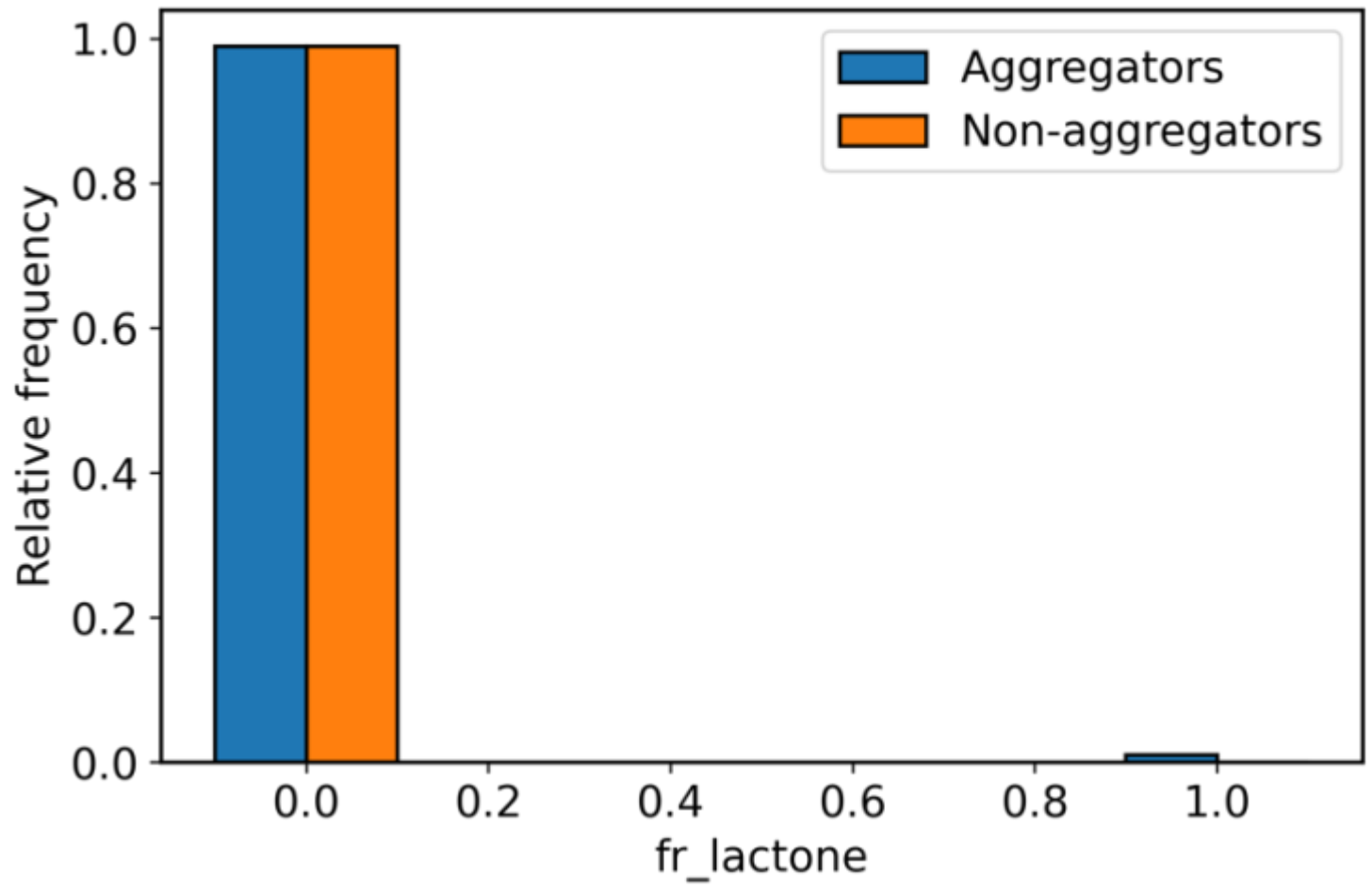

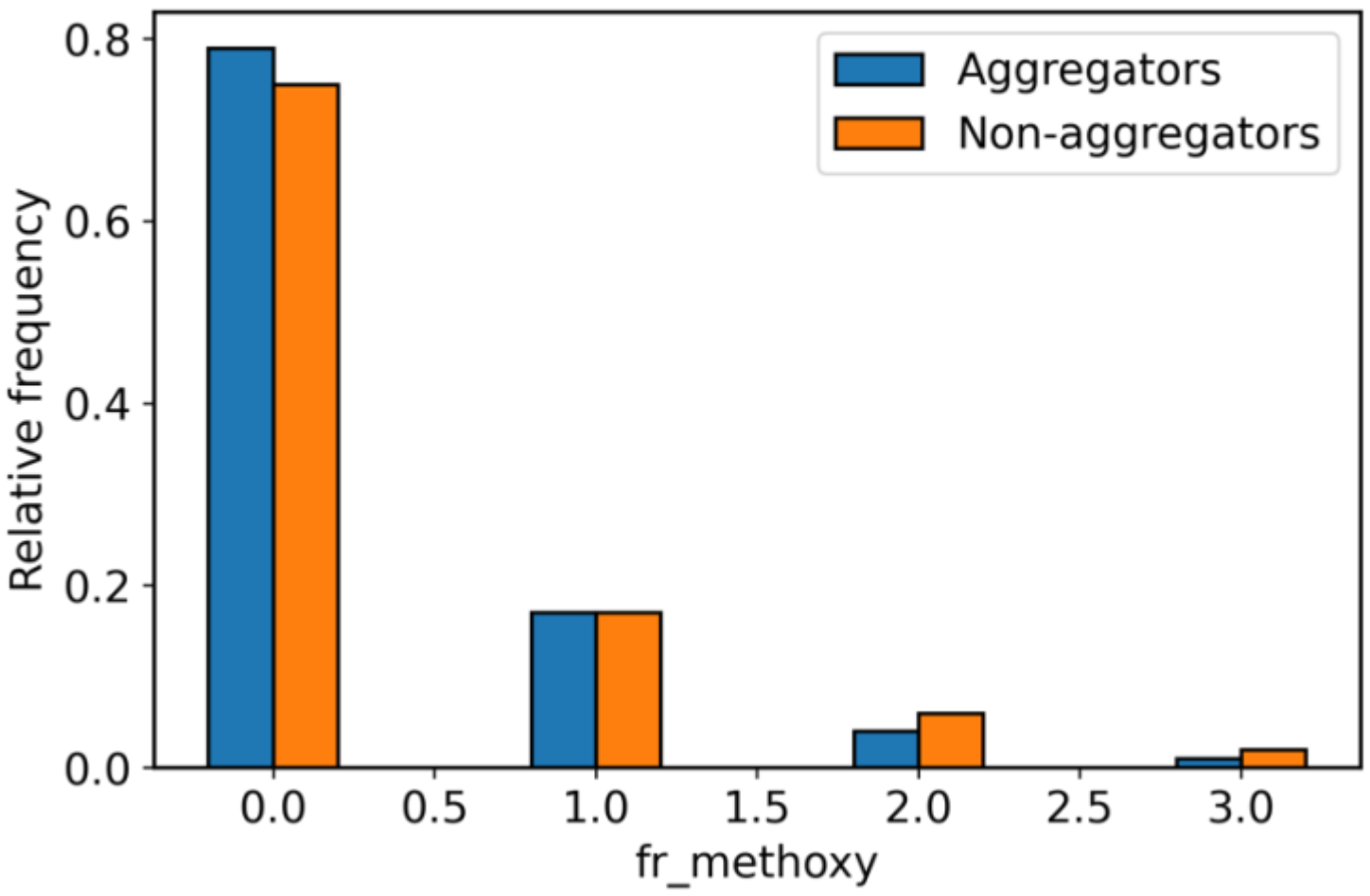

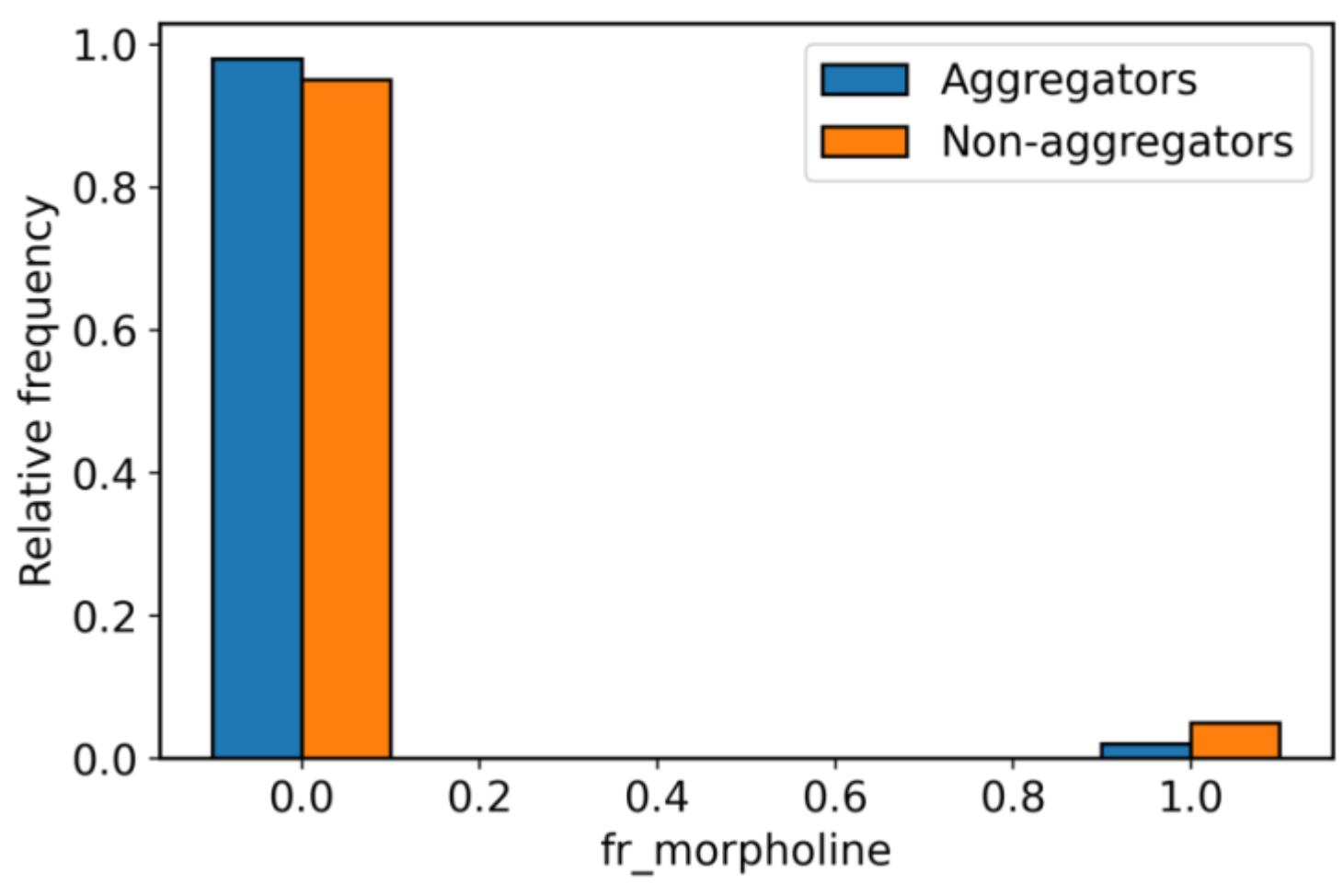

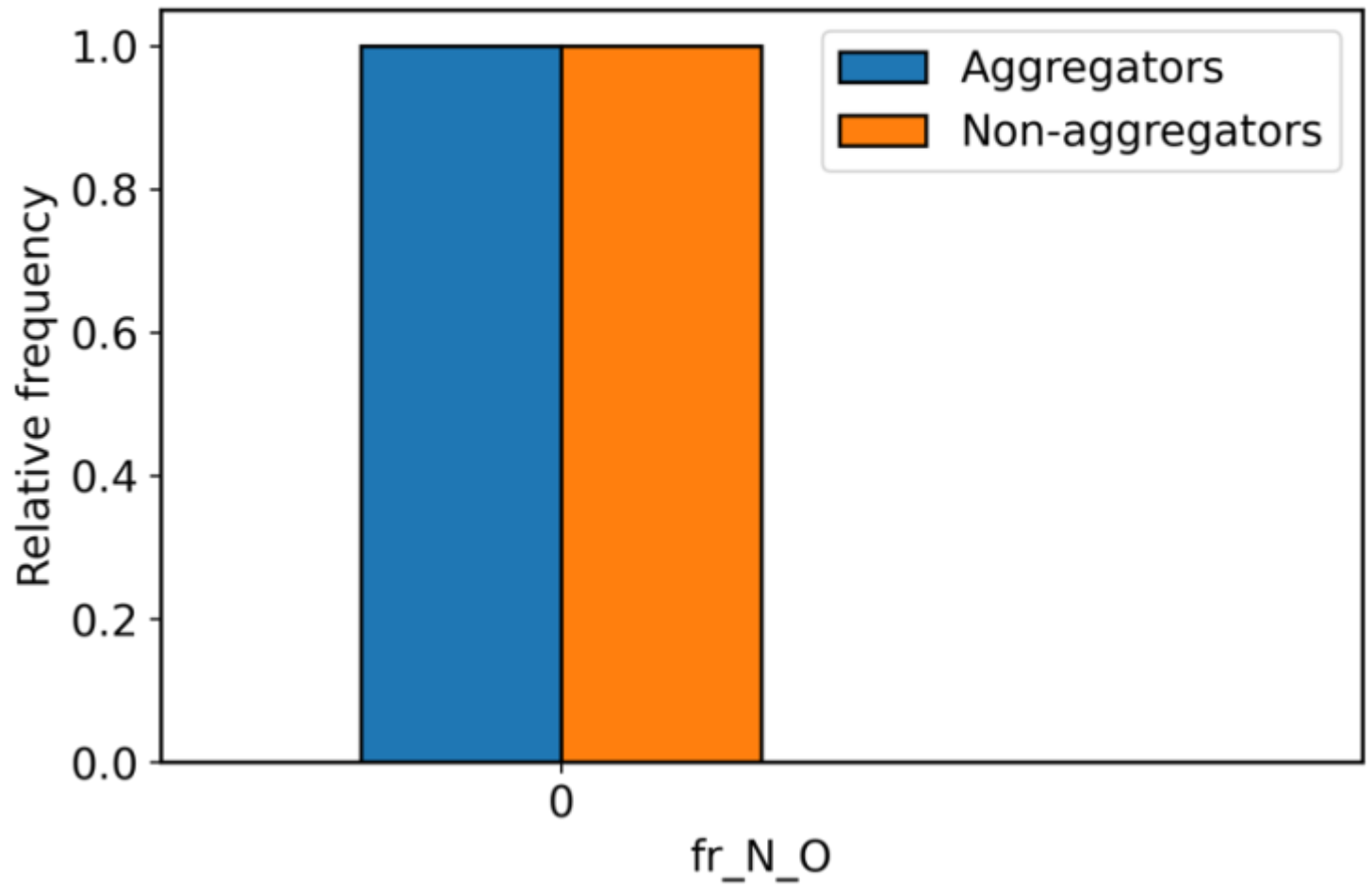

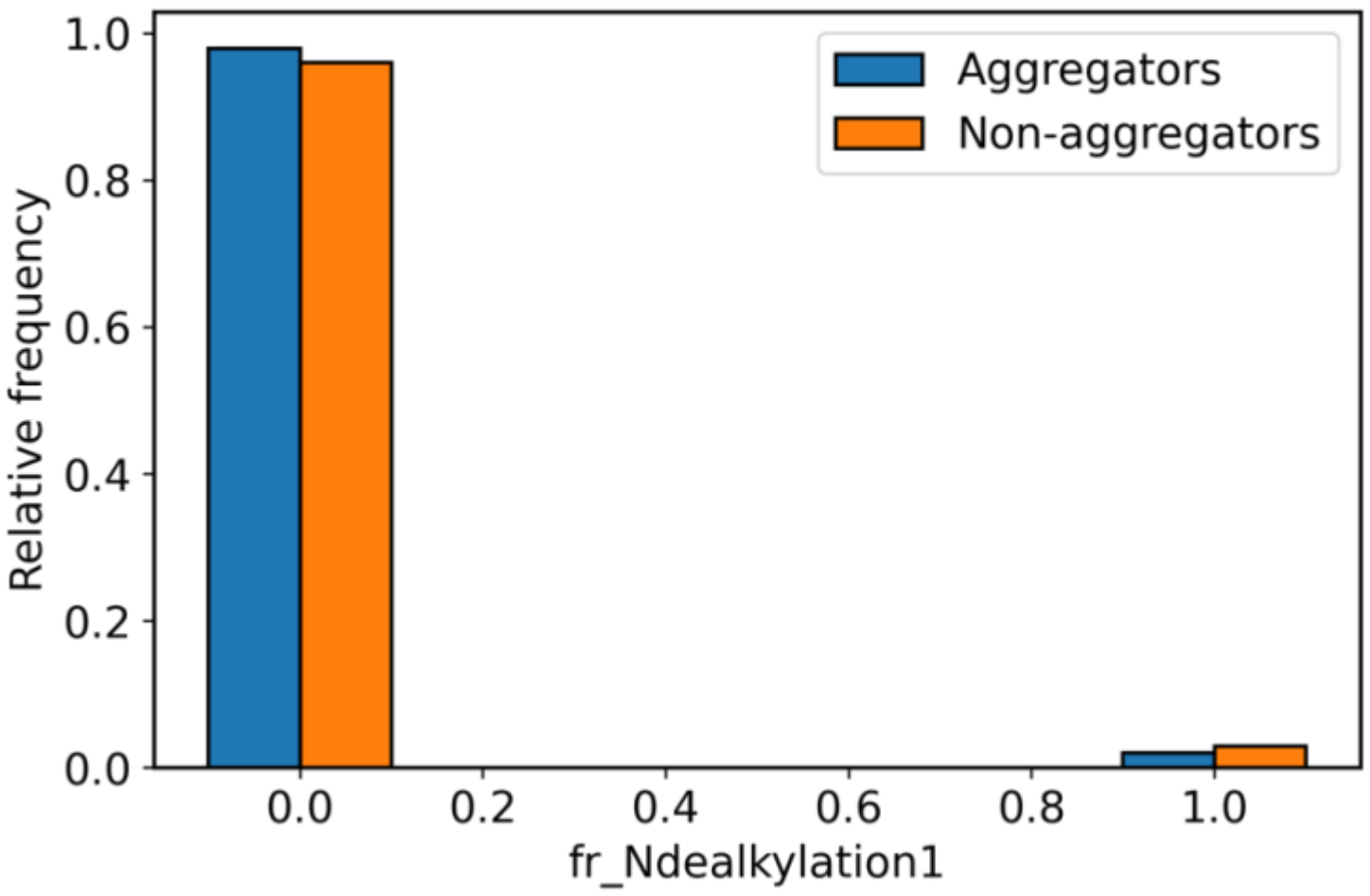

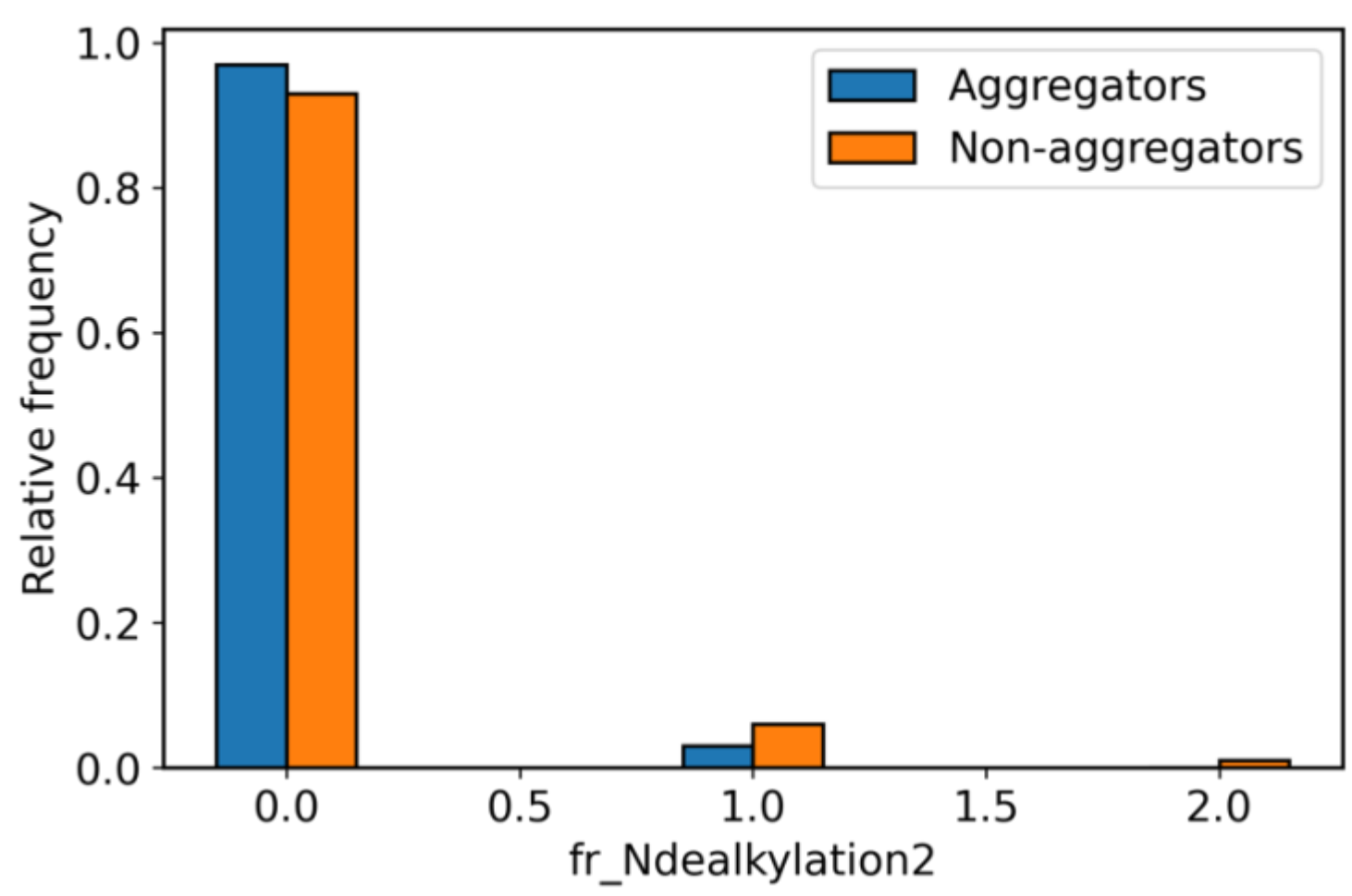

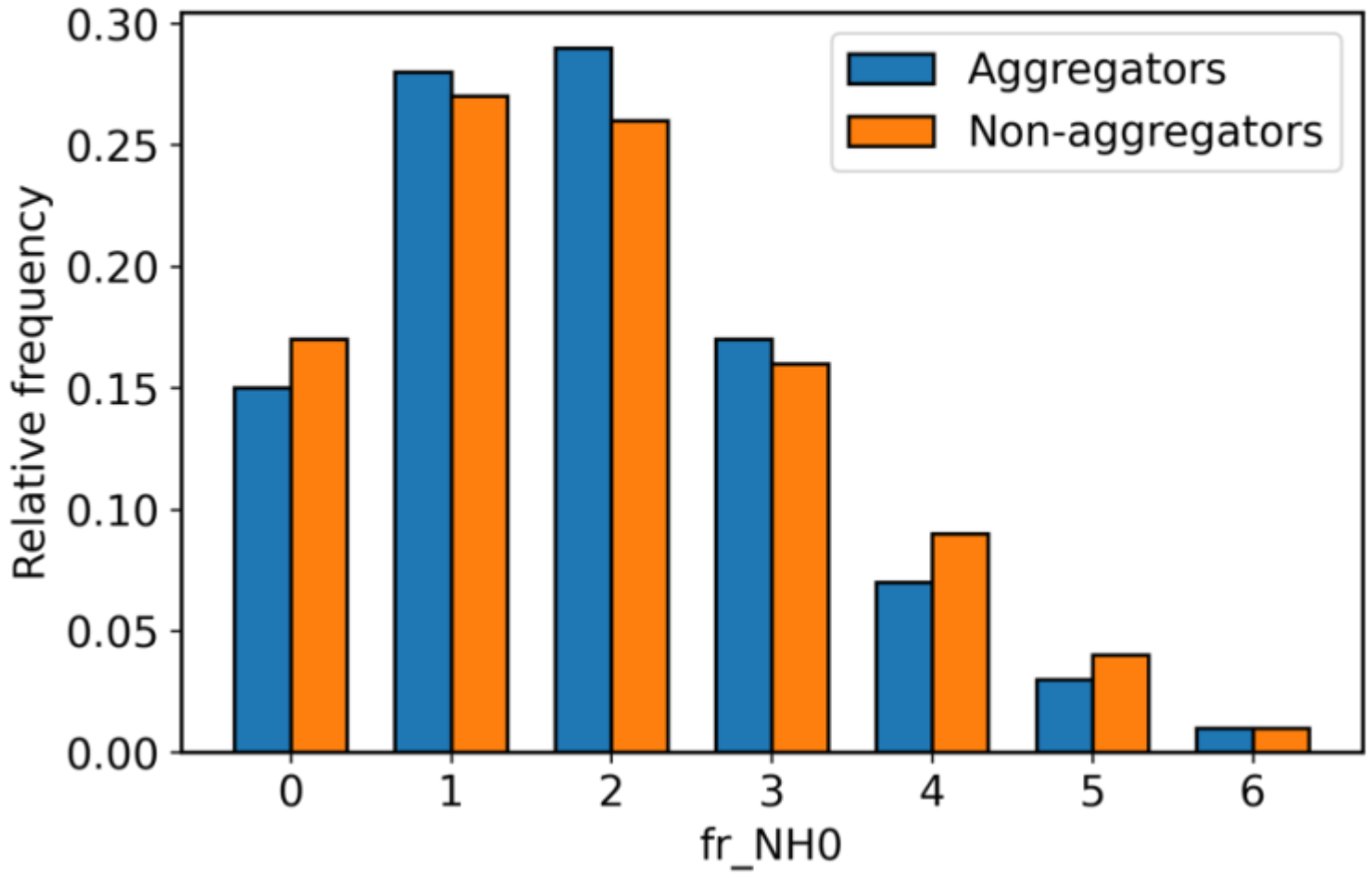

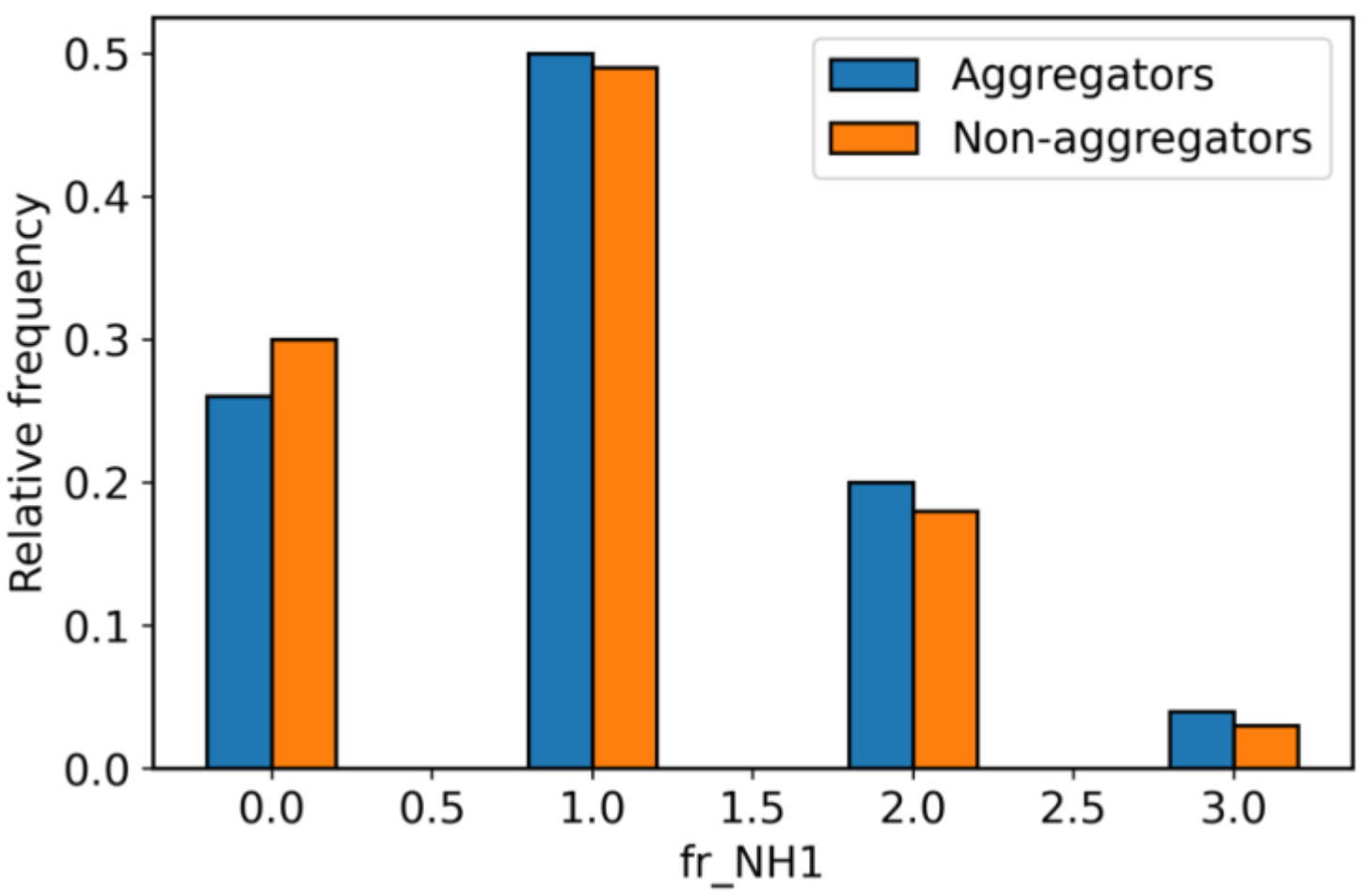

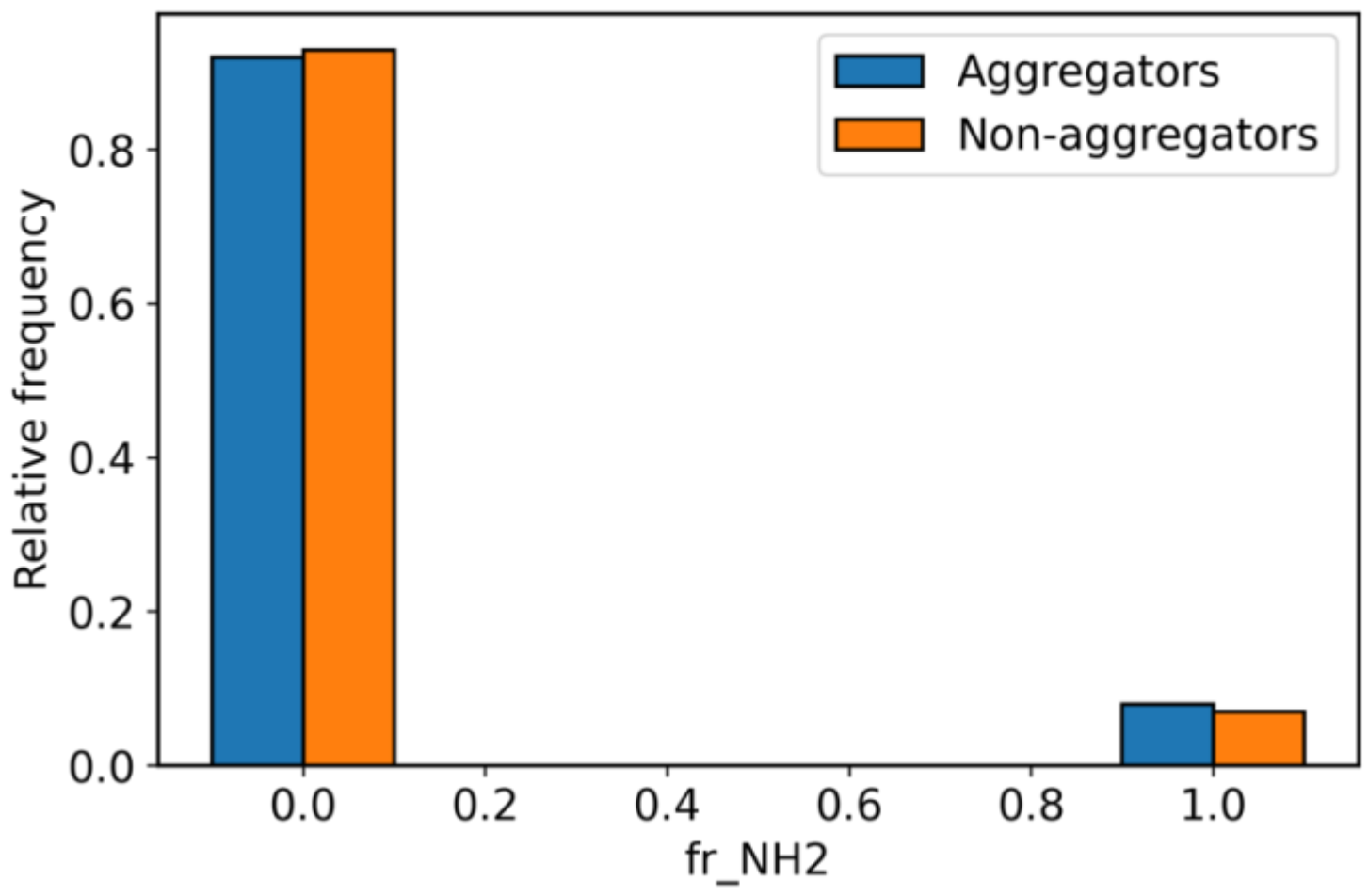

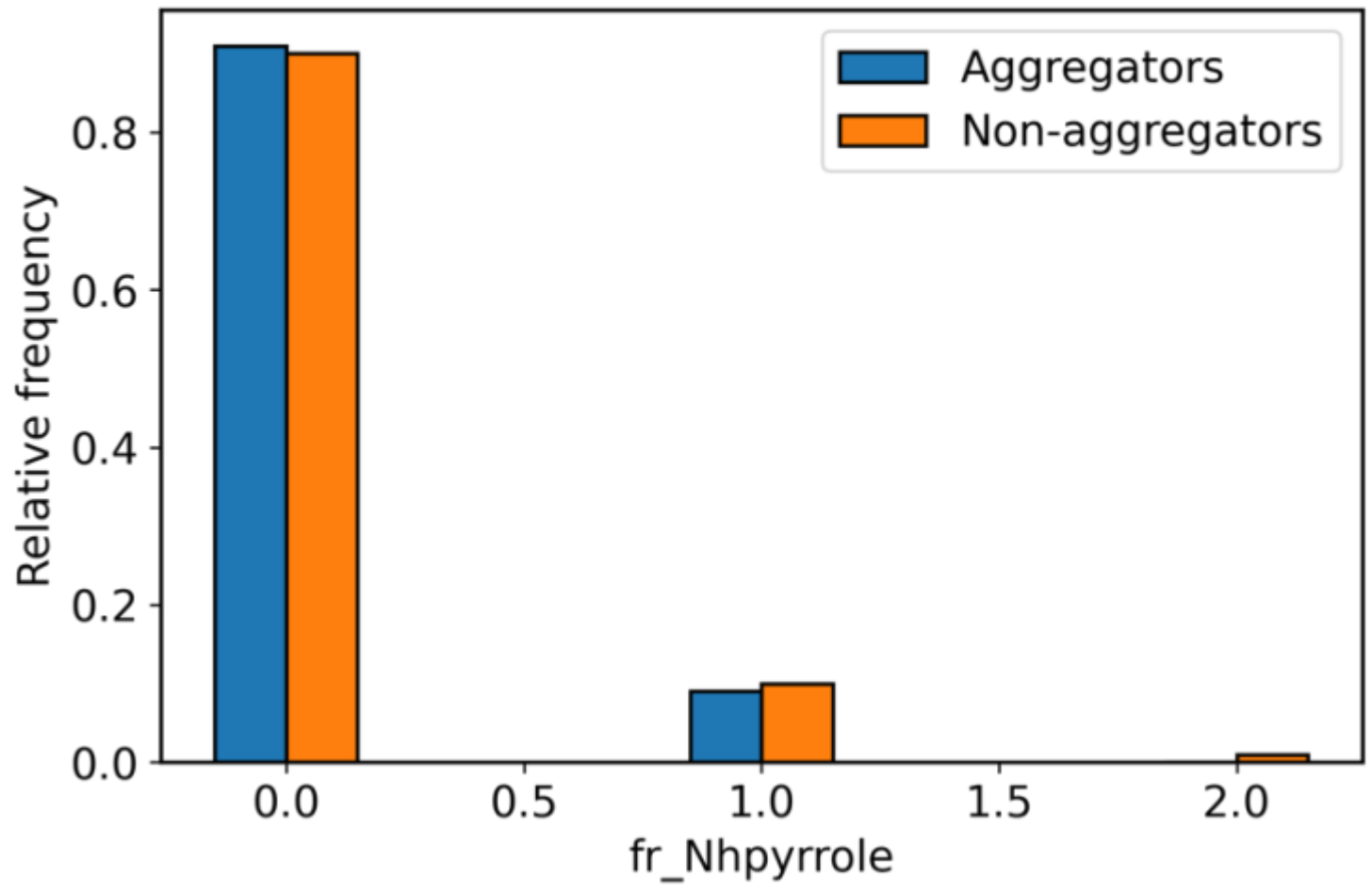

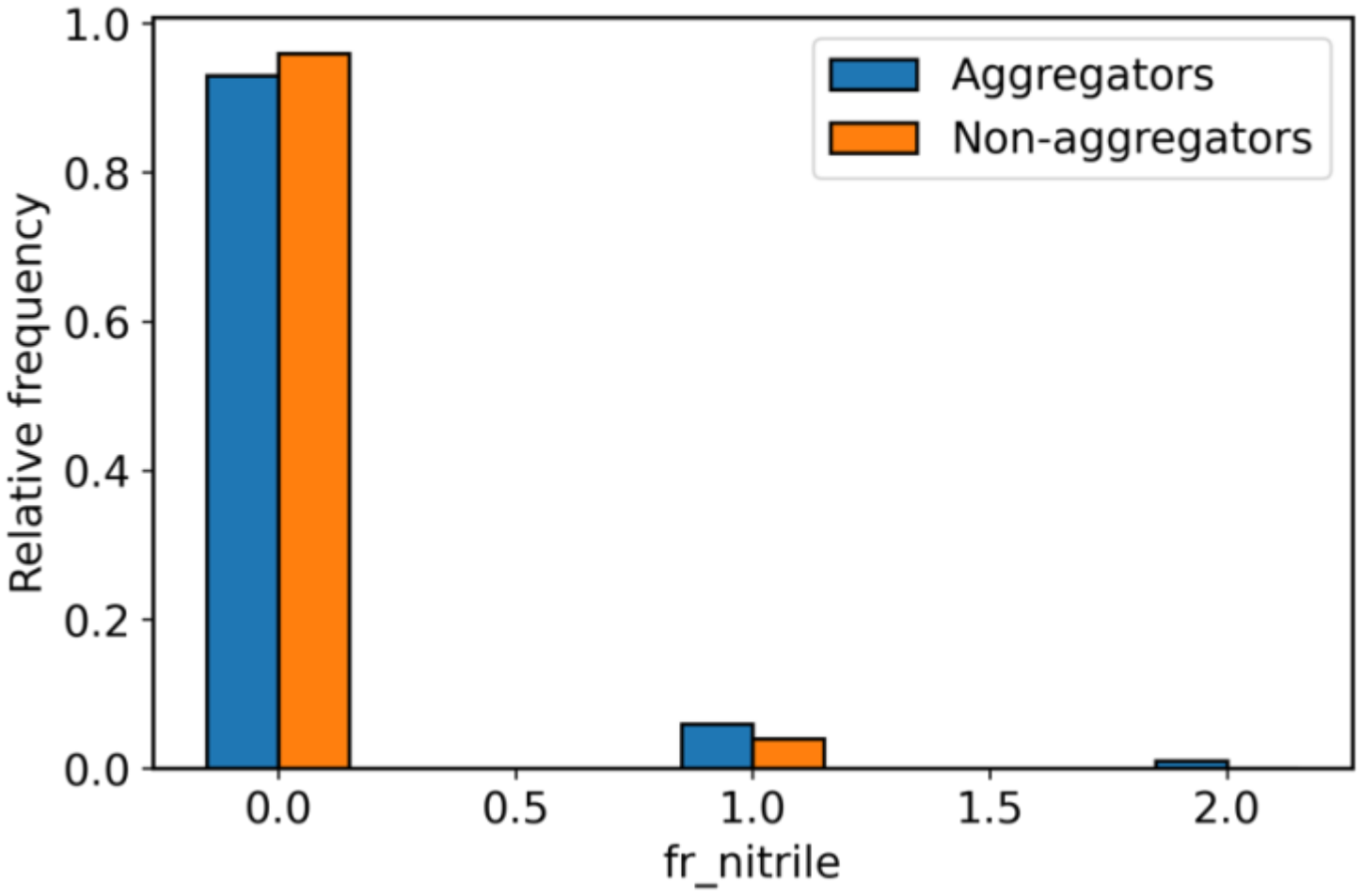

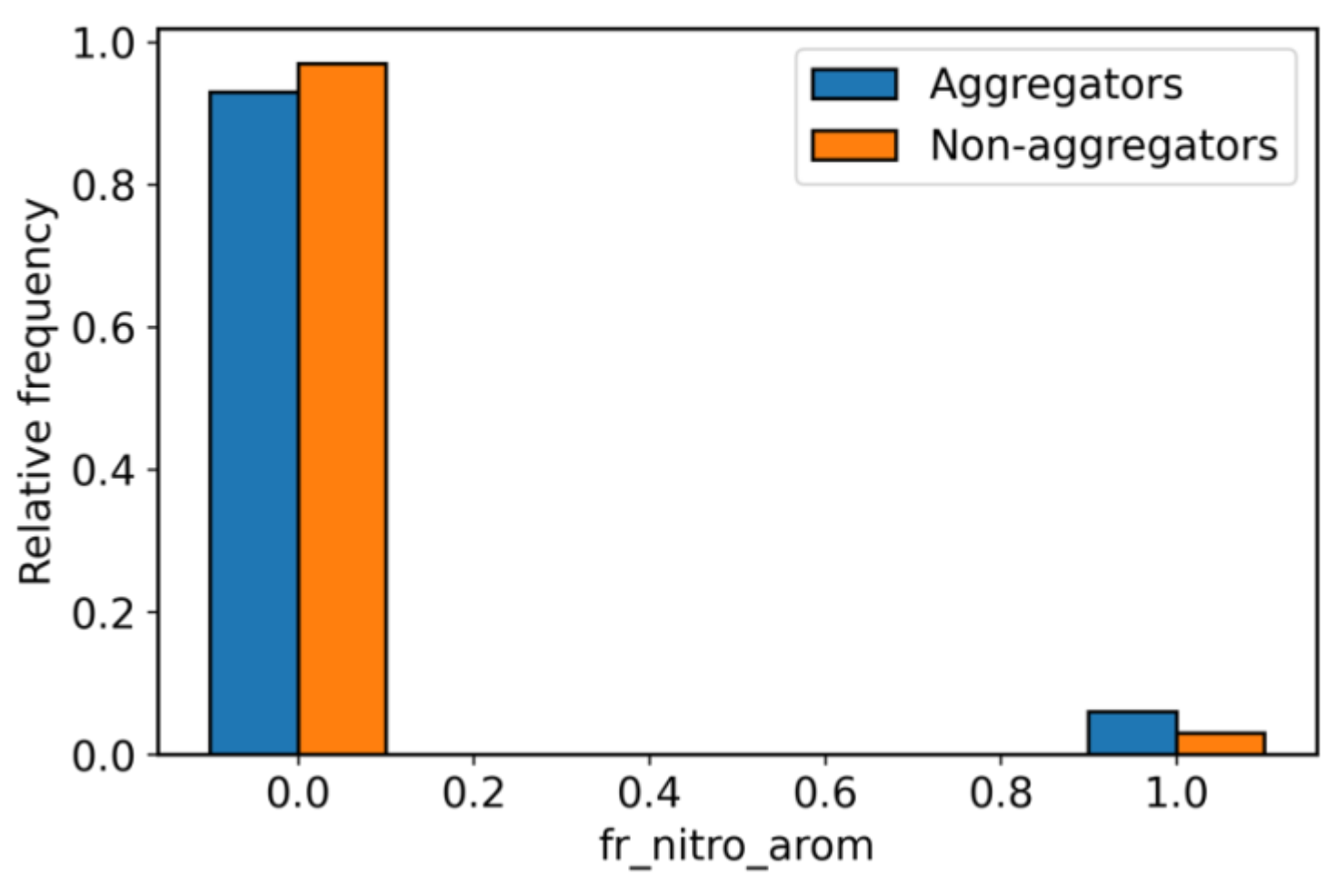

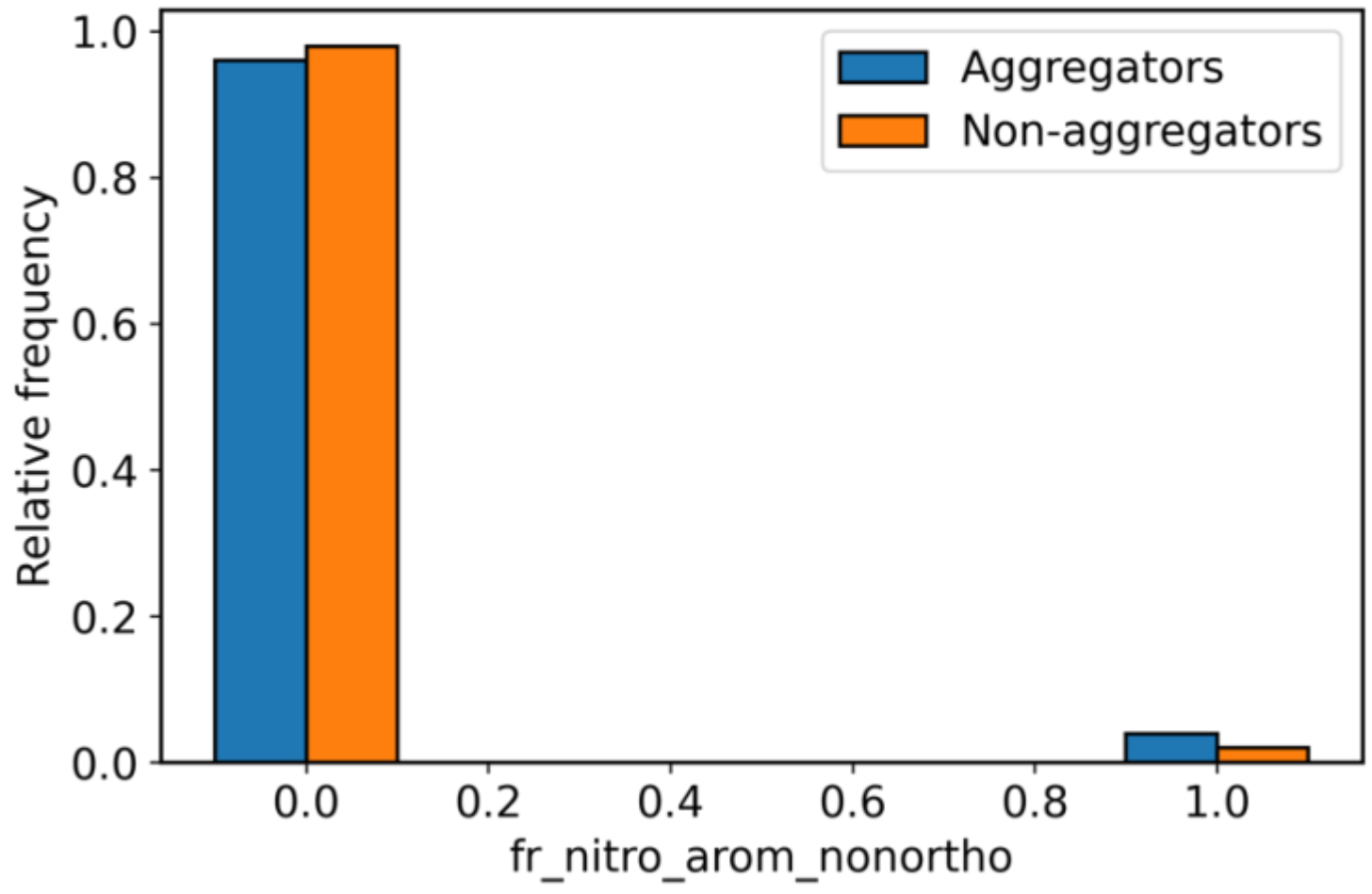

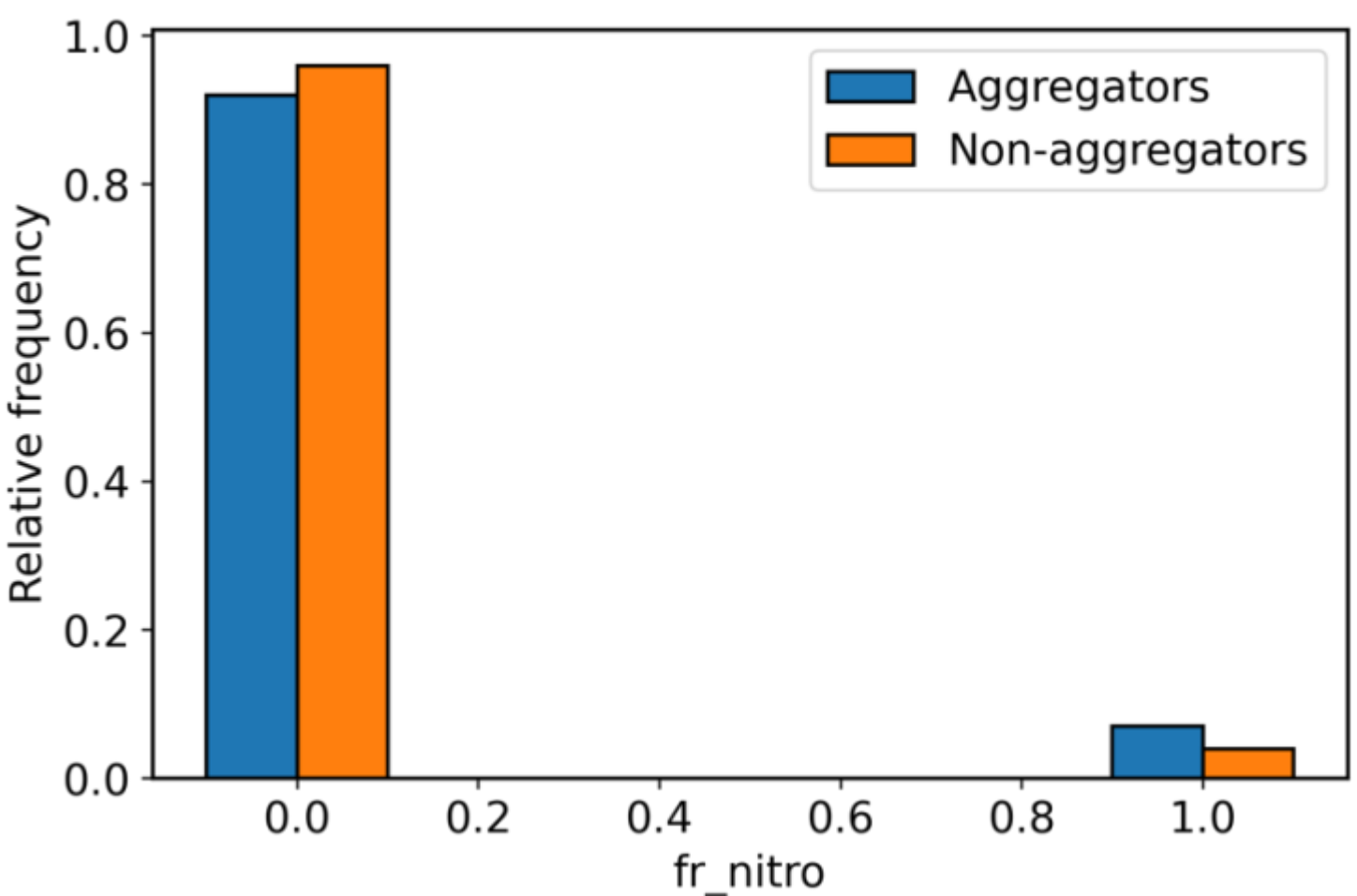

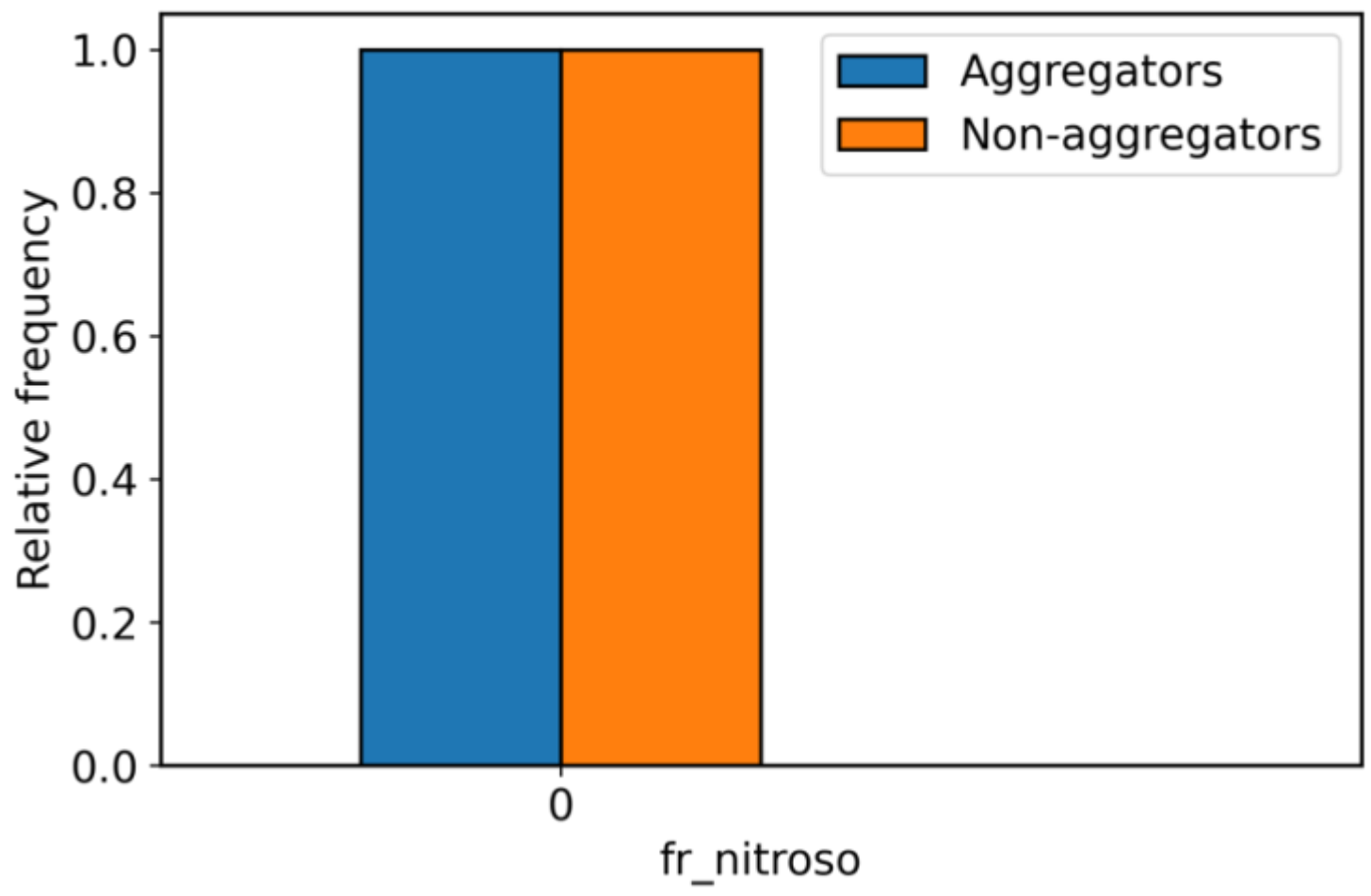

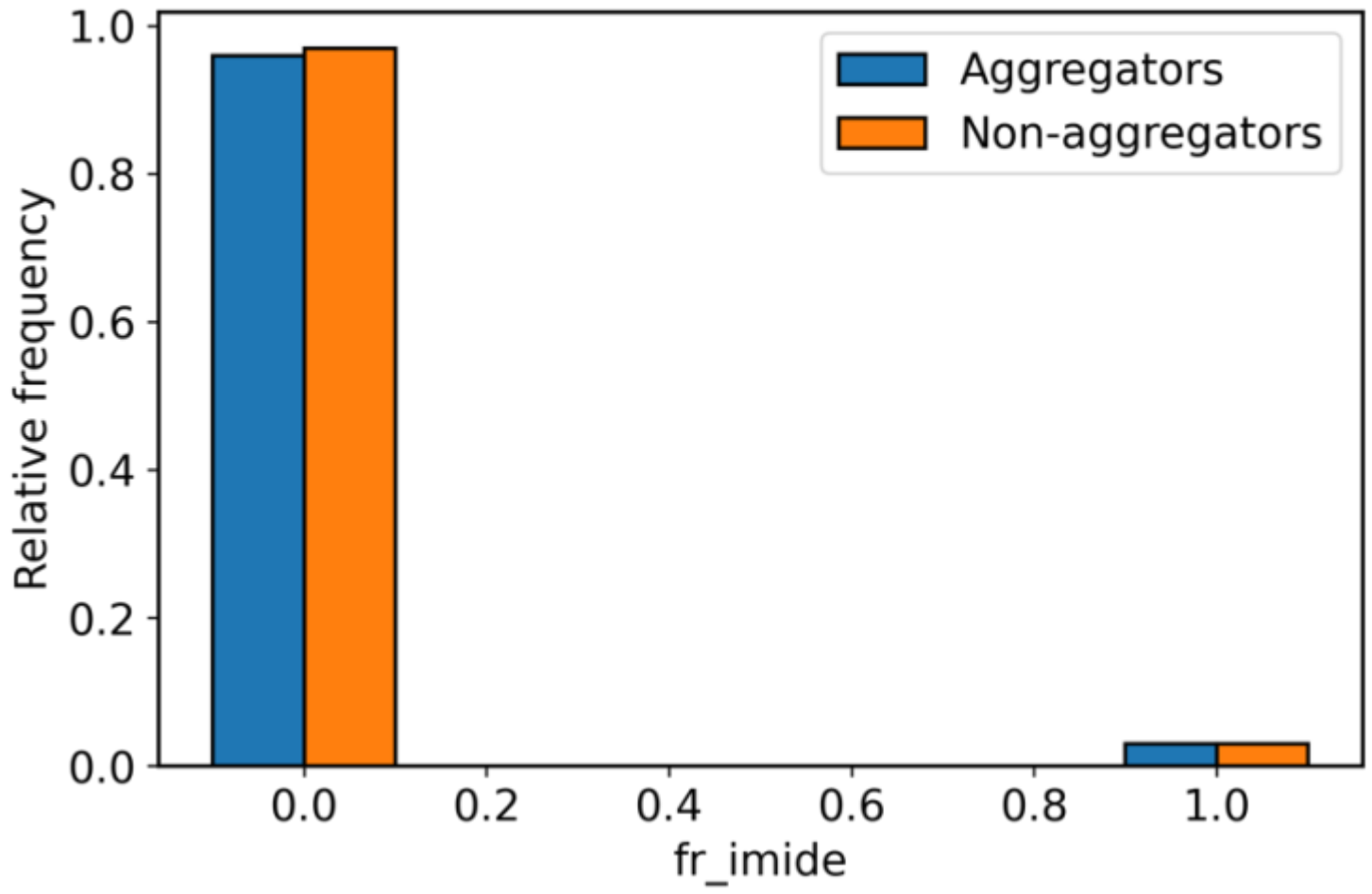

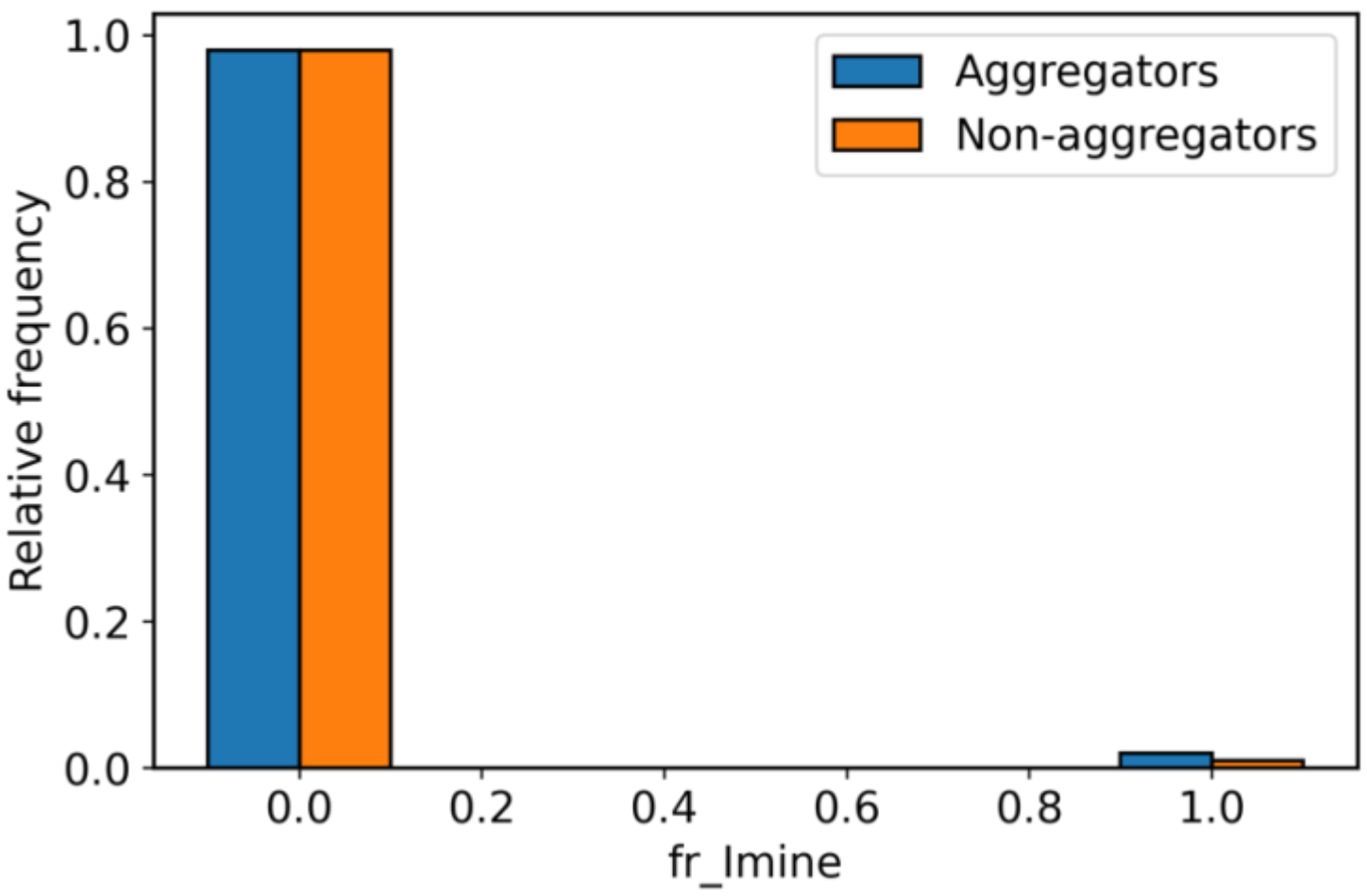

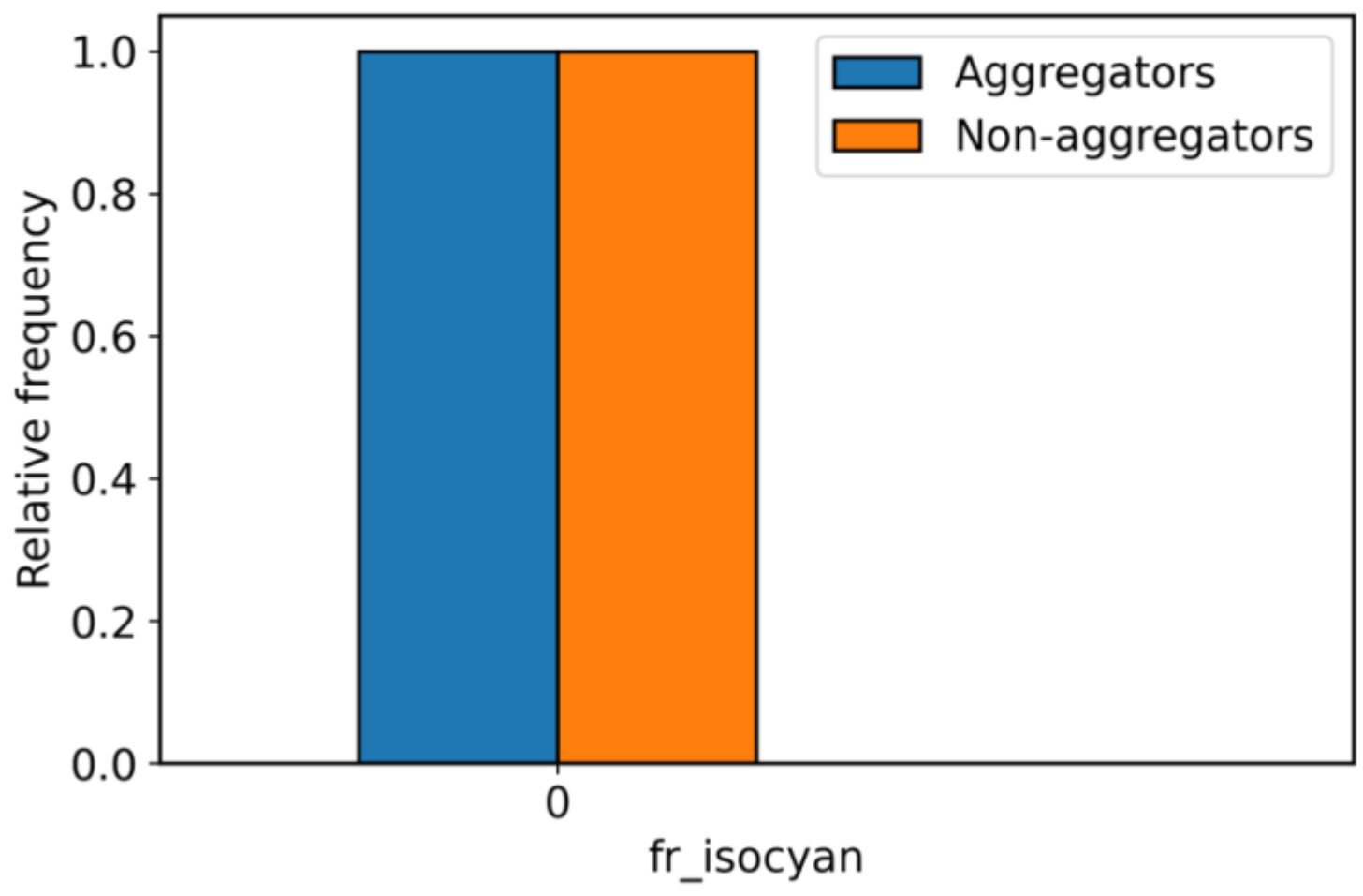

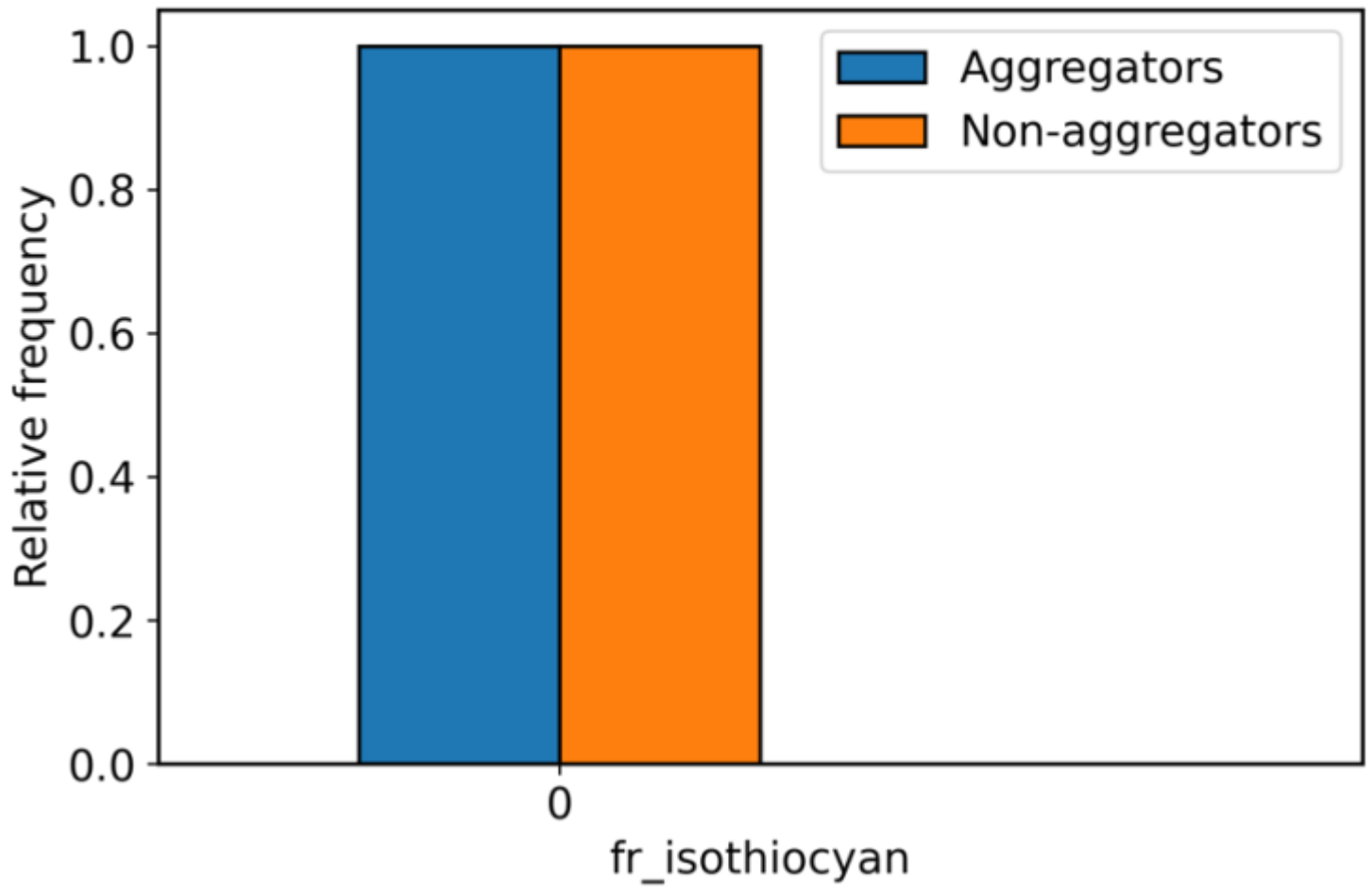

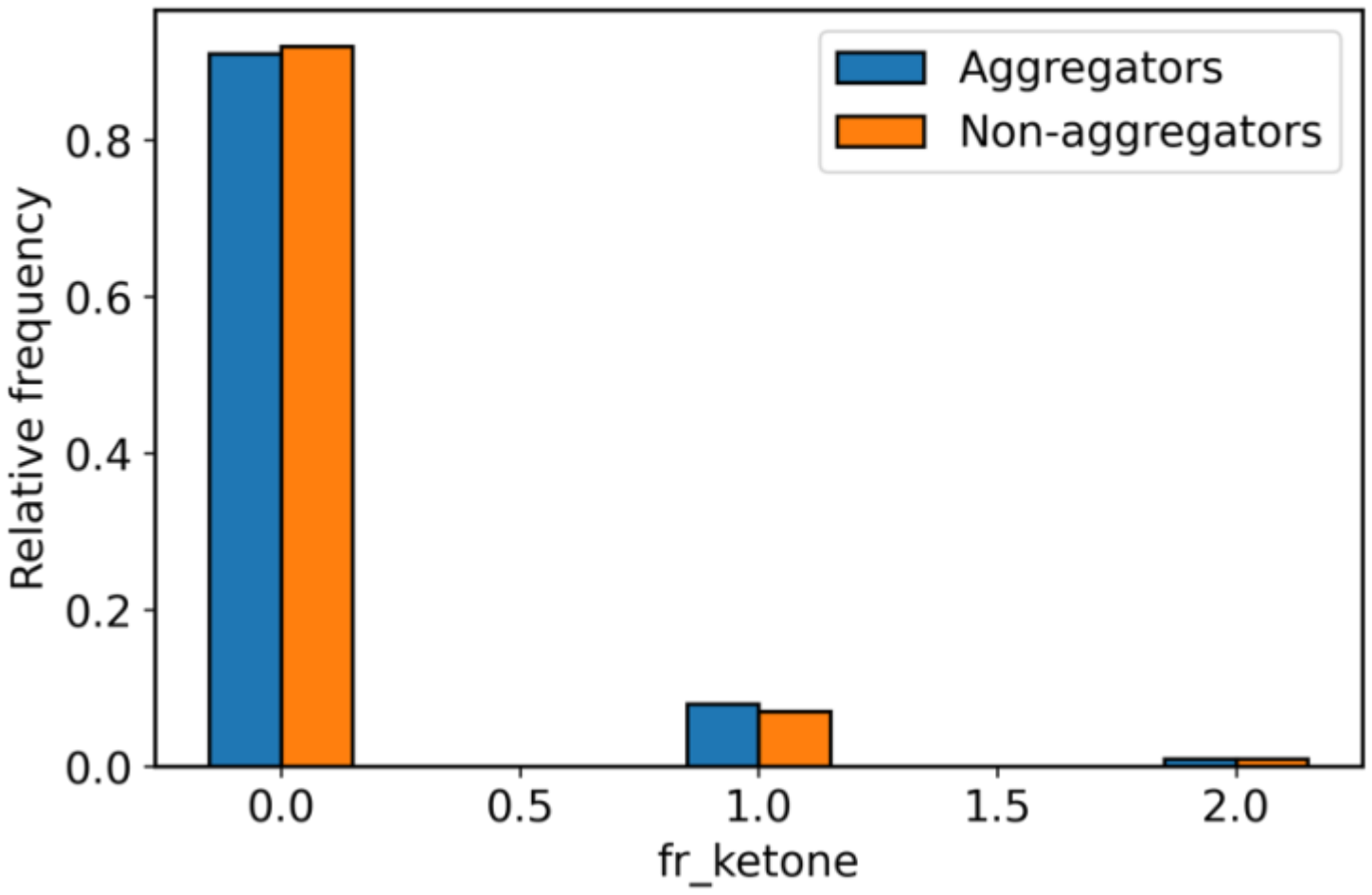

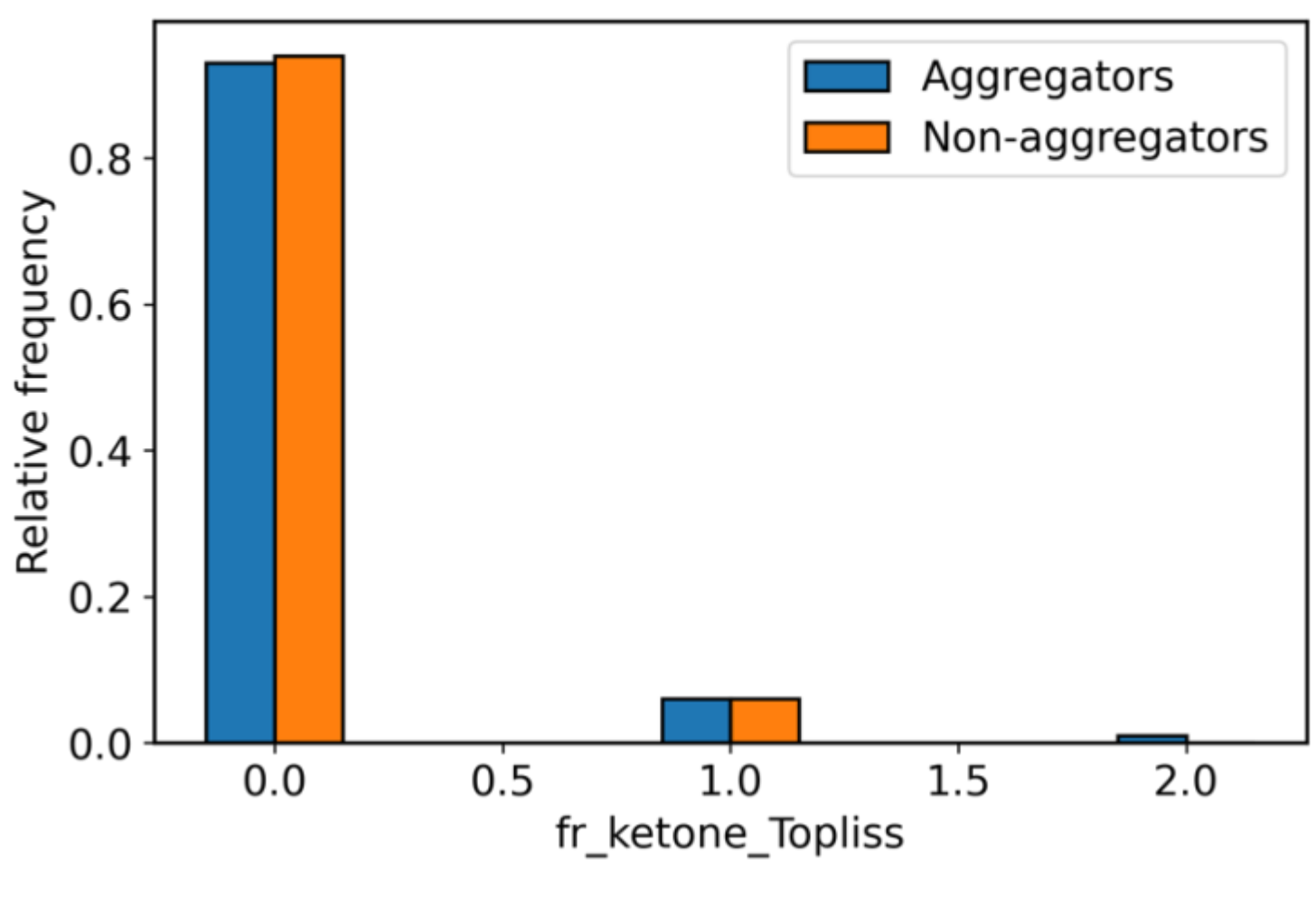

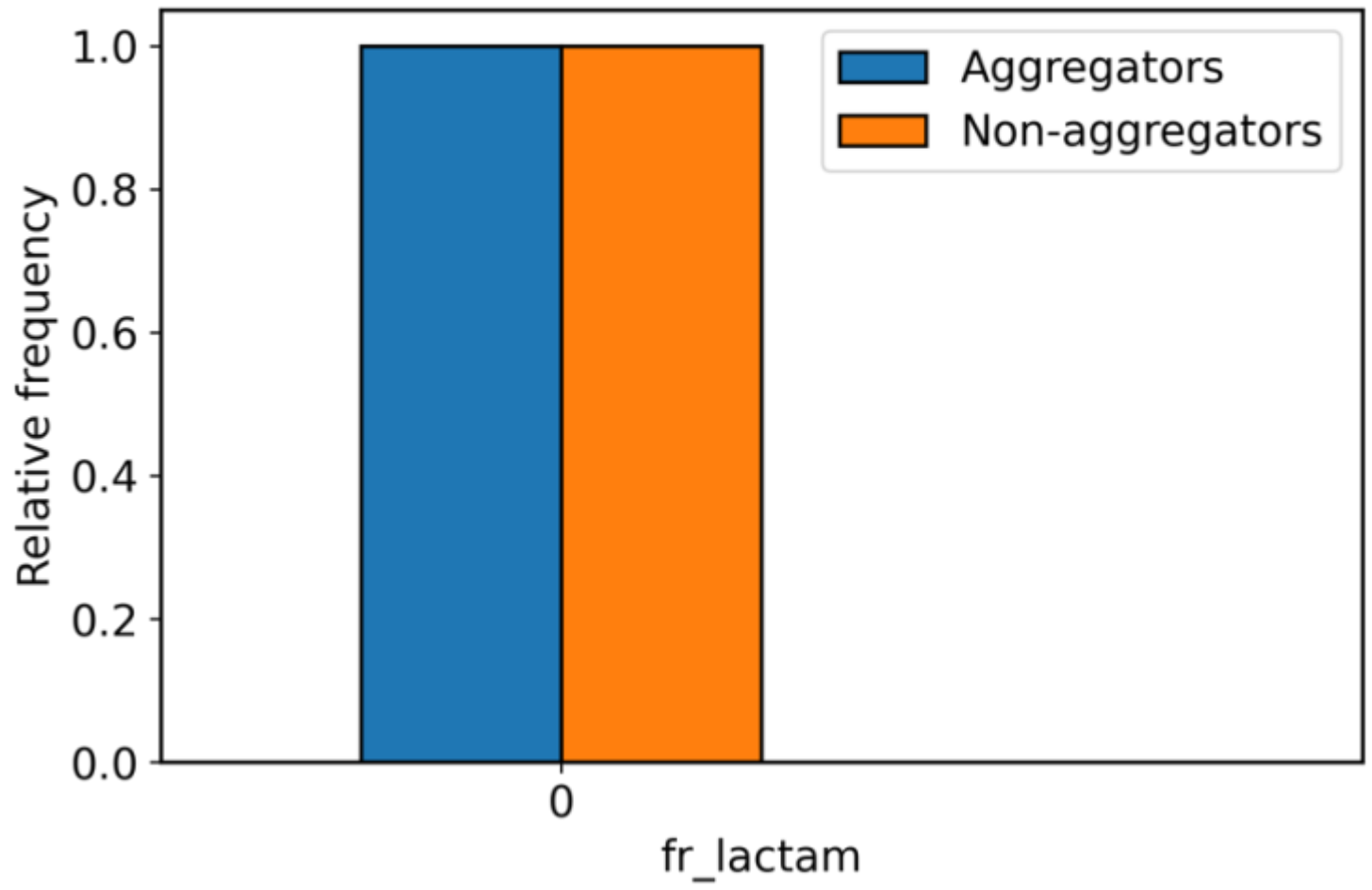

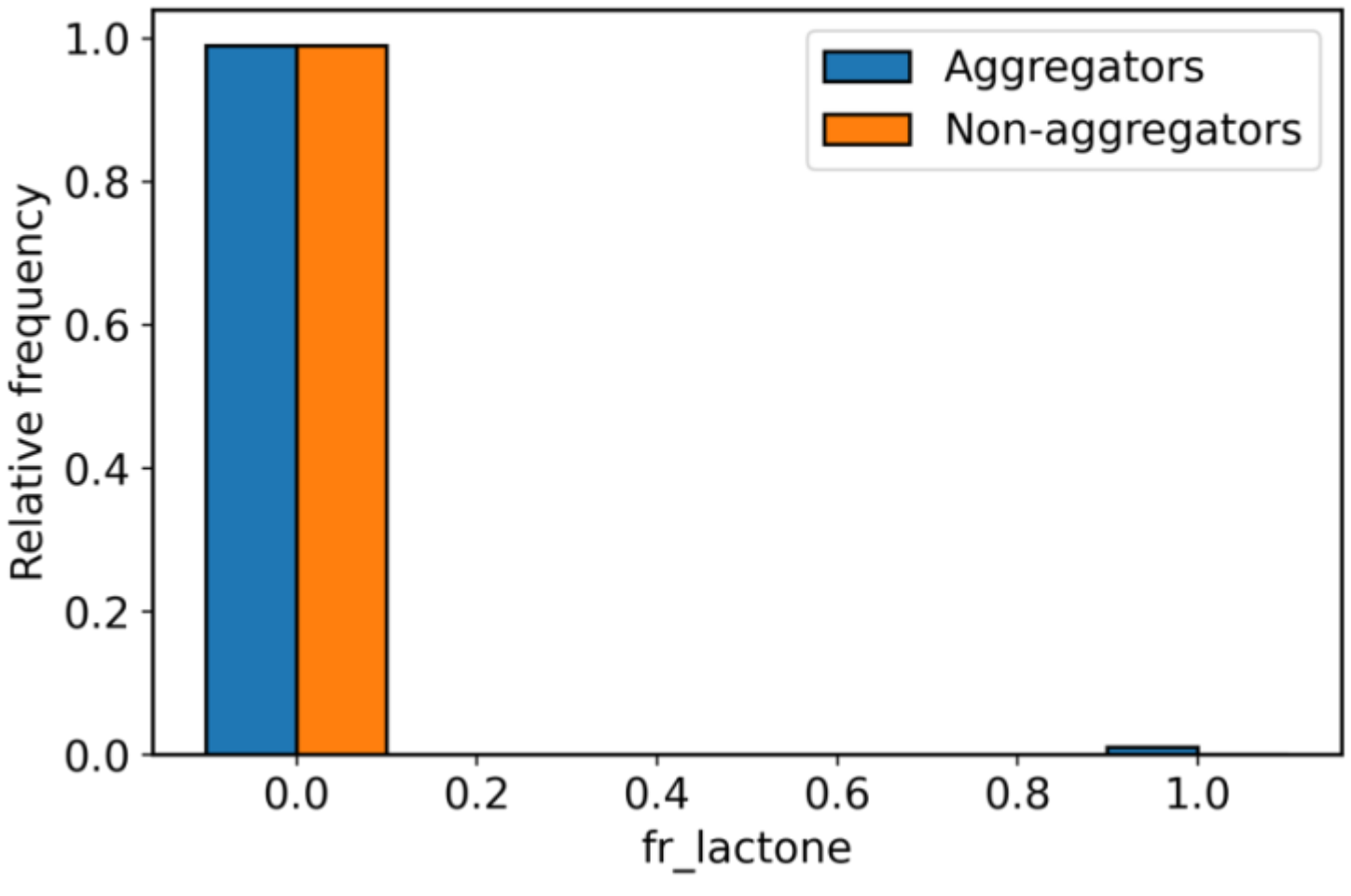

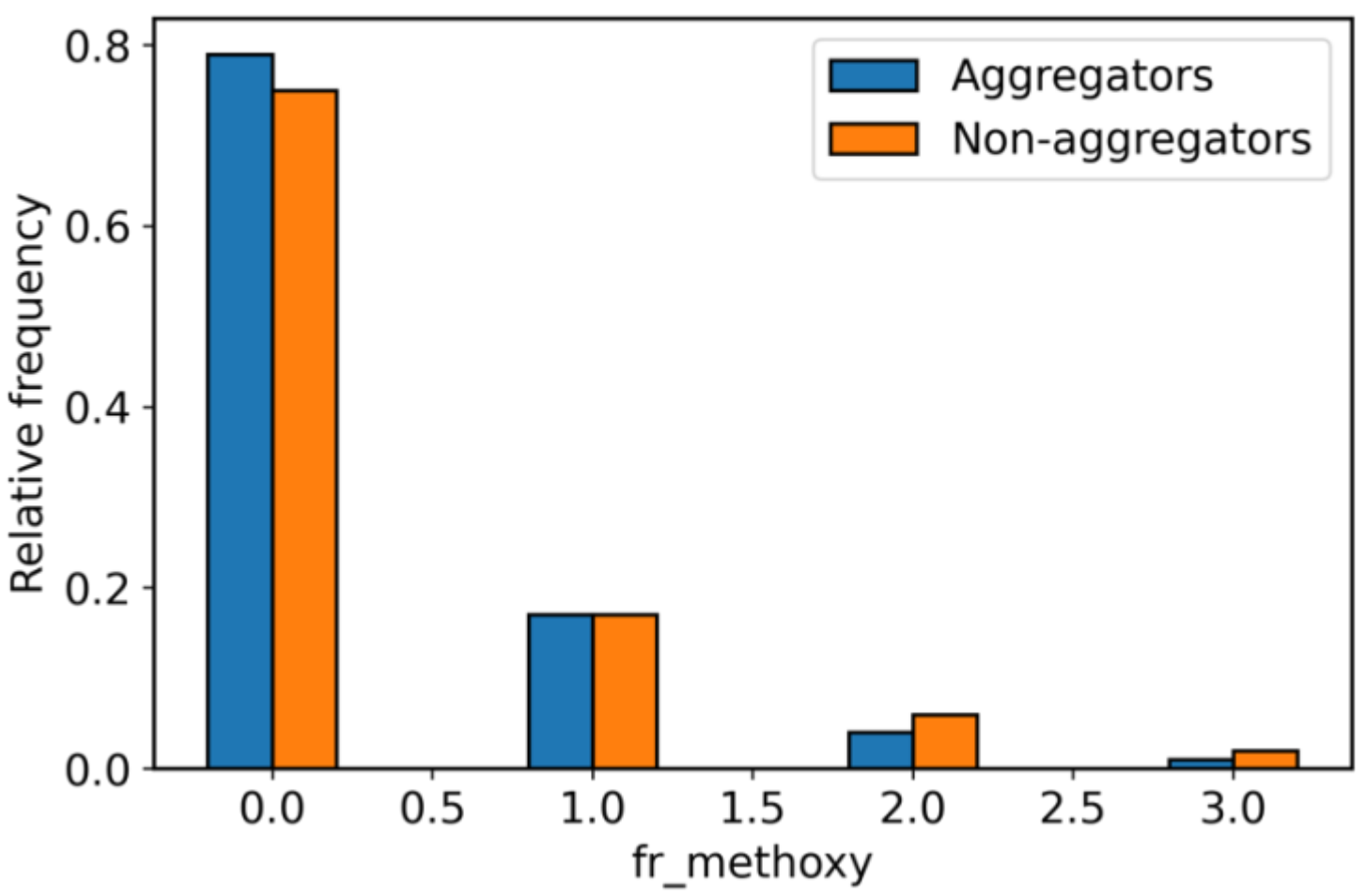

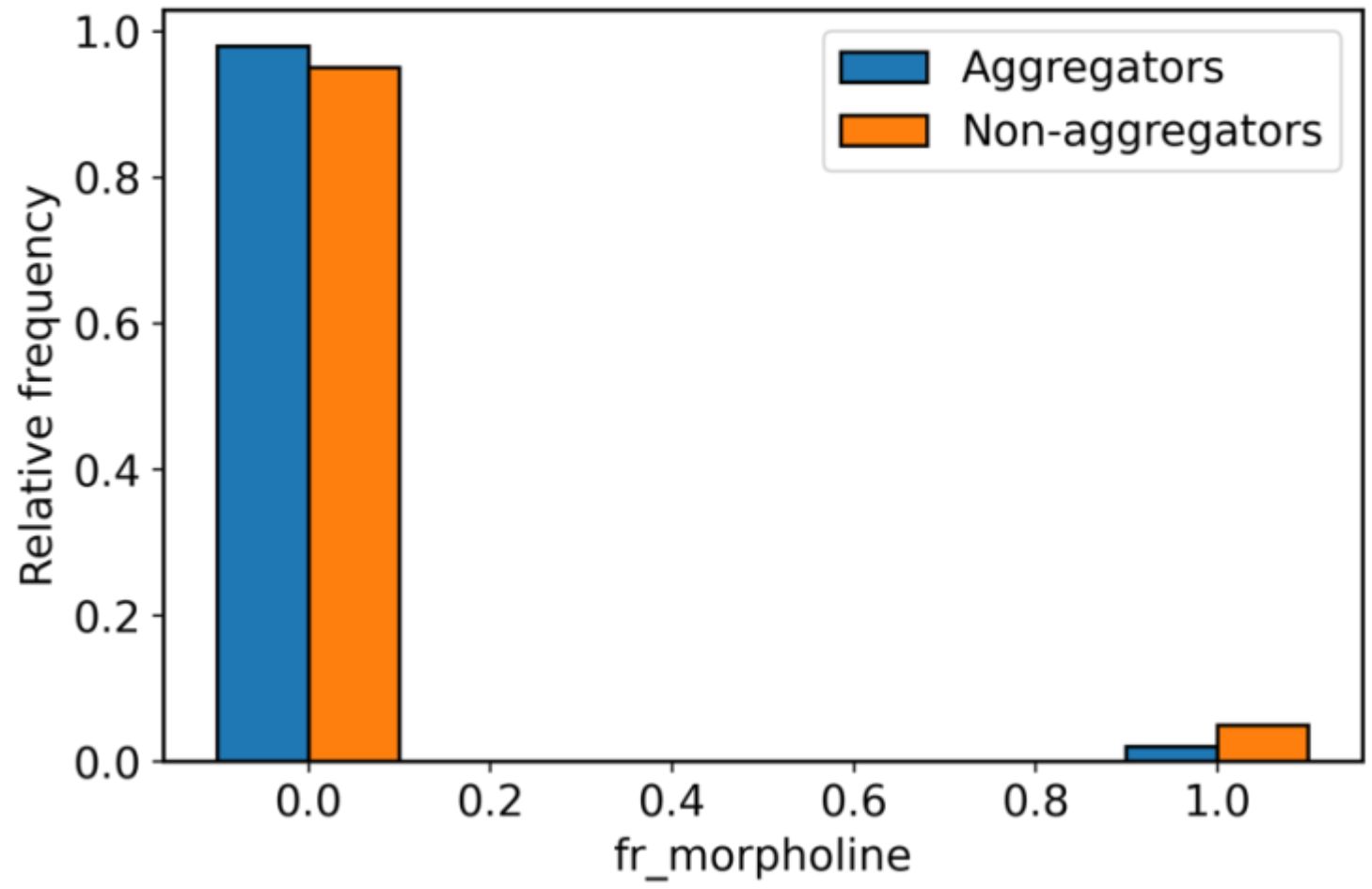

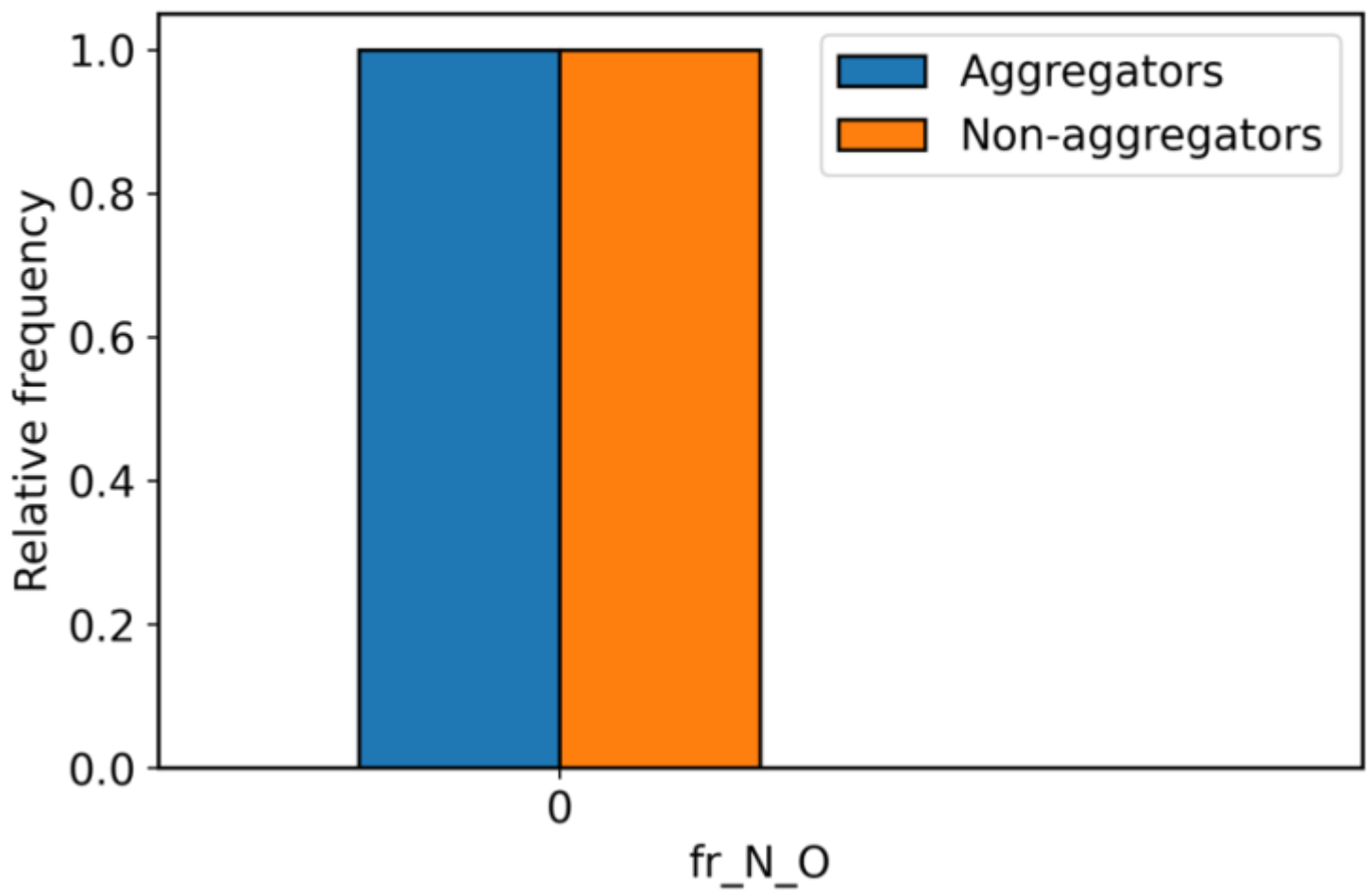

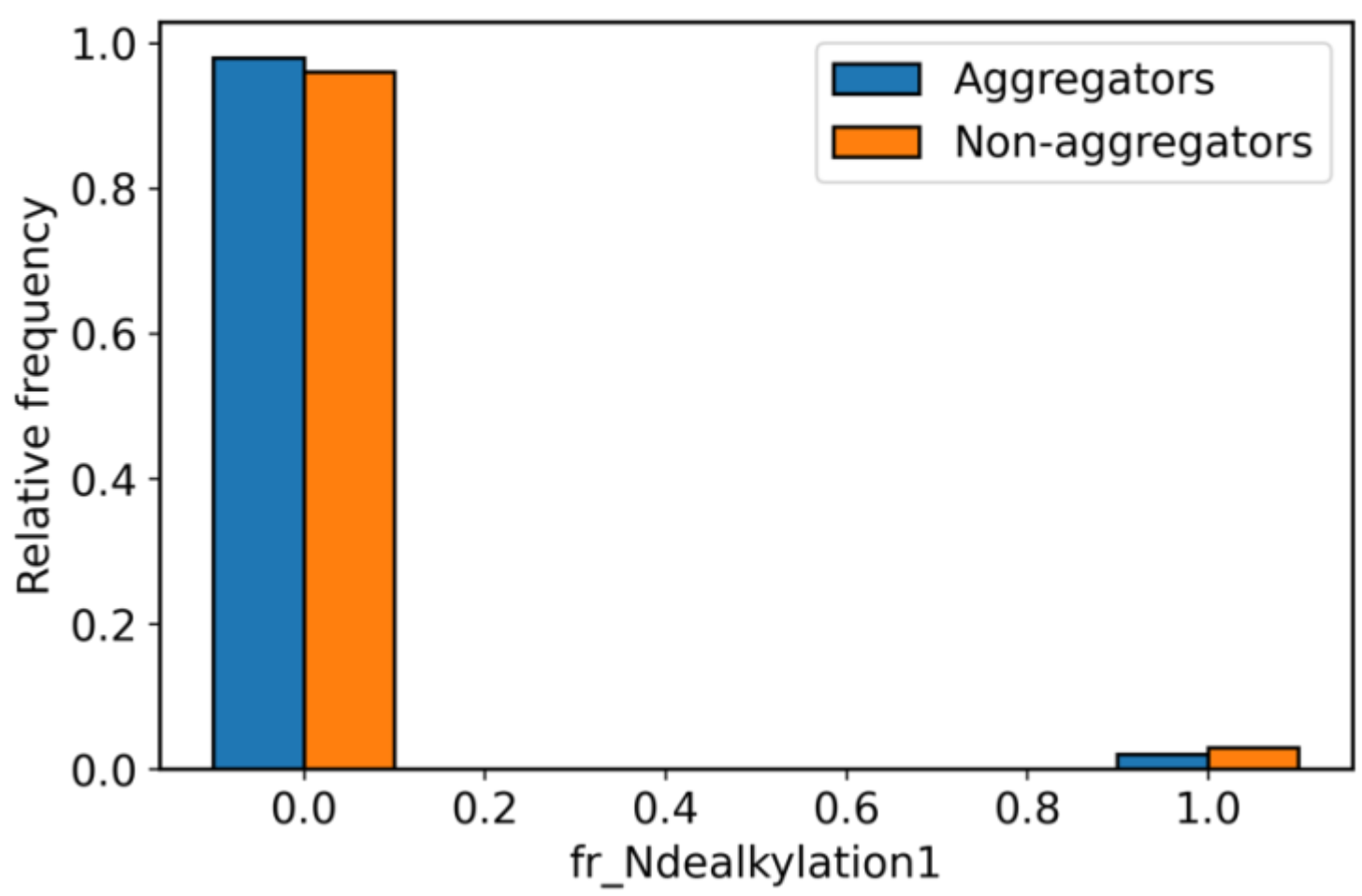

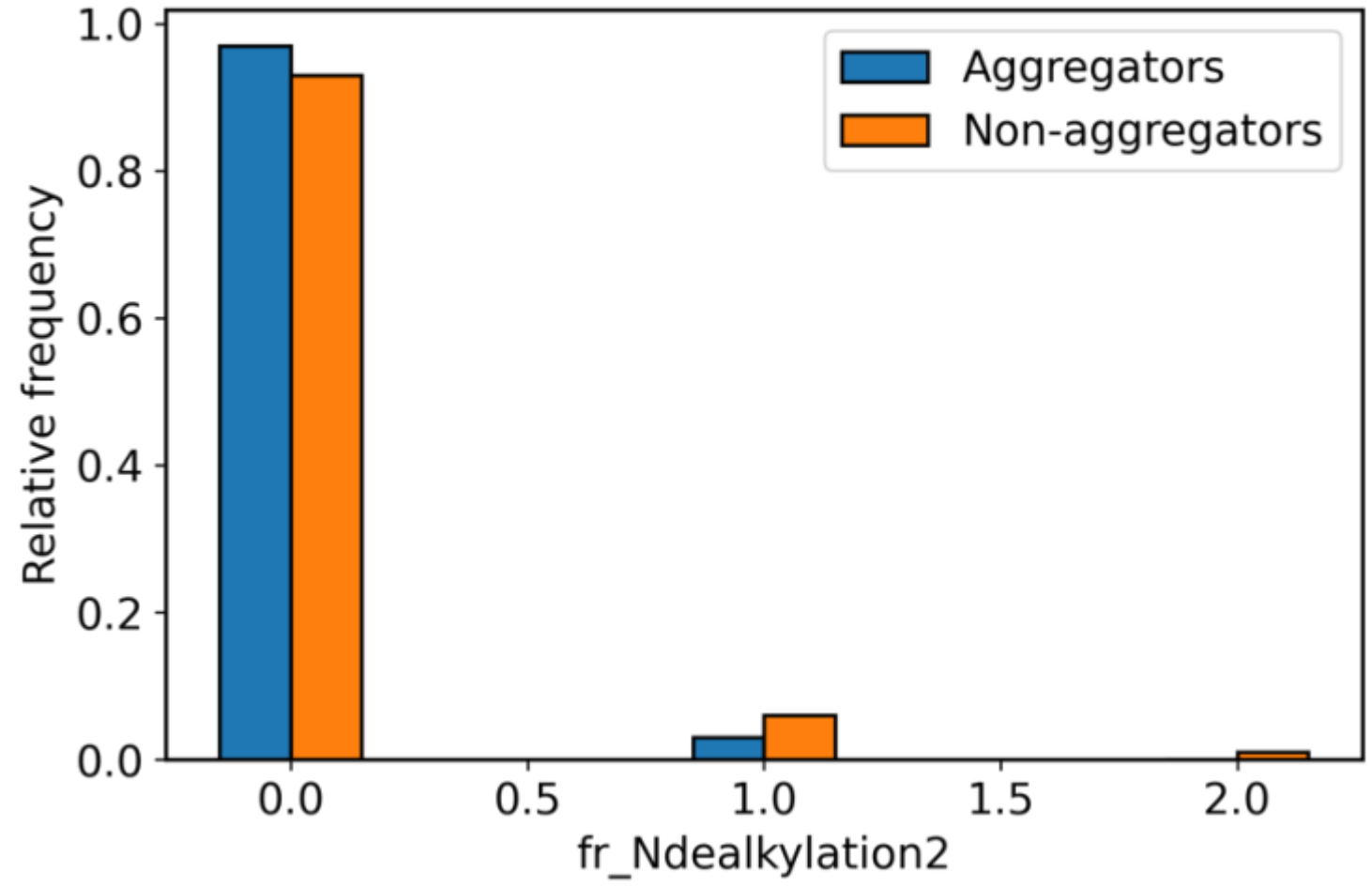

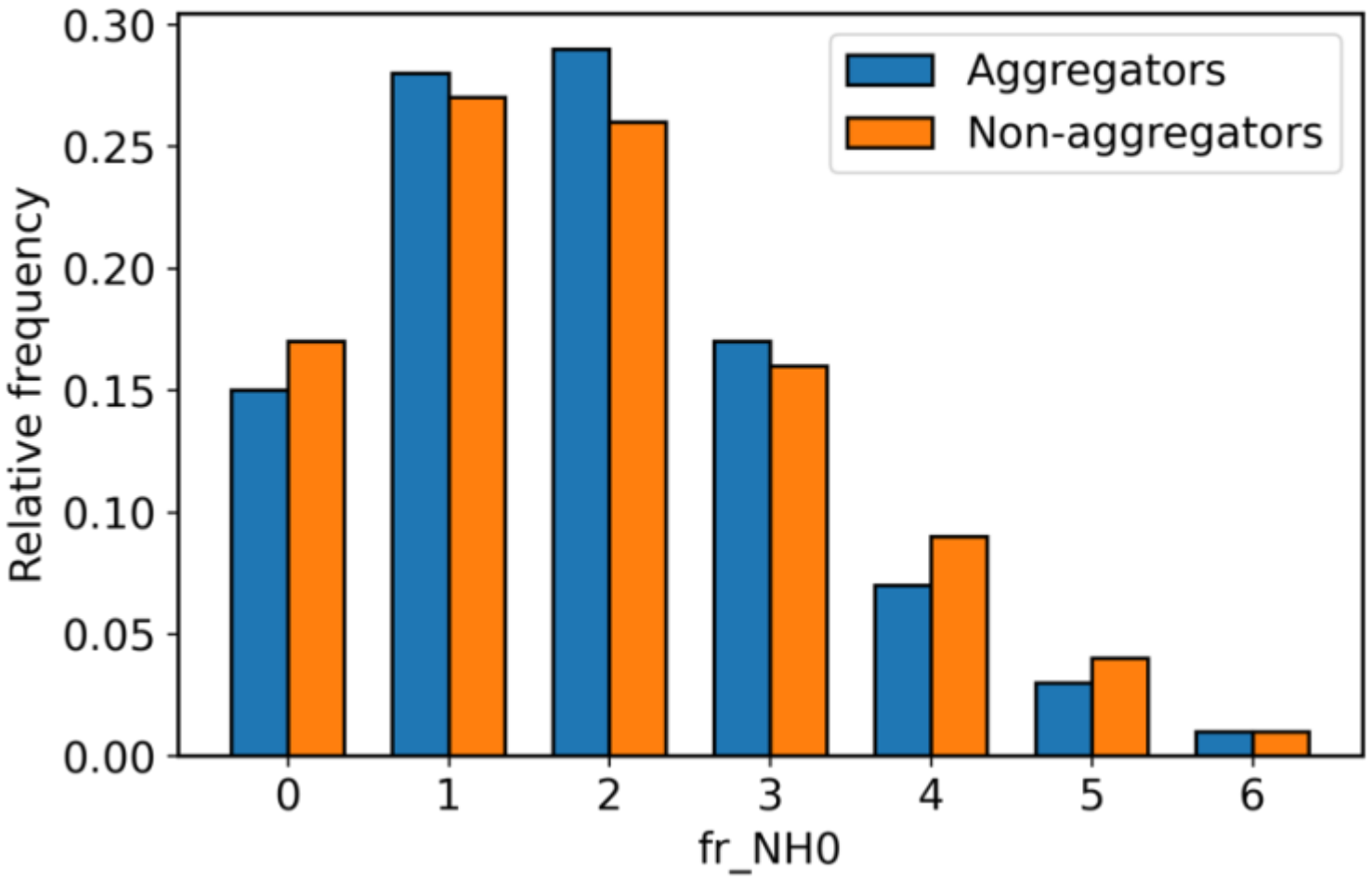

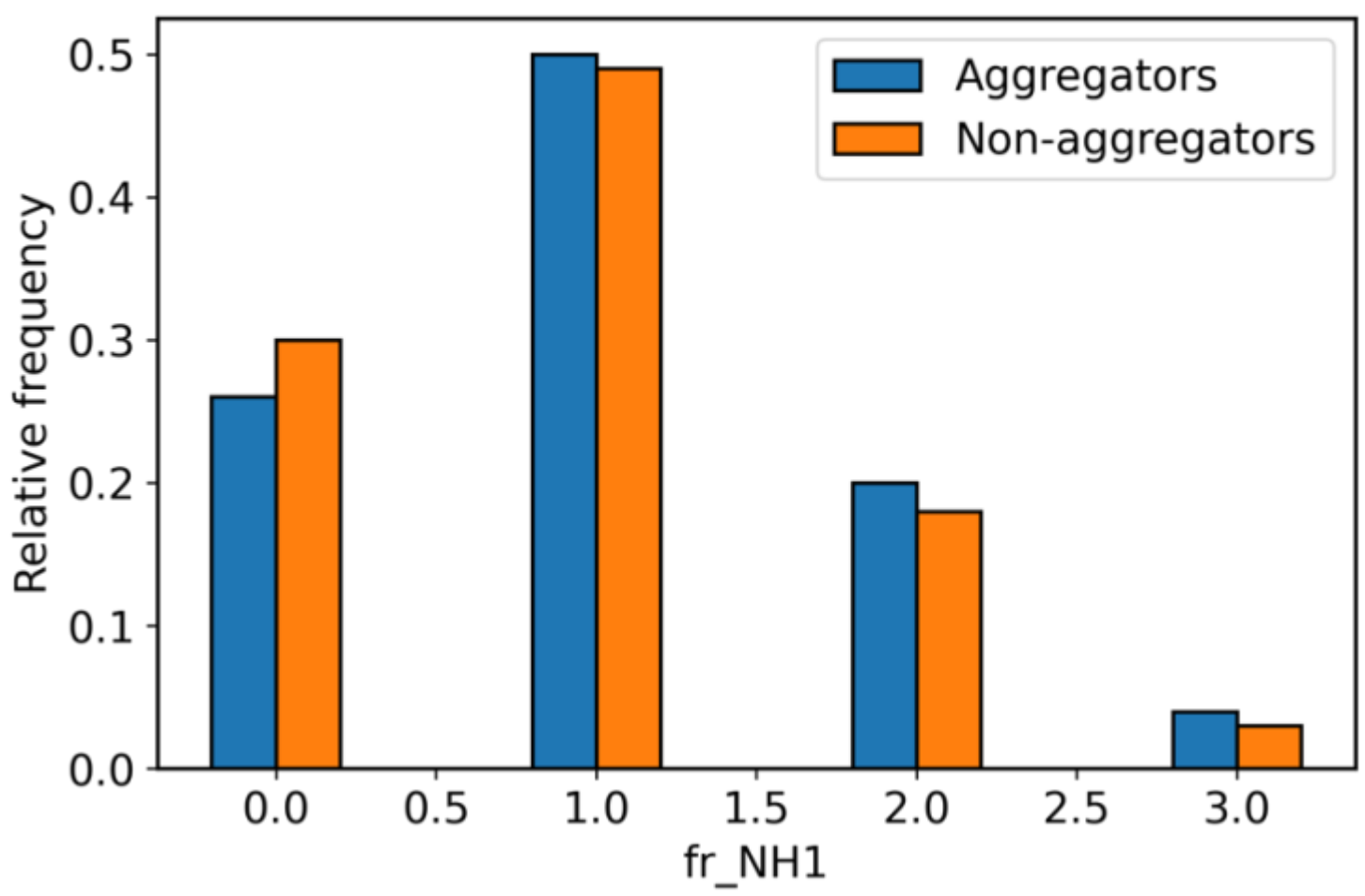

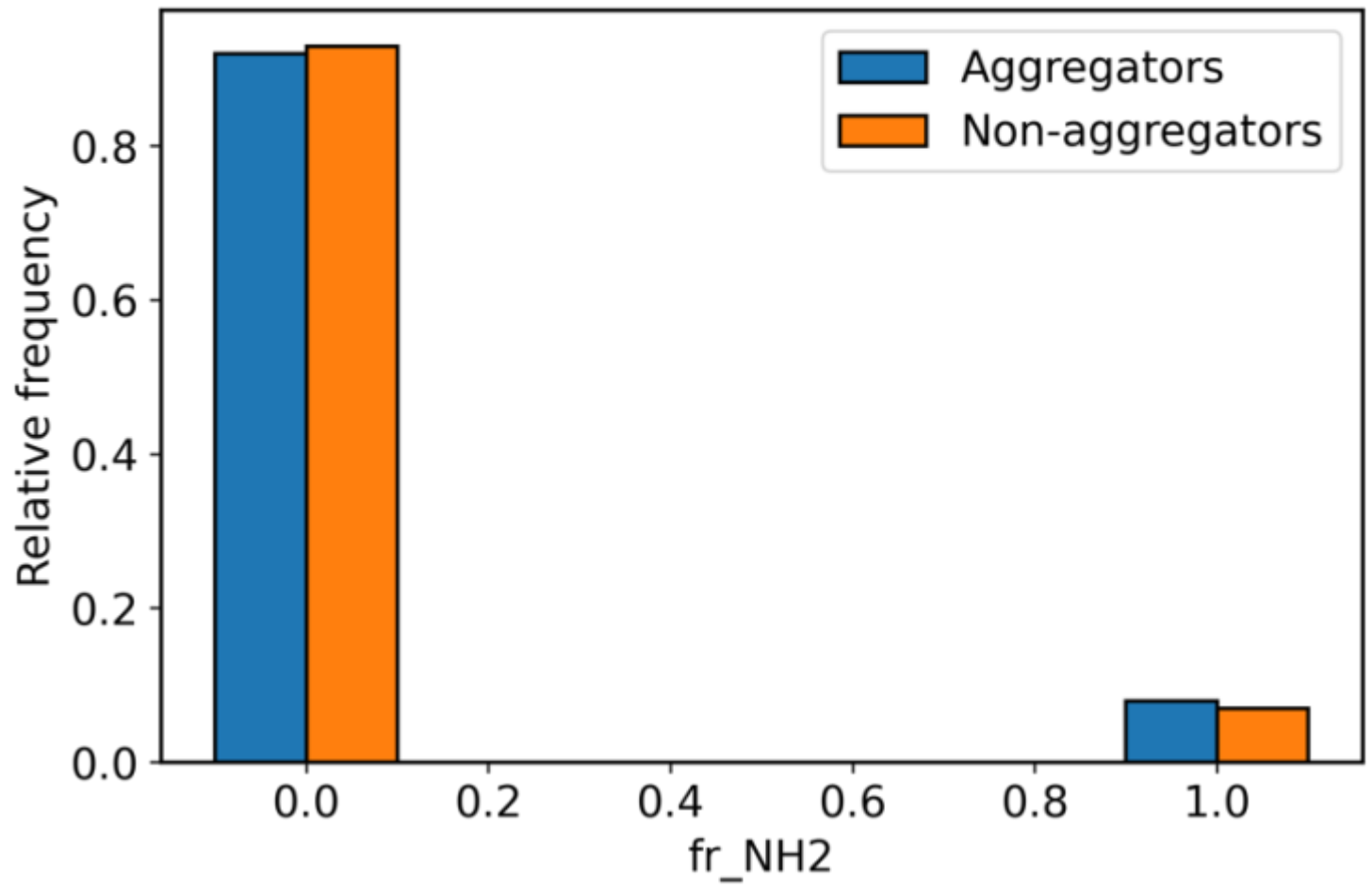

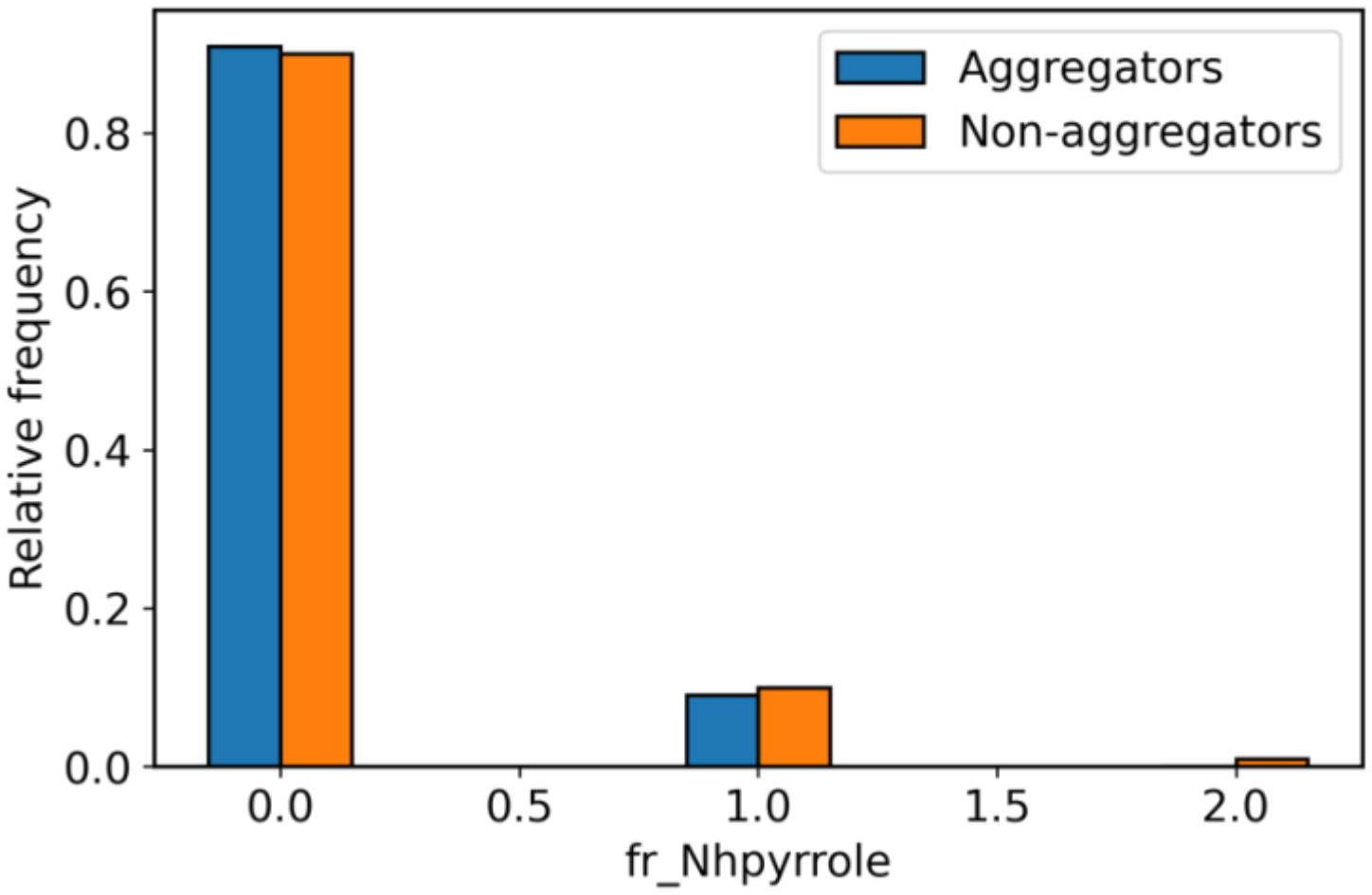

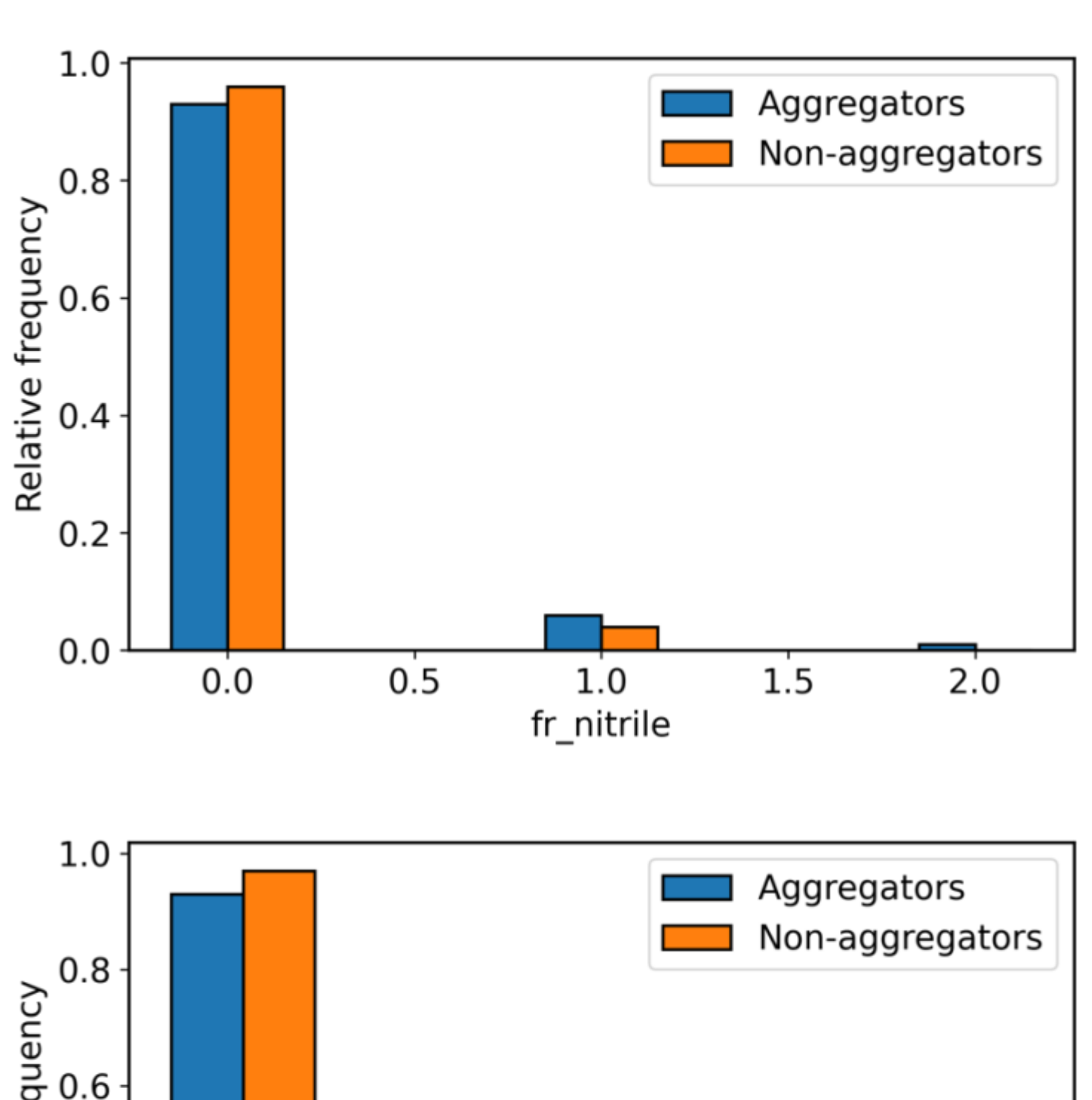

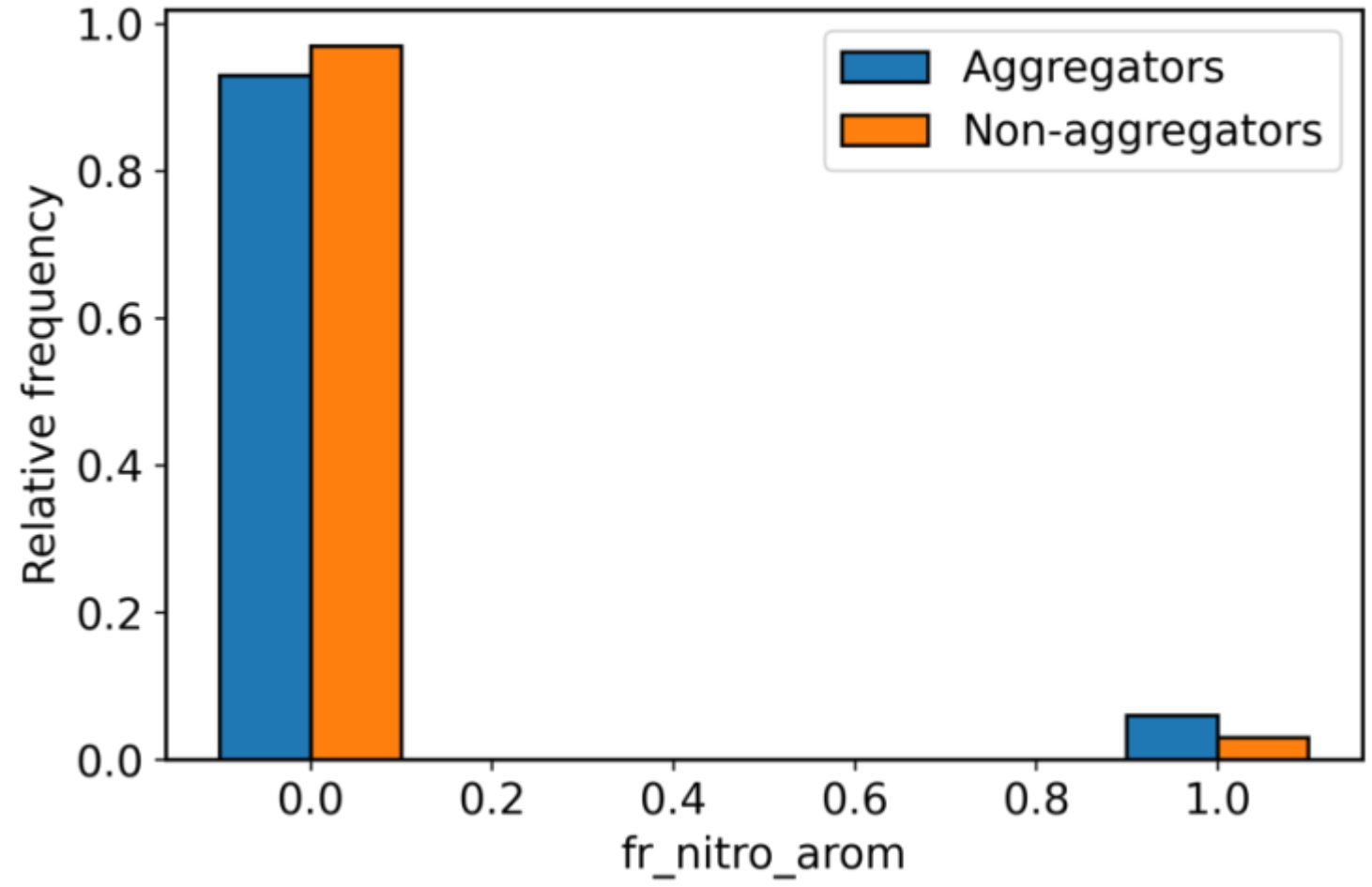

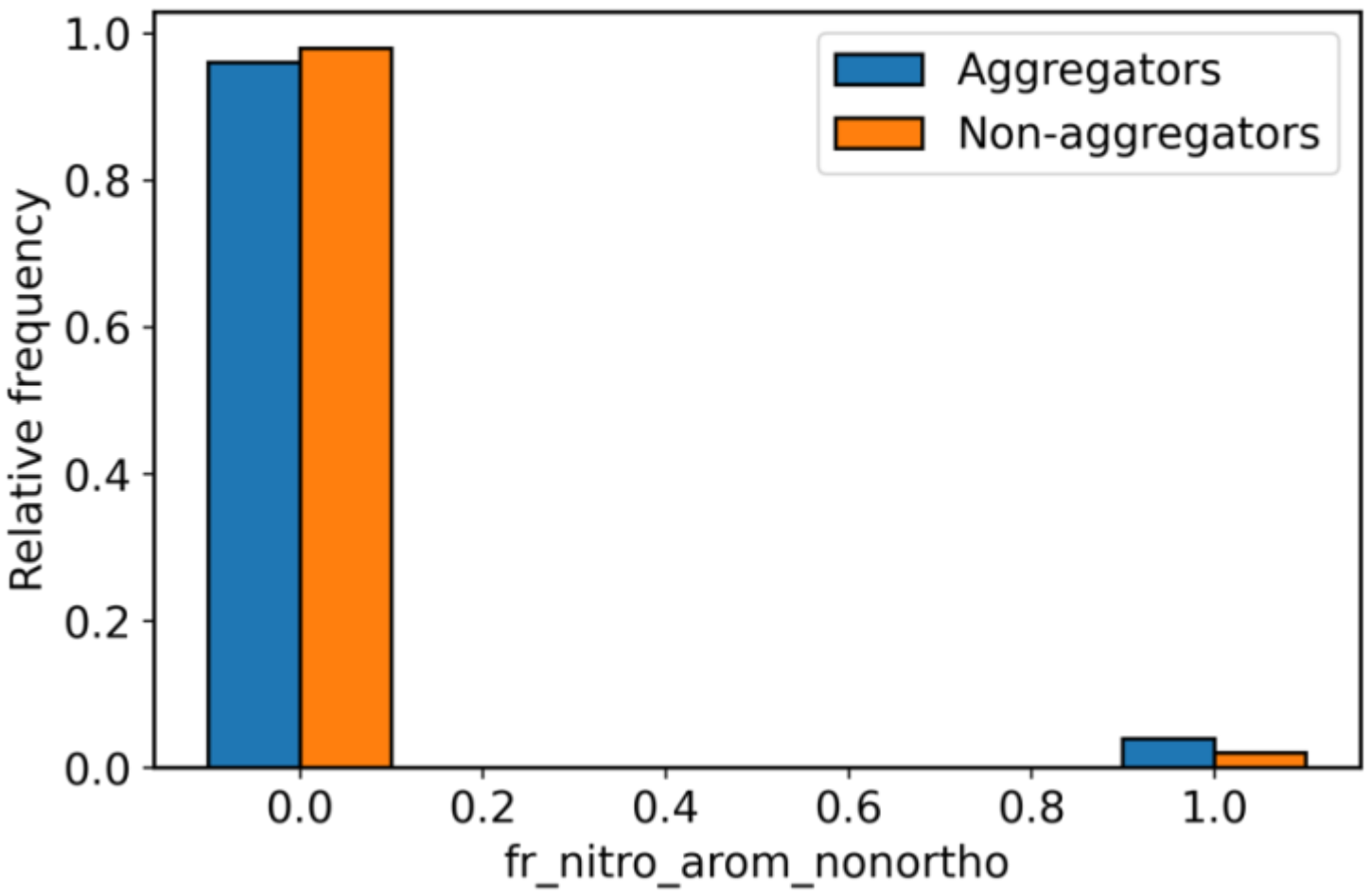

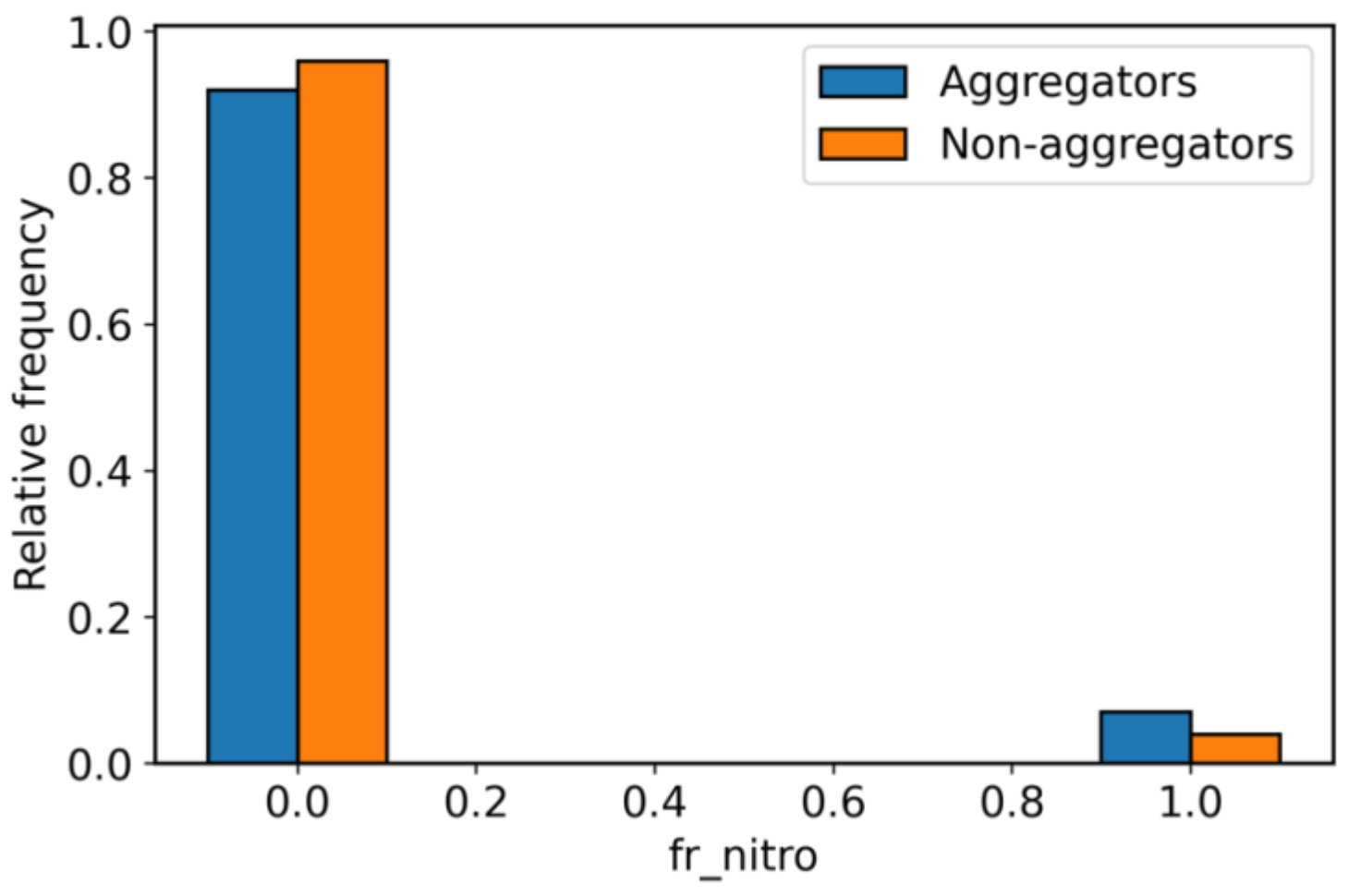

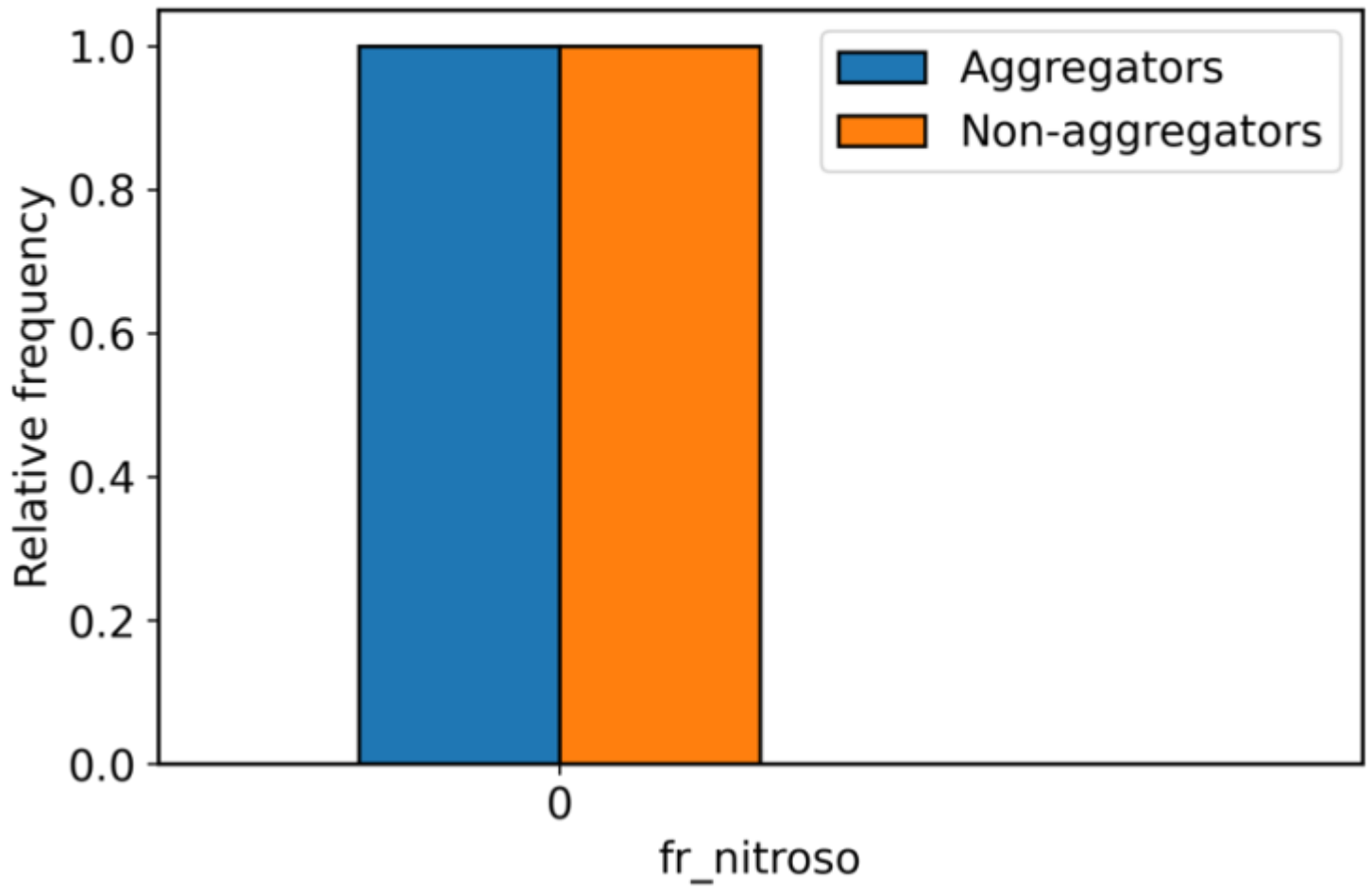

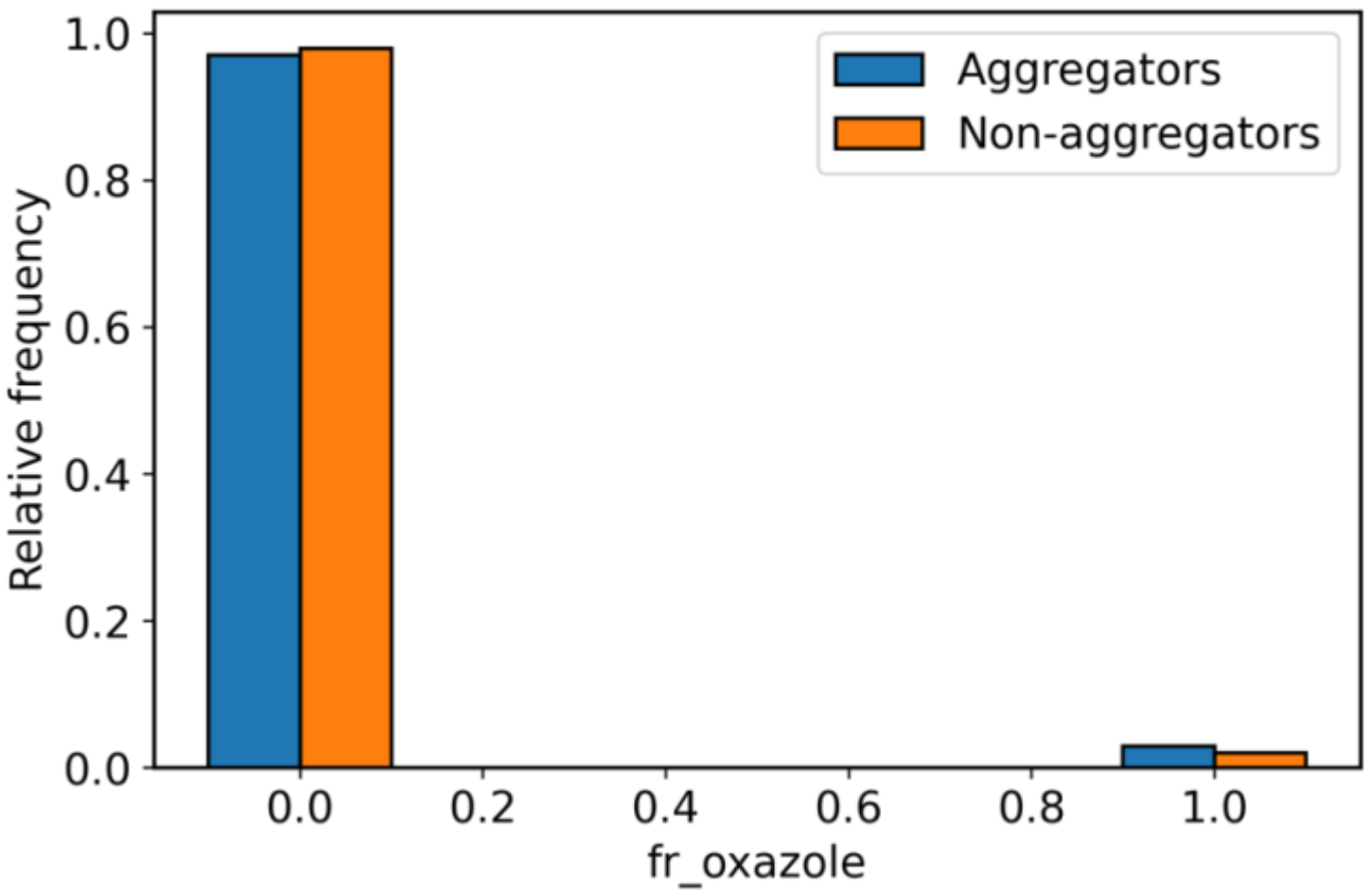

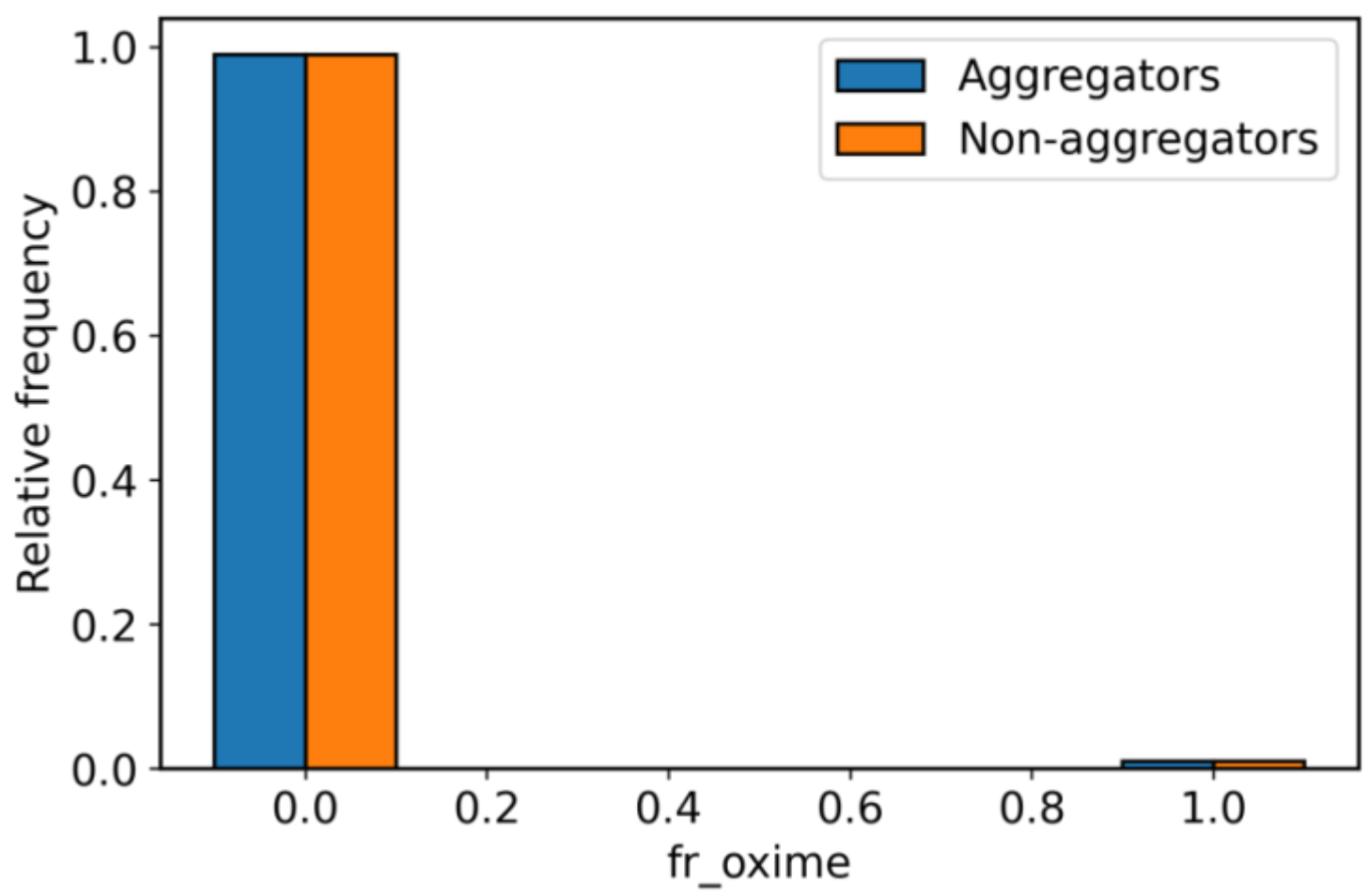

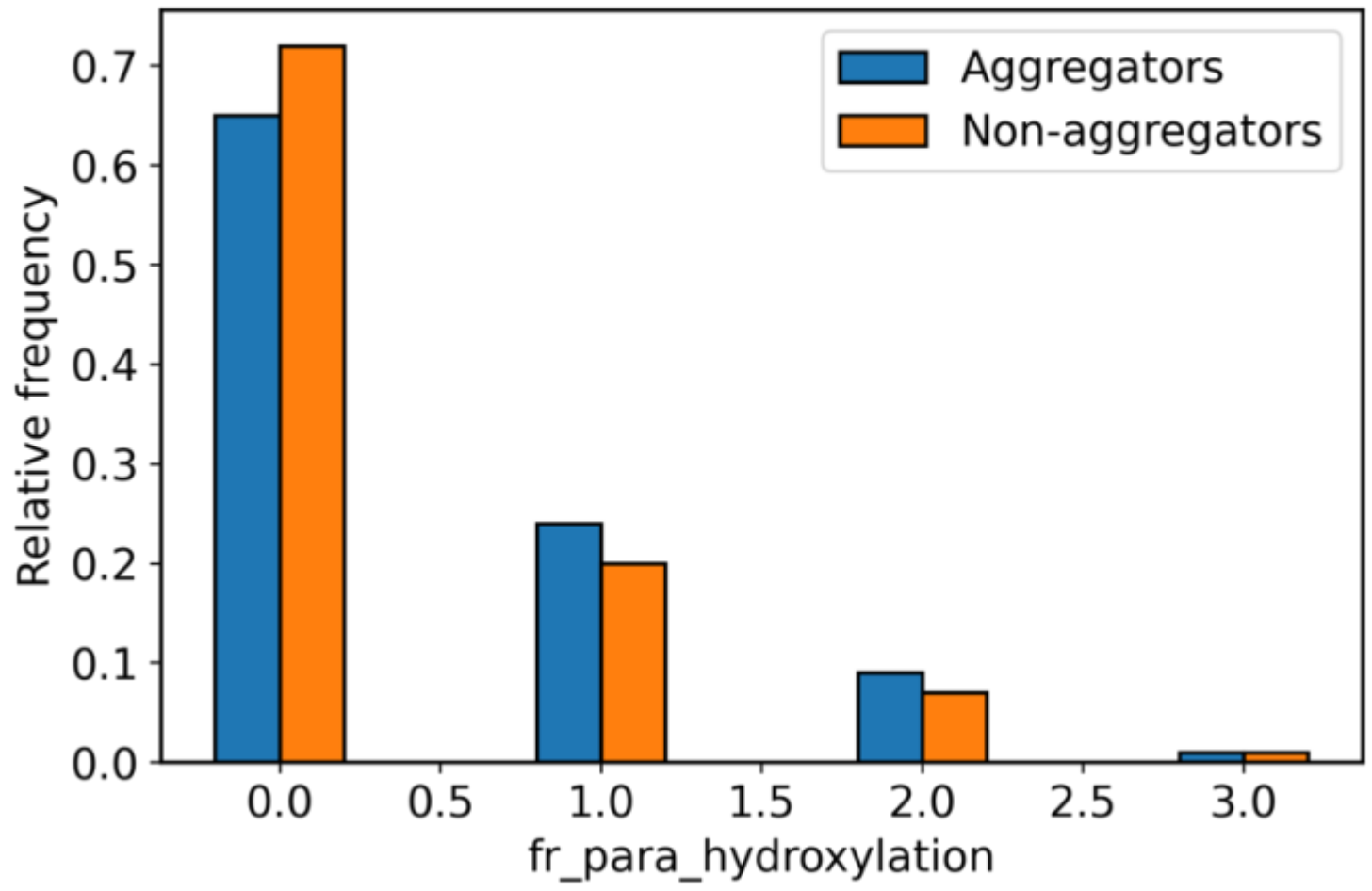

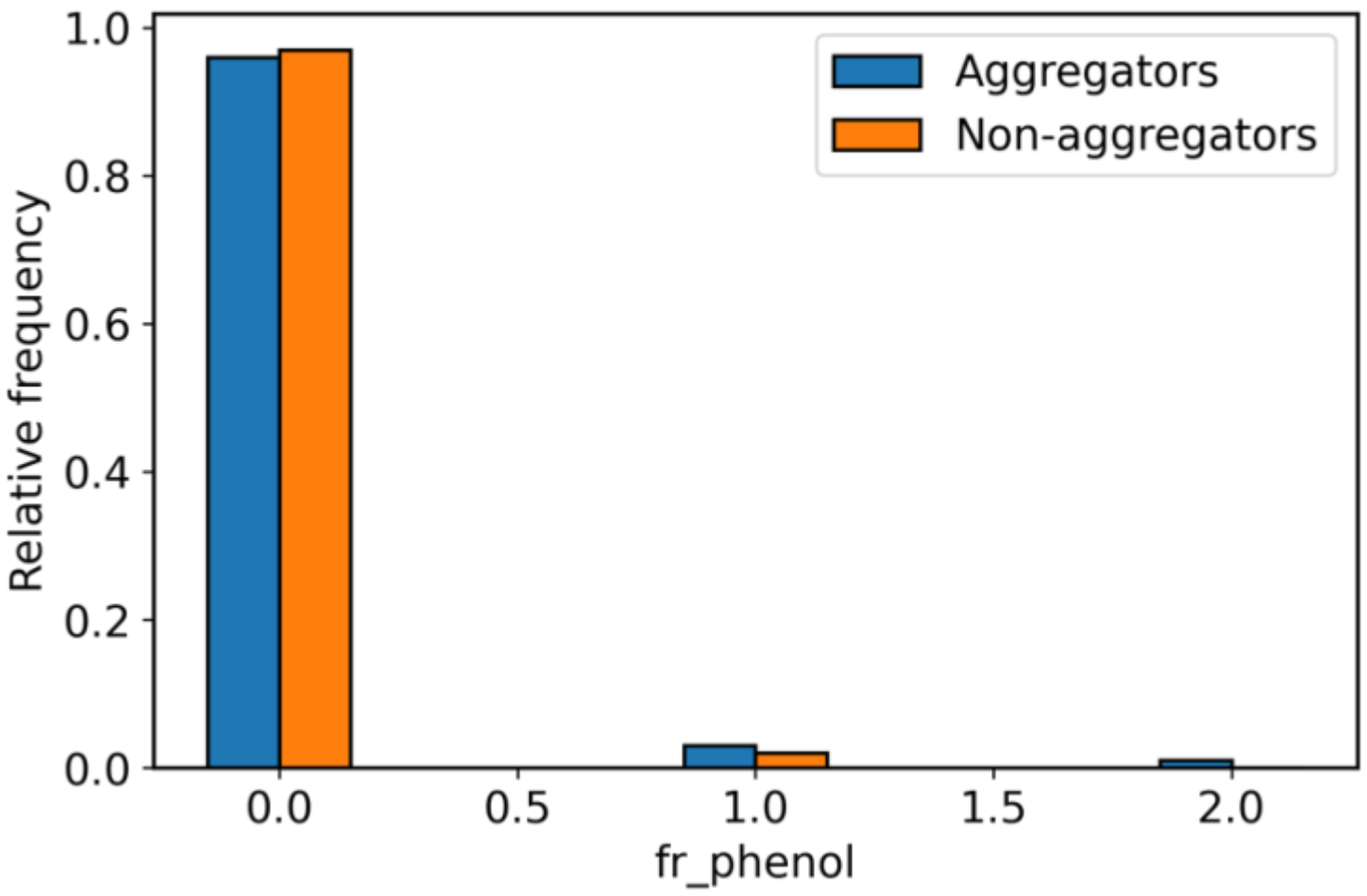

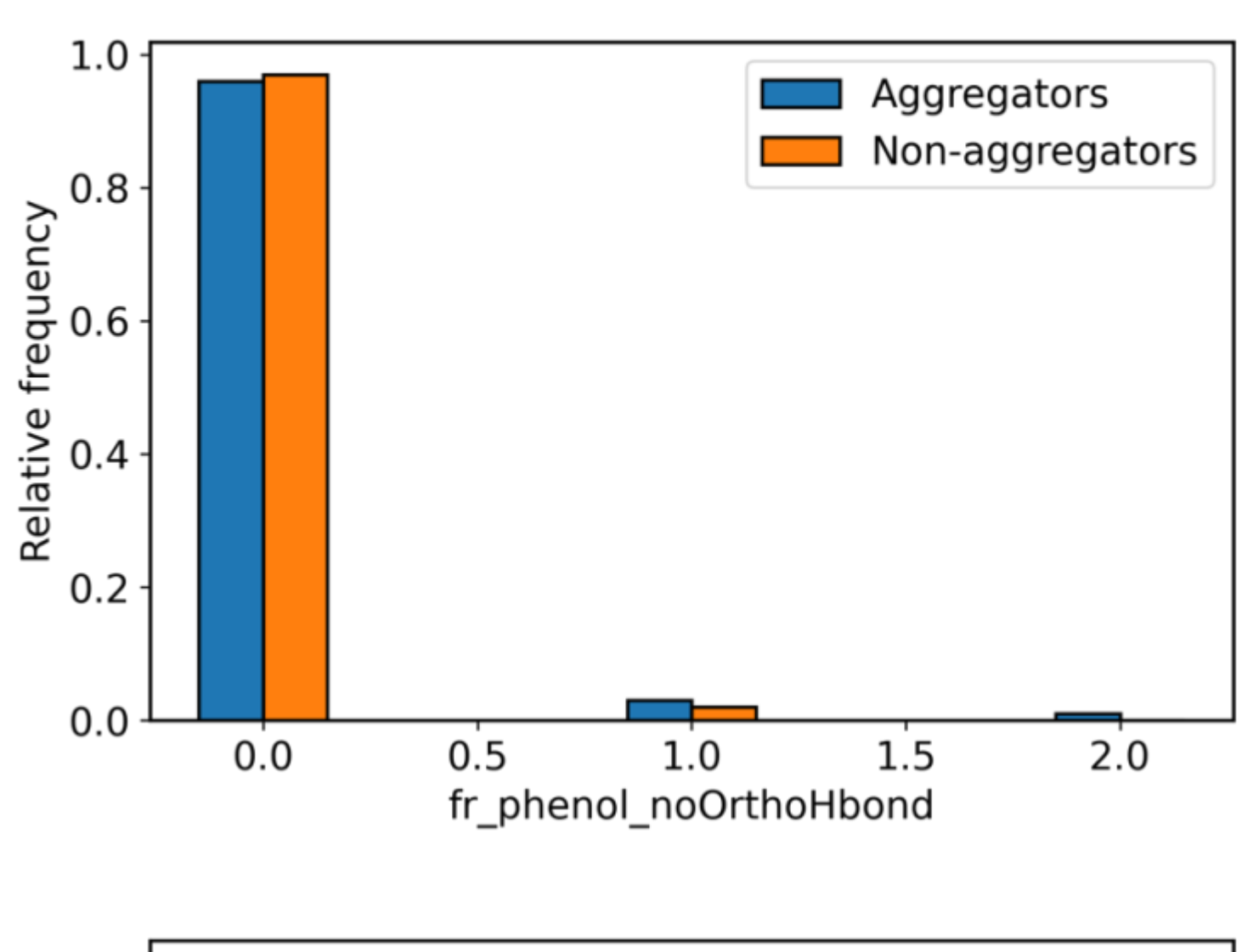

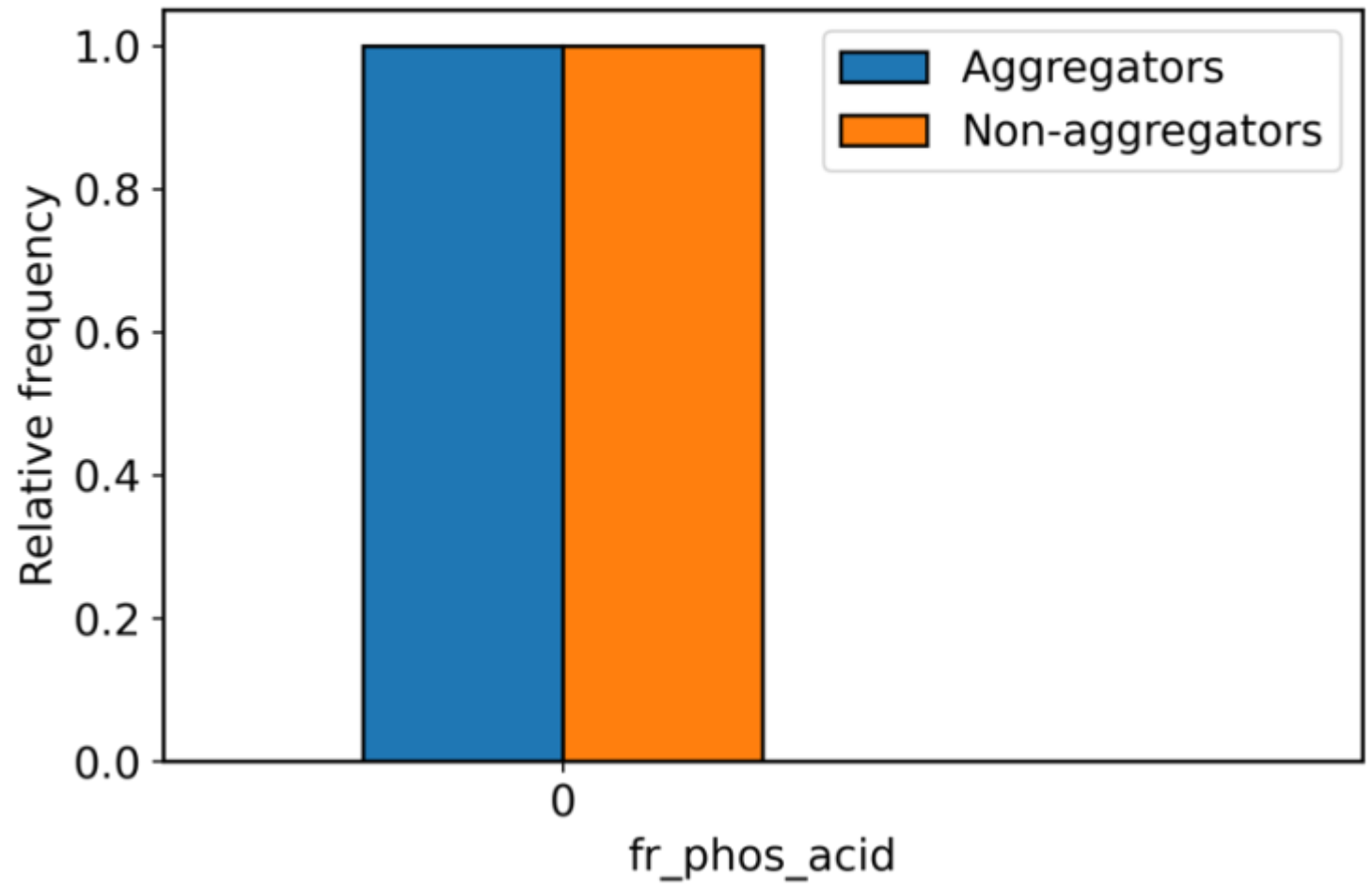

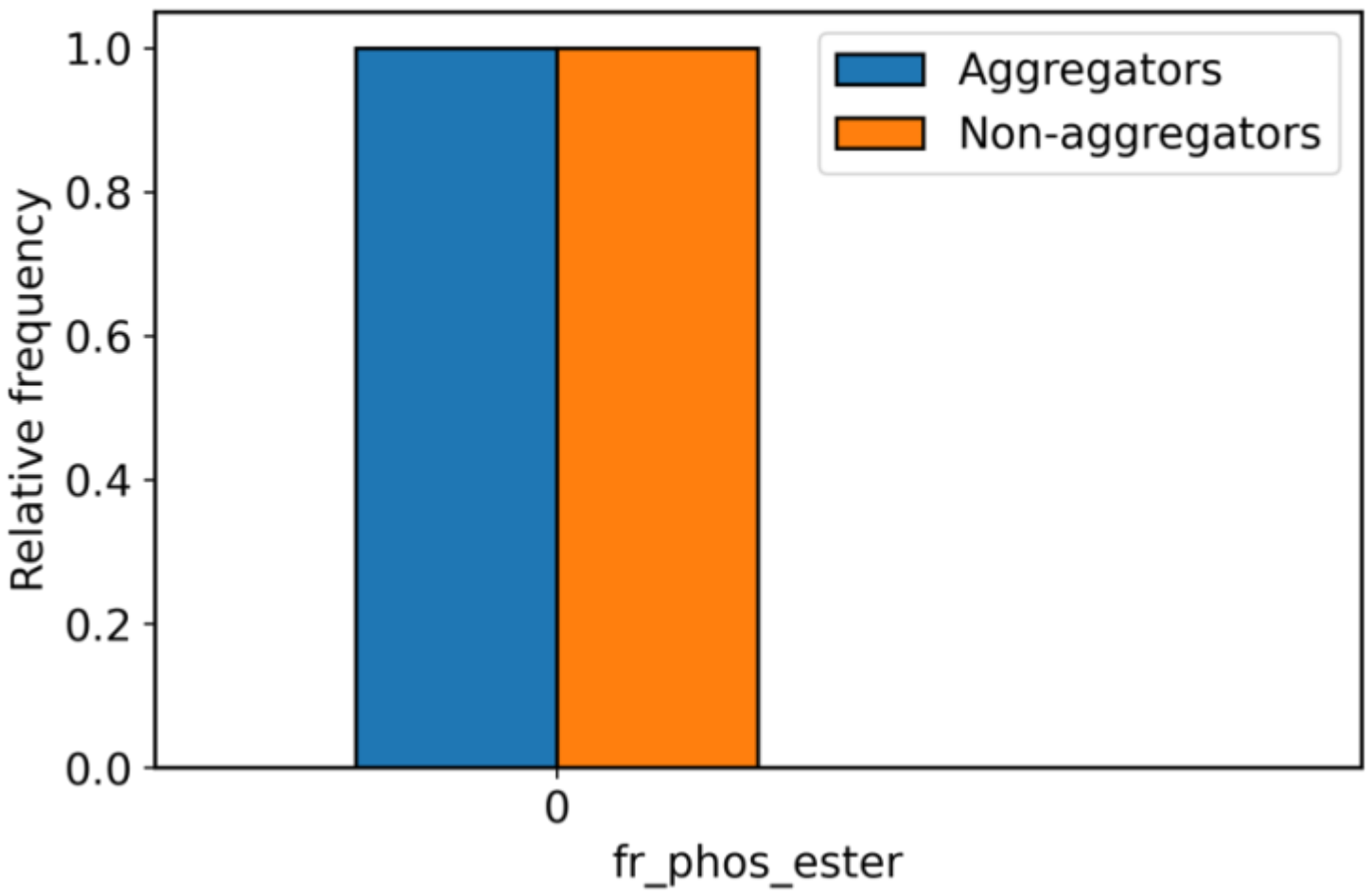

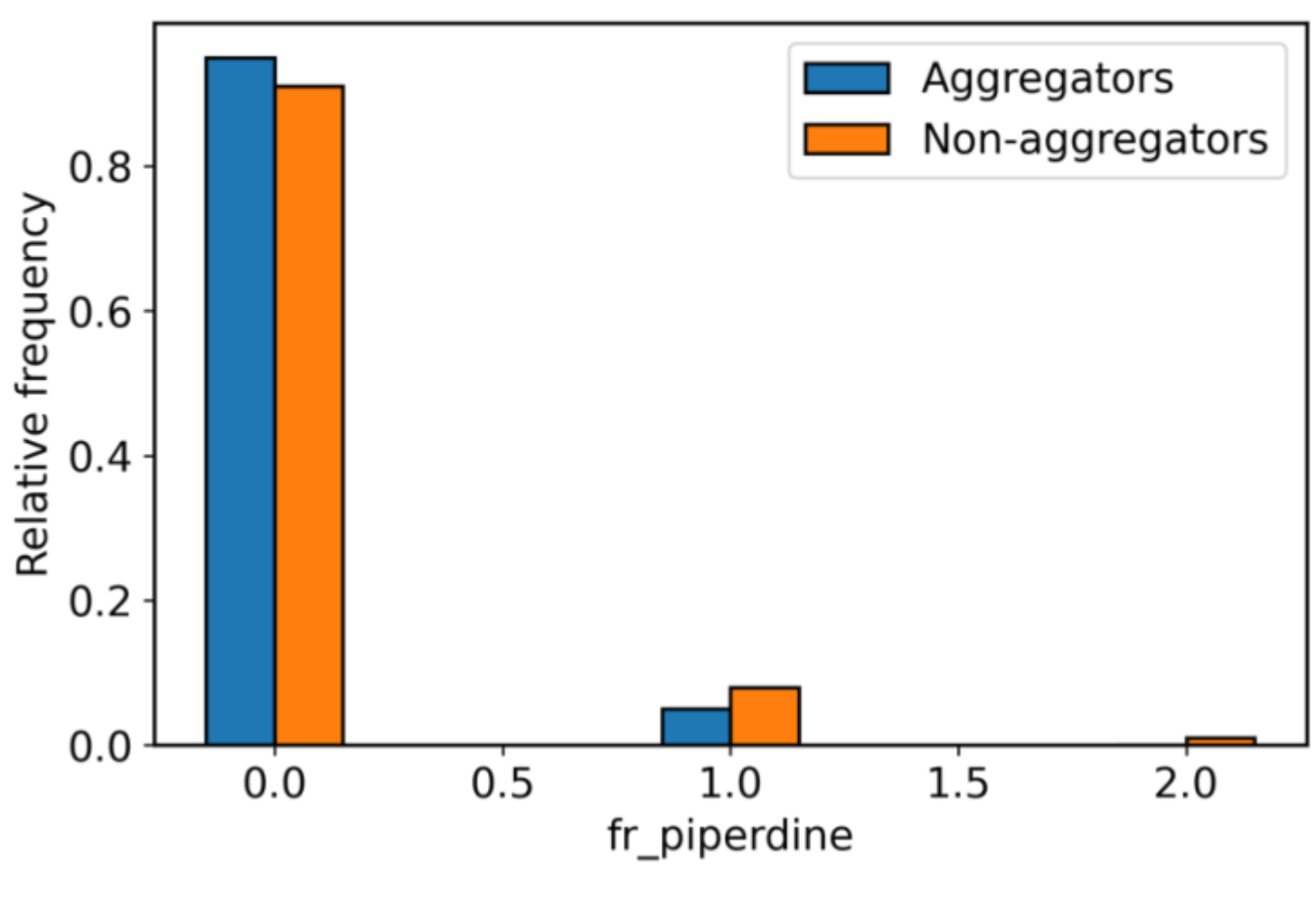

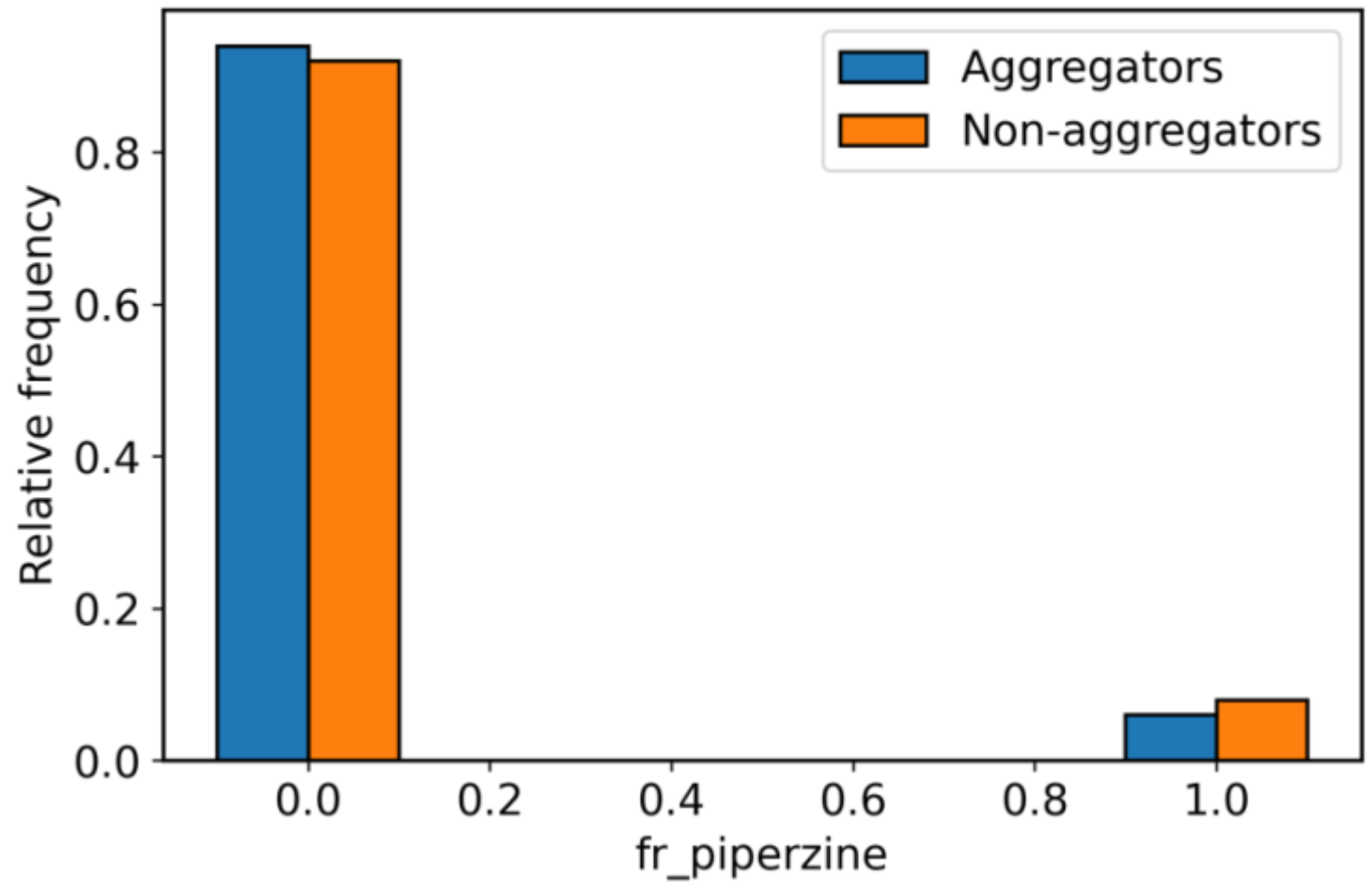

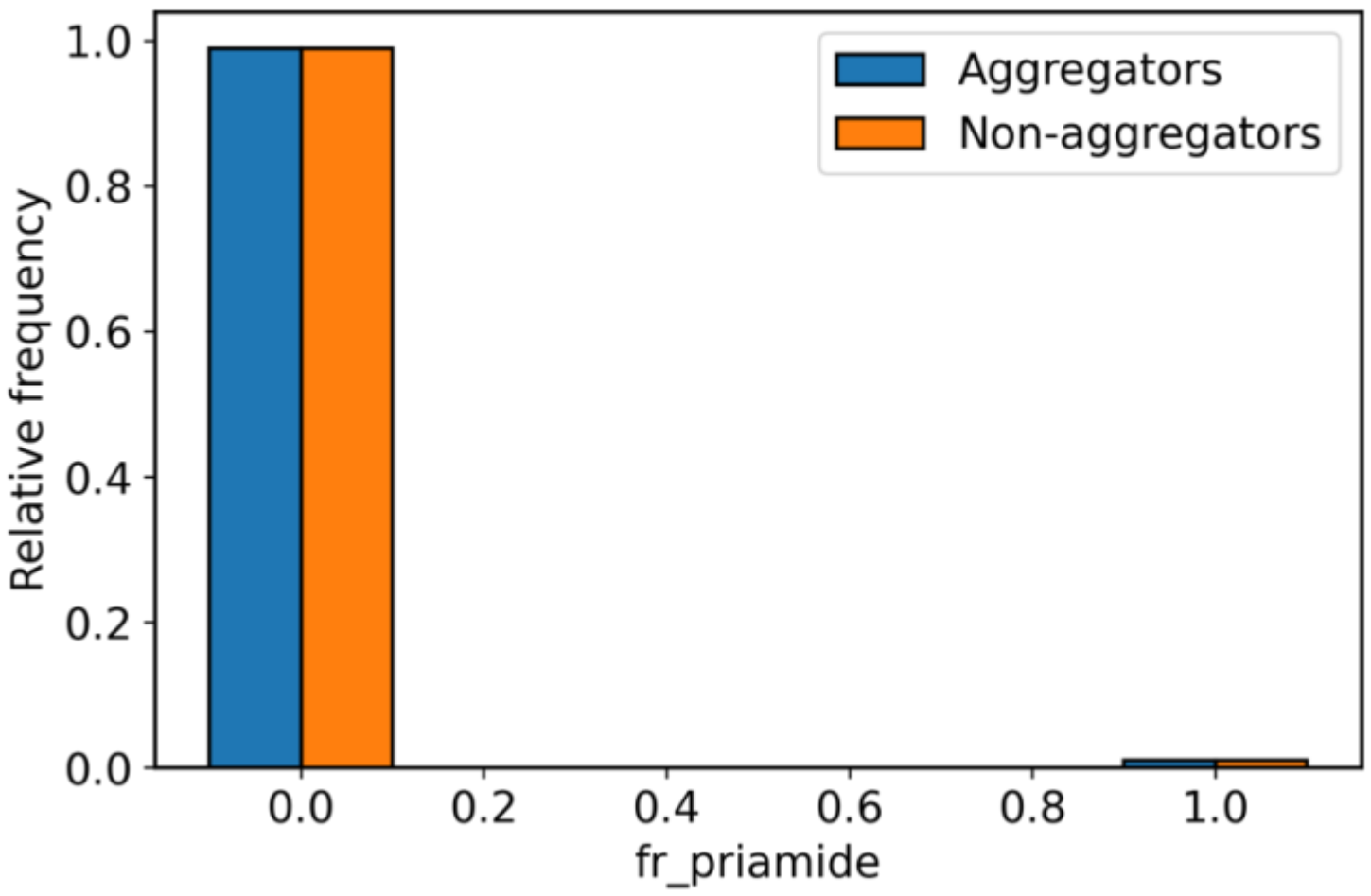

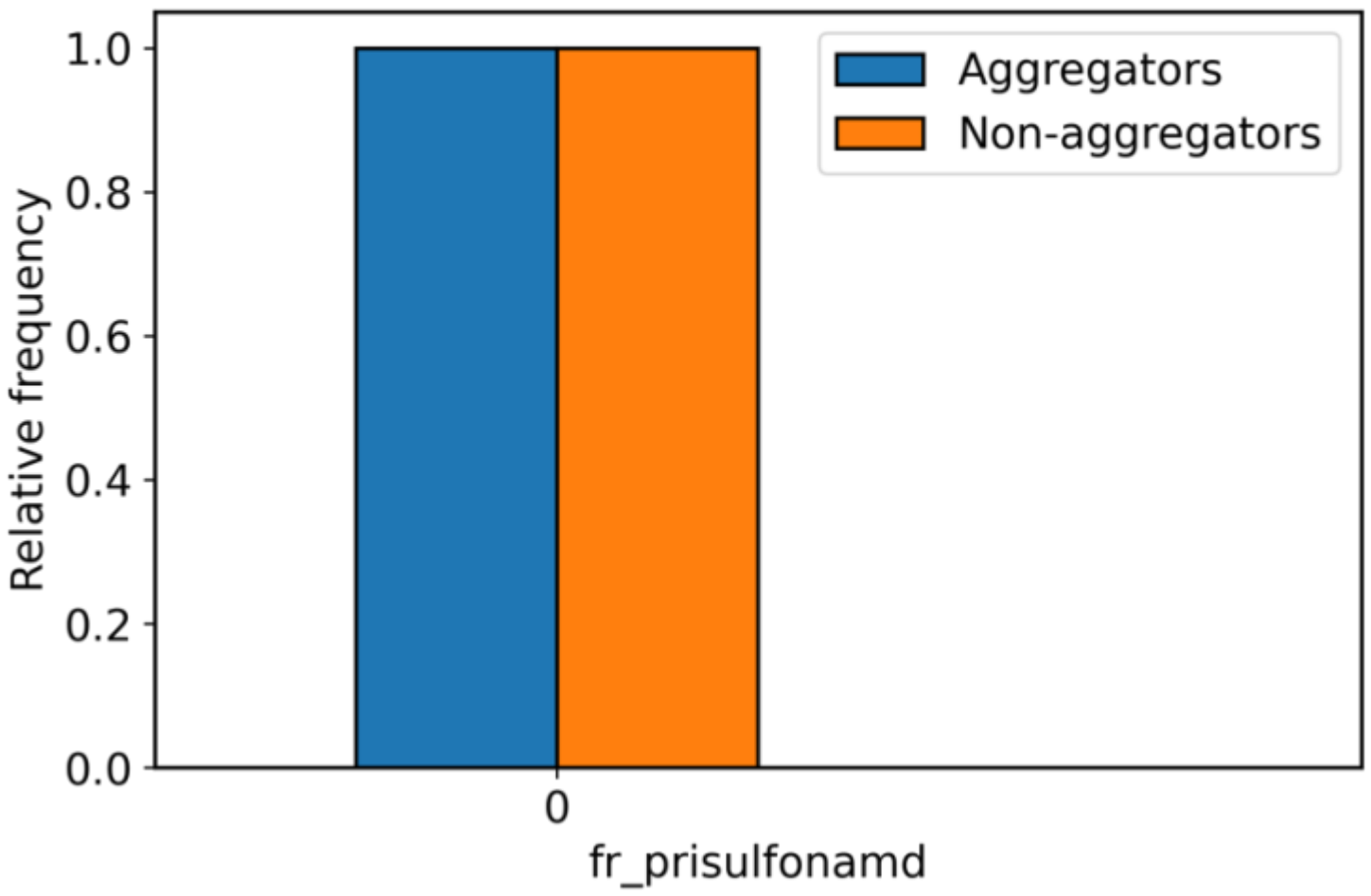

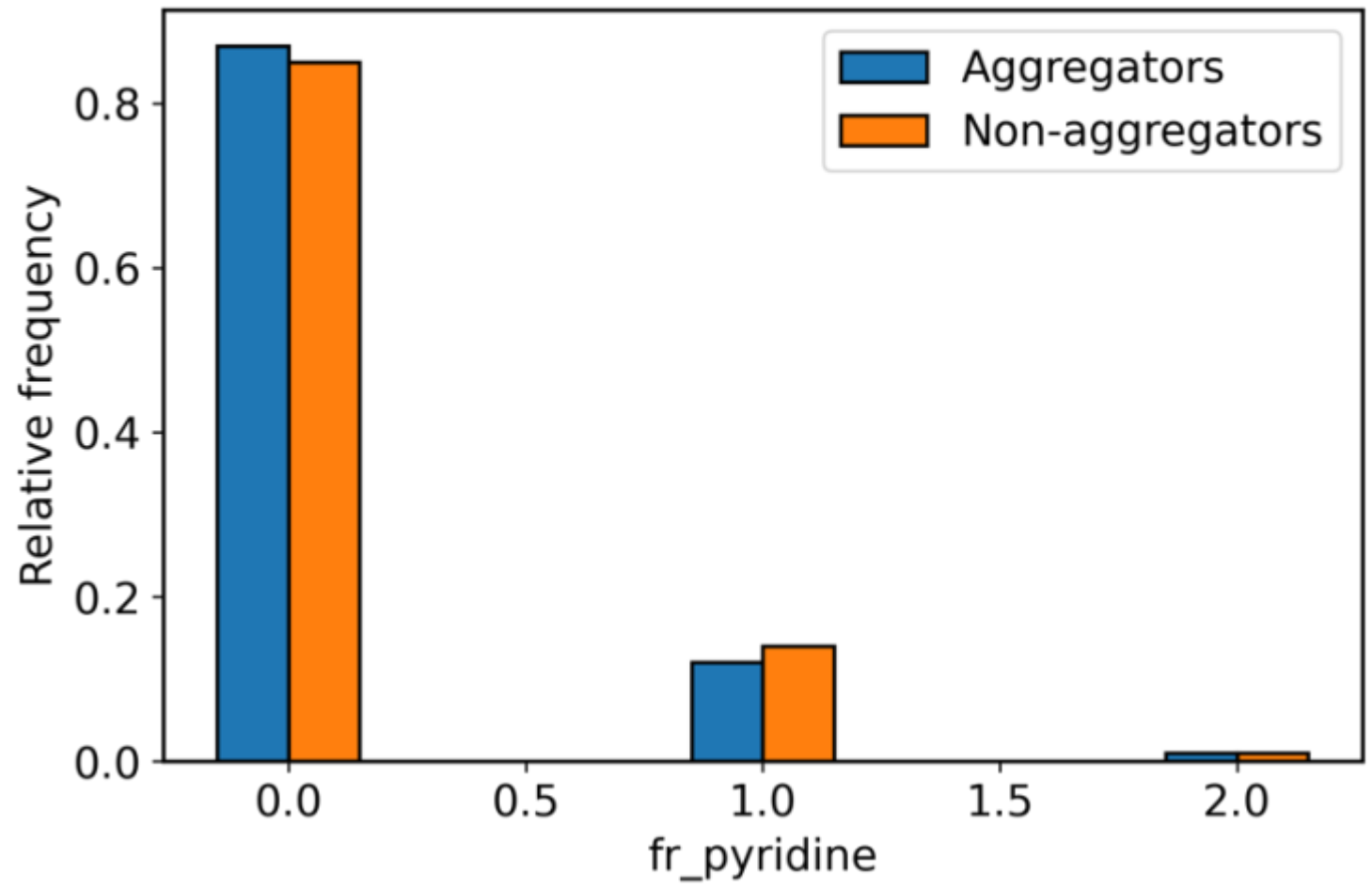

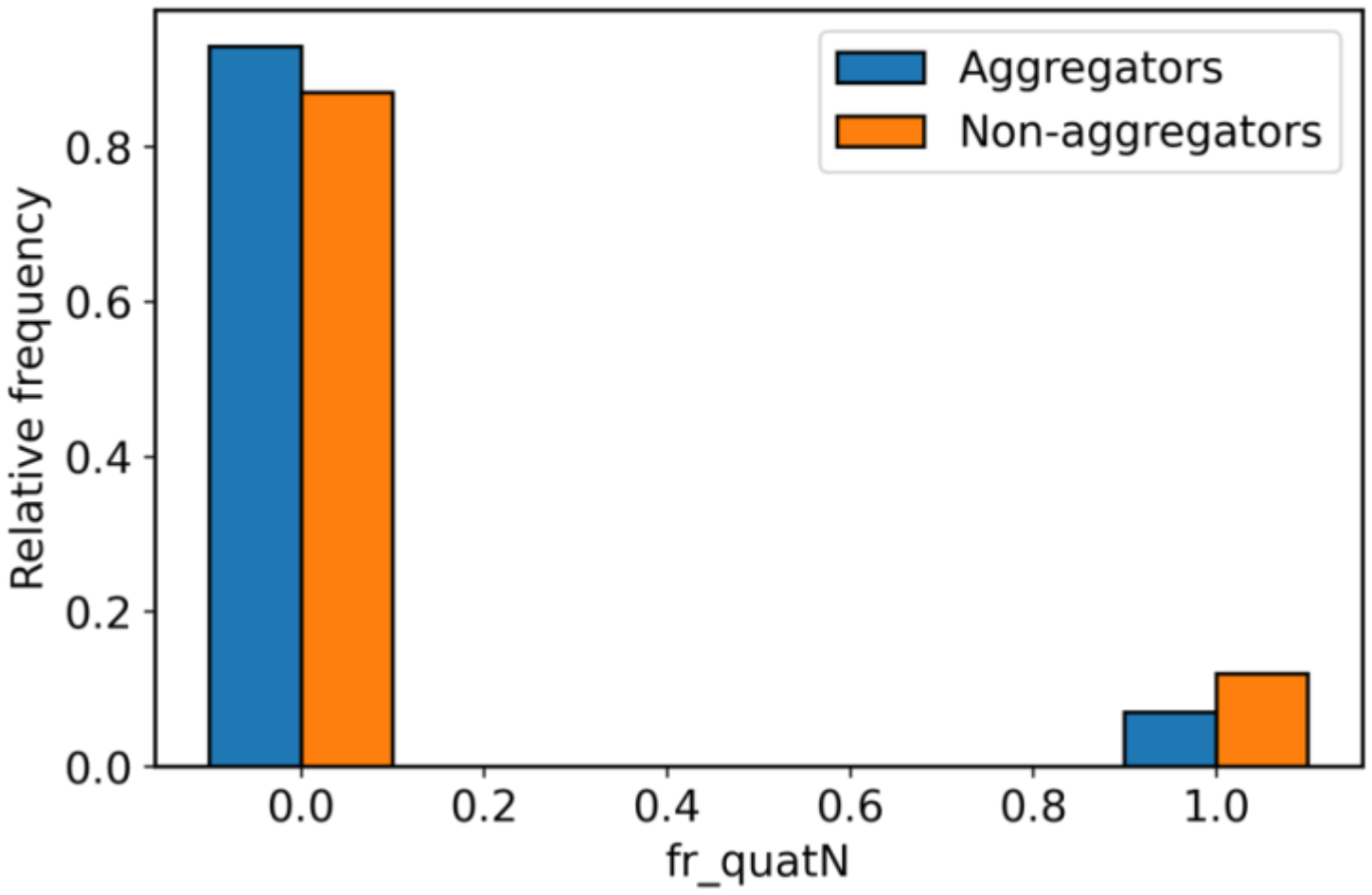

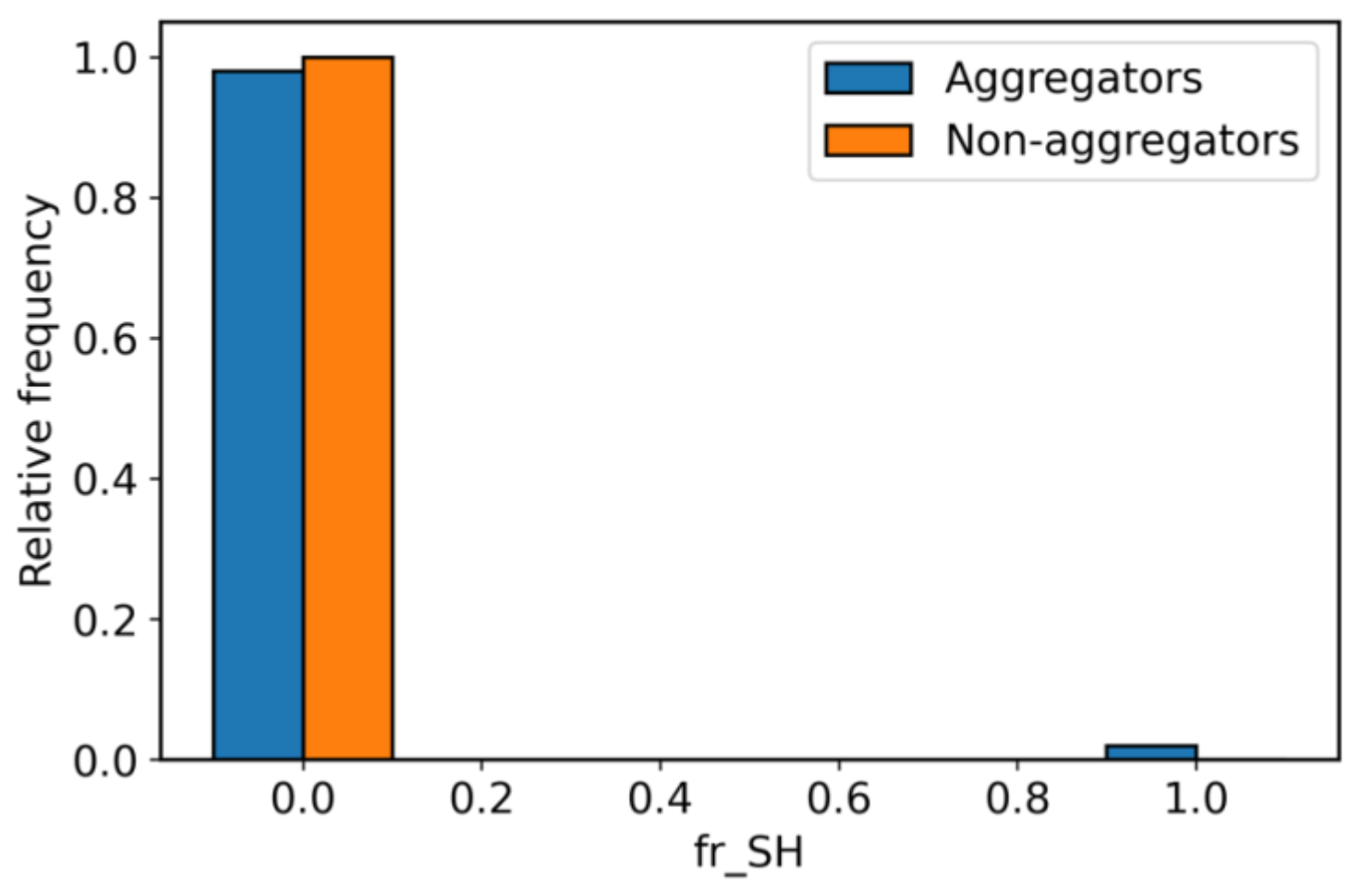

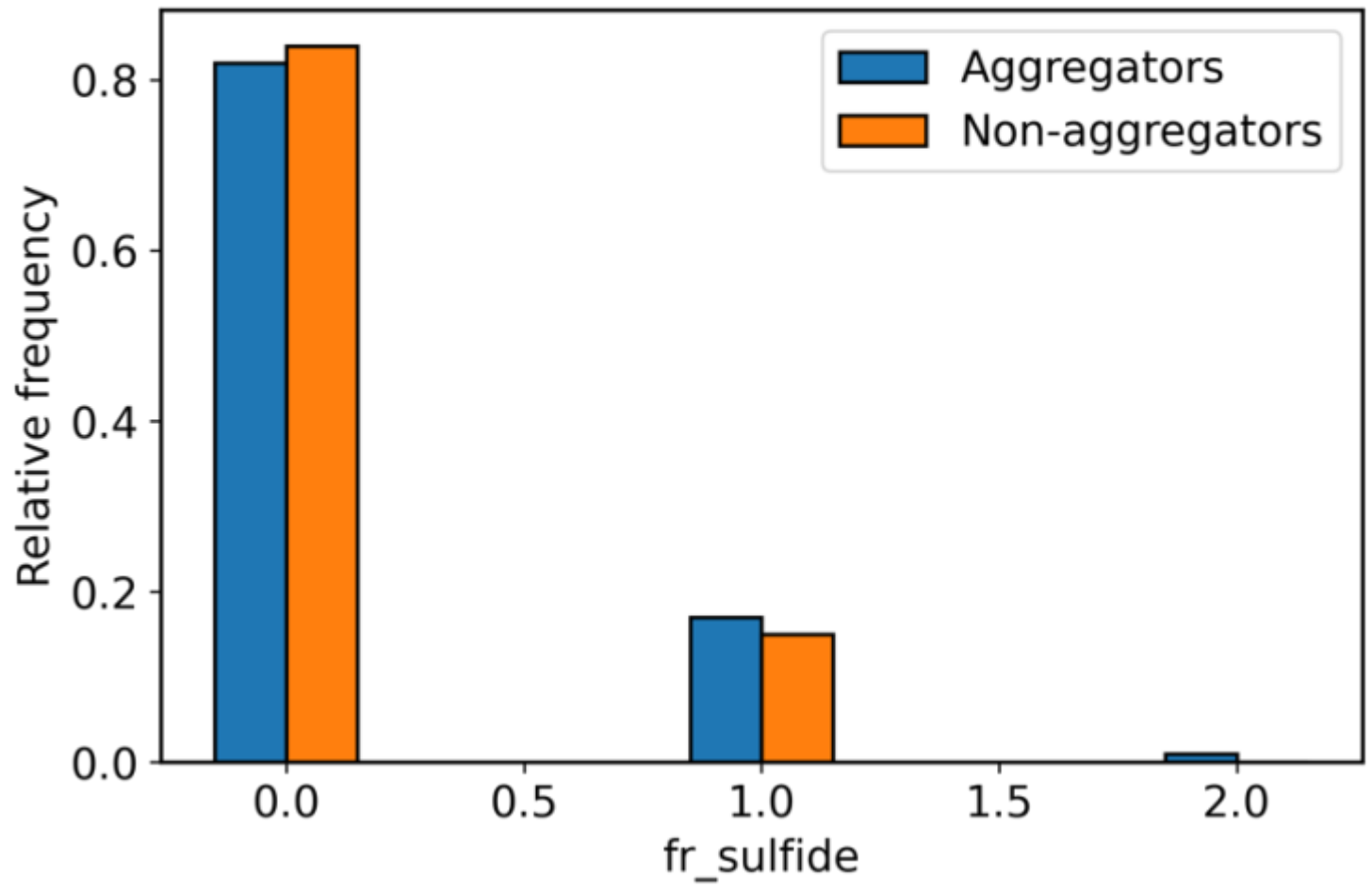

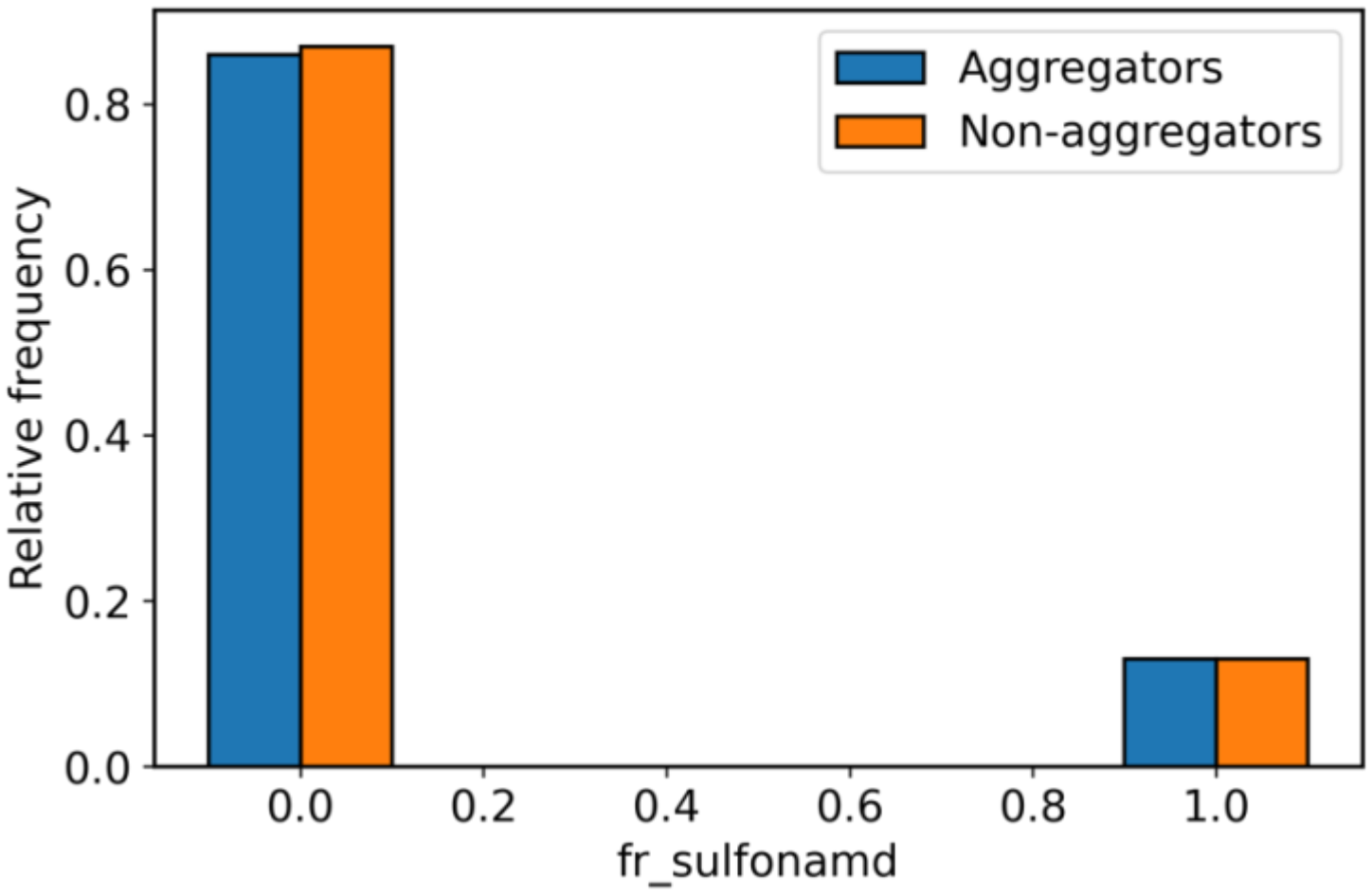

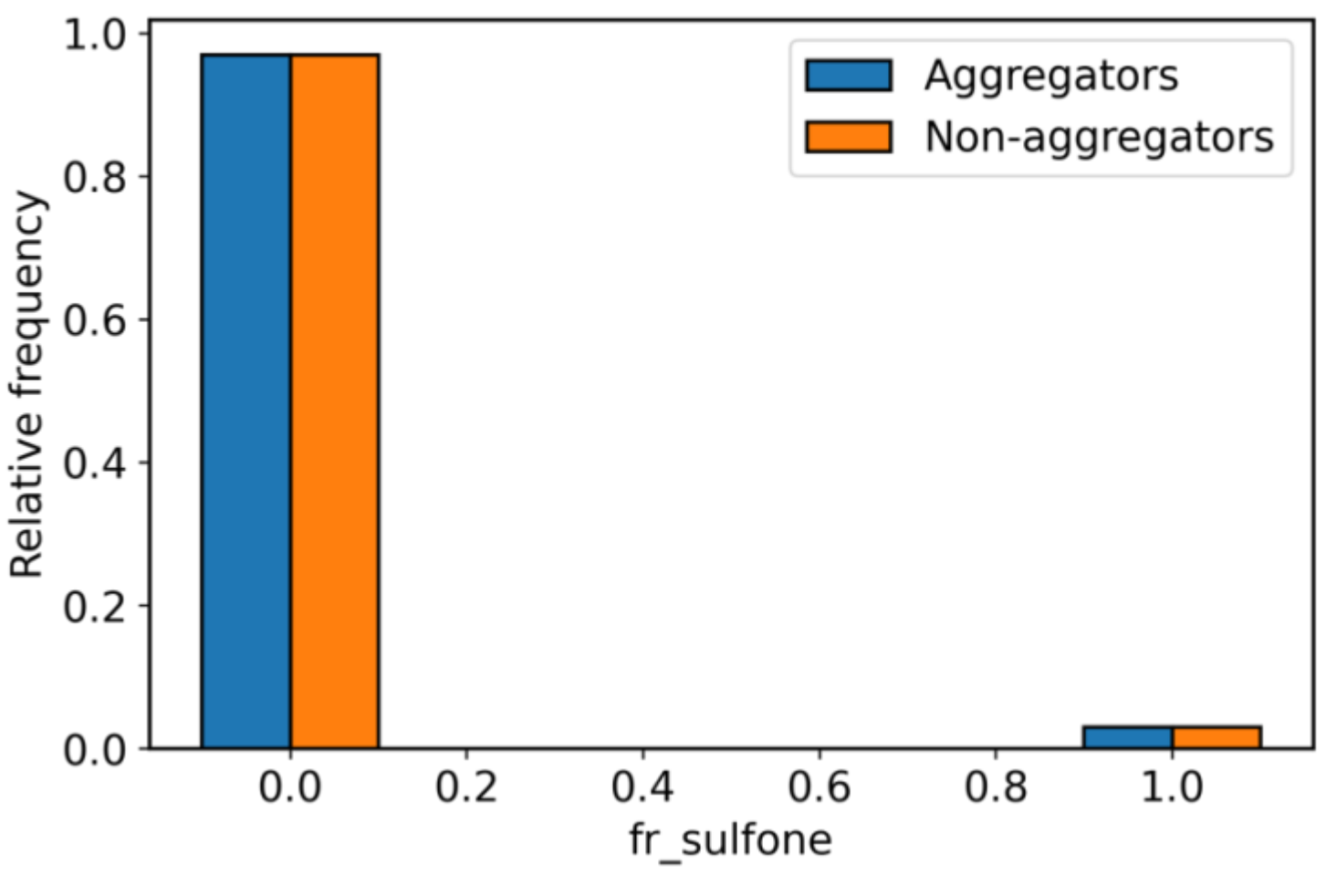

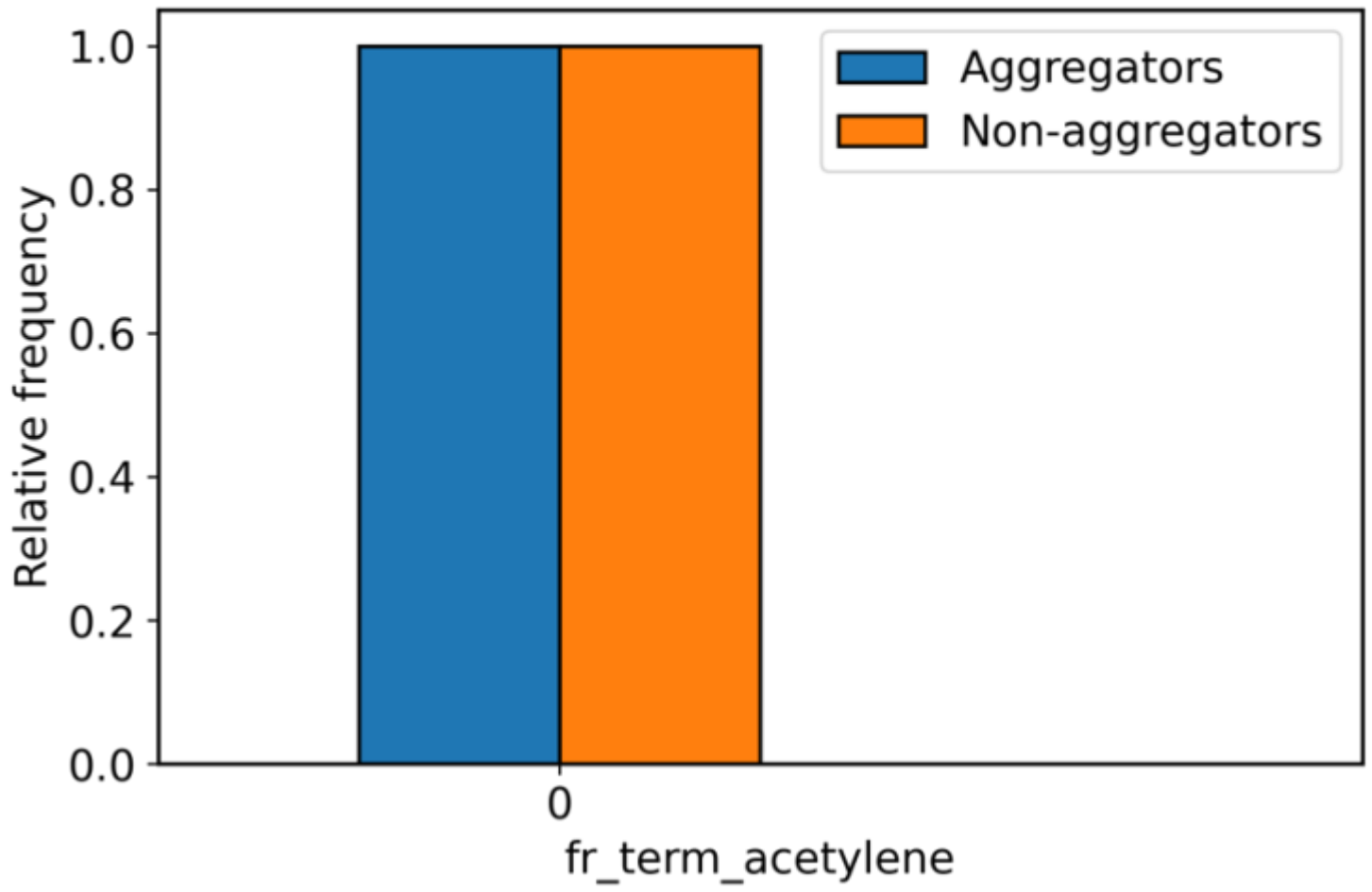

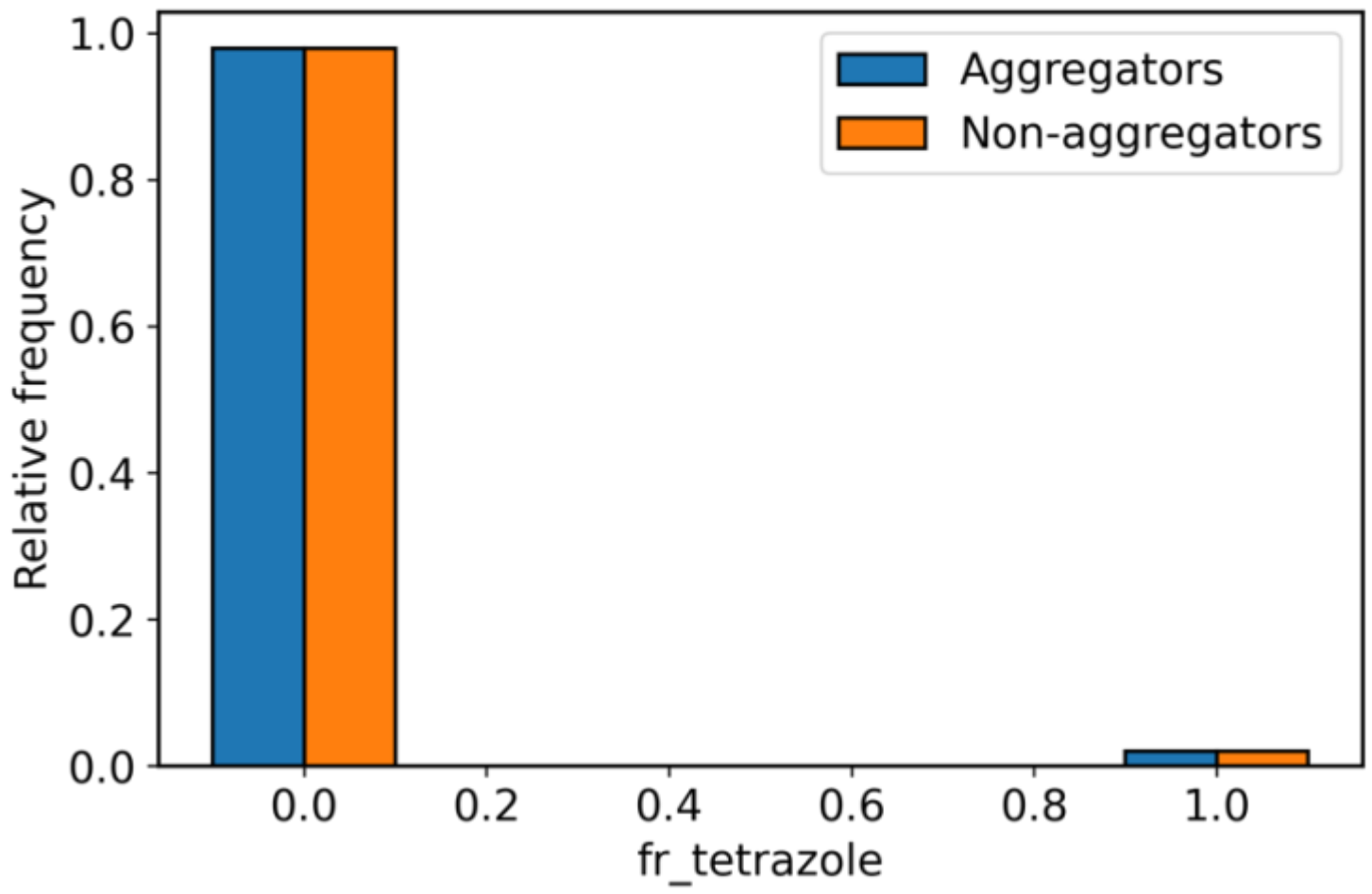

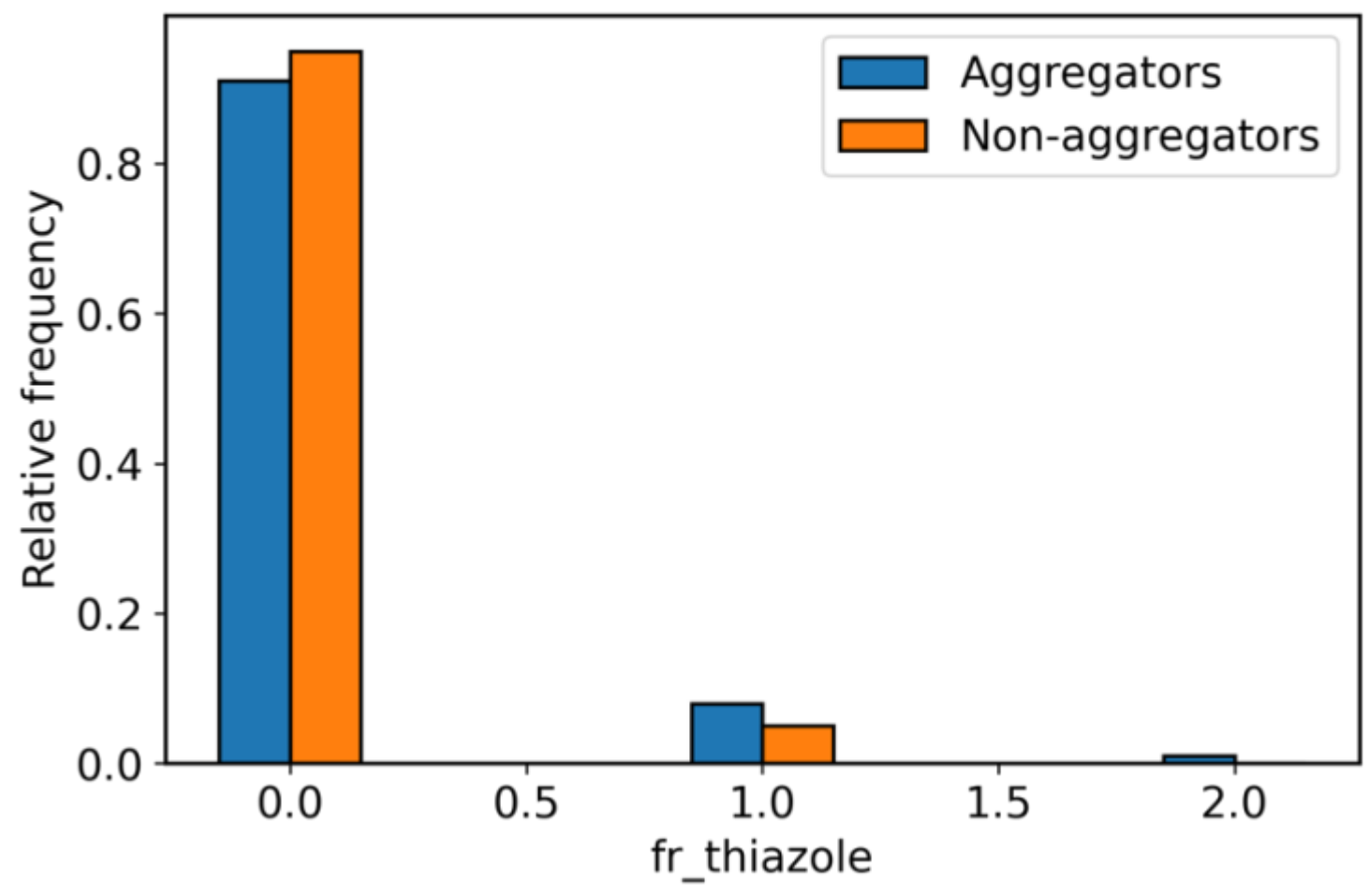

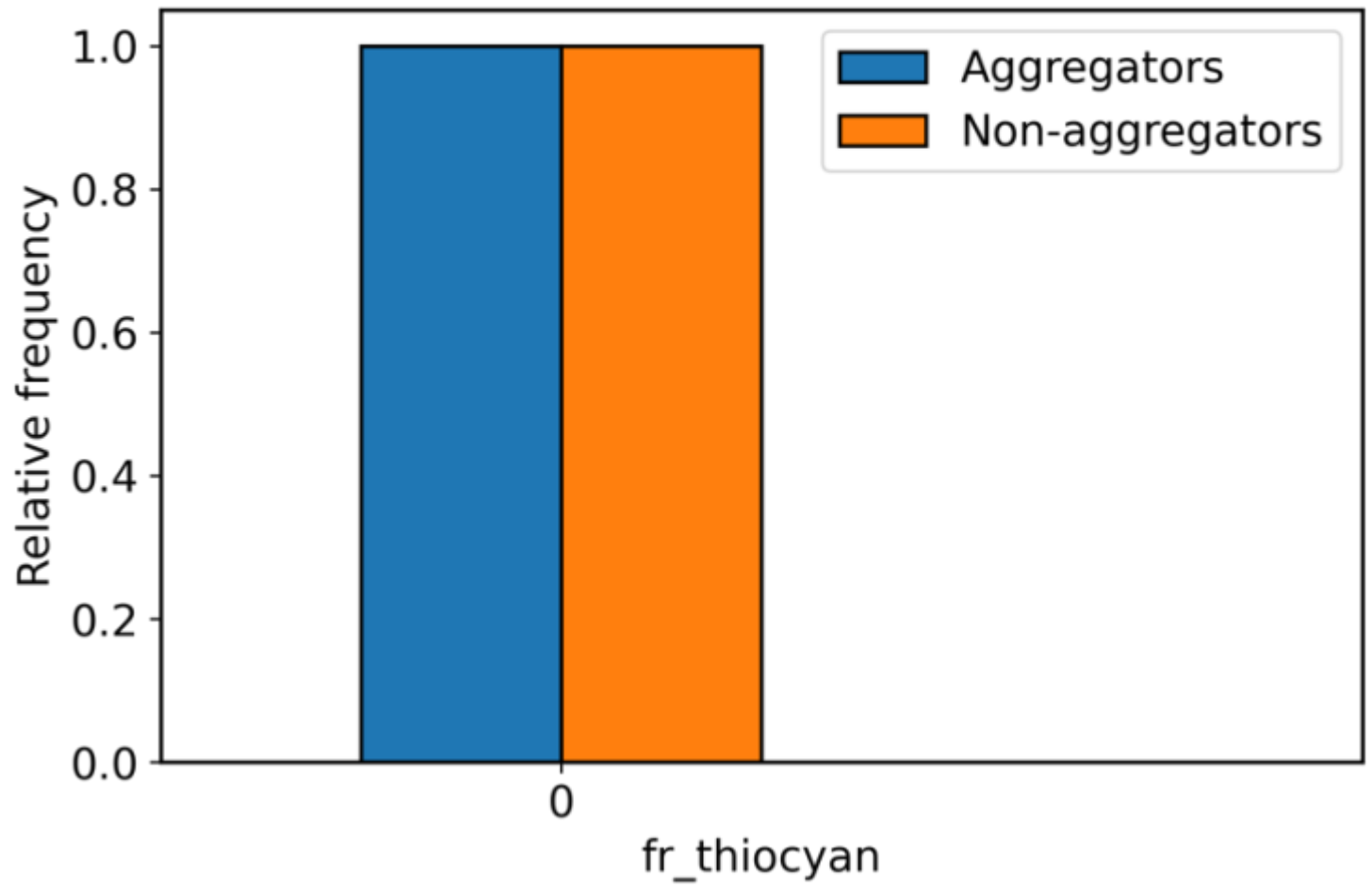

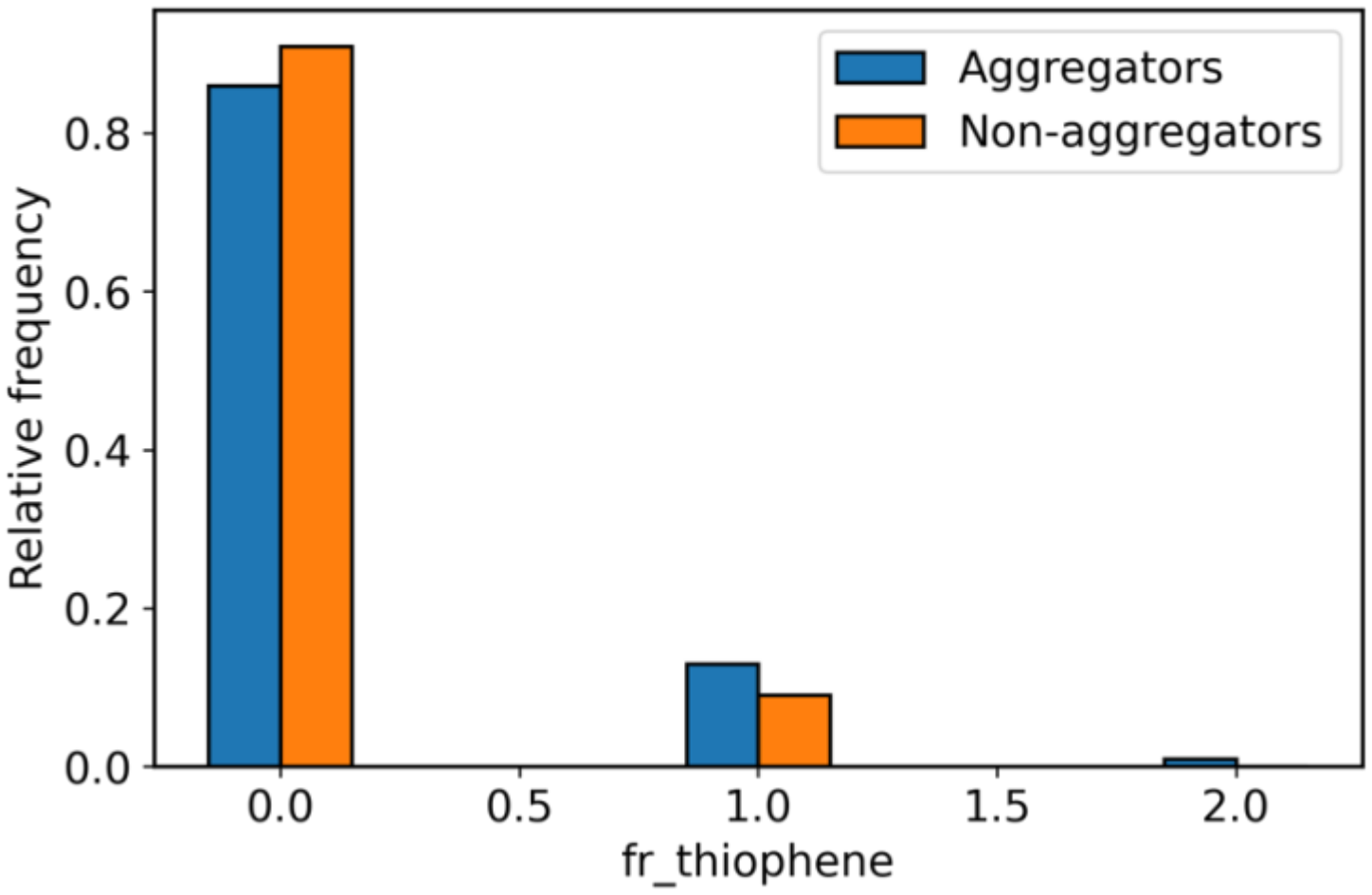

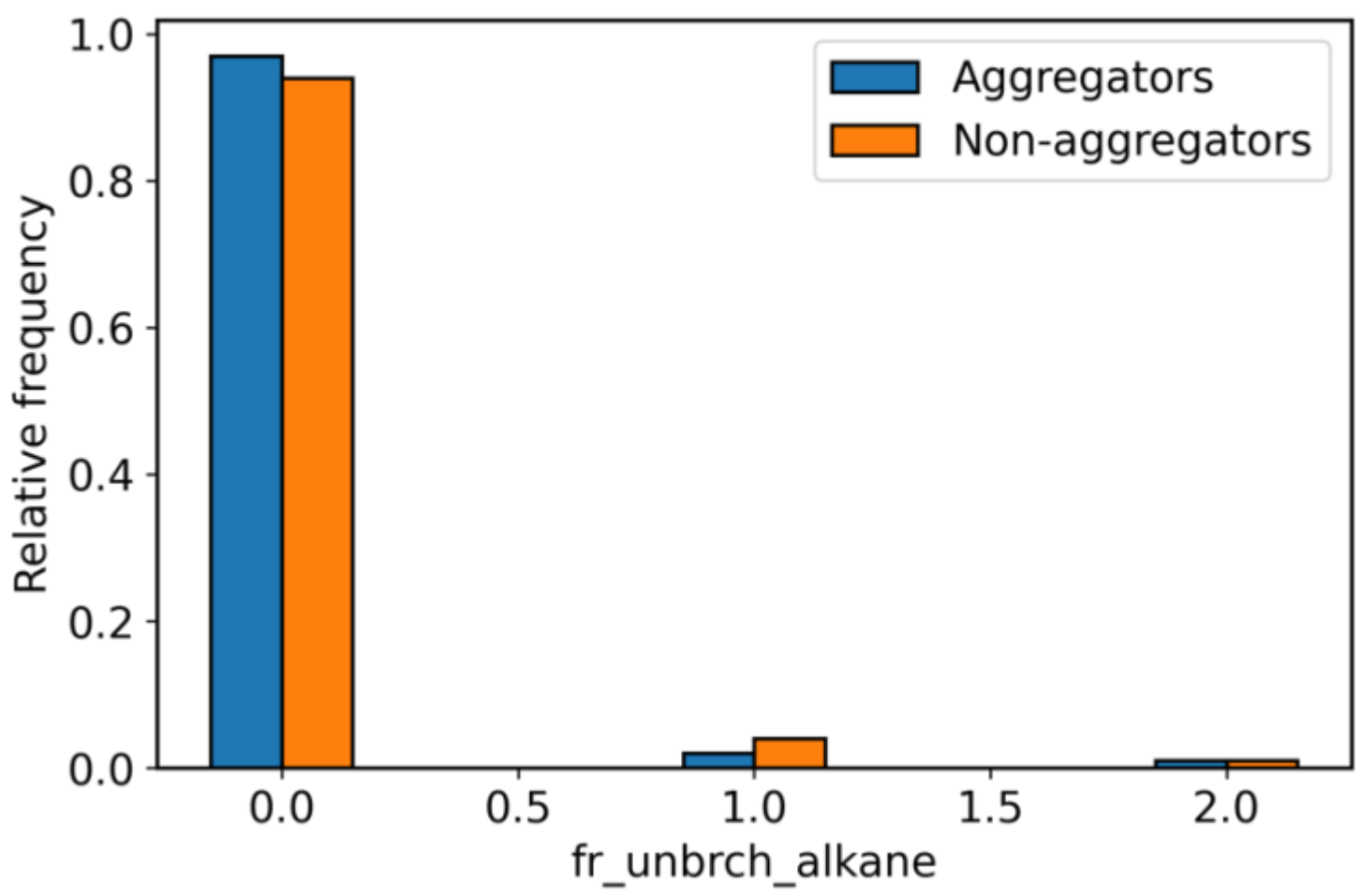

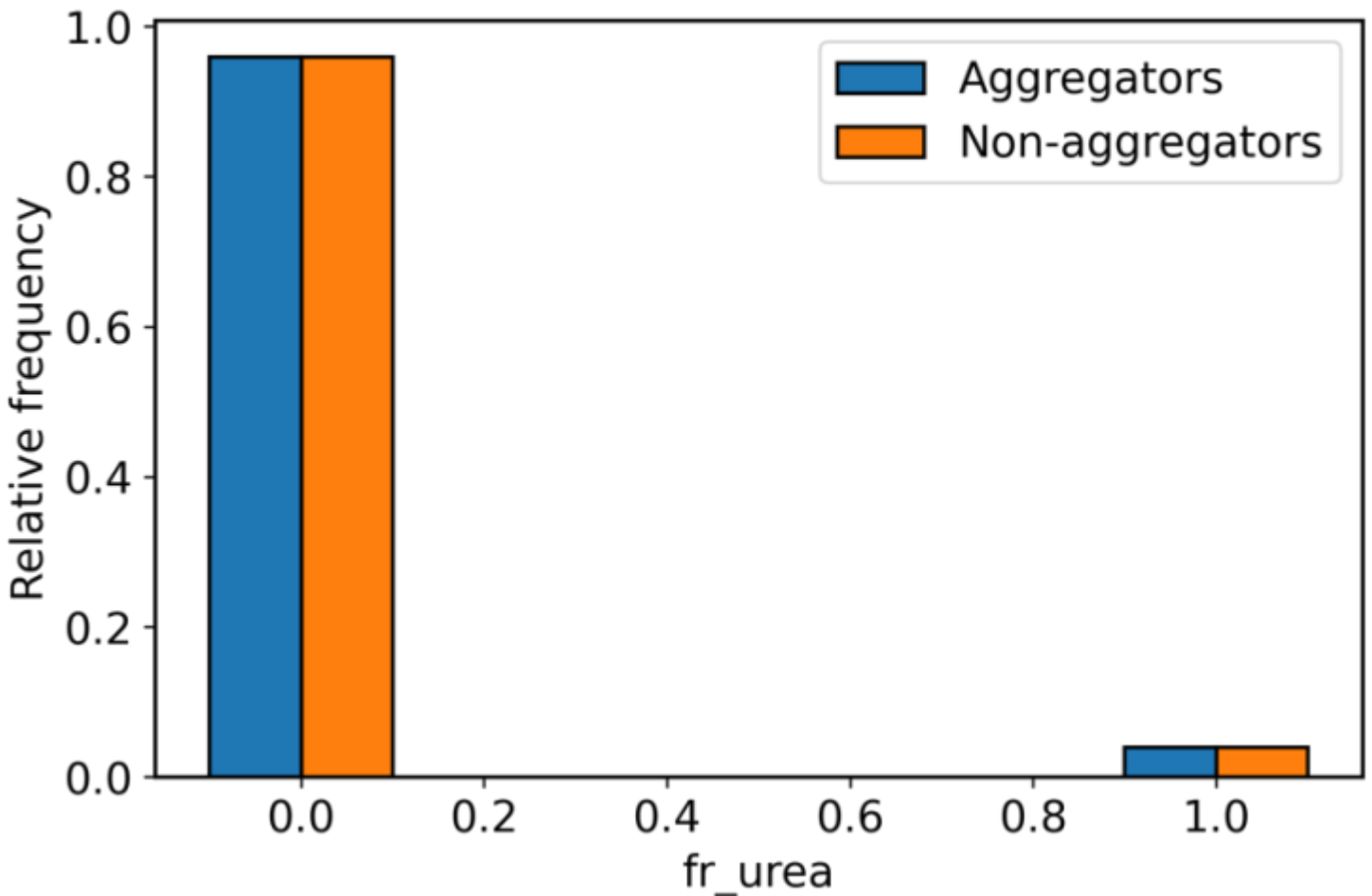

## SXVII. DFT energies and coordinates for pyridine derivatives E-P, monomers

### Pyridine Derivative E (monomer)

HF = -362.8645756 hartrees
Zero-point correction= 0.121380 (Hartree/Particle)
Thermal correction to Gibbs Free Energy= 0.090214
Sum of electronic and zero-point Energies= -362.743195
Sum of electronic and thermal Free Energies= -362.774362

Standard orientation:

| Center Number | Atomic Number | Atomic Type | Coordinates (Angstroms) X | Y | Z |
|---|---|---|---|---|---|
| 1 | 6 | 0 | 1.074043 | 1.179809 | -0.000217 |
| 2 | 6 | 0 | 1.135810 | -0.202853 | -0.000302 |
| 3 | 7 | 0 | 0.025020 | -0.958332 | -0.000250 |
| 4 | 6 | 0 | -1.146969 | -0.353964 | -0.000010 |
| 5 | 6 | 0 | -1.314992 | 1.028263 | 0.000198 |
| 6 | 6 | 0 | -0.172323 | 1.797108 | 0.000070 |
| 7 | 6 | 0 | 2.441372 | -0.933267 | 0.000202 |
| 8 | 8 | 0 | -2.255757 | -1.129714 | 0.000036 |
| 9 | 1 | 0 | 1.984328 | 1.763503 | -0.000405 |
| 10 | 1 | 0 | -2.305729 | 1.460347 | 0.000371 |
| 11 | 1 | 0 | -0.246379 | 2.877455 | 0.000149 |
| 12 | 1 | 0 | 3.282285 | -0.241361 | -0.005866 |
| 13 | 1 | 0 | 2.514035 | -1.580427 | -0.876305 |
| 14 | 1 | 0 | 2.518742 | -1.570020 | 0.883989 |
| 15 | 1 | 0 | -1.978014 | -2.054026 | -0.000110 |

**Pyridine Derivative F (monomer)**

HF = -362.8622972 hartrees

Zero-point correction= 0.121878 (Hartree/Particle)

Thermal correction to Gibbs Free Energy= 0.091163

Sum of electronic and zero-point Energies= -362.740420

Sum of electronic and thermal Free Energies= -362.771134

Standard orientation:

| Center Number | Atomic Number | Atomic Type | Coordinates (Angstroms) X | Y | Z |
|---|---|---|---|---|---|
| 1 | 6 | 0 | -0.535836 | 1.433915 | 0.000005 |
| 2 | 6 | 0 | 0.589768 | 0.635227 | -0.000016 |
| 3 | 6 | 0 | 0.359069 | -0.747592 | -0.000032 |
| 4 | 7 | 0 | -0.831818 | -1.312778 | -0.000018 |
| 5 | 6 | 0 | -1.902224 | -0.507789 | 0.000008 |
| 6 | 6 | 0 | -1.806921 | 0.864458 | 0.000013 |
| 7 | 6 | 0 | 1.981712 | 1.185083 | -0.000003 |
| 8 | 8 | 0 | 1.441034 | -1.560332 | 0.000028 |
| 9 | 1 | 0 | -0.420114 | 2.511552 | 0.000004 |
| 10 | 1 | 0 | -2.868256 | -0.998567 | -0.000002 |
| 11 | 1 | 0 | -2.697521 | 1.477406 | 0.000024 |
| 12 | 1 | 0 | 1.960944 | 2.274463 | -0.000254 |
| 13 | 1 | 0 | 2.537439 | 0.850624 | -0.878791 |
| 14 | 1 | 0 | 2.537252 | 0.851037 | 0.879065 |
| 15 | 1 | 0 | 1.131306 | -2.474222 | 0.000013 |

**Pyridine Derivative G (monomer)**

HF = -323.5421013 hartrees

Zero-point correction= 0.094012 (Hartree/Particle)

Thermal correction to Gibbs Free Energy= 0.065222

Sum of electronic and zero-point Energies= -323.448089

Sum of electronic and thermal Free Energies= -323.476880

Standard orientation:

| Center Number | Atomic Number | Atomic Type | Coordinates (Angstroms) X | Y | Z |
|---|---|---|---|---|---|
| 1 | 6 | 0 | 1.187548 | 1.151260 | -0.000010 |
| 2 | 6 | 0 | -0.187921 | 1.222827 | 0.000011 |
| 3 | 6 | 0 | -0.897108 | 0.023041 | 0.000074 |
| 4 | 7 | 0 | -0.331427 | -1.170086 | 0.000023 |
| 5 | 6 | 0 | 1.006844 | -1.216660 | -0.000025 |
| 6 | 6 | 0 | 1.807559 | -0.095558 | -0.000007 |
| 7 | 8 | 0 | -2.247495 | 0.083992 | -0.000038 |
| 8 | 1 | 0 | 1.777082 | 2.059338 | -0.000025 |
| 9 | 1 | 0 | -0.716273 | 2.166049 | -0.000011 |
| 10 | 1 | 0 | 1.446286 | -2.207289 | -0.000011 |
| 11 | 1 | 0 | 2.884178 | -0.191728 | -0.000009 |
| 12 | 1 | 0 | -2.592861 | -0.817168 | -0.000061 |

**Pyridine Derivative H (monomer)**

HF = -362.8610997 hartrees

Zero-point correction= 0.121393 (Hartree/Particle)

Thermal correction to Gibbs Free Energy= 0.090072

Sum of electronic and zero-point Energies= -362.739707

Sum of electronic and thermal Free Energies= -362.771028

Standard orientation:

| Center Number | Atomic Number | Atomic Type | Coordinates (Angstroms) X | Y | Z |
|---|---|---|---|---|---|
| 1 | 6 | 0 | 1.340844 | -0.011829 | -0.000206 |
| 2 | 6 | 0 | 0.569177 | -1.156474 | -0.000223 |
| 3 | 7 | 0 | -0.771337 | -1.165455 | -0.000062 |
| 4 | 6 | 0 | -1.387842 | -0.001730 | 0.000077 |
| 5 | 6 | 0 | -0.724111 | 1.223428 | -0.000048 |
| 6 | 6 | 0 | 0.651394 | 1.204139 | -0.000188 |
| 7 | 8 | 0 | -2.742698 | 0.004197 | 0.000207 |
| 8 | 6 | 0 | 2.839239 | -0.069977 | 0.000269 |
| 9 | 1 | 0 | 1.046957 | -2.130463 | -0.000451 |
| 10 | 1 | 0 | -1.285375 | 2.147619 | -0.000112 |
| 11 | 1 | 0 | 1.203015 | 2.137306 | -0.000357 |
| 12 | 1 | 0 | -3.047638 | -0.911258 | 0.000115 |
| 13 | 1 | 0 | 3.190447 | -1.102085 | -0.006831 |
| 14 | 1 | 0 | 3.250115 | 0.422713 | 0.884055 |
| 15 | 1 | 0 | 3.251213 | 0.435440 | -0.875728 |

**Pyridine Derivative I (monomer)**

HF = -346.8086801 hartrees

Zero-point correction= 0.132951 (Hartree/Particle)

Thermal correction to Gibbs Free Energy= 0.101973

Sum of electronic and zero-point Energies= -346.675729

Sum of electronic and thermal Free Energies= -346.706707

Standard orientation:

| Center Number | Atomic Number | Atomic Type | Coordinates (Angstroms) X | Y | Z |
|---|---|---|---|---|---|
| 1 | 6 | 0 | -0.393047 | 0.765701 | 0.000003 |
| 2 | 6 | 0 | -0.586779 | -0.618718 | 0.000086 |
| 3 | 6 | 0 | 0.544534 | -1.427814 | 0.000038 |
| 4 | 6 | 0 | 1.826001 | -0.895117 | -0.000024 |
| 5 | 6 | 0 | 1.991315 | 0.482034 | -0.000050 |
| 6 | 6 | 0 | 0.882457 | 1.313265 | -0.000011 |
| 7 | 6 | 0 | -1.970117 | -1.201501 | -0.000034 |
| 8 | 8 | 0 | -1.444611 | 1.642765 | 0.000148 |
| 9 | 1 | 0 | 0.409678 | -2.503550 | 0.000091 |
| 10 | 1 | 0 | 2.686778 | -1.551415 | 0.000077 |
| 11 | 1 | 0 | 2.983935 | 0.914841 | -0.000118 |
| 12 | 1 | 0 | 0.992422 | 2.390674 | -0.000040 |
| 13 | 1 | 0 | -2.537985 | -0.886963 | 0.880583 |
| 14 | 1 | 0 | -1.927314 | -2.289609 | 0.001300 |
| 15 | 1 | 0 | -2.536700 | -0.888947 | -0.882178 |
| 16 | 1 | 0 | -2.280107 | 1.165745 | -0.000943 |

**Pyridine Derivative J (monomer)**

HF = -362.8520864 hartrees

Zero-point correction= 0.121082 (Hartree/Particle)

Thermal correction to Gibbs Free Energy= 0.090076

Sum of electronic and zero-point Energies= -362.731005

Sum of electronic and thermal Free Energies= -362.762010

Standard orientation:

| Center Number | Atomic Number | Atomic Type | Coordinates (Angstroms) X | Y | Z |
|---|---|---|---|---|---|
| 1 | 6 | 0 | 0.379007 | 0.768323 | 0.000009 |
| 2 | 6 | 0 | 0.550582 | -0.621485 | -0.000008 |
| 3 | 7 | 0 | -0.499666 | -1.441593 | -0.000089 |
| 4 | 6 | 0 | -1.729378 | -0.926049 | -0.000095 |
| 5 | 6 | 0 | -1.975473 | 0.433416 | 0.000068 |
| 6 | 6 | 0 | -0.897015 | 1.300603 | 0.000142 |
| 7 | 6 | 0 | 1.923985 | -1.218113 | 0.000107 |
| 8 | 8 | 0 | 1.431419 | 1.635069 | -0.000281 |
| 9 | 1 | 0 | -2.549558 | -1.634361 | -0.000250 |
| 10 | 1 | 0 | -2.990191 | 0.808159 | 0.000223 |
| 11 | 1 | 0 | -1.031388 | 2.375060 | 0.000318 |
| 12 | 1 | 0 | 2.490816 | -0.906377 | 0.882214 |
| 13 | 1 | 0 | 2.491094 | -0.906382 | -0.881809 |
| 14 | 1 | 0 | 1.858194 | -2.303842 | 0.000081 |
| 15 | 1 | 0 | 2.267100 | 1.158175 | 0.000750 |

**Pyridine Derivative K (monomer)**

HF = -342.9931098 hartrees
Zero-point correction= 0.133682 (Hartree/Particle)
Thermal correction to Gibbs Free Energy= 0.102389
Sum of electronic and zero-point Energies= -342.859428
Sum of electronic and thermal Free Energies= -342.890721

Standard orientation:

| Center Number | Atomic Number | Atomic Type | Coordinates (Angstroms) X | Y | Z |
|---|---|---|---|---|---|
| 1 | 6 | 0 | -1.123560 | 1.159906 | 0.003281 |
| 2 | 6 | 0 | -1.147458 | -0.223658 | -0.000323 |
| 3 | 7 | 0 | -0.026565 | -0.960459 | -0.004884 |
| 4 | 6 | 0 | 1.152729 | -0.340465 | -0.008966 |
| 5 | 6 | 0 | 1.267036 | 1.058025 | -0.002368 |
| 6 | 6 | 0 | 0.112076 | 1.802118 | 0.004249 |
| 7 | 6 | 0 | -2.439721 | -0.981083 | 0.003454 |
| 8 | 7 | 0 | 2.273881 | -1.129504 | -0.065893 |
| 9 | 1 | 0 | -2.046868 | 1.722931 | 0.008522 |
| 10 | 1 | 0 | 2.244379 | 1.522748 | -0.001456 |
| 11 | 1 | 0 | 0.164807 | 2.884155 | 0.011720 |
| 12 | 1 | 0 | -3.295394 | -0.307087 | 0.014053 |
| 13 | 1 | 0 | -2.496410 | -1.631034 | 0.879237 |
| 14 | 1 | 0 | -2.508681 | -1.618473 | -0.880705 |
| 15 | 1 | 0 | 3.133385 | -0.706782 | 0.244452 |
| 16 | 1 | 0 | 2.146951 | -2.085776 | 0.223657 |

**Pyridine Derivative L (monomer)**

HF = -287.6127912 hartrees

Zero-point correction= 0.117078 (Hartree/Particle)

Thermal correction to Gibbs Free Energy= 0.087096

Sum of electronic and zero-point Energies= -287.495713

Sum of electronic and thermal Free Energies= -287.525696

Standard orientation:

| Center Number | Atomic Number | Atomic Type | Coordinates (Angstroms) X | Y | Z |
|---|---|---|---|---|---|
| 1 | 7 | 0 | -0.253353 | -1.187072 | 0.000026 |
| 2 | 6 | 0 | -0.874170 | -0.001494 | 0.000029 |
| 3 | 6 | 0 | -0.165744 | 1.195669 | 0.000025 |
| 4 | 6 | 0 | 1.218005 | 1.167855 | -0.000005 |
| 5 | 6 | 0 | 1.860185 | -0.059095 | -0.000028 |
| 6 | 6 | 0 | 1.078760 | -1.201045 | 0.000005 |
| 7 | 6 | 0 | -2.371484 | -0.018724 | -0.000024 |
| 8 | 1 | 0 | -0.700965 | 2.136309 | 0.000062 |
| 9 | 1 | 0 | 1.785605 | 2.090046 | -0.000018 |
| 10 | 1 | 0 | 2.939080 | -0.135200 | -0.000049 |
| 11 | 1 | 0 | 1.546760 | -2.179531 | 0.000000 |
| 12 | 1 | 0 | -2.780490 | 0.990782 | -0.000088 |
| 13 | 1 | 0 | -2.744981 | -0.545936 | 0.880225 |
| 14 | 1 | 0 | -2.744847 | -0.545962 | -0.880326 |

**Pyridine Derivative M (monomer)**

HF = -323.5303203 hartrees

Zero-point correction= 0.093878 (Hartree/Particle)

Thermal correction to Gibbs Free Energy= 0.065012

Sum of electronic and zero-point Energies= -323.436442

Sum of electronic and thermal Free Energies= -323.465308

Standard orientation:

| Center Number | Atomic Number | Atomic Type | Coordinates (Angstroms) X | Y | Z |
|---|---|---|---|---|---|
| 1 | 6 | 0 | -1.157173 | 1.159073 | -0.000013 |
| 2 | 6 | 0 | 0.226419 | 1.200190 | 0.000019 |
| 3 | 6 | 0 | 0.914839 | -0.002887 | 0.000080 |
| 4 | 6 | 0 | 0.190164 | -1.189431 | 0.000165 |
| 5 | 7 | 0 | -1.137993 | -1.228003 | 0.000009 |
| 6 | 6 | 0 | -1.796917 | -0.068131 | -0.000066 |
| 7 | 8 | 0 | 2.271079 | -0.092432 | -0.000369 |
| 8 | 1 | 0 | -1.735221 | 2.073611 | -0.000030 |
| 9 | 1 | 0 | 0.765872 | 2.140048 | -0.000002 |
| 10 | 1 | 0 | 0.721045 | -2.135175 | 0.000202 |
| 11 | 1 | 0 | -2.878791 | -0.123575 | -0.000157 |
| 12 | 1 | 0 | 2.660415 | 0.787685 | 0.001759 |

**Pyridine Derivative N (monomer)**

HF = -323.5356016 hartrees

Zero-point correction= 0.094179 (Hartree/Particle)

Thermal correction to Gibbs Free Energy= 0.065366

Sum of electronic and zero-point Energies= -323.441422

Sum of electronic and thermal Free Energies= -323.470236

Standard orientation:

| Center Number | Atomic Number | Atomic Type | Coordinates (Angstroms) X | Y | Z |
|---|---|---|---|---|---|
| 1 | 6 | 0 | 1.187888 | -1.115011 | 0.000060 |
| 2 | 6 | 0 | -0.188948 | -1.207232 | 0.000005 |
| 3 | 6 | 0 | -0.926563 | -0.029382 | -0.000007 |
| 4 | 6 | 0 | -0.244612 | 1.181668 | 0.000006 |
| 5 | 6 | 0 | 1.136599 | 1.152756 | 0.000072 |
| 6 | 7 | 0 | 1.866520 | 0.035787 | -0.000054 |
| 7 | 8 | 0 | -2.273433 | -0.116151 | 0.000009 |
| 8 | 1 | 0 | 1.780065 | -2.023220 | -0.000134 |
| 9 | 1 | 0 | -0.684174 | -2.168882 | -0.000066 |
| 10 | 1 | 0 | -0.780960 | 2.122165 | -0.000064 |
| 11 | 1 | 0 | 1.685633 | 2.087735 | -0.000135 |
| 12 | 1 | 0 | -2.664920 | 0.764101 | -0.000106 |

**Pyridine Derivative O (monomer)**

HF = -362.8514983 hartrees

Zero-point correction= 0.121303 (Hartree/Particle)

Thermal correction to Gibbs Free Energy= 0.089963

Sum of electronic and zero-point Energies= -362.730196

Sum of electronic and thermal Free Energies= -362.761535

Standard orientation:

| Center Number | Atomic Number | Atomic Type | Coordinates (Angstroms) X | Y | Z |
|---|---|---|---|---|---|
| 1 | 6 | 0 | -0.659984 | 1.196633 | -0.000023 |
| 2 | 6 | 0 | -1.331886 | -0.021244 | -0.000005 |
| 3 | 7 | 0 | -0.663305 | -1.178075 | -0.000023 |
| 4 | 6 | 0 | 0.665806 | -1.153972 | -0.000012 |
| 5 | 6 | 0 | 1.404249 | 0.020837 | 0.000006 |
| 6 | 6 | 0 | 0.721993 | 1.225784 | -0.000024 |
| 7 | 6 | 0 | -2.828807 | -0.091618 | 0.000030 |
| 8 | 8 | 0 | 2.767141 | 0.030299 | 0.000070 |
| 9 | 1 | 0 | -1.221981 | 2.121579 | -0.000059 |
| 10 | 1 | 0 | 1.180713 | -2.110305 | -0.000064 |
| 11 | 1 | 0 | 1.267264 | 2.161288 | -0.000053 |
| 12 | 1 | 0 | -3.187723 | -0.628530 | 0.880647 |
| 13 | 1 | 0 | -3.187796 | -0.628596 | -0.880501 |
| 14 | 1 | 0 | -3.270817 | 0.904116 | 0.000017 |
| 15 | 1 | 0 | 3.098125 | -0.873934 | -0.000213 |

**Pyridine Derivative P (monomer)**

HF = -362.8578173 hartrees

Zero-point correction= 0.121853 (Hartree/Particle)

Thermal correction to Gibbs Free Energy= 0.090789

Sum of electronic and zero-point Energies= -362.735965

Sum of electronic and thermal Free Energies= -362.767028

Standard orientation:

| Center Number | Atomic Number | Atomic Type | Coordinates (Angstroms) X | Y | Z |
|---|---|---|---|---|---|
| 1 | 7 | 0 | -1.133748 | 1.142852 | 0.000017 |
| 2 | 6 | 0 | -1.177048 | -0.197204 | 0.000074 |
| 3 | 6 | 0 | -0.032333 | -0.978461 | 0.000005 |
| 4 | 6 | 0 | 1.209329 | -0.357415 | -0.000020 |
| 5 | 6 | 0 | 1.264984 | 1.030250 | -0.000008 |
| 6 | 6 | 0 | 0.067797 | 1.718222 | -0.000020 |
| 7 | 6 | 0 | -2.534432 | -0.829657 | -0.000029 |
| 8 | 8 | 0 | 2.314132 | -1.135487 | 0.000002 |
| 9 | 1 | 0 | -0.098577 | -2.058539 | -0.000005 |
| 10 | 1 | 0 | 2.211993 | 1.554512 | -0.000014 |
| 11 | 1 | 0 | 0.080455 | 2.802824 | -0.000006 |
| 12 | 1 | 0 | -3.098213 | -0.513476 | -0.880016 |
| 13 | 1 | 0 | -3.097999 | -0.514213 | 0.880358 |
| 14 | 1 | 0 | -2.468895 | -1.916752 | -0.000491 |
| 15 | 1 | 0 | 3.104638 | -0.584836 | 0.000025 |

## SXVIII. DFT energies and coordinates for pyridine derivatives E-P, Dimers

### Pyridine Derivative E (dimer)

HF = -725.7489444 hartrees
Zero-point correction= 0.244125 (Hartree/Particle)
Thermal correction to Gibbs Free Energy= 0.200360
Sum of electronic and zero-point Energies= -725.504820
Sum of electronic and thermal Free Energies= -725.548584

Standard orientation:

| Center Number | Atomic Number | Atomic Type | Coordinates (Angstroms) X | Y | Z |
|---|---|---|---|---|---|
| 1 | 6 | 0 | 4.061995 | 0.321359 | -0.000187 |
| 2 | 6 | 0 | 2.832251 | 0.949331 | -0.000275 |
| 3 | 7 | 0 | 1.684963 | 0.246270 | -0.000000 |
| 4 | 6 | 0 | 1.732458 | -1.079565 | 0.000364 |
| 5 | 6 | 0 | 2.933579 | -1.791618 | 0.000486 |
| 6 | 6 | 0 | 4.104306 | -1.070476 | 0.000207 |
| 7 | 6 | 0 | 2.699732 | 2.439637 | -0.000663 |
| 8 | 8 | 0 | 0.585232 | -1.761317 | 0.000654 |
| 9 | 1 | 0 | 4.970030 | 0.908015 | -0.000407 |
| 10 | 1 | 0 | 2.918593 | -2.872463 | 0.000793 |
| 11 | 1 | 0 | 5.056750 | -1.585788 | 0.000293 |
| 12 | 1 | 0 | 3.676762 | 2.919722 | -0.000850 |
| 13 | 1 | 0 | 2.145636 | 2.771613 | 0.879649 |
| 14 | 1 | 0 | 2.145527 | 2.771148 | -0.881083 |
| 15 | 1 | 0 | -0.207398 | -1.146647 | 0.000279 |
| 16 | 6 | 0 | -4.062013 | -0.321277 | -0.000019 |
| 17 | 6 | 0 | -2.832305 | -0.949321 | -0.000415 |
| 18 | 7 | 0 | -1.684980 | -0.246322 | -0.000358 |
| 19 | 6 | 0 | -1.732408 | 1.079521 | 0.000109 |
| 20 | 6 | 0 | -2.933480 | 1.791641 | 0.000547 |
| 21 | 6 | 0 | -4.104243 | 1.070559 | 0.000478 |
| 22 | 6 | 0 | -2.699880 | -2.439636 | -0.000866 |
| 23 | 8 | 0 | -0.585167 | 1.761234 | 0.000152 |
| 24 | 1 | 0 | -4.970083 | -0.907879 | -0.000089 |
| 25 | 1 | 0 | -2.918443 | 2.872487 | 0.000925 |
| 26 | 1 | 0 | -5.056657 | 1.585925 | 0.000807 |
| 27 | 1 | 0 | -3.676940 | -2.919661 | -0.001425 |
| 28 | 1 | 0 | -2.146134 | -2.771701 | 0.879637 |
| 29 | 1 | 0 | -2.145367 | -2.771128 | -0.881095 |
| 30 | 1 | 0 | 0.207376 | 1.146464 | 0.000030 |

**Pyridine Derivative F (dimer)**

HF = -725.7426051 hartrees

Zero-point correction= 0.244454 (Hartree/Particle)

Thermal correction to Gibbs Free Energy= 0.200633

Sum of electronic and zero-point Energies= -725.498151

Sum of electronic and thermal Free Energies= -725.541972

Standard orientation:

| Center Number | Atomic Number | Atomic Type | Coordinates (Angstroms) X | Y | Z |
|---|---|---|---|---|---|
| 1 | 6 | 0 | -4.184701 | -0.527431 | -0.000368 |
| 2 | 6 | 0 | -3.383706 | 0.593863 | 0.000638 |
| 3 | 6 | 0 | -1.994688 | 0.373487 | 0.000239 |
| 4 | 7 | 0 | -1.450769 | -0.835386 | -0.001040 |
| 5 | 6 | 0 | -2.255976 | -1.907241 | -0.002016 |
| 6 | 6 | 0 | -3.624407 | -1.804867 | -0.001729 |
| 7 | 6 | 0 | -3.929382 | 1.987111 | 0.002080 |
| 8 | 8 | 0 | -1.186605 | 1.435581 | 0.001236 |
| 9 | 1 | 0 | -5.261843 | -0.406887 | -0.000098 |
| 10 | 1 | 0 | -1.760076 | -2.870677 | -0.003047 |
| 11 | 1 | 0 | -4.242841 | -2.691243 | -0.002527 |
| 12 | 1 | 0 | -5.018729 | 1.970028 | 0.002184 |
| 13 | 1 | 0 | -3.592319 | 2.542296 | -0.876095 |
| 14 | 1 | 0 | -3.592121 | 2.540564 | 0.881271 |
| 15 | 1 | 0 | -0.221928 | 1.150901 | 0.000396 |
| 16 | 6 | 0 | 4.184704 | 0.527392 | -0.000172 |
| 17 | 6 | 0 | 3.383684 | -0.593883 | 0.000787 |
| 18 | 6 | 0 | 1.994681 | -0.373465 | 0.000258 |
| 19 | 7 | 0 | 1.450790 | 0.835424 | -0.001100 |
| 20 | 6 | 0 | 2.256019 | 1.907261 | -0.002025 |
| 21 | 6 | 0 | 3.624447 | 1.804846 | -0.001610 |
| 22 | 6 | 0 | 3.929321 | -1.987149 | 0.002310 |
| 23 | 8 | 0 | 1.186595 | -1.435550 | 0.001186 |
| 24 | 1 | 0 | 5.261842 | 0.406817 | 0.000202 |
| 25 | 1 | 0 | 1.760144 | 2.870709 | -0.003121 |
| 26 | 1 | 0 | 4.242908 | 2.691203 | -0.002370 |
| 27 | 1 | 0 | 5.018670 | -1.970099 | 0.002524 |
| 28 | 1 | 0 | 3.592334 | -2.542341 | -0.875889 |
| 29 | 1 | 0 | 3.591955 | -2.540576 | 0.881477 |
| 30 | 1 | 0 | 0.221961 | -1.150762 | 0.000362 |

**Pyridine Derivative G (dimer)**

HF = -647.10185 hartrees

Zero-point correction= 0.189083 (Hartree/Particle)

Thermal correction to Gibbs Free Energy= 0.148965

Sum of electronic and zero-point Energies= -646.912767

Sum of electronic and thermal Free Energies= -646.952885

Standard orientation:

| Center Number | Atomic Number | Atomic Type | Coordinates (Angstroms) X | Y | Z |
|---|---|---|---|---|---|
| 1 | 6 | 0 | 4.215814 | -0.362025 | -0.000207 |
| 2 | 6 | 0 | 3.183105 | -1.270080 | -0.000388 |
| 3 | 6 | 0 | 1.873030 | -0.782724 | -0.000240 |
| 4 | 7 | 0 | 1.596610 | 0.516276 | 0.000028 |
| 5 | 6 | 0 | 2.614816 | 1.387828 | 0.000193 |
| 6 | 6 | 0 | 3.934492 | 1.003815 | 0.000101 |
| 7 | 8 | 0 | 0.862969 | -1.653632 | -0.000429 |
| 8 | 1 | 0 | 5.241057 | -0.710095 | -0.000297 |
| 9 | 1 | 0 | 3.355844 | -2.337385 | -0.000624 |
| 10 | 1 | 0 | 2.336194 | 2.435158 | 0.000423 |
| 11 | 1 | 0 | 4.721429 | 1.744415 | 0.000260 |
| 12 | 1 | 0 | -0.022543 | -1.178162 | -0.000241 |
| 13 | 6 | 0 | -4.215820 | 0.362005 | 0.000336 |
| 14 | 6 | 0 | -3.183121 | 1.270073 | 0.000228 |
| 15 | 6 | 0 | -1.873038 | 0.782734 | 0.000095 |
| 16 | 7 | 0 | -1.596606 | -0.516261 | 0.000011 |
| 17 | 6 | 0 | -2.614798 | -1.387826 | 0.000106 |
| 18 | 6 | 0 | -3.934481 | -1.003832 | 0.000291 |
| 19 | 8 | 0 | -0.862976 | 1.653645 | -0.000071 |
| 20 | 1 | 0 | -5.241067 | 0.710061 | 0.000467 |
| 21 | 1 | 0 | -3.355877 | 2.337375 | 0.000257 |
| 22 | 1 | 0 | -2.336162 | -2.435152 | 0.000049 |
| 23 | 1 | 0 | -4.721409 | -1.744439 | 0.000383 |
| 24 | 1 | 0 | 0.022555 | 1.178199 | -0.000050 |

**Pyridine Derivative H (dimer)**

HF = -725.7400057 hartrees

Zero-point correction= 0.244121 (Hartree/Particle)

Thermal correction to Gibbs Free Energy= 0.199376

Sum of electronic and zero-point Energies= -725.495885

Sum of electronic and thermal Free Energies= -725.540630

Standard orientation:

| Center Number | Atomic Number | Atomic Type | Coordinates (Angstroms) X | Y | Z |
|---|---|---|---|---|---|
| 1 | 6 | 0 | 4.058992 | -0.350325 | 0.001215 |
| 2 | 6 | 0 | 2.809500 | -0.929164 | 0.001002 |
| 3 | 7 | 0 | 1.660487 | -0.235948 | -0.000143 |
| 4 | 6 | 0 | 1.715680 | 1.087055 | -0.001115 |
| 5 | 6 | 0 | 2.930113 | 1.778362 | -0.001079 |
| 6 | 6 | 0 | 4.094259 | 1.049054 | 0.000071 |
| 7 | 8 | 0 | 0.573796 | 1.782904 | -0.002235 |
| 8 | 6 | 0 | 5.312884 | -1.172300 | 0.002754 |
| 9 | 1 | 0 | 2.702909 | -2.008786 | 0.001746 |
| 10 | 1 | 0 | 2.931849 | 2.859693 | -0.001961 |
| 11 | 1 | 0 | 5.049142 | 1.562332 | 0.000100 |
| 12 | 1 | 0 | -0.216970 | 1.164897 | -0.001662 |
| 13 | 1 | 0 | 5.081981 | -2.237767 | 0.001403 |
| 14 | 1 | 0 | 5.920583 | -0.956643 | 0.884073 |
| 15 | 1 | 0 | 5.923827 | -0.955032 | -0.875903 |
| 16 | 6 | 0 | -4.058999 | 0.350309 | 0.001080 |
| 17 | 6 | 0 | -2.809518 | 0.929170 | 0.001064 |
| 18 | 7 | 0 | -1.660496 | 0.235971 | -0.000154 |
| 19 | 6 | 0 | -1.715670 | -1.087035 | -0.001402 |
| 20 | 6 | 0 | -2.930088 | -1.778362 | -0.001556 |
| 21 | 6 | 0 | -4.094244 | -1.049070 | -0.000330 |
| 22 | 8 | 0 | -0.573787 | -1.782881 | -0.002559 |
| 23 | 6 | 0 | -5.312903 | 1.172265 | 0.002663 |
| 24 | 1 | 0 | -2.702947 | 2.008794 | 0.002028 |
| 25 | 1 | 0 | -2.931809 | -2.859693 | -0.002639 |
| 26 | 1 | 0 | -5.049118 | -1.562365 | -0.000456 |
| 27 | 1 | 0 | 0.216960 | -1.164854 | -0.001821 |
| 28 | 1 | 0 | -5.920823 | 0.956212 | 0.883731 |
| 29 | 1 | 0 | -5.923619 | 0.955373 | -0.876246 |
| 30 | 1 | 0 | -5.082016 | 2.237736 | 0.001831 |

**Pyridine Derivative I (dimer)**

HF = -693.629853 hartrees

Zero-point correction= 0.266898 (Hartree/Particle)

Thermal correction to Gibbs Free Energy= 0.221096

Sum of electronic and zero-point Energies= -693.362955

Sum of electronic and thermal Free Energies= -693.408757

Standard orientation:

| Center | Atomic | Atomic | Coordinates (Angstroms) | | |
|---|---|---|---|---|---|
| Number | Number | Type | X | Y | Z |
| 1 | 6 | 0 | 1.786359 | -0.775085 | 0.065089 |
| 2 | 6 | 0 | 1.461894 | -0.092730 | -1.110882 |
| 3 | 6 | 0 | 1.542817 | 1.295002 | -1.094008 |
| 4 | 6 | 0 | 1.938227 | 1.990959 | 0.039719 |
| 5 | 6 | 0 | 2.265678 | 1.289729 | 1.190568 |
| 6 | 6 | 0 | 2.191259 | -0.093994 | 1.203835 |
| 7 | 6 | 0 | 1.039994 | -0.840658 | -2.341392 |
| 8 | 8 | 0 | 1.718712 | -2.138915 | 0.148243 |
| 9 | 1 | 0 | 1.284848 | 1.838543 | -1.995832 |
| 10 | 1 | 0 | 1.986797 | 3.072198 | 0.023163 |
| 11 | 1 | 0 | 2.575650 | 1.817127 | 2.084126 |
| 12 | 1 | 0 | 2.431962 | -0.657883 | 2.096413 |
| 13 | 1 | 0 | 1.836347 | -1.499939 | -2.698524 |
| 14 | 1 | 0 | 0.790816 | -0.147638 | -3.143667 |
| 15 | 1 | 0 | 0.158979 | -1.462121 | -2.155378 |
| 16 | 1 | 0 | 1.338117 | -2.504483 | -0.656514 |
| 17 | 6 | 0 | -1.786840 | -0.774253 | -0.065572 |
| 18 | 6 | 0 | -1.461987 | -0.092833 | 1.110830 |
| 19 | 6 | 0 | -1.541953 | 1.294967 | 1.094786 |
| 20 | 6 | 0 | -1.936815 | 1.991872 | -0.038547 |
| 21 | 6 | 0 | -2.264677 | 1.291561 | -1.189839 |
| 22 | 6 | 0 | -2.191188 | -0.092201 | -1.203939 |
| 23 | 6 | 0 | -1.040820 | -0.841793 | 2.340962 |
| 24 | 8 | 0 | -1.720128 | -2.138087 | -0.149524 |
| 25 | 1 | 0 | -1.283687 | 1.837790 | 1.996957 |
| 26 | 1 | 0 | -1.984648 | 3.073134 | -0.021342 |
| 27 | 1 | 0 | -2.574231 | 1.819710 | -2.083097 |
| 28 | 1 | 0 | -2.432187 | -0.655389 | -2.096878 |
| 29 | 1 | 0 | -0.790963 | -0.149416 | 3.143583 |
| 30 | 1 | 0 | -0.160448 | -1.464106 | 2.154736 |
| 31 | 1 | 0 | -1.837857 | -1.500428 | 2.697765 |
| 32 | 1 | 0 | -1.339863 | -2.504337 | 0.655078 |

**Pyridine Derivative J (dimer)**

HF = -725.7165687 hartrees

Zero-point correction= 0.243553 (Hartree/Particle)

Thermal correction to Gibbs Free Energy= 0.198781

Sum of electronic and zero-point Energies= -725.473016

Sum of electronic and thermal Free Energies= -725.517787

Standard orientation:

| Center Number | Atomic Number | Atomic Type | Coordinates (Angstroms) X | Y | Z |
|---|---|---|---|---|---|
| 1 | 6 | 0 | 1.709948 | -0.244003 | -0.834157 |
| 2 | 6 | 0 | 1.699566 | 0.832250 | 0.062230 |
| 3 | 7 | 0 | 1.728853 | 0.629728 | 1.378232 |
| 4 | 6 | 0 | 1.753961 | -0.619784 | 1.843388 |
| 5 | 6 | 0 | 1.771172 | -1.729099 | 1.020751 |
| 6 | 6 | 0 | 1.752637 | -1.537166 | -0.349927 |
| 7 | 6 | 0 | 1.665568 | 2.239232 | -0.446810 |
| 8 | 8 | 0 | 1.685148 | -0.066142 | -2.185061 |
| 9 | 1 | 0 | 1.766447 | -0.732672 | 2.921157 |
| 10 | 1 | 0 | 1.795304 | -2.725023 | 1.442119 |
| 11 | 1 | 0 | 1.759875 | -2.370566 | -1.041037 |
| 12 | 1 | 0 | 0.771016 | 2.420240 | -1.049450 |
| 13 | 1 | 0 | 2.534914 | 2.451577 | -1.075567 |
| 14 | 1 | 0 | 1.662845 | 2.938282 | 0.386489 |
| 15 | 1 | 0 | 1.600385 | 0.867590 | -2.403294 |
| 16 | 6 | 0 | -1.715122 | -0.301841 | 0.819122 |
| 17 | 6 | 0 | -1.707295 | 0.822630 | -0.015805 |
| 18 | 7 | 0 | -1.722804 | 0.692741 | -1.341625 |
| 19 | 6 | 0 | -1.730619 | -0.528963 | -1.875116 |
| 20 | 6 | 0 | -1.742774 | -1.682726 | -1.115326 |
| 21 | 6 | 0 | -1.738808 | -1.566799 | 0.263591 |
| 22 | 6 | 0 | -1.689706 | 2.200304 | 0.569417 |
| 23 | 8 | 0 | -1.708273 | -0.199989 | 2.177954 |
| 24 | 1 | 0 | -1.732871 | -0.582452 | -2.957579 |
| 25 | 1 | 0 | -1.751745 | -2.653670 | -1.591929 |
| 26 | 1 | 0 | -1.743230 | -2.436740 | 0.908108 |
| 27 | 1 | 0 | -0.799453 | 2.356340 | 1.185500 |
| 28 | 1 | 0 | -2.563733 | 2.370424 | 1.204472 |
| 29 | 1 | 0 | -1.689814 | 2.943206 | -0.225018 |
| 30 | 1 | 0 | -1.638448 | 0.721012 | 2.448482 |

**Pyridine Derivative K (dimer)**

HF = -685.996633 hartrees

Zero-point correction= 0.268847 (Hartree/Particle)

Thermal correction to Gibbs Free Energy= 0.223201

Sum of electronic and zero-point Energies= -685.727786

Sum of electronic and thermal Free Energies= -685.773432

Standard orientation:

| Center Number | Atomic Number | Atomic Type | Coordinates (Angstroms) X | Y | Z |
|---|---|---|---|---|---|
| 1 | 6 | 0 | 4.132845 | 0.399997 | 0.398900 |
| 2 | 6 | 0 | 2.916887 | 0.979765 | 0.086255 |
| 3 | 7 | 0 | 1.835244 | 0.249548 | -0.220169 |
| 4 | 6 | 0 | 1.920867 | -1.083006 | -0.224311 |
| 5 | 6 | 0 | 3.118904 | -1.748281 | 0.085991 |
| 6 | 6 | 0 | 4.222089 | -0.990900 | 0.395961 |
| 7 | 6 | 0 | 2.730565 | 2.466432 | 0.073869 |
| 8 | 7 | 0 | 0.799292 | -1.773718 | -0.583374 |
| 9 | 1 | 0 | 4.987586 | 1.016393 | 0.641300 |
| 10 | 1 | 0 | 3.155035 | -2.829880 | 0.077397 |
| 11 | 1 | 0 | 5.159259 | -1.476628 | 0.640154 |
| 12 | 1 | 0 | 3.653528 | 2.985029 | 0.329843 |
| 13 | 1 | 0 | 1.956733 | 2.757344 | 0.787725 |
| 14 | 1 | 0 | 2.404446 | 2.799826 | -0.913670 |
| 15 | 1 | 0 | 0.772176 | -2.748423 | -0.333431 |
| 16 | 1 | 0 | -0.085656 | -1.273475 | -0.501115 |
| 17 | 6 | 0 | -4.132788 | -0.399893 | 0.399140 |
| 18 | 6 | 0 | -2.916948 | -0.979745 | 0.086170 |
| 19 | 7 | 0 | -1.835349 | -0.249590 | -0.220538 |
| 20 | 6 | 0 | -1.920865 | 1.082972 | -0.224611 |
| 21 | 6 | 0 | -3.118773 | 1.748326 | 0.086009 |
| 22 | 6 | 0 | -4.221929 | 0.991009 | 0.396250 |
| 23 | 6 | 0 | -2.730754 | -2.466425 | 0.073659 |
| 24 | 7 | 0 | -0.799274 | 1.773559 | -0.583849 |
| 25 | 1 | 0 | -4.987519 | -1.016225 | 0.641740 |
| 26 | 1 | 0 | -3.154837 | 2.829927 | 0.077434 |
| 27 | 1 | 0 | -5.159007 | 1.476797 | 0.640680 |
| 28 | 1 | 0 | -1.957003 | -2.757462 | 0.787555 |
| 29 | 1 | 0 | -2.404584 | -2.799763 | -0.913880 |
| 30 | 1 | 0 | -3.653778 | -2.984965 | 0.329529 |
| 31 | 1 | 0 | -0.771973 | 2.748243 | -0.333847 |
| 32 | 1 | 0 | 0.085602 | 1.273170 | -0.501605 |

**Pyridine Derivative L (dimer)**

HF = -575.2356486 hartrees

Zero-point correction= 0.235277 (Hartree/Particle)

Thermal correction to Gibbs Free Energy= 0.192884

Sum of electronic and zero-point Energies= -575.000372

Sum of electronic and thermal Free Energies= -575.042764

Standard orientation:

| Center Number | Atomic Number | Atomic Type | Coordinates (Angstroms) X | Y | Z |
|---|---|---|---|---|---|
| 1 | 7 | 0 | -2.217707 | 0.621007 | 0.732002 |
| 2 | 6 | 0 | -1.681106 | 0.830250 | -0.474569 |
| 3 | 6 | 0 | -1.310220 | -0.223933 | -1.305386 |
| 4 | 6 | 0 | -1.500371 | -1.524795 | -0.879497 |
| 5 | 6 | 0 | -2.050941 | -1.742319 | 0.373557 |
| 6 | 6 | 0 | -2.385752 | -0.638825 | 1.135807 |
| 7 | 6 | 0 | -1.522986 | 2.251110 | -0.919939 |
| 8 | 1 | 0 | -0.870327 | -0.015752 | -2.272175 |
| 9 | 1 | 0 | -1.213297 | -2.356674 | -1.510330 |
| 10 | 1 | 0 | -2.210879 | -2.740724 | 0.757431 |
| 11 | 1 | 0 | -2.814276 | -0.770460 | 2.123506 |
| 12 | 1 | 0 | -2.463578 | 2.616212 | -1.341515 |
| 13 | 1 | 0 | -0.755024 | 2.342428 | -1.687713 |
| 14 | 1 | 0 | -1.265740 | 2.893590 | -0.078653 |
| 15 | 7 | 0 | 2.218318 | 0.619488 | -0.732085 |
| 16 | 6 | 0 | 1.681880 | 0.829412 | 0.474436 |
| 17 | 6 | 0 | 1.309948 | -0.224302 | 1.305390 |
| 18 | 6 | 0 | 1.498890 | -1.525402 | 0.879704 |
| 19 | 6 | 0 | 2.049313 | -1.743628 | -0.373297 |
| 20 | 6 | 0 | 2.385188 | -0.640565 | -1.135700 |
| 21 | 6 | 0 | 1.525127 | 2.250480 | 0.919625 |
| 22 | 1 | 0 | 0.870198 | -0.015561 | 2.272123 |
| 23 | 1 | 0 | 1.211006 | -2.356917 | 1.510647 |
| 24 | 1 | 0 | 2.208332 | -2.742240 | -0.757013 |
| 25 | 1 | 0 | 2.813629 | -0.772749 | -2.123361 |
| 26 | 1 | 0 | 2.465933 | 2.614552 | 1.341620 |
| 27 | 1 | 0 | 0.756921 | 2.342716 | 1.687047 |
| 28 | 1 | 0 | 1.269005 | 2.893205 | 0.078184 |

**Pyridine Derivative M (dimer)**

HF = -647.0718874 hartrees

Zero-point correction= 0.188920 (Hartree/Particle)

Thermal correction to Gibbs Free Energy= 0.146832

Sum of electronic and zero-point Energies= -646.882968

Sum of electronic and thermal Free Energies= -646.925055

Standard orientation:

| Center Number | Atomic Number | Atomic Type | Coordinates (Angstroms) X | Y | Z |
|---|---|---|---|---|---|
| 1 | 6 | 0 | -3.248368 | 1.334241 | 0.671556 |
| 2 | 6 | 0 | -2.077852 | 0.693589 | 0.302449 |
| 3 | 6 | 0 | -2.165350 | -0.587877 | -0.228099 |
| 4 | 6 | 0 | -3.430720 | -1.158660 | -0.357786 |
| 5 | 7 | 0 | -4.553631 | -0.543686 | -0.003975 |
| 6 | 6 | 0 | -4.459351 | 0.686697 | 0.503368 |
| 7 | 8 | 0 | -1.100529 | -1.312378 | -0.628741 |
| 8 | 1 | 0 | -3.219652 | 2.332754 | 1.087676 |
| 9 | 1 | 0 | -1.116998 | 1.177524 | 0.423531 |
| 10 | 1 | 0 | -3.516447 | -2.158875 | -0.769275 |
| 11 | 1 | 0 | -5.386935 | 1.170612 | 0.784735 |
| 12 | 1 | 0 | -0.238191 | -0.822839 | -0.491613 |
| 13 | 6 | 0 | 2.812570 | 1.703487 | -0.526679 |
| 14 | 6 | 0 | 3.860407 | 0.940433 | -0.046687 |
| 15 | 6 | 0 | 3.614991 | -0.382055 | 0.292951 |
| 16 | 6 | 0 | 2.328789 | -0.883350 | 0.140149 |
| 17 | 7 | 0 | 1.327730 | -0.140674 | -0.323323 |
| 18 | 6 | 0 | 1.559737 | 1.128207 | -0.651668 |
| 19 | 8 | 0 | 4.639664 | -1.137266 | 0.762913 |
| 20 | 1 | 0 | 2.964278 | 2.738010 | -0.802838 |
| 21 | 1 | 0 | 4.855446 | 1.351693 | 0.066024 |
| 22 | 1 | 0 | 2.107149 | -1.913434 | 0.399504 |
| 23 | 1 | 0 | 0.716546 | 1.695454 | -1.026486 |
| 24 | 1 | 0 | 4.333901 | -2.031493 | 0.949129 |

**Pyridine Derivative N (dimer)**

HF = -647.0835497 hartrees

Zero-point correction= 0.189088 (Hartree/Particle)

Thermal correction to Gibbs Free Energy= 0.146406

Sum of electronic and zero-point Energies= -646.894462

Sum of electronic and thermal Free Energies= -646.937144

Standard orientation:

| Center Number | Atomic Number | Atomic Type | Coordinates (Angstroms) X | Y | Z |
|---|---|---|---|---|---|
| 1 | 6 | 0 | -2.119885 | -1.276398 | 0.513302 |
| 2 | 6 | 0 | -3.442682 | -0.913047 | 0.636397 |
| 3 | 6 | 0 | -3.855059 | 0.280351 | 0.052346 |
| 4 | 6 | 0 | -2.918438 | 1.056083 | -0.623425 |
| 5 | 6 | 0 | -1.619130 | 0.600275 | -0.687841 |
| 6 | 7 | 0 | -1.211687 | -0.544098 | -0.136665 |
| 7 | 8 | 0 | -5.148516 | 0.633091 | 0.166328 |
| 8 | 1 | 0 | -1.765327 | -2.200501 | 0.954332 |
| 9 | 1 | 0 | -4.144731 | -1.538103 | 1.170776 |
| 10 | 1 | 0 | -3.201895 | 1.991123 | -1.089026 |
| 11 | 1 | 0 | -0.864720 | 1.178671 | -1.208820 |
| 12 | 1 | 0 | -5.308017 | 1.469209 | -0.285559 |
| 13 | 6 | 0 | 4.599815 | -0.125386 | -0.027004 |
| 14 | 6 | 0 | 3.586245 | -1.018458 | -0.303870 |
| 15 | 6 | 0 | 2.263766 | -0.597882 | -0.166383 |
| 16 | 6 | 0 | 2.042937 | 0.715690 | 0.249109 |
| 17 | 6 | 0 | 3.135999 | 1.521869 | 0.498551 |
| 18 | 7 | 0 | 4.407571 | 1.136043 | 0.371129 |
| 19 | 8 | 0 | 1.280225 | -1.460691 | -0.437662 |
| 20 | 1 | 0 | 5.631105 | -0.444117 | -0.132187 |
| 21 | 1 | 0 | 3.808840 | -2.028260 | -0.622730 |
| 22 | 1 | 0 | 1.041081 | 1.103156 | 0.375844 |
| 23 | 1 | 0 | 2.974080 | 2.544571 | 0.821526 |
| 24 | 1 | 0 | 0.363309 | -1.057147 | -0.301824 |

**Pyridine Derivative O (dimer)**

HF = -725.7160373 hartrees

Zero-point correction= 0.243620 (Hartree/Particle)

Thermal correction to Gibbs Free Energy= 0.197338

Sum of electronic and zero-point Energies= -725.472417

Sum of electronic and thermal Free Energies= -725.518699

Standard orientation:

| Center Number | Atomic Number | Atomic Type | Coordinates (Angstroms) X | Y | Z |
|---|---|---|---|---|---|
| 1 | 6 | 0 | 4.095584 | -0.315932 | 0.618126 |
| 2 | 6 | 0 | 3.876804 | -0.136934 | -0.743299 |
| 3 | 7 | 0 | 2.637518 | -0.054568 | -1.234791 |
| 4 | 6 | 0 | 1.602954 | -0.144510 | -0.403767 |
| 5 | 6 | 0 | 1.737061 | -0.318965 | 0.971049 |
| 6 | 6 | 0 | 3.023149 | -0.407262 | 1.485226 |
| 7 | 6 | 0 | 5.012627 | -0.029821 | -1.715612 |
| 8 | 8 | 0 | 0.679498 | -0.408238 | 1.809143 |
| 9 | 1 | 0 | 5.107342 | -0.382743 | 0.997315 |
| 10 | 1 | 0 | 0.614226 | -0.078435 | -0.844620 |
| 11 | 1 | 0 | 3.171333 | -0.545376 | 2.549307 |
| 12 | 1 | 0 | 4.986815 | 0.931566 | -2.233241 |
| 13 | 1 | 0 | 4.944704 | -0.812172 | -2.474518 |
| 14 | 1 | 0 | 5.974100 | -0.122406 | -1.211474 |
| 15 | 1 | 0 | -0.186190 | -0.259283 | 1.327159 |
| 16 | 6 | 0 | -3.199707 | 1.474170 | -0.518017 |
| 17 | 6 | 0 | -2.020006 | 1.247683 | 0.181487 |
| 18 | 7 | 0 | -1.703113 | 0.016099 | 0.588208 |
| 19 | 6 | 0 | -2.510999 | -1.005277 | 0.319682 |
| 20 | 6 | 0 | -3.700992 | -0.849673 | -0.373249 |
| 21 | 6 | 0 | -4.050207 | 0.422402 | -0.800002 |
| 22 | 6 | 0 | -1.056937 | 2.347034 | 0.507319 |
| 23 | 8 | 0 | -4.531375 | -1.890936 | -0.649182 |
| 24 | 1 | 0 | -3.448365 | 2.476346 | -0.841132 |
| 25 | 1 | 0 | -2.201305 | -1.985602 | 0.667888 |
| 26 | 1 | 0 | -4.973594 | 0.577280 | -1.343837 |
| 27 | 1 | 0 | -0.092804 | 2.166230 | 0.026136 |
| 28 | 1 | 0 | -0.882225 | 2.397390 | 1.583576 |
| 29 | 1 | 0 | -1.435107 | 3.310943 | 0.170460 |
| 30 | 1 | 0 | -4.160747 | -2.708547 | -0.300278 |

**Pyridine Derivative P (dimer)**

HF = -725.73 hartrees

Zero-point correction= 0.244123 (Hartree/Particle)

Thermal correction to Gibbs Free Energy= 0.197382

Sum of electronic and zero-point Energies= -725.485876

Sum of electronic and thermal Free Energies= -725.532618

Standard orientation:

| Center Number | Atomic Number | Atomic Type | Coordinates (Angstroms) X | Y | Z |
|---|---|---|---|---|---|
| 1 | 7 | 0 | 1.378599 | -0.931279 | -0.173263 |
| 2 | 6 | 0 | 1.846184 | -0.362454 | 0.947706 |
| 3 | 6 | 0 | 3.046844 | 0.323984 | 0.967596 |
| 4 | 6 | 0 | 3.784976 | 0.430615 | -0.205028 |
| 5 | 6 | 0 | 3.298869 | -0.159154 | -1.366852 |
| 6 | 6 | 0 | 2.093388 | -0.821070 | -1.292434 |
| 7 | 6 | 0 | 1.001567 | -0.499794 | 2.174914 |
| 8 | 8 | 0 | 4.948200 | 1.108643 | -0.164192 |
| 9 | 1 | 0 | 3.409270 | 0.776545 | 1.880802 |
| 10 | 1 | 0 | 3.845406 | -0.096515 | -2.298630 |
| 11 | 1 | 0 | 1.673807 | -1.288601 | -2.175902 |
| 12 | 1 | 0 | 0.805082 | -1.552917 | 2.382802 |
| 13 | 1 | 0 | 0.036896 | -0.009924 | 2.023499 |
| 14 | 1 | 0 | 1.486845 | -0.053705 | 3.041098 |
| 15 | 1 | 0 | 5.359609 | 1.118228 | -1.035778 |
| 16 | 7 | 0 | -4.099926 | 0.890727 | -0.086556 |
| 17 | 6 | 0 | -2.815464 | 1.267970 | -0.141891 |
| 18 | 6 | 0 | -1.769599 | 0.359602 | -0.192246 |
| 19 | 6 | 0 | -2.046049 | -1.007237 | -0.181047 |
| 20 | 6 | 0 | -3.381319 | -1.401447 | -0.126062 |
| 21 | 6 | 0 | -4.348654 | -0.420353 | -0.081728 |
| 22 | 6 | 0 | -2.543303 | 2.741309 | -0.147097 |
| 23 | 8 | 0 | -1.093607 | -1.946056 | -0.221551 |
| 24 | 1 | 0 | -0.748122 | 0.712277 | -0.239923 |
| 25 | 1 | 0 | -3.647863 | -2.449855 | -0.116748 |
| 26 | 1 | 0 | -5.393888 | -0.707845 | -0.038134 |
| 27 | 1 | 0 | -2.955517 | 3.205060 | 0.751430 |
| 28 | 1 | 0 | -3.023510 | 3.212139 | -1.007375 |
| 29 | 1 | 0 | -1.475101 | 2.950595 | -0.187440 |
| 30 | 1 | 0 | -0.165013 | -1.544151 | -0.209996 |

**SXIX. $^{1}$H and $^{13}$C NMR spectra for methylclioquinol**

$^{1}$H NMR (400 MHz, DMSO) $\delta$ (ppm): 9.03 (1H, dd, J = 1.60, Ar, 4.16 Hz), 8.55 (1H, dd, Ar, J = 1.60, 8.60 Hz), 8.15 (1H, s, Ar), 7.77 (1H, dd, Ar, J = 4.16, 8.60 Hz), 4.06 (3H, s, OCH3), 3.32 (residual $H_2O$). $^{13}$C NMR (400 MHz, DMSO) $\delta$ (ppm): 155.71, 151.07, 141.66, 134.58, 133.15, 126.73, 125.61, 123.38, 91.19, 61.82.

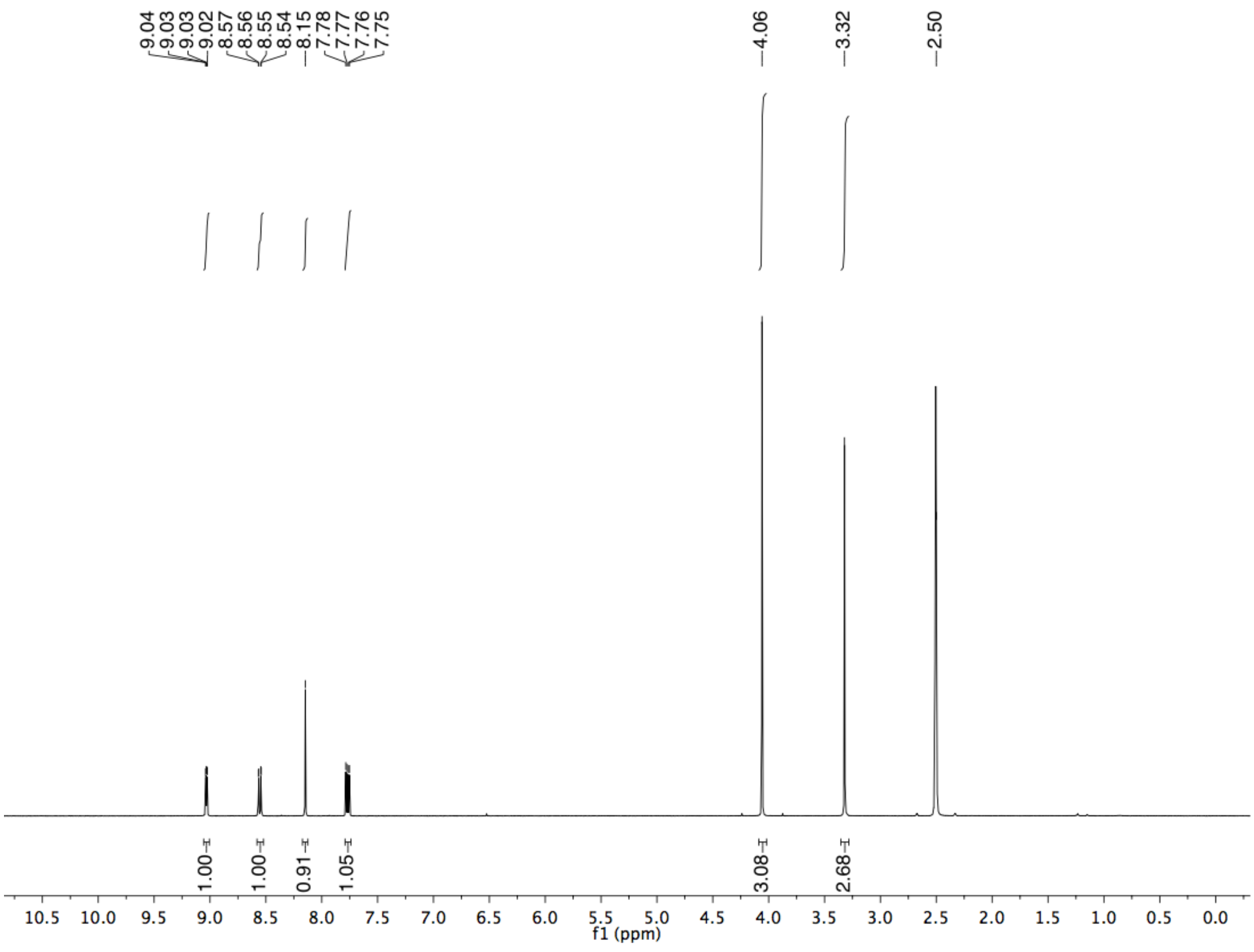

**Figure S7.** $^{1}$H NMR spectrum of methylclioquinol in DMSO-d6.

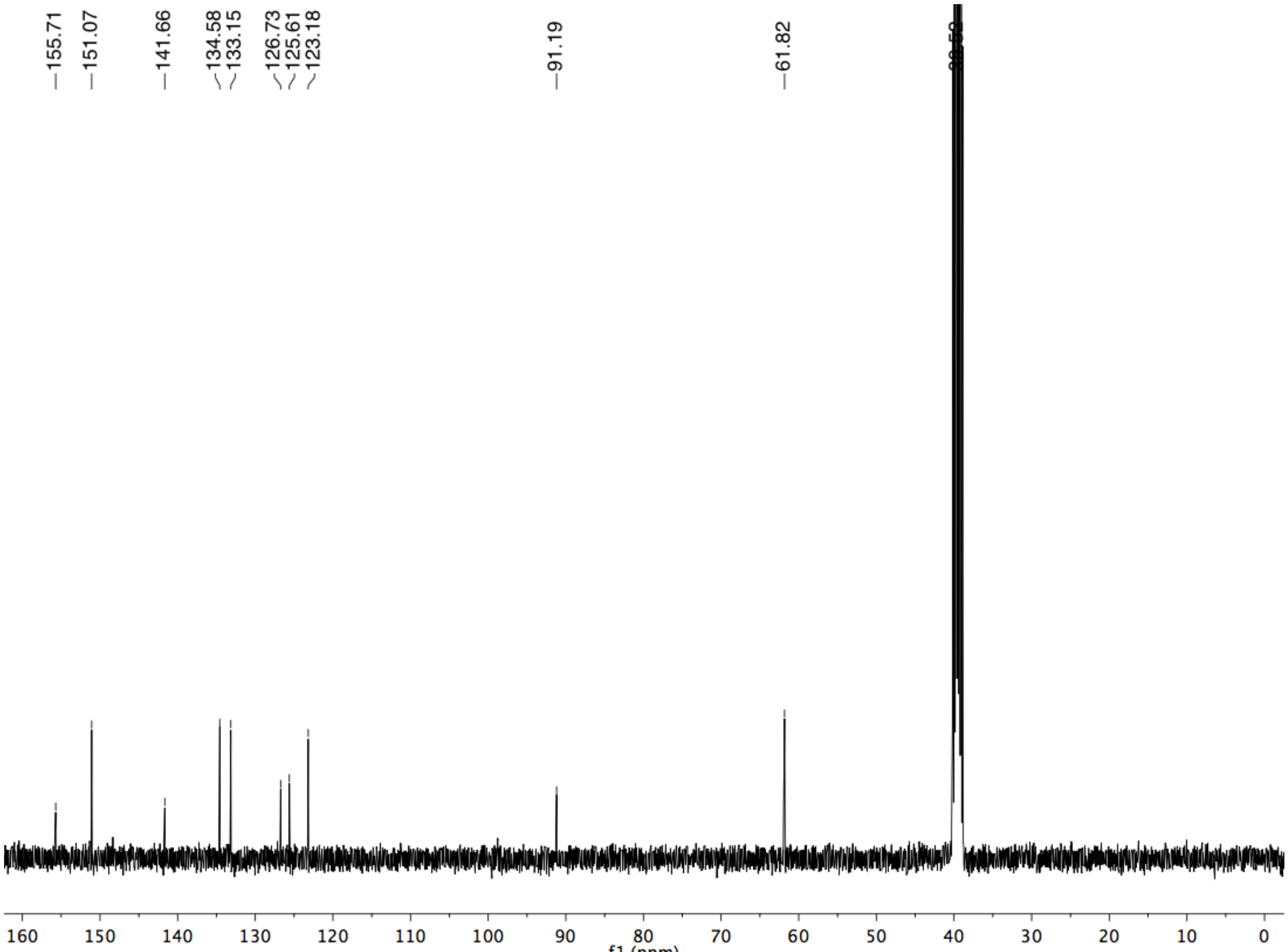

**Figure S8.** $^{13}$C NMR spectrum of methylclioquinol in DMSO-d6.